\pdfoutput=1
\documentclass[12pt,aps,pre,nofootinbib]{revtex4}
\usepackage[isolatin]{inputenc}
\usepackage[OT1]{fontenc}
\usepackage{graphicx}
\usepackage{amssymb,amsmath}
\usepackage{hyperref}

\newcommand{\nn}{\nonumber}

\renewcommand{\epsilon}{\varepsilon}
\newcommand{\red}{\color{red}}

\graphicspath{{./figures/}{./figures/open-diag/}{./figures-new/}}
\allowdisplaybreaks[1]

\newcommand{\rme}{{\mathrm{e}}}
\newcommand{\rmd}{{\mathrm{d}}}

\newcommand{\largebilderscale}{0.7}
\newcommand{\bilderscale}{0.35}

\newcommand{\fig}[2]{\includegraphics[width=#1\columnwidth]{#2}}

\newlength{\bilderlength} 
\newcommand{\usebilderscale}{\bilderscale}
\newcommand{\bilderskip}{\hspace*{0.8ex}}

\newcommand{\largediagram}[1]{%
\renewcommand{\usebilderscale}{\largebilderscale}%
\raisebox{1mm}{\diagram{#1}}\renewcommand{\usebilderscale}{\bilderscale}}
\newcommand{\diagram}[1]{%
\settowidth{\bilderlength}{\bilderskip%
\includegraphics[scale=\usebilderscale]{#1}\bilderskip}%
\parbox{\bilderlength}{\bilderskip%
\includegraphics[scale=\usebilderscale]{#1}\bilderskip}}

\newcommand{\rB}{{\rvec_{\!\scriptscriptstyle{\mathrm{B}}}}}
\newcommand{\drB}{{{\dot{\rvec}}_{\!\scriptscriptstyle{\mathrm{B}}}}}
\newcommand{\rR}{{\rvec_{\!\scriptscriptstyle{\mathrm{R}}}}}
\newcommand{\drR}{{{\dot{\rvec}}_{\!\scriptscriptstyle{\mathrm{R}}}}}
\newcommand{\trR}{{\tilde \rvec_{\!\scriptscriptstyle{\mathrm{R}}}}}
\newcommand{\sB}{{s_{\scriptscriptstyle{\mathrm{B}}}}}

\newcommand{\sR}{{s_{\scriptscriptstyle{\mathrm{R}}}}}
\newcommand{\tsR}{{\tilde s_{\scriptscriptstyle{\mathrm{R}}}}}
\newcommand{\PhiB}{{\Phi_{\scriptscriptstyle{\mathrm{B}}}}}
\newcommand{\PhiR}{{\Phi_{\scriptscriptstyle{\mathrm{R}}}}}

\newcommand{\PsiR}{{\Psi_{\scriptscriptstyle{\mathrm{R}}}}}
\newcommand{\fB}{f_{\scriptscriptstyle{\mathrm{B}}}}
\newcommand{\fR}{f_{\scriptscriptstyle{\mathrm{R}}}}
\newcommand{\gB}{g_{\scriptscriptstyle{\mathrm{B}}}}
\newcommand{\gR}{g_{\scriptscriptstyle{\mathrm{R}}}}
\newcommand{\tgR}{\tilde g_{\scriptscriptstyle{\mathrm{R}}}}
\newcommand{\tgB}{\tilde g_{\scriptscriptstyle{\mathrm{B}}}}
\newcommand{\tuB}{\tilde u_{\scriptscriptstyle{\mathrm{B}}}}
\newcommand{\tvB}{\tilde v_{\scriptscriptstyle{\mathrm{B}}}}
\newcommand{\trB}{\tilde\rvec_{\scriptscriptstyle{\mathrm{B}}}}
\newcommand{\tLB}{\tilde L_{\scriptscriptstyle{\mathrm{B}}}}
\newcommand{\tZ}{\tilde{\mathbb{Z}}(\gR)}
\newcommand{\tZg}{\tilde{\mathbb{Z}}_g(\gR)}
\newcommand{\qB}{{\qvec_{\scriptscriptstyle{\mathrm{B}}}}}
\newcommand{\qR}{{\qvec_{\scriptscriptstyle{\mathrm{R}}}}}
\newcommand{\SR}{{{\cal S}_{{\mathrm{R}}}}}
\newcommand{\SB}{{{\cal S}_{{\mathrm{B}}}}}
\newcommand{\tSB}{{{\tilde{\cal S}}_{{\mathrm{B}}}}}
\newcommand{\LR} {L_{\scriptscriptstyle{\mathrm{R}}}}
\newcommand{\sss}{\scriptscriptstyle}
\newcommand{\uR}{{u_{\scriptscriptstyle{\mathrm{R}}}}}
\newcommand{\vR}{{v_{\scriptscriptstyle{\mathrm{R}}}}}

\begin{document}

\title{\sffamily\bfseries\large Field Theory of the RNA Freezing
Transition} 
\author{\sffamily\bfseries\normalsize Fran{\c c}ois David$^{(1)}$ and Kay J\"org Wiese$^{(2)}$}
\address{$^{(1)}${Institut de Physique Th\'eorique\\
\mbox{\textrm{CNRS,~URA~2306,~F-91191~Gif-sur-Yvette,France}}\\
CEA, IPhT, F-91191 Gif-sur-Yvette, France}
\\ \label{a12}
$^{(2)}$Laboratoire de Physique Th\'eorique de l'Ecole Normale
Sup\'erieure, \\
\mbox{\textrm{Ecole Normale Superieure, unite mixte CNRS UMR 8549,}}\\
24 rue Lhomond, 75005 Paris, France
}
\email{francois.david@cea.fr, wiese@lpt.ens.fr}
\date{\small\today}

\begin{abstract}
Folding of RNA is subject to  a competition between entropy, relevant at high temperatures, and the random, or random looking, sequence, determining the low-temperature phase. It is known from numerical simulations that for random as well as biological sequences, high- and low-temperature phases are different, e.g.\ the exponent $\rho$ describing the pairing probability between two bases is $\rho= \frac 32$ in the high-temperature phase, and $\rho\approx \frac43$ in the low-temperature (glass) phase. Here, we present,  for  random sequences, a field theory of the phase transition separating high- and low-temperature phases. We establish  the existence of the latter by showing that the underlying theory is renormalizable to all orders in perturbation theory. We test this result via an explicit 2-loop calculation, which yields $\rho\approx 1.36$ at the transition, as well as diverse other critical exponents, including the response to an applied external force (denaturation transition). 
\end{abstract}

\maketitle
\newcommand{\clean}{}


\section{Introduction}

\subsection{Random RNA}
{T}{ogether} with DNA and proteins, RNA plays a key role in biology.
As such it is important to understand its spatial
conformations. While for protein the lowest-energy fold depends
strongly on the chemical constitution, and is only tractable
numerically, the problem for RNA is simpler, due to a clear separation
in energy scales between primary structure (the sequence), secondary
structure (pairing of bases in a fold) and tertiary structure
(embedding of a fold in 3-d space). 

RNA molecules consist of 4 bases -- adenine, guanine, cytidine and uracil -- which are attached to a flexible sugar-phosphate backbone. In contrast to duplex DNA molecules (where uracil is replaced with thymine), there does not exist an independent complementary strand, and the RNA molecule folds back on itself.
Experimentally important (see e.g.\ \cite{TinocoBustamante1999}) is the observation, that  topologically intertwined pairings such as knots and pseudo-knots do not seem to play a crucial role for the structure, though they are present \cite{XayaphoummineBucherIsambert2005}. Therefore, for many problems and for many practical purposes, the folding configuration may be considered as topologically planar, which graphically amounts to the rule to draw the
sequence and the pairings on the plane without self-intersection (figure \ref{RNA-figs}).
This approximation makes the problem of RNA folding considerably simpler, since it allows for instance a recursive calculation of the partition function of a RNA strand in a polynomial time (as a function of the length of the strand). A lot of work has now been invested to find the most efficient algorithm \cite{ZukerStiegler1981,FernandezColubri1998,BompfunewererBackofenBernhartHertelHofackerStadlerWill2008}.  
The planar approximation is also the starting point for the study of more general configurations, by performing expansions in terms of the topological number of the latter. Such studies may involve beautiful mathematical tools like random-matrix theory \cite{OrlandZee2002}.

The folding of planar configurations of RNA strands is a fascinating subject in itself, with a lot of attention from physicists and mathematicians (besides biophysicists and biochemists). In particular, planar folded configurations are topologically equivalent to tree-like configurations, and the statistics and combinatorics of trees is a vast subject of its own.
The homopolymer problem (all bases identical) was already solved in 1968 by de Gennes \cite{deGennes1968}.  
In this simple case the pairing-probability  ${\cal P}$ of two RNA-bases decays with the distance $\ell$ between the two bases along the backbone according to the scaling law ${\cal P}(\ell)\propto\ell^{-3/2}$. Irrespective of the embedding in 3-dimensional space, the statistics of the configurations is that of so-called ``generic trees", or equivalently of the mean-field approximation for branched polymers.
For Real RNA molecules however, the  optimal fold depends on the sequence. 
Most studies, in particular numerical ones, focus on the configuration space and on the statistics and dynamics of folding for specific (and biologically relevant) sequences \cite{XayaphoummineBucherIsambert2005,KineFold}. 

Since the pioneering work of Bundschuh and Hwa \cite{BundschuhHwa1999,BundschuhHwa2000,BundschuhHwa2002a,BundschuhHwa2002}, several authors have studied the statistical physics of 
RNA secondary structures for \emph{random sequences}  and \emph{random bond energy} models \cite{KrzakalaMezardMueller2002,HuiTang2006}.
One motivation is  to understand the relative role of general sequence disorder and of specific biological sequences in the behavior of long RNA strands, and whether some properties are generic irrespective of the details of the sequence.
In addition the physics of random RNA sequences is interesting in its own, as a highly nontrivial example of (seemingly 1D) disordered systems, where ordering (due to attractive pairing interactions) and frustration (due to the sequence disorder and the topological constraint of planarity) coexist.
A key feature of the above models is that
there appears be a continuous freezing transition between a weak-disorder phase, at large scales indistinguishable from the homopolymer
case,  and a strong-disorder or glass phase with non-trivial scaling, and of
possible biological relevance since the conformation and properties of
RNA depends on the sequence disorder, i.e.\ on the primary structure.
Not much is known about the transition, even from numerical work; indeed its localization is non-trivial  \cite{HuiTang2006}. 
Better studied numerically is the glass phase  at strong disorder, or equivalently zero temperature
\cite{BundschuhHwa1999,BundschuhHwa2000,BundschuhHwa2002a,BundschuhHwa2002,KrzakalaMezardMueller2002,Monthus2007}.
However, the nature of the freezing transition and of the low-temperature phase are still poorly understood, and contradictory results are reported \cite{Monthus2007}. It is e.g.\ disputed whether replica-symmetry breaking exists in the latter 
\cite{PagnaniParisiRicciTersenghi2000,Hartmann2001,PagnaniParisiRicci-Tersenghi2001}.
The glass phase appears in the  solution of \cite{BundschuhHwa1999,BundschuhHwa2000,BundschuhHwa2002a,BundschuhHwa2002} for the partition function for
$n=2$ replicas (instead of $n=0$ relevant for the disordered
system) and in numerical simulations 
\cite{BundschuhHwa1999,BundschuhHwa2000,BundschuhHwa2002a,BundschuhHwa2002,KrzakalaMezardMueller2002,Monthus2007}. 
 One feature which seems to be robust is  the pairing probability ${\cal P}(\ell)\propto\ell^{-\rho_{\mathrm{glass}}}$ with $\rho_{\mathrm{glass}}\approx \frac 43$, independent of the disorder, be it sequence disorder, or random-pairing energies \cite{KrzakalaMezardMueller2002,HuiTang2006}. 

To better interpret the numerics, finite-size effects  have to be understood. A first step in this direction was the recent analytic solution of a simplified hierarchical model \cite{DavidHagendorfWiese2007b}, corresponding to a broad distribution of pairing energies, and with a pairing exponent $\rho=(\sqrt{17}-3)/2=0.5615...$. 

Pulling a DNA molecule at both ends has become an important experimental technique, which may one day allow to identify the DNA sequence by its force-extension characteristics \cite{BaldazziBraddeCoccoMarinariMonasson2007}. For RNA, the problem is more complicated, since folded RNA is not a linear strand, thus the sequence of base-pair openings is not clear in advance. There is a rapidly increasing bibliography on the subject \cite{ManosasRitort2005,LiphardtOnoaSmithTinocoBustamante2001}. Remarkably, RNA pulling  gives one of the first direct tests \cite{LiphardtDumontSmithJrBustamante2002} of Jarzynski's equality \cite{Jarzynski1997,Jarzynski1997b}.  
Averaged quantities can more easily be estimated and measured, either for homopolymers  or numerically for disordered sequences \cite{KrzakalaMezardMueller2002}. Efforts have been undertaken to include experimentally relevant details, as the elasticity of the free RNA strands
\cite{GerlandBundschuhHwa2001,GerlandBundschuhHwa2003}. 

\subsection{The field-theory approach}
This paper is devoted to a renormalization group study of the freezing transition and of the force-induced denaturation transition of RNA with random pairing energies.
In \cite{LaessigWiese2005} L\"assig and Wiese (LW) pioneered a field theoretical approach for the freezing transition for this model. 
They proposed a continuum formulation for the perturbative weak-disorder expansion of random RNA.
Its starting point (the free theory) is the homopolymer model.
They analyzed the divergences of this expansion at first order in the disorder strength, and  they showed their model to be renormalizable at first order in perturbation theory. Assuming scaling  at the freezing transition, they showed that this transition can be described by an UV stable fixed point at finite disorder strength, and that the coupling (disorder strength) and the length of the RNA strand (number of bases) have to be renormalized at one-loop order.
This allowed them to calculate the critical exponents (to be described later) for the freezing transition.
Using a ``locking argument" (see below), the scaling exponents for random RNA in the {\em strong disorder} phase were estimated, in good agreement with
numerics \cite{BundschuhHwa1999,BundschuhHwa2000,BundschuhHwa2002a,BundschuhHwa2002,KrzakalaMezardMueller2002}.

It is important to understand if this approach defines a consistent theory beyond first order (if possible to all orders), and if the estimates of \cite{LaessigWiese2005} for the
scaling exponents are reliable. 
Indeed the diagrammatics in the LW model is of a new type, although it bears similarities with the diagrammatics of the Edwards model for polymers, i.e.\ self-avoiding random walks, and self-avoiding polymerized membranes, with non-local interactions. It is not at all obvious if the (now standard) field-theoretic renormalization formalism, leading to the renormalization-group picture, is valid for this kind of model.
It is the purpose of this article to show this, and to present applications of this field-theoretical formalism.
We shall introduce a formulation of the LW model in terms of interacting random walks in $d=3$ dimensions, and field-theory tools developed for self-avoiding membranes
\cite{DDG1,DDG2,DDG3,DDG4,WieseHabil,WieseDavid1995,DavidWiese1996,Wiese1996a,Wiese1997a,Wiese1997b,WieseDavid1997,DavidWiese1998,WieseKardar1998a,WieseKardar1998b,DavidWiese2004}.
We show that this model is  consistent and renormalizable to all orders of the weak-disorder perturbative expansion, and deduce that the LW model is indeed renormalizable.
Our formulation is in fact more convenient  for explicit calculations than the original LW formulation.
It allows us to derive new scaling relations between exponents, and to calculate critical exponents at second order. A  short summary of this approach and  its results at second order has already been published \cite{DavidWiese2006}.
Our formulation allows also to treat the related problem of the denaturation transition of RNA strands induced by an external pulling force.
The modelisation of this effect and the principle of the renormalization group calculation has been presented in \cite{DavidHagendorfWiese2007a}  by the two authors and C. Hagendorf, but the details of the second order calculation are presented for the first time here.

\subsection{Organization of the article}
The article is organized as follows:
In section \ref{s:LW-theory}, we discuss basic properties of RNA
molecules and their folding, the equivalent description of these foldings in terms of trees, arch systems, and random-height models in subsection \ref{ss:RNAfold}. 
We then present the L\"assig-Wiese field theory for RNA folding, firstly for the free theory (no disorder) in subsection \ref{ss:LWfree}, and secondly for the interacting theory (with disorder) in subsection \ref{Random RNA}. The perturbative expansion of the interacting theory, its short-distance (UV) singularities and its renormalization are briefly discussed in subsection \ref{ss:LWpert}.

In section  \ref{s:RW} we introduce our representation of the model in terms of interacting random walks.
The basic idea relating random planar foldings to random walks in 3-dimensional space is recalled in subsection \ref{ss:RWbasics}.
The representation of a free folded RNA strand (no disorder) in terms of a closed random walk and the precise concepts and notations are given in subsection \ref{s:3B}. We then generalize this representation to open random walks, since this will prove convenient for renormalization.
In order to take into account the planarity of the folding, we  introduce auxiliary ``dressing fields'', before taking a large-$N$ limit ($N$ is the number of components of these fields), as detailed in subsection \ref{s:3C}.
The disordered (``interacting'') model is introduced in subsection \ref{ss:RepRWdiag}.
Since the disorder in the random pairing energies is quenched, we introduce $n$ replicas. The average over the disorder gives an effective non-local interaction between replicas, given by the so called replica-overlap operator ${\Psi}$. Finally,  one  takes the $n\to 0$ limit (``replica trick'').
The principles of the perturbative expansion and its diagrammatics are given. 
The model and its diagrammatics are easily extendible to an interacting {\em open} RW (open strand), subsection \ref{ss:soRW}, and to multiple interacting RWs (multiple strands), subsection \ref{s:3F}.  This is required in order to extract all renormalizations without going to 3-loop order.

Section \ref{s:renormalization} deals with the short-distance UV divergences, and their renormalization. Defining the model in (fictitious) $d$ dimensions, with $d=3$ relevant for RNA folding, dimensional analysis shows that $d$  may be used as an analytic regularisation parameter (dimensional regularization), and that the model is expected to be renormalizable for $d=2$. This  is briefly explained in subsection \ref{ss:UVdimanalysis}.
The crucial tool to analyze the short-distance singularities  is the Multilocal Operator Product Expansion (MOPE), which generalizes the standard Wilson OPE for local field theories. It  is an extension of the MOPE introduced in \cite{DDG3,DDG4} for self-avoiding manifold models, and used in \cite{DDG3,DDG4,WieseHabil,WieseDavid1995,DavidWiese1996,Wiese1996a,Wiese1997a,Wiese1997b,WieseDavid1997,DavidWiese1998,WieseKardar1998a,WieseKardar1998b,DavidWiese2004} for  interacting tethered membranes and polymers.
This MOPE and its structure for the different relevant (local and multilocal) operators is introduced and discussed in subsection \ref{ss.mope}.
In subsection \ref{ss:Renblty} we use the MOPE formalism to analyze the UV divergences of our model for random RNA folding, and show that it is indeed renormalizable (for $\epsilon =d-2\to 0$).
In subsection \ref{ss:rsBsRO} we discuss the general structure of the counterterms and of the renormalized action. We show that UV finiteness requires a renormalization of the coupling constant $g$ (as expected), a renormalization of the field $\vec{\mathbf{r}}$ (which represents the position of the random walk in the $d$-dimensional fictitious space), plus an additional  renormalization for a boundary operator in the case of open RWs, which is crucial for the consistency of the model.
The definition of the renormalization-group beta-functions and of the anomalous dimensions of the operators is given in subsection \ref{MSanDim}.
Two slightly different renormalization schemes, denoted $\mathrm{MS}$ and $\overline{\mathrm{MS}}$, and based on the standard minimal subtraction scheme (subtraction of poles in $\epsilon=d-2$)  are introduced in subsection \ref{ss:MSandMSbar}. They will be used for the explicit calculations.
In subsection \ref{ss:RenPhi} we discuss renormalization of the so-called contact operator ${\Phi}$, and show that its anomalous dimension is not independent of the renormalization of $\vec{\mathbf{r}}$,  thanks to a new scaling relation that we derive with the help of the multi-strand model.
Finally, in subsection \ref{ss:LSscheme} we apply our results to show that the L\"assig-Wiese model for random RNA folding is indeed renormalizable, and we make precise the relation between our renormalization of $g$ and of $\vec{\mathbf{r}}$ and the renormalization of the coupling constant $g$ and of the RNA strand length $L$ in the LW model. This was the first initial motivation of our study.

We then compute at second order (two loops) the renormalization-group functions of the random RNA folding model, and the scaling exponents for the freezing transition.
In section \ref{s:2loops} we give explicitly all diagrams and integrals.
In subsection \ref{ss:2Lpresent} we present the principle of our calculation for RNA strands of fixed length. 
In subsection \ref{sss:GCanS} we present the calculation for another ensemble, the so called ``grand canonical'' ensemble, where the length of the RNA strand is a fluctuating variable distributed with an exponential distribution involving a chemical potential $\tau$. This is reminiscent of the two ensembles present for self-avoiding polymers (fixed length), and the $n=0$ field theory the after deGennes mapping (Laplace transform) \cite{DeGennes1972}. 
The calculations require the evaluation of diagrams to two loops in perturbation theory. This is done in subsection \ref{ss:1stdiag} for one-strand configurations, in subsection \ref{ss:2stdiag} for two-strand configurations, and in subsection \ref{ss:PhiDiag} for diagrams involved in the renormalization of the contact operator ${\Phi}$.

In section \ref{s:RG2L} we apply our two-loop calculations to the freezing transition.
In subsection \ref{ss:Poles2L} we sum the results of section \ref{s:2loops} and  compute the ultraviolet poles in $\epsilon=d-2$  for the partition functions at two loops, for an arbitrary number of replicas $n$. This determines  the counterterms, the beta function and  the anomalous dimensions, in the $\mathrm{MS}$ scheme (subsection \ref{ss:CT2L}), in the $\overline{\mathrm{MS}}$ scheme (subsection \ref{ss:CTMS'}), and in the grand-canonical ensemble (subsection \ref{ss:CTGrCa}).
In subsection \ref{FPexp2L} we study the RG flow for $\epsilon>0$. We show that the two-loop calculation confirms the existence of an UV stable fixed point (i.e.\ a phase transition) at positive coupling, for $n=0$ as well as for $n>0$. The $n=0$  fixed point describes the freezing transition induced by strong enough disorder in the random RNA model. We compute the critical exponents at second order in $\epsilon$, and check that the results are consistent between the different schemes and the different ensembles.

Finally, in section \ref{s:tension} we generalize our approach to an applied external
force pulling on the RNA strand. This problem was first studied by the two authors and C. Hagendorf in \cite{DavidHagendorfWiese2007a}, where it was shown that our model could be extended to describe the denaturation transition induced  by an external pulling force, and where a one-loop calculation was performed.
In subsection \ref{ss:TenModel} we recall the model, and in subsection \ref{ss:PertTen} its diagrammatics, while in subsection \ref{ss:RenTen} we derive its renormalizability, and define the form of the renormalized action and of the RG functions.
In subsection \ref{ss:2LTen} we present new results, namely the details of the two-loop calculation of the counterterms and of the RG functions.
In subsection \ref{ss:FrDenTen} we discuss the physical meaning of our calculations for the influence of an applied force on the freezing transition, and on the nature of the denaturation transition for weak and strong disorder.

Section \ref{s:conclusion} offers conclusions and further perspectives.
 \clean

\section{The L\"assig-Wiese field theory}
\label{s:LW-theory} 

\subsection{RNA folding representations}
\label{ss:RNAfold}

\subsubsection{Pairing configurations}
\begin{figure}[h]
\raisebox{3cm}{(a)}\fig{0.45}{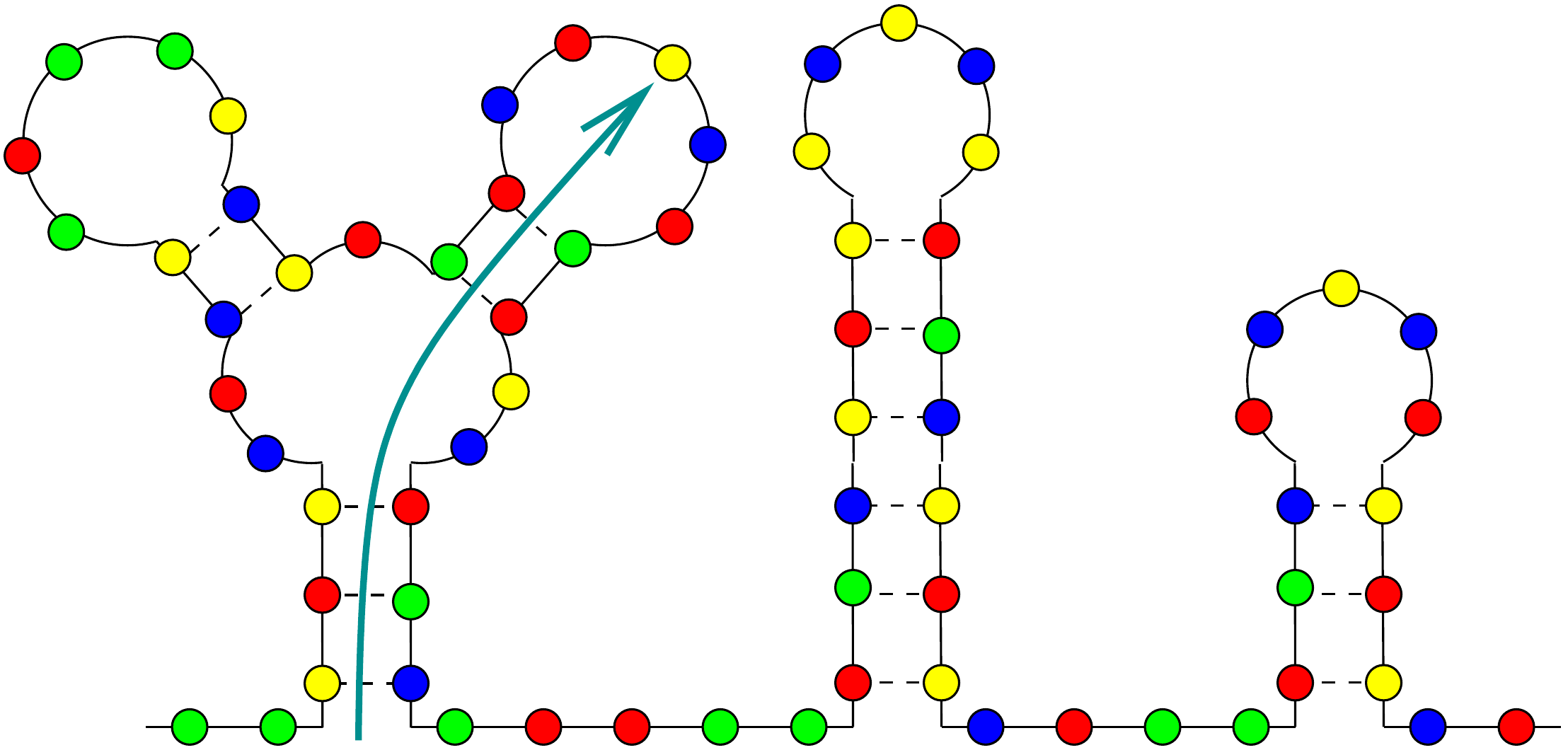}
\medskip \\
\raisebox{2.8cm}{(b)}\fig{0.9}{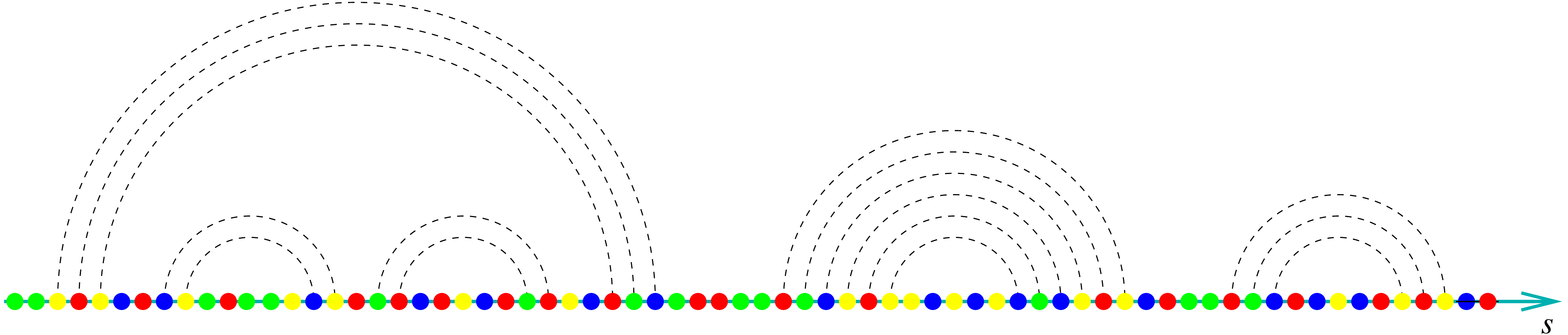}\medskip \\
\raisebox{1.8cm}{(c)}\fig{0.9}{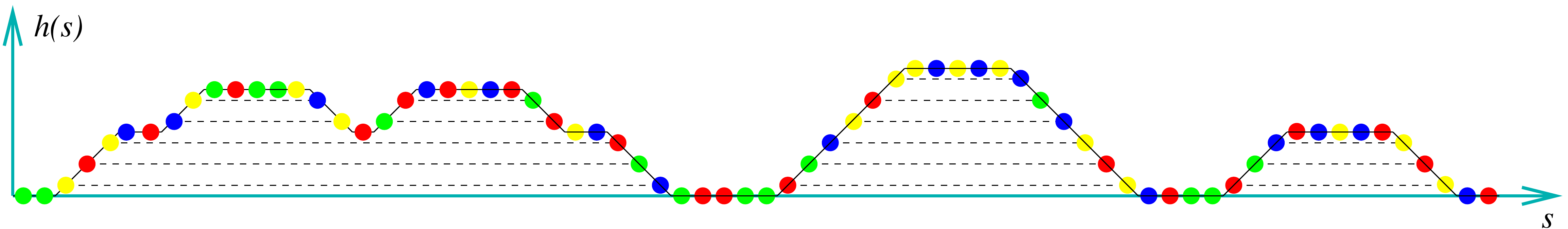}
\caption{(a) an (open) RNA molecule. Bases are represented as discs
(color-coded for the bases); pairings are indicated with dashed black
lines. (b) the same configuration, redrawn s.t.\ bases lie on a line;
allowed configurations consist of non-intersecting rainbow
configurations. (c) the same configuration, redrawn in the height
picture: each time a circle in (b) starts to the right, $h (s)$ is
increased. Each time it comes from the left, $h (s)$ is decreased. The
such constructed height function $h (s)$ of base $s$ is the minimal
number of bonds that have to be opened in (a) to reach base $s$. It is
also equivalent to the number of arcs over base $s$ in (b). }
\label{RNA-figs}
\end{figure}

As explained above, RNA molecules consist of 4 bases, adenin, guanin, cystein and thymin, which are attached to a sugar phosphate backbone. 
In contrast to DNA molecules, there does not exist a complementary strand, and the RNA molecule has to fold back onto itself. 
For the reasons alluded above, we consider that an allowed RNA-fold is an RNA configuration, which can be drawn in the plane without
self-intersections, see figure \ref{RNA-figs}~(a). 
Equivalently, this can be redrawn as a set of arch-diagrams, given on figure \ref{RNA-figs}~(b), or as a height diagram with the constraint that the height is zero at the both ends, and changes by zero or one between neighbors, figure \ref{RNA-figs}~(c). This will be explained below.

\newcommand{\pairing}{{\bf \Phi}}

In order to describe the LW model, we need to more precise.
We consider a strand with $L$ bases, i.e.\ of length $L$.
We label successive bases by integers $i=1,\ldots,L$, and denote the pairing between two different base $i$ and $j$ by the {\em ordered} pair $(i,j)$.
A planar pairing configuration $\pairing$ is given by the collection of $N$ pairings
$$\pairing=\{(i_{{1}},i_2),(i_3,i_4),\cdots (i_{2N-1},i_{2N})\}$$
such that all the $i_a$'s are different and such that the corresponding configuration is planar, i.e.\ no knot or pseudoknot configurations are allowed. This implies that for any two pairings in $\pairing$ we have
\begin{equation}
\label{T1}
(i_1,i_2)\,,  (i_3,i_4)\quad \implies \quad\text{either}\ \begin{cases}
      i_2<i_3 , \\
       i_4<i_1 ,\\
      i_1<i_3<   i_4<i_2,\\
      i_3<i_1< i_2<i_4.
\end{cases}
\end{equation}
\begin{figure}[h]
\begin{center}
\includegraphics[width=2.5in]{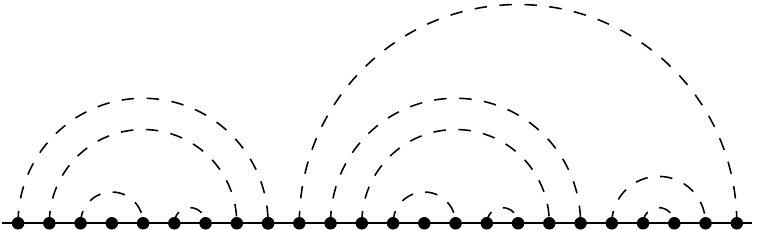}\qquad
\includegraphics[width=2.5in]{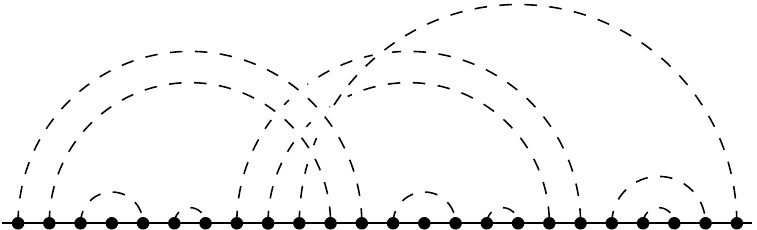}
\caption{Planar (left) and non-planar (right) pairings}
\label{T2}
\end{center}
\end{figure}
A pairing configuration is \emph{compact} if all bases are paired, that is if $L=2N$.

Any planar pairing configuration can be represented by an arch system.
Associating to each interval $\mathbf{i}=]i,i+1[$ the number of arches i.e.\ the {\em height}  $h\mathbf{(i})$ above it, each planar configuration is in one-to-one correspondence with a path $\mathbf{i}\to h(\mathbf{i})$ over the non-negative integers with increment $h(\mathbf{i}+1)-h(\mathbf{i})=0,\,\pm 1$ for general configurations (``Motzkin paths'') or $h(\mathbf{i}+1)-h(\mathbf{i})=\pm 1$ for compact  configurations (``Dyck paths'').

Finally, to each pairing configuration $\pairing$ we associate  the pairing function $\Phi(i,j)$ which is defined by
\begin{equation}
\label{phifun}
\Phi(i,j)=\begin{cases}
      1& \text{if\ }(i,j)\in\pairing \text{, i.e.\ if $i$ and $j$ are paired}, \\
      0& \text{otherwise}.
\end{cases}
\end{equation}
Defining by $\eta(i,j)$ the pairing energy between to bases $i$ and $j$, 
the energy of a folded configuration is 
\begin{equation}
\label{Edisalp1}
{E}[\Phi;\eta]=\sum_{1\le i< j\le L} \eta(i,j)\,\Phi(i,j)\ .
\end{equation}
\subsubsection{Scaling exponents}

Irrespective of the precise statistics of pairings, these representations allow to define two scaling exponents, $\zeta$ and $\rho$, which play an important role in the study of RNA folding. 
First, the average height $\overline{\left< h
\right>}$ scales with the size $L$ of the RNA-molecule as 
\begin{equation}\label{zeta-def}
\overline{\left< h \right>} \sim L^{\zeta}\ .
\end{equation}
Here, we denote by $\left< \ldots \right>$ thermal averages, and by an overbar $\overline{\rule{0mm}{1.3ex}\ldots}$ disorder averages. 

The probability that bases $i$ and $j$ are paired, scales like 
\begin{equation}\label{ph1}
\overline{\left< \Phi(i,j) \right>} \sim |i-j|^{-\rho}\ .
\end{equation}
provided that $1\ll |i-j|\ll L$.
These exponents are not independent \cite{LaessigWiese2005}. Note
that the height at position $k$ is the number of rainbow-arches
starting before and ending after $k$
\begin{equation}\label{ph2}
h (k) = \sum_{0<i\le k}\ \sum_{k<j\le L}  \Phi(i,j)
\end{equation}
Summing over all $k$ on both sides and taking thermal and disorder averages, yields from scaling, assuming that $1<\rho< 2$ 
\begin{equation}\label{ph3}
L^{1+\zeta} \sim L^{3-\rho}\ .
\end{equation}
This yields the important scaling relation
\begin{equation}\label{rho+zeta=2}
 \rho + \zeta =2 \ .
\end{equation}
The pairing statistics depends on the set of pairing energies $\eta(i,j)$. For ``homo-polymers'', i.e.\ a uniform $\eta(i,j)=\eta<0$ for all $i,j$, de Gennes  \cite{DeGennes1972} has shown that $\rho=\rho_0=\frac 3 2$ and $\zeta=\zeta_0=\frac 12$. This can be understood from the fact that the height $h(s)$ is a random walk in time $s$, constrained to remain positive (see subsection \ref{count}). These exponents are also relevant for random RNA in the high-temperature phase. 

We define the ``pair contact'' probability and the exponent $\theta$ by
\begin{equation}\overline{\left< \Phi(i,j)\right>^2}\sim |i-j|^{-\theta}\ .
\end{equation}
We expect that in the high-temperature phase $\theta= 2 \rho=3$, whereas in the low-temperature phase this relation is not satisfied. In the glass phase, and if the partition function is dominated by a single or a few configurations,  $\theta_{\mathrm{glass}}= \rho_{\mathrm{glass}}$. Since 
\begin{equation}
\left[ \overline{\left< \Phi(i,j)\right>}\right]^2 \le 
\overline{\left< \Phi(i,j)\right>^2} \le \overline{\left< \Phi(i,j)\right>}\ , 
\end{equation}
it follows that in all cases 
\begin{equation}\label{10}
\rho\le \theta \le 2 \rho\ .
\end{equation}
We expect that upon lowering the temperature, there will be a phase transition with different universal exponents $\rho^*$ and $\theta^*$. Finally, in the low-temperature phase, there is a third set of exponents $\rho_{\mathrm{glass}}$ and $\zeta_{\mathrm{glass}}$. All these exponents must satisfy relation (\ref{10}). 

\subsubsection{Free energy, finite-size scaling, and divergence of specific heat}
\label{s:free-energy}

General scaling analysis yields that close to a fixed point $g^*$ of the renormalization-group beta function  $\beta  (g)$, i.e.\ close to a phase transition
\begin{equation}\label{a33b}
|g-g^*| \sim \xi^{-\beta ' (g^*)}  \ ,
\end{equation}
where $\xi$ is the correlation length. As will be explained later, the coupling $g$ comes with the pair-contact operator, and thus
\begin{equation}\label{a34b}
\beta ' (g_{c}) \equiv 2-\theta^{*}
\ .
\end{equation}
Since  close to the transition $g$ varies continuously with $T$, this gives 
\begin{equation}\label{a35b}
\xi \sim |T-T_{c}|^{-\nu^*}\ , \qquad \nu^* = \frac{1}{2-\theta^{*}}\ .
\end{equation}
The free energy scales like the inverse correlation volume, i.e.\ in
one dimension like $1/\xi$ 
\footnote{A simple model for the last equation is as follows: Suppose that the system is correlated over a size $\xi$. Then there are $(L/\xi)^{d}$ independent uncorrelated degrees of freedom. If they have Ising character (2 states), then 
$
\delta {\cal F} = -k_{B} T \ln Z =  -k_{B} T \ln
\left( 2^{(\frac L \xi)^{d}}\right) =  -k_{B} T \left(\frac L \xi\right)^{\!\!d} \ln 2
$.}
\begin{equation}\label{a36b}
\delta {\cal F} \sim L/\xi
\ .
\end{equation}Using (\ref{a35b}), 
the divergence of the specific heat becomes:
\begin{equation}
c = \frac{\rmd^{2}}{\rmd T^{2}} {\cal F} \sim |T-T_{c}|^{\frac{2 \theta^{*}  -3} {2-\theta^{*} }}\ .
\end{equation}
Thus for   $1<\theta^{*}< 3/2$, this phase transition is of  second order.

For our model, it is difficult to extract the correlation length $\xi$ from a simulation or experiment, since there is no scale at which a correlation function starts to fall off exponentially. Rather, $\xi$ is the scale, where the contact probability $\left< \Phi(i,j) \right> \sim |i-j|^{-\rho}$ crosses over from $\rho=\frac 32$ to $\rho=\rho^*$ or $\rho=\rho_{\mathrm{glass}}$, which will turn out to be $\rho^*\approx \rho_{\mathrm{glass}}\approx \frac43$.

\subsection{The free theory}
\label{ss:LWfree}

We now recall the formulation of the LW continuum theory in the case where there is no disorder.

\subsubsection{Counting configurations}\label{count}
In the absence of disorder, all pairing configurations are assumed to be equiprobable, with the topological constraint that they must be planar configurations.
If we restrict ourselves to the case of compact configurations, the number of planar pairings for a strand of length $L=2N$ (number of Dyck paths of length $2N$) is given by the {\em Catalan number}
\begin{equation}
\label{SumCat}
C_N:=\sum_{\text{planar\,compact}\,\pairing} 1={1\over N+1}\begin{pmatrix}
      2N    \\
      N  
\end{pmatrix}
\mathop{\simeq}_{N\to\infty} N^{-3/2} {4}^N\,\pi^{-1/2}\ .
\end{equation}
In the general case (Motzkin paths), or in more realistic models where there is a weight for forming an arch, 
the number of planar configurations obeys a similar asymptotics
\begin{equation}
\label{Z0def}
Z_0(L)=\sum_{\text{planar}\,\pairing} 1=
\mathop{\simeq}_{L\to\infty} L^{-\rho_0} c_{\scriptscriptstyle{0}}^L\,a_{\scriptscriptstyle{0}}
\quad \text{with}\quad\rho_0=3/2\ , 
\end{equation}
where $c_{\scriptscriptstyle{0}}$ and $a_{\scriptscriptstyle{0}}$ are non-universal constants.
The exponent $\rho_0=3/2$, which governs the power-law correction factor $L^{-3/2}$, is a universal scaling exponent playing an essential role in the problem. The  value of this exponent can also be understood from the observation, that in the height formulation of the problem, the planar pairing ensemble becomes a random walk ensemble on the half line $h(i)\in\mathbb{N}$. The exponent $\rho_0=3/2$ is then nothing but the exponent for the probability for the first return to the origin at large time for a random walker in one dimension.

$Z_0(L)$ is  the partition function for  planar pairings of a RNA strand with length $L$, when the energy for every possible pairing is the same, and when the only constraint comes from the planarity condition. This problem is similar to the problem of folding for an homopolymer considered by de Gennes in 1968 \cite{deGennes1968}.

Let us now consider a strand with length $L$ and impose the constraint that there are $P$ fixed planar sub-pairings  in the configuration. This collection of sub-pairings is denoted
\begin{equation}
\label{PsiColl}
{\bf\Omega}=\{(i_1,j_1),(i_2,j_2),\cdots,(i_P,j_P)\}\ .
\end{equation}
These $P$ sub-pairings divide the structure (with length $L$) into $P+1$ substructures with backbone lengths $L_0$, $L_1$, $\ldots$, $L_P$ such that 
\begin{equation}
\label{SumLen}
L_0+L_1+\cdots +L_P=L-2P\ .
\end{equation}
The number of planar configurations with the  substructure ${\bf \Omega}$ fixed is denoted $Z_0(L|{\bf \Omega})$, and is given by the product of the number of configurations in each substructure, 
\begin{equation}
\label{Z0Psi}
Z_0(L|{\bf\Omega})=\sum_{\pairing\supset{\bf\Omega}} 1\ =\ \sum_\Phi \Phi(i_1,i_2)\cdots\Phi(i_P,j_P)=Z_0(L_0)Z_0(L_1)\cdots Z_0(L_P)\ .
\end{equation}
We write the sum over configurations as an \emph{unnormalized expectation value} of an observable $\mathbf{O}$ as
\begin{equation}
\label{UnNormE}
\sum_\pairing\ \mathbf{O}\ ={\langle \mathbf{O} \rangle}_0\ ,
\end{equation}
so that the partition function is the e.v.\ of the ``unity operator" $\mathbf{1}$
\begin{equation}
\label{Z0=1}
Z_0(L)={\langle\,\mathbf{1}\,\rangle}_0
\quad,
\end{equation}
while the number of configurations with a fixed planar substructure ${\bf \Omega}$ is the e.v.\ of the operator 
\begin{equation}
\label{T4}
\Omega(i_1,\cdots, j_p)=\Phi(i_1,i_2)\cdots\Phi(i_P,j_P)
\end{equation}
and reads
\begin{equation}
\label{ZPsinot}
Z_0(L|{\bf \Omega})
={\langle\,\Omega\,\rangle}_0
={\langle \Phi(i_1,j_1)\cdots\Phi(i_P,j_P) \rangle}_0
\end{equation}
With these notations the number of configurations with a fixed planar substructure behaves in the large-size limit
($L\to\infty$, all $L_n/L$ fixed and of $\mathcal{O}(1)$)
\begin{equation}
\label{ZPsiLarL}
Z_0(L|{\bf \Omega})
={\langle\,\Omega\,\rangle}_0
\ \mathop{\simeq}_{L\to\infty}\  L_0^{-\rho_0}L_1^{-\rho_0}\cdots L_P^{-\rho_0}\,c_{\scriptscriptstyle{0}}^{L-2P}\,a_{\scriptscriptstyle{0}}^{P+1}
\end{equation}
\begin{figure}[t]
\begin{center}
\includegraphics[width=2in]{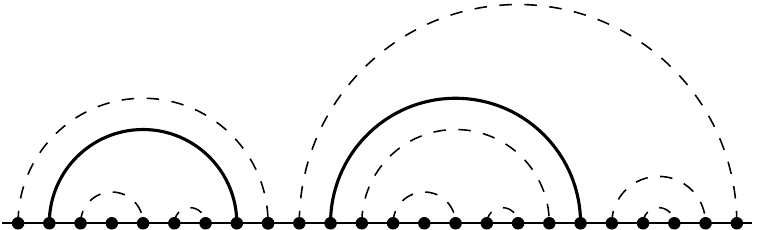}
\quad
\includegraphics[width=2in]{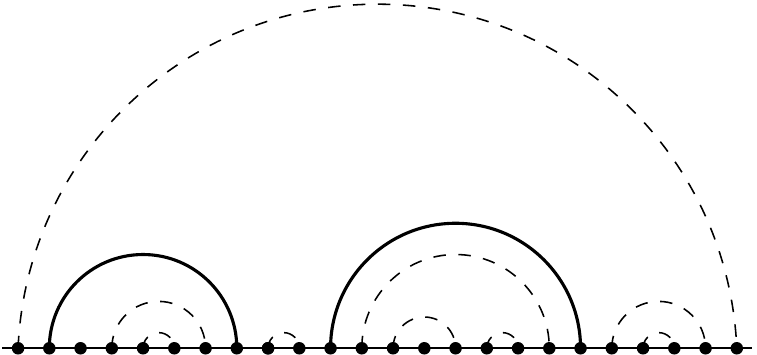}\\
\ \\
\includegraphics[width=2in]{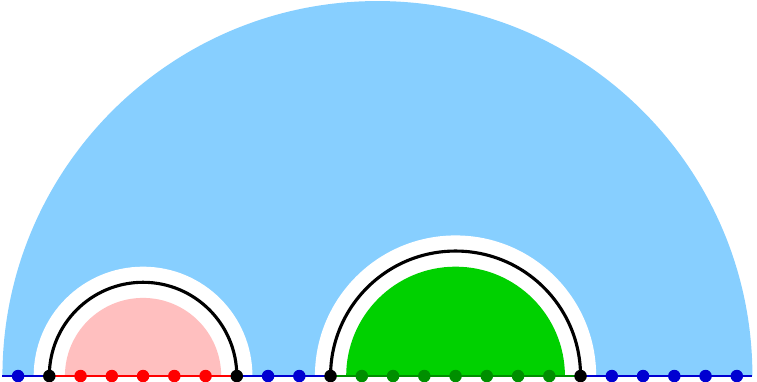}
\caption{Two planar configurations with the same fixed substructure ${\bf \Omega}$, and graphical representation of $Z(L,{\bf \Omega})$.}
\label{T5}
\end{center}
\end{figure}

\subsubsection{Continuum theory}
In the rest of this article, we are interested in the scaling behavior for long RNA strands, and take the limit $L\to\infty$.
We consider RNA in presence of disorder induced by the heterogeneity of the base sequence (primary structure), and will construct a perturbation theory in the strength of the disorder.
In this perturbation theory, each term involves the expectation value for a product of a finite number of $\Phi(i,j)$.
Our starting point is a continuum free theory where:
\begin{enumerate}
  \item The length of the strand $L$ is rescaled to be finite.
  \item The positions $i_a$ of the bases become a continuous variables $s_a$,
\begin{equation}
\label{iTos}
 i\in\{1,2,\cdots , L\}\ \to s\in [0,L]\ .
\end{equation}
  \item The non-universal factor $c_0^La_0$ in the partition function $Z_0(L)$ is absorbed in the normalization of the expectation value ${\langle\ \rangle}_0$ so that the continuum partition function of the strand with length $L$ is 
  \begin{equation}
\label{Z0Cont}
Z_0(L)={\langle\ \mathbf{1}\ \rangle}_0=L^{-\rho_{\scriptscriptstyle{0}}}\quad,\qquad\rho_0=3/2
\end{equation}
\item Similarly, the non-universal factor $a_0/c_0^2$ is absorbed in the normalization of the operator $\Phi(i,j)$, s.t.\  in the continuum limit the operator $\Omega$ is defined by its expectation value
\begin{equation}
\label{Z0PsiC}
Z_0(L|{\bf \Omega})
={\langle\,\Omega\,\rangle}_0
={\langle \Phi(s_1,t_1)\cdots\Phi(s_P,t_P) \rangle}_0
\end{equation}
with
\begin{equation}
\label{Z0PsiE}
{\langle \Phi(s_1,t_1)\cdots\Phi(s_P,t_P) \rangle}_0\ =\ \begin{cases}
    L_0^{-\rho_0}L_1^{-\rho_0}\cdots L_P^{-\rho_0}  & \text{if $\Omega$ is a planar structure }, \\
    0  & \text{otherwise}.
\end{cases}
\end{equation}
\end{enumerate}

\subsubsection{Diagrammatic representation}
\begin{figure}
\begin{center}
\includegraphics[width=2in]{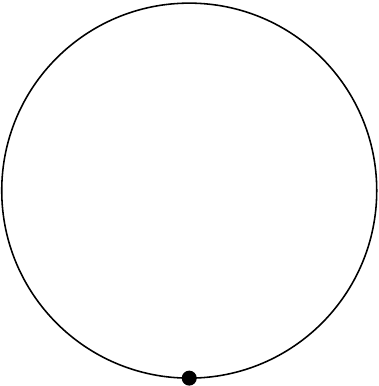}
\raisebox{12.ex}{$\ =\sum\limits_\Phi\ $}
\includegraphics[width=2in]{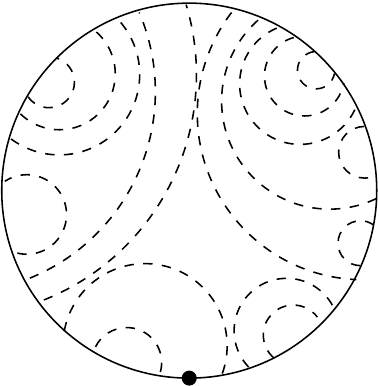}\\
\ \\
\includegraphics[width=2in]{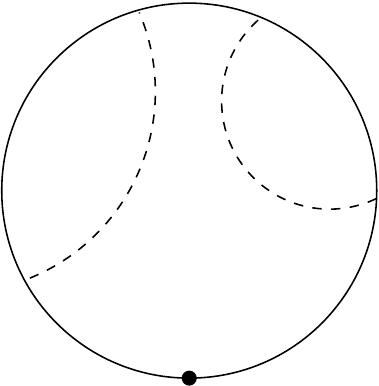}
\raisebox{12.ex}{$\ =\sum\limits_\Phi\ $}
\includegraphics[width=2in]{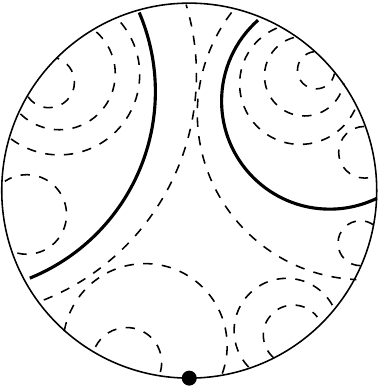}
\caption{Equivalent diagrammatic representation as closed arch structures.}
\label{FigClo}
\end{center}
\end{figure}

A convenient diagrammatic representation of the  operator $\Phi$ in the free theory is the following. We represent the partition function for a strand of length $L$, $Z_0(L)$, by a single line with length $L$. This single line represents the whole (normalized) sum over all planar pairings between points on the line.
\begin{equation}
{\langle\mathbf{1}\rangle}_0=\ \raisebox{.2 ex}{\includegraphics[width=1.5in]{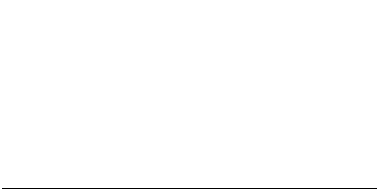}}\ =\sum\limits_\Phi\quad \includegraphics[width=1.5in]{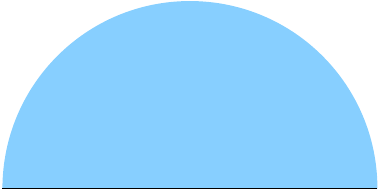}\ .
\label{FigZ0}
\end{equation}
The operator $\Phi(s,t)$ is then represented by a dashed arch over the line joining points $s$ and $t$ (it is a bi-local vertex joining $s$ and $t$):
\begin{equation}
\left<\Phi(s,t)\right>_{0}=\raisebox{1.5 ex}{\largediagram{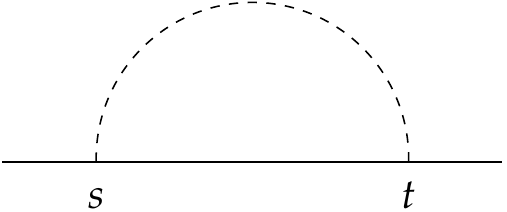}}\ .
\label{FigPhi}
\end{equation}
The partition function for a strand with fixed planar substructure ${\bf \Omega}$, $Z_0(L | {\bf \Omega})={\langle\Omega\rangle}_0$, is the expectation value of a product of $\Phi$ operators and is depicted by the corresponding planar collection of arches over the line. If the substructure $\Omega$ is  non-planar,  the expectation value $\langle\Omega\rangle_0$ is zero, according to (\ref{Z0PsiE}).
\begin{equation}
\langle\Omega\rangle_0\ =\  \includegraphics[width=2in]{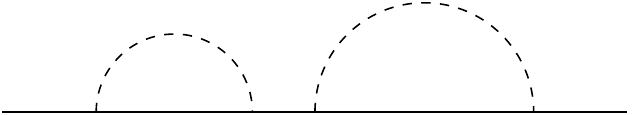}
\ =\sum\limits_{\Phi \supset \Omega}\ 
\includegraphics[width=2in]{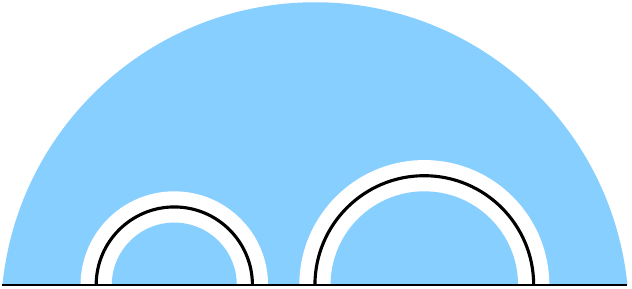} \ .
\label{FigPsi}
\end{equation}
The expectation value of $\Omega$ depends only on the sub-backbone lengths, hence on the distances between the end-points of the arches  considered to be on a closed circle. Both endpoints of the strand are identified, since this does not change the statistics. There is formally no difference between open and closed RNA strands, since we are interested in the secondary structure, not in the tertiary structure, thus steric effects are absent.
An alternative diagrammatic representation for the partition function and the $\Phi$ operators is to depict $Z_0$ as a closed loop with a marked point which depicts the endpoints of the strand. Similarly, the partition function for a strand with a fixed planar substructure $\Omega$, $Z_0(L | {\bf \Omega})={\langle\Omega\rangle}_0$ is depicted as a closed planar arch system. This is represented on figure \ref{FigClo}.

\subsection{Random RNA, disorder and the L\"assig-Wiese field theory}
\label{Random RNA}
\subsubsection{Random RNA}
The field theory approach initiated in  \cite{LaessigWiese2005} by L\"assig and Wiese is based on the random RNA model proposed by Bundschuh and Hwa \cite{BundschuhHwa1999,BundschuhHwa2002a}.
In this model one assumes that to each pair $(i,j)$ is associated a pairing energy $\eta(i,j)$, and that the total energy $E$ for a pairing configuration $\Phi$ is the sum of the pairing energies associated to each pair. 
With our notations the configurational energy $E$ may be written as
\begin{equation}
\label{Edisor}
E[\Phi;\eta]=\sum_{1\le i<j\le L}\eta(i,j)\,\Phi(i,j)\ ,
\end{equation}
with $\Phi(i,j)$ the contact function defined by (\ref{phifun}). Given the collection of pairing energies ${{\eta}}=\{\eta({i,j});\,i<j\}$, the partition function for the RNA strand at finite temperature is the sum over all planar configurations
\begin{equation}
\label{ZeDef}
Z_{\eta}=\sum_\Phi\exp\left({-\beta\,E[\Phi;\eta]}\right)\ ,
\end{equation}
with $\beta=1/k_BT$ the usual Boltzmann factor.
For fixed pairing energies $\eta$, the partition function of a strand of length $L$ can be computed recursively (see e.g.\ \cite{BundschuhHwa2002a,KrzakalaMezardMueller2002}) in a time $T=\mathcal{O}(L^3)$.

While biological sequences are highly structured in order to fulfill their biological function, here we consider 
\emph{random RNA sequences}. 
 While this may, or may not be realistic for real RNA, it is at least an important benchmark against which to compare experimental results for biologically functional RNA. 
 
However, the random-sequence model is still not amenable to an analytical treatment. We therefore  assume that the pairing energies are independent Gaussian random variables. This approximation neglects correlations between the random pairing energies. Numerically it seems that at least in the low-temperature phase, these correlations do not affect the large-distance properties \cite{KrzakalaMezardMueller2002}. We have to leave to future research  to develop an analytical handle on this problem. 

We choose $\eta(i,j)$ to be a random variable with probability distribution 
\begin{equation}
\label{DistrE}
P(\eta(i,j))\ =\ {1\over \sqrt{2\sigma^2\pi}}\,\exp\left({-{1\over 2\sigma^2}[\eta(i,j)-\eta_0]^2}\right)
\ .
\end{equation}
$\eta_0<0$ is the mean pairing energy and $\sigma$ its variance. 
The randomness in the pairing-energy distribution amounts to the introduction of quenched disorder in the system. 
The average over the Gaussian disorder $\eta$ is denoted  by the horizontal overline $\overline{\vphantom{m}\qquad}$, so that (with the ordering $i<j$ and $k<l$)
\begin{equation}
\label{AverE2}
\overline{\eta(i,j)\eta(k,l)}=\eta_0^2+\sigma^2\,\delta_{ik}\delta_{jl}
\quad.
\end{equation}
The averaged free energy for the system is
\begin{equation}
\label{AverF}
\overline{F}=-\frac1{\beta}\,\overline{\log{Z_\eta}}
\quad,
\end{equation}
and the expectation value for an observable $\mathbf{O}$ is
\begin{equation}
\label{AverO}
\overline{\langle\mathbf{O}\rangle}:=\overline{{1\over Z_\eta}\sum_\Phi\mathbf{O}\,\exp\left(-\beta\,E[\Phi;\eta]\right)}
\quad.
\end{equation}
For instance the probability that the bases $i$ and $j$ are paired  is
\begin{equation}
\label{AverPhi}
\overline{\langle \Phi(i,j)\rangle}\ =\  \overline{{1\over Z_\eta}\sum_\Phi\Phi(i,j)\,\exp\left(-\beta\,E[\Phi;\eta]\right)}
\quad,
\end{equation}
and  the probability for a given pairing substructure $\Omega$ to occur in the random pairing-energy ensemble is
\begin{equation}
\label{AverPsi}
\overline{\langle \Omega\rangle}\ =\  \overline{{1\over Z_\eta}\sum_\Phi\Omega\,\exp\left(-\beta\,E[\Phi;\eta]\right)}
\ =\  \overline{{1\over Z_\eta}\sum_{\Phi\supset\Omega}\,\exp\left(-\beta\,E[\Phi;\eta]\right)}
\quad.
\end{equation}

\subsubsection{Weak-disorder expansion and replicas}

The idea of \cite{LaessigWiese2005} is to study the model by a perturbative  weak-disorder expansion, and to extract its large-length $L$ scaling behavior at finite (and if possible large) disorder by renormalization group techniques.
The perturbation expansion can be constructed in the discrete model by expanding in powers of the disorder $\eta(i,j)$ and using (\ref{AverE2}) so that we get a perturbative expansion in powers of the effective disorder strength (coupling constant)
\begin{equation}
\label{gdef}
g=\beta^2\,\sigma^2
\ .
\end{equation}
The quenched average over the disorder is done by the standard replica trick, which is well-defined for a perturbative expansion.
One considers $n$ replicas of the system, labeled by an index $\alpha=1,2,\cdots, n$. Finally, one has to take the limit $n\to 0$. 
The pairing configuration $\Phi_\alpha$ of the replica $\alpha$ is given by the pairing function $\Phi_\alpha (i,j)$
\begin{equation}
\label{phifun}
\Phi_\alpha(i,j)=\begin{cases}
      1& \text{if\ }(i,j)\in\Phi_\alpha\ \text{i.e.\ if $i$ and $j$ are paired for the replica $\alpha$}, \\
      0& \text{otherwise}.
\end{cases}
\quad.
\end{equation}
Since the disorder is quenched, all replicas see the same pairing energy $\eta(i,j)$, and the configurational energy for a replica ensemble $\Phi=\{\Phi_\alpha\}$ is
\begin{equation}
\label{Edisalp}
\mathcal{E}[\Phi;\eta]=\sum_{\alpha=1}^{n}E[\Phi_\alpha;\eta]=\sum_{\alpha=1}^{n}\ \sum_{1\le i< j\le L} \eta(i,j)\,\Phi_\alpha(i,j)
\quad.
\end{equation}
The average over the disorder  gives the partition function for the $n$-times replicated system 
\begin{equation}
\label{ZnRep}
\mathcal{Z}=\overline{\left(Z_\eta\right)^n}=\overline{\sum_{\{\Phi_\alpha\}}\,\exp\left(-\beta\,\mathcal{E}[\Phi;\eta]\right)}
\ .
\end{equation}
The average over the disorder can be taken explicitly since the disorder is Gaussian. 
From (\ref{AverE2}) one has
\begin{equation}
\label{GauAve}
\overline{\exp\left(-\beta\sum_{\alpha=1}^{n}\sum_{i<j}\eta(i,j)\Phi_\alpha(i,j)\right)}
=
\exp\left({g\over 2}\,\sum_{\alpha,\beta=1}^{n}\sum_{i<j}\Phi_\alpha(i,j)\,\Phi_\beta(i,j)\right)
\end{equation}
with $g$ given by (\ref{gdef}).
One obtains an effective attractive interaction between replicas, and one can rewrite the system in terms of an effective ``Hamiltonian" 
\begin{equation}
\label{Heff1}
\beta\mathcal{H}_{\mathrm{tot}}[\Phi]=\beta\mathcal{H}_{0}[\Phi]+\beta\mathcal{H}[\Phi]=\beta\eta_{0}\sum_{\alpha=1}^{n}\sum_{i<j}\Phi_{\alpha}(i,j)-{g\over 2}\,\sum_{\alpha,\beta=1}^{n}\sum_{i<j}\Phi_\alpha(i,j)\Phi_\beta(i,j)
\quad.
\end{equation}
The first contribution, proportional to $\beta \eta_{0}$ is the one present for a homopolymer, which we can solve analytically. The second term, proportional to the disorder, contains two contributions:   the diagonal contribution 
 $\alpha=\beta$ 
\begin{equation}
\label{sum2sum}
\sum_{i<j}\Phi_\alpha(i,j)\Phi_\alpha(i,j)=\sum_{i<j}\Phi_\alpha(i,j)\quad ,
\end{equation}
leading to a change of $\eta_{0}$
\begin{equation}
\eta_{0}^{\mathrm{eff}} = \eta_{0}-\frac{g}{2\beta}\ ,\qquad \beta\mathcal{H}_{0}[\Phi]= \beta\eta_{0}^{\mathrm{eff}}\sum_{\alpha=1}^{n}\sum_{i<j}\Phi_{\alpha}(i,j)\ ;
\end{equation}
and an off-diagonal part  $\mathcal{H}[\Phi]$ 
\begin{equation}
\label{Heffr2}
\mathcal{H}[\Phi]=-g\,\sum_{\alpha <\beta}\sum_{i<j}\Psi_{\alpha\beta}(i,j)
\quad,\qquad
\Psi_{\alpha\beta}(i,j):=\Phi_\alpha(i,j)\Phi_\beta(i,j)
\ ,
\end{equation}
where 
$\Psi_{\alpha\beta}(i,j)$ with $\alpha\neq \beta$ is the {\em pair contact} or  \emph{overlap operator}. It gives the probability that the bases $i$ and $j$ are paired both in replica $\alpha$ and $\beta$.
The partition function and the e.v.\ of observables are now 
\begin{equation}
\label{Z&Orep}
\mathcal{Z}=\sum_\Phi \exp(-\mathcal{H}_{\mathrm{tot}}[\Phi])
\quad,\quad\ 
\overline{\langle\mathbf{O}\rangle}=
\lim_{n\to 0}\langle\mathbf{O}[\Phi]\rangle
=\lim_{n\to 0}\left({1\over \mathcal{Z}}\sum_\Phi \mathbf{O}[\Phi]\,\exp\left(-\mathcal{H}_{\mathrm{tot}}[\Phi]\right)\right)
\ .
\end{equation}
The idea is to do perturbation theory in ${\cal H}$, using the solvable theory with ${\cal H}_{0}$ as reference.

\subsubsection{Continuum limit}
The L\"assig-Wiese field theory \cite{LaessigWiese2005} is obtained by taking the continuum limit of this model.
It is defined in terms of the continuum pairing operators for each replica $\Phi_\alpha(u,v)$, $u,\,v\in[0,L]$, and the {\em overlap} or {\em pair-contact} operator $\Psi_{\alpha\beta}(u,v)$.
One starts from the free theory for $n$ independent non-interacting replicas. 
The partition function for a bundle of $n$ free replicas is
\begin{equation}
\label{ Z0n}
\mathcal{Z}_{0}(L)=\sum_{\{\Phi_\alpha\}}1=
\prod_\alpha {\langle\mathbf{1}\rangle}_0={Z_0(L)}^n=L^{-n\,\rho_0}
\quad.
\end{equation}
We represent diagrammatically this partition function by a collection of $n$ lines, or by a fat bundle.
\begin{equation}
\mathcal{Z}_0\ = \includegraphics[width=2in]{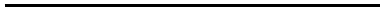}\ =\ \prod\limits_\alpha\  
\includegraphics[width=2.5in]{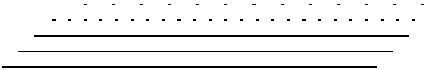}
\end{equation}
The expectation value for a product $\Omega$ of $n$ (different) operators $\Omega_\alpha$ living in replica $\alpha$ factorizes into
\begin{equation}
\label{ZPsirep}
\mathcal{Z}_0(L|{\bf \Omega})=\sum_{\{\Phi_\alpha\}}\prod_\alpha\Omega_\alpha
={\langle\Omega\rangle}_0
=\prod_\alpha{\langle\Omega_\alpha\rangle}_0=\prod_\alpha Z_0(L|{\bf \Omega}_\alpha)
\quad.
\end{equation}
We represent it diagrammatically by the collection of the $n$ planar arch structures relative to each $\Omega_\alpha$
\begin{equation}
\mathcal{Z}_0(L|{\bf\Omega})\ =\ \includegraphics[width=2.5in]{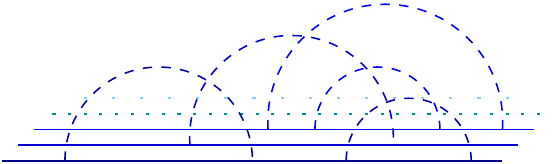}\ .
\label{FigPsin}
\end{equation}
The continuum model with disorder is given by the theory with an effective disorder Hamiltonian corresponding to an attractive 2-replica interaction which is the continuum limit of the discrete effective Hamiltonian (\ref{Heffr2})
\begin{equation}
\label{HintRep}
\mathcal{H}[\Phi]=-g\sum_{\alpha <\beta}\iint_{0\le u<v\le L}\Psi_{\alpha\beta}(u,v)
\quad,\qquad\Psi_{\alpha\beta}(u,v):=\Phi_\alpha(u,v)\,\Phi_\beta(u,v)
\quad.
\end{equation}
The partition function for $n$ replica of a strand with length $L$ with disorder is the continuum version of (\ref{ZnRep}). 
Therefore it is  given by
\begin{equation}
\label{ZintRep}
\mathcal{Z}(L)={\langle \exp\left(-\mathcal{H}[\Phi]\right)\rangle}_0
\quad.
\end{equation}
It will be expanded in powers of the coupling constant $g$, each term of order $g^k$ being of the form
${\langle {\mathcal{H}[\Phi]}^k\rangle}_0$ and can be computed using (\ref{ZPsirep}).
Similarly the partition function for $n$ replicas with a given set of substructures $\Omega$ (i.e.\ the ``expectation value" for the operator $\Omega$) with disorder is
\begin{equation}
\label{ZOmRep}
\mathcal{Z}(L|\Omega)=\overline{\langle\Omega\rangle}={\langle\Omega\,\exp\left(-\mathcal{H}[\Phi]\right)\rangle}_0
\quad.
\end{equation}
It will be computed as a formal power-series expansion in $g$ in terms of the ${\langle \Omega\,{\mathcal{H}[\Phi]}^k\rangle}_0$.
The details of these perturbative expansions will be discussed and studied in the next sections.

\subsection{Perturbative expansion for the L\"assig-Wiese theory}
\label{ss:LWpert}
\begin{figure}[t]
\begin{center}
\includegraphics[width=3.5in]{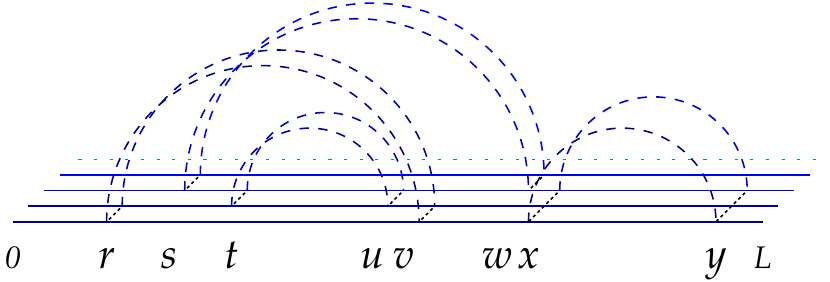}
\caption{Example of a product of four $\Psi$-operators, consistent with planarity. }
\label{FigPab3}
\end{center}
\end{figure}

According to the LW diagrammatics, the continuum overlap operator $\Psi_{\alpha\beta}(u,v)$ is represented by a double arch between points $u$ and $v$ on the two lines for replicas $\alpha$ and $\beta$
\begin{equation}\label{59}
\Psi_{\alpha\beta}(u,v)=\parbox{2.in}{\includegraphics[width=2.in]{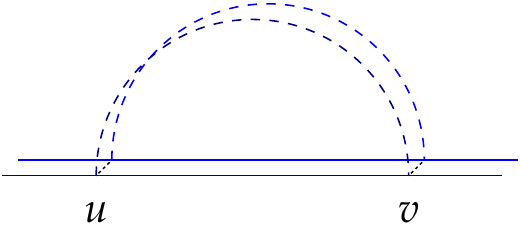}}\ .
\end{equation}The perturbative expansion in $g$  of the partition function  $\mathcal{Z}(L)$ involves integrals of e.v.\ of products of $\Psi$ operators
\begin{equation}
\label{IntPPsi}
\iiint_{\{u_i,\,v_i\}} {\langle \Psi_{\alpha_1\beta_1}(u_1,v_1) \cdots  \Psi_{\alpha_k\beta_k}(u_k,v_k) \rangle}_0
\end{equation}
which is represented as a set of $k$ double arches between the replicas. An example is given on figure \ref{FigPab3}.
At a given order $g^k$, the number of different replicas coupled by the $\Psi$-arches $n_{\mathrm{diff}}$ is bounded by $2\le n_{\mathrm{diff}} \le 2k$.
Let us consider a configuration $\Psi\cdots\Psi$ associated to replica pairs $(\alpha_i,\beta_i)$ and base pairs $(u_i,v_i)$; for each replica among the $n_{\mathrm{diff}}$ coupled replica, and consider the corresponding reduced system of $\Phi$-arches. The e.v.\ ${\langle \Psi_{\alpha_1\beta_1}(u_1,v_1) \cdots  \Psi_{\alpha_k\beta_k}(u_k,v_k) \rangle}_0$ is non-zero if and only if for each replica $\alpha$ the reduced system of  $\Phi_\alpha\cdots\Phi_\alpha$ is \emph{planar.}
For each replica $\alpha$ the e.v.\ of the product $\Phi_\alpha\cdots\Phi_\alpha$ is the product over the $p_\alpha+1$ cycles  of their backbone lengths, to the power $-\rho_0=-3/2$
\begin{equation}
\label{PPhiL}
\langle\underbrace{\Phi_\alpha\cdots\Phi_\alpha}_{p_\alpha} {\rangle}_0 =\left(\prod_{j=0}^{p_\alpha}\ell_{j,\alpha}\right)^{-\rho_0}
\end{equation}
The e.v.\ of the product $\Psi\cdots\Psi$ is now the product of the previous terms for each of the $n_{\mathrm{diff}}$ coupled replicas, times the product of the free partition function $ \langle\mathbf{1}{\rangle}_0=L^{-3/2}$ for the $n-n_{\mathrm{diff}}$ uncoupled replicas.
\begin{equation}
\label{PPsiL}
\langle\!\!\langle\underbrace{\Psi\cdots\Psi}_{k}  {\rangle\!\!\rangle}_0 =
\prod_{\alpha=1}^{n_{\mathrm{diff}}}\left(\prod_{j=0}^{p_\alpha}\ell_{j,\alpha}\right)^{-\rho_0}\ \left(L^{-\rho_0}\right)^{n-n_{\mathrm{diff}}}
\end{equation}
An example is given on figure \ref{FigPab3}.

We expect the scaling dimensions of $\Phi$ and $\Psi$ to change in the presence of disorder. It is the aim of this article to calculate these changes. In a perturbative field theory, the latter can usually be extracted from the divergences of the diagrams, as e.g.\ the one given on figure \ref{FigPab3}. We therefore have to achieve two things: Calculate these divergences, but {\em even more
importantly}, find which quantities they renormalize. In a standard field theory, this  task is not difficult: The needed renormalizations are associated to the marginal and relevant operators present in the original theory, or generated by the perturbation expansion. Here, and up to now, we only have a perturbation theory, but no field theoretic action to renormalize, so we do not know which quantities will need renormalization! In the following, we will construct such a field theoretic representation, which will tell us which quantities to renormalize. In a second step, we will then calculate the necessary diagrams. 

Before doing so, let us as an example calculate the diagram drawn in equation \ref{59}, to see that indeed there are divergences:
\begin{equation}
\largediagram{PsiArch} = \int_0^{L}\rmd u\, \int_{u}^{L}\rmd v\, |v-u|^{-2\rho_{0}} \left(L-|v-u|\right)^{-2\rho_{0}} =\frac12 L^{2-4\rho_{0}} \frac{\Gamma(1-2\rho_{0})^{2}}{\Gamma(2-4\rho_{0})}
\end{equation}
The diagram has a pole in $1/(1-2\rho_{0})$, renormalizing the free energy. It has also a pole in $1/(1-\rho_{0})$; the latter can be interpreted as a  renormalization of the length of the RNA molecule. However, it is not at all obvious why, and how to do this, thus a proper representation as an action is necessary.
 
 \clean

\newcommand{\rvec}{\mathbf{r}}
\newcommand{\qvec}{\mathbf{q}}
\newcommand{\cS}{\mathcal{S}}
\newcommand{\cZ}{\mathcal{Z}}
\newcommand{\cZo}{\mathcal{Z}^{{(1)}}}
\newcommand{\cZt}{\mathcal{Z}^{{(2)}}}
\newcommand{\cZp}{\mathcal{Z}^{{(p)}}}
\newcommand{\tcZo}{{\widetilde{\mathcal{Z}}}^{\scriptscriptstyle{(1)}}}

\newcommand{\ssst}[1]{{{{#1}}}}

\renewcommand{\imath}{\mathrm{i}}
\newcommand{\emath}{\mathrm{e}}

\section{The Random-Walk representation}
\label{s:RW}
\subsection{Basic ideas}
\label{ss:RWbasics}
The diagrammatic expansion of the model bears  strong similarities with the diagrammatic expansion of the Edwards model \cite{Edwards1965,DesCloizeauxJannink}, which describes 3-dimensional random walks with a weak repulsive interaction upon contact, and which has been widely used for polymers and self-avoiding membranes. 

Here is an heuristic explanation for this similarity:
Planar pairing configurations for a RNA strand are in one-to-one correspondence with planar arch systems over a linear strand, which are themselves in one-to-one correspondence with discrete paths (Dyck or Motzkin paths) on the half-line of integers $\mathbb{N}$. 
In particular the (normalized) partition function of the free strand with length $L$ is nothing but the probability of first return to the origin at time $t=L$ for a random walk on $\mathbb{N}$, or $\mathbb{Z}$, which scales at large times as the continuous random walk on $\mathbb{R}$ (the Wiener process), that is as $t^{-3/2}$.
Now for several observables, the one-dimensional random walk on the half-line $\mathbb{R}_+$ behaves as the three-dimensional random walk on the full space $\mathbb{R}^3$. 
In particular, the \emph{first-return probability} to the origin for a RW in one dimension scales as the \emph{total return probability} to the origin for a RW in three dimensions.
As a consequence, the pairing operator $\Phi(i,j)$ has a natural representation in the 3d RW picture as the so-called \emph{contact operator} $\delta^3(\rvec(u)-\rvec(v))$ (probability of contact at times $u$ and $v$ for the random walk $\rvec(t)$). Similarly, many observables and many questions about scaling can easily be represented or formulated in the RW picture. In particular, the \emph{analytic regularization} used in \cite{LaessigWiese2005}, where the contact exponent $\rho_0=3/2$ is analytically continued to $\rho\in[1,3/2]$ and used as an UV regularization parameter to construct an $\epsilon$-expansion for the RG equations and the scaling exponents, is nothing but the classical \emph{dimensional regularization} where the dimension  $d_0=3$ of space for the RW is analytically continued to $d\in[2,3]$, and used to construct  a $d=2+\epsilon$ expansion.

There is however an important difference. The planarity constraint for the pairings implies that the product of several pairing operators vanishes if the resulting configuration is not planar. This is a global topological constraint that cannot be represented by local operators in a RW representation.
To implement this constraint, we shall introduce additional matrix-like degrees of freedom in the RW representation which allow to deal with the topology of the diagrammatics and to take the planar limit as a large-$N$ limit (where $N$ is the dimension of the ``internal space" associated to these additional degrees of freedom).
This is a usual trick in QFT and in statistical mechanics. In particular it has been introduced in \cite{OrlandZee2002} for the problem of RNA secondary structure enumeration and statistics.

Thus we construct in this section a quite involved RW-like representation of the LW model, which involves ``generalized random walks" in a $d\times N\times n$ dimensional space, where $d=2\rho$ is the dimension of space, $N$ is the dimension of internal space, and $n$ the dimension of replica space. We are interested in the limit $d\to 3$ ($\rho_0=3/2$), $N\to\infty$ (planar limit) and $n\to 0$ (limit of quenched disorder).

This representation turns out to be very powerful. It allows to  apply the mathematical tools developed in the renormalization of polymers and self-avoiding membranes, in particular the so-called Multilocal Operator Product Expansion (MOPE).
Although the random RNA model is mapped only on the closed RW subsector of the RW model, there are other observables, associated to open random walks, which have no interpretation in terms of RNA observables, but which are much easier to study and to compute. They allow a more direct calculation of some of the renormalisation group functions and of the scaling exponents for the random RNA model.

\subsection{The simple RW model}
\label{s:3B}
In the rest of this article, we normalize the Dirac ``$\delta$-function" in $\mathbb{R}^d$ as
\begin{equation}
\label{tdelta}
{\tilde\delta}^d(\rvec)=(4\pi)^{d/2}\,\delta^d(\rvec)
\quad.
\end{equation}
With this normalization most of the annoying factors involving powers of $4\pi$ disappear in the calculations.

\subsubsection{Closed Random Walk}
We start from  the random-walk process in $\mathbb{R}^d$ in the time interval $t\in[0,L]$, described by the random variable $\rvec (t)$ ($\rvec=\{r^\mu;\,\mu=1\cdots d\}$). The Euclidean action for the RW is (with proper normalization)
\begin{equation}
\label{S0r}
\cS_0[\rvec]=\int_0^L \!\!\mathrm{d}t\,{1\over 4}\,{\dot{\rvec}(t)}^2\ ,
\end{equation}
and the functional measure is the standard Feynman-Kac measure \footnote{The measure 
has dimension $[t]^{d/2}=[\rvec]^{d}$.} (in Euclidean time) for the quantum particle with mass $m=1/2$ 
\begin{equation}
\label{DmeasR}
\mathcal{D}[\rvec]=\prod_t{\mathrm{d}^d\rvec(t)\over (4\pi a)^{d/2}}\quad,\qquad a=\text{UV regulator}\ \simeq\ {1\over \delta(t=0)}
\ .
\end{equation} 
The partition function for the closed RW (periodic boundary conditions) is defined as
\begin{align}
\label{Z0rCl}
Z_0^{(\mathrm{closed})}(L)&=\int \mathcal{D}[\rvec]\,\exp\bigl({-\cS_0[\rvec]}\bigr)\,{\tilde\delta}^d(\rvec(0)-\rvec(L))
\ .
\end{align}
To extract the infinite factor from the translational zero mode in $\mathbb{R}^d$, we formally write
\begin{equation}
\label{VolSpa}
\mathrm{Vol}(\mathbb{R}^d)=\int \mathrm{d}^d\rvec=(2\pi)^d\,\delta^d(\mathbf{q}=\mathbf{0})
\end{equation}
(with $\mathbf{q}$ the momentum in the conjugate space of $\mathbb{R}^d$), and 
\begin{align}
\label{Z0c}
Z_0^{(\mathrm{closed})}(L)&= (2\pi)^d\,\delta^d(\mathbf{q}=\mathbf{0}) \,
 {\widetilde Z}_0^{(\mathrm{closed})}(L)\ .
\end{align}
The normalized closed partition function ${\widetilde  Z}_0^{(\mathrm{closed})}(L)$ is nothing but the heat kernel given by
\begin{align}
\label{Zb0c}
\widetilde {Z}_0^{(\mathrm{closed})}(L)&=\langle 0|\exp\left(L\Delta_\rvec\right) |0\rangle=
\int_{\rvec(0)=\mathbf{0}} \hskip -2.em\mathcal{D}[\rvec]\,\exp\bigl({-\cS_0[\rvec]}\bigr)
\ {\tilde\delta}^d(\rvec(0)-\rvec(L))
=L^{-d/2}\ .
\end{align}
Note the disappearance of the usual $(4\pi)^{-d/2}$ factor, thanks to the normalization (\ref{tdelta}) for the Dirac  $\tilde\delta$ distribution in the definition of the partition function.

For $d=3$, $\widetilde  Z_0^{\mathrm{(closed)}}(L)$ is equal to the partition function $Z_0(L)$ of the free closed RNA strand in the continuum limit of the LW model, as defined by (\ref{Z0Cont}) (hence the similar notation). 
Therefore we also denote it as the unnormalized expectation value
\begin{equation}
\label{Zb0c2}
\widetilde {Z}_0^{(\mathrm{closed})}(L)=\langle\mathbf{1}\rangle_0^{\mathrm{closed}}
\quad,
\end{equation}
and represent it as a single line with length $L$, as for the RNA model, 
\begin{equation}
\widetilde {Z}_0^{(\mathrm{closed})}(L)\ =\ 
\raisebox{.3 ex}{\includegraphics[width=1.5in]{single-line}}
\ =\ 
\raisebox{-4.0 ex}{\includegraphics[width=1.5in]{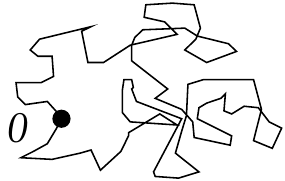}}
\label{s3-2}
\end{equation}
The normalization for the action (\ref{S0r}) was chosen  such that the propagator (the IR-finite 2-point function) becomes
\begin{equation}
\label{rprop}
{\langle {[r^\mu(u)-r^\mu(v)] [r^\nu(u)-r^\nu(v)]}\rangle}_0=\delta^{\mu\nu} |u-v|
\quad.
\end{equation}

\subsubsection{Open Random Walk}
Although the observables for the random RNA pairing model are related to observables for a \emph{closed random walk}, we also consider \emph{open random walks with free boundary conditions}.
For the open RW it is convenient to define the generating function 
\begin{equation}
\label{Z0oqq}
Z_0^{\mathrm{(open)}}(\qvec_1,\qvec_2;L)=\int \mathcal{D}[\rvec]\,\exp\left({-\cS_0[\rvec]}\right)
\,\exp\left(\imath(\qvec_1\rvec(0)+\qvec_2\rvec(L))\right)\ .
\end{equation}
It is the Fourier transform of the partition function for an open RW with fixed boundaries,
\begin{equation}
\label{Z0orr}
Z_0^{\mathrm{(open)}}(\rvec_1,\rvec_2;L)=\int_{\rvec(0)=\rvec_1,\,\rvec(L)=\rvec_2}\hskip-4.em \mathcal{D}[\rvec]\,\exp\left({-\cS_0[\rvec]}\right)\ .
\end{equation}
$\qvec_1$ and $\qvec_2$ are the momenta flowing through both end points $t=0$ and $t=L$ of the RW.

\noindent With our normalization for the measure, and using translational invariance, it is given by
\begin{align}
\label{Z0orre}
Z_0^{\mathrm{(open)}}(\rvec_1,\rvec_2;L)&=
\langle\rvec_1|\exp\left(L\,\Delta_\rvec\right)|\rvec_2\rangle
\nonumber\\
&=
(4\pi L)^{-d/2}\,\exp\left({-{(\rvec_1-\rvec_2)^2\over 4L}}\right)
\end{align}
hence
\begin{equation}
\label{Z0oqqe}
Z_0^{\mathrm{(open)}}(\qvec_1,\qvec_2;L)=
(2\pi)^d\,\delta^d(\qvec_1+\qvec_2)\,
\widetilde  Z_0^{\mathrm{(open)}}(\qvec_1;L)
\end{equation}
with
\begin{equation}
\label{Z0obq}
\widetilde  Z_0^{\mathrm{ (open)}}(\qvec_1;L)=\exp(-L\,\qvec_1^2)\ .
\end{equation}
These notations will be useful later.
We represent diagrammatically the open RW  function as a single line with length $L$ with bars at its end points (if necessary for clarity), 
\begin{equation}
\widetilde  Z_0^{\mathrm{(open)}}(\qvec;L)\ =\ 
\includegraphics[width=1.5in]{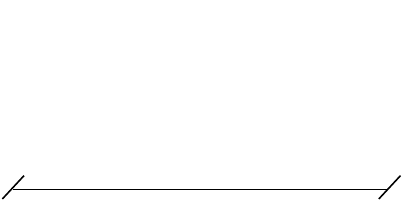}
\ =\ 
\parbox{3in}{\includegraphics[width=3in]{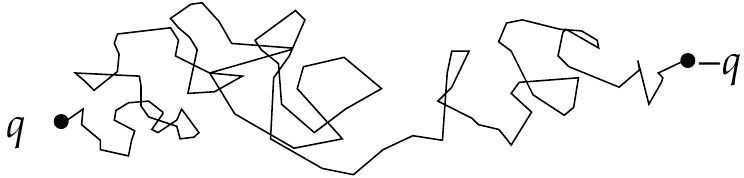}}
\end{equation}

\subsubsection{Contact operator}

\begin{figure}[!h]
\begin{center}
\includegraphics[width=4.4 in]{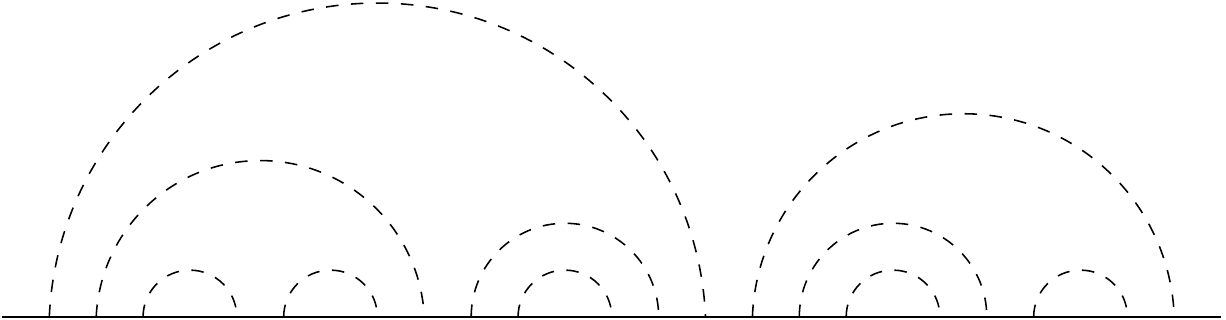}
\\$0\ \ u_1\, u_2\, u_3\ \  \ \ v_3\, u_4\ \  \ \ v_4\, v_2\, u_5\, u_6\  \  \ v_6\, v_5\,v_1\, u_7\, u_8\, u_9\  \ \ v_9\, v_8\, u_{10}\  \ v_{10}\, v_{7}\ \ L$\\
\vskip 2.ex
\includegraphics[width=2.8in]{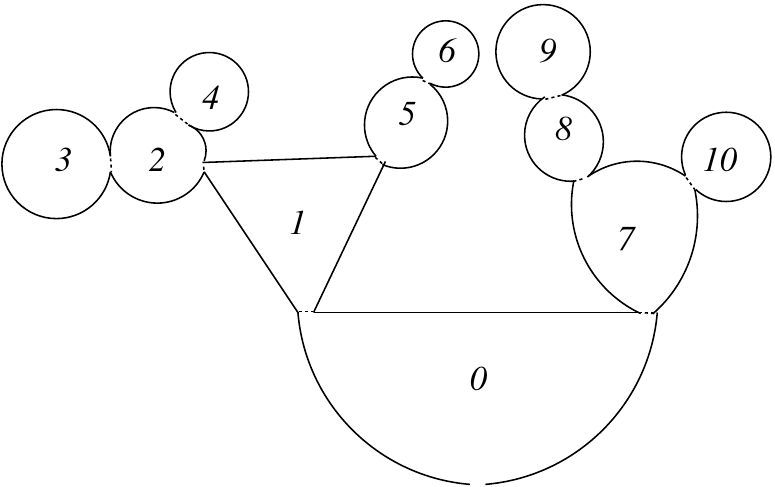}\\
\caption{An arch system (top) , and the corresponding $\phi^4$ cactus-like diagram (bottom) with its loops.
The length of the $k=11$ loops are $\ell_0=L-(v_1-u_1)-(v_7-u_7)$, $\ell_1=(u_2-u_1)+(u_5-v_2) +(v_1-v_5)$, $\ell_2=(u_3-u_2)+(u_4-v_3)+(v_2-v_4)$, $\ell_3=(v_3-u_3)$, $\ell_4=(v_4-u_4)$, $\ell_5=(u_6-u_5)+(v_5-v_6)$, $\ell_6=(v_6-u_6)$, $\ell_7=(u_8-u_7)+(u_{10}-v_8)+(v_7-v_{10})$, $\ell_8=(u_9-u_8)+(v_8-v_9)$, $\ell_9=(v_9-u_9)$ and $\ell_{10}=(v_{10}-u_{10})$.
The Symanzik polynomial $P_L$ is the product of the length for each loop $P_L=\prod_{i=0,10}\ell_i$}
\label{KDelFig}
\end{center}
\end{figure}

The contact operator $\Xi(u,v)$ is defined as
\begin{equation}
\label{DefDel}
\Xi(u,v)=\tilde\delta^d(\rvec(u)-\rvec(v))=(4\pi)^{d/2}\,\delta^d(\rvec(u)-\rvec(v))\ .
\end{equation}
Again, the factor of $(4\pi)^{d/2}$ is a normalization factor simplifying the calculations.
We represent it diagrammatically as an arch joining the points $u$ and $v$.
The partition function with one contact operator inserted is thus 
\begin{eqnarray}
\label{1DelPF}
{\langle \Xi(u,v)\rangle}_0^{\mathrm{closed}}&=&\raisebox{-.6 ex}{\largediagram{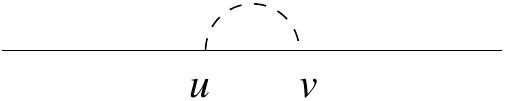}} \nn\\
&=&
\int_{\rvec(0)=\rvec(L)=\mathbf{0}} \hskip -3.5em\mathcal{D}[\rvec]\,\exp\bigl({-\cS_0[\rvec]}\bigr)\,\Xi(u,v)
=  (L-|u-v|)^{-d/2}\,|u-v|^{-d/2}
\end{eqnarray}
This is nothing but the product of the sizes of all loops, raised to the power of $-d/2$.
More generally, the partition function with $K$ contact operators inserted is
\begin{align}
\label{KDelPF}
{\langle\Xi(u_1,v_1) \cdots  \Xi(u_K,v_K)  \rangle}_0^{\mathrm{closed}}&=
\int_{\rvec(0)=\rvec(L)=\mathbf{0}} \hskip -3.5em\mathcal{D}[\rvec]\,\exp\bigl({-\cS_0[\rvec]}\bigr)\,\Xi(u_1,v_1)\cdots\Xi(u_K,v_K)
\nonumber\\
&= \, P_L[u_i,v_i]^{-d/2}
\end{align}
where $P_L$ is the Symanzik polynomial of the $K+1$ loop ($\phi^4$-like) diagram obtained by  contracting to a 4-vertex each arch associated to a contact operator, or equivalently the product of all loop sizes, raised to the power of $-d/2$.
An example of such an arch system (with $K=10$) is depicted in figure~\ref{KDelFig}, together with the corresponding $\phi^4$  diagram.

The main problem with this simple RW representation it that the product of several  contact operators  $\Delta$ does not vanish if the corresponding arch configurations are non-planar. 
As a consequence, the  model of interacting RWs with a contact interaction given by the action
\begin{equation}
\label{s3-5}
\cS[\rvec]= {1\over 4} \int_{0<t<L} \hskip -1.5em\dot{\rvec}(t)^2-g\iint_{0<s<t<L}\hskip -2 em 
{\tilde\delta}^d(\rvec(s)-\rvec(t))
\end{equation}
(the attractive Edwards model) has a perturbative expansion which contains, besides the planar contributions which correspond to terms in the expansion for a RNA pairing model, many more non-planar contributions. Thus the complete action must be more complicated. (Note that (\ref{s3-5}) is of course only a toy model, since it does not contain the replica part necessary to treat the disorder.)

\subsection{The dressed planar RW model}
\label{s:3C}
In order to classify the pairing configurations according to their topology in the RW representation, and to keep  only the planar configurations, we introduce additional matrix-like degrees of freedom and modify the action  accordingly.

\subsubsection{Auxiliary fields}
\begin{figure}[t]
\begin{center}
${\langle\tilde\gamma(t_1)\gamma(t_2)\rangle}_0\ =\ t_1$\includegraphics[width=1.5in]{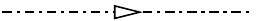}$\ t_2$
\\
\vskip 4ex
$\widetilde {Z}_0^{(\mathrm{closed})}(L)\ =\ $
\includegraphics[width=1.5in]{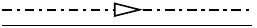}
\\
\vskip 4ex
$\widetilde {Z}_0^{(\mathrm{open})}(\qvec,L)\ =\ $
\includegraphics[width=1.5in]{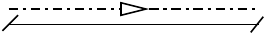}
\caption{The auxiliary field propagator and the dressed closed and open RW partition functions}
\label{AuxilProp}
\end{center}
\end{figure}

First we add a conjugate pair of auxiliary $N$-component  fields $\gamma_a(t)$ and $\tilde\gamma_a(t)$ with a dynamical It\^o like action
\begin{equation}
\label{Sgtg}
\cS_0[\gamma,\tilde\gamma]=\sum_{a=1}^N \int_{0<t<L} \tilde\gamma_a(t)\dot\gamma_a(t)
\end{equation}
$a=1,\ldots N$ is a color index which will play its role later.
The action is such that the propagator for these auxiliary fields is the causal Heaviside function $\theta$
\begin{equation}
\label{Prgtg}
\langle\tilde\gamma_a(t_1)\gamma_b(t_2)\rangle_0=\delta_{ab}\,\theta(t_2-t_1)=\begin{cases}
      \delta_{ab}& \text{if}\ t_2>t_1, \\
      0& \text{otherwise}\ ,
\end{cases}
\end{equation}
while
\begin{equation}
\label{Prgg}
{\langle\tilde\gamma_a(t_1)\tilde\gamma_b(t_2)\rangle}_0={\langle\gamma_a(t_1)\gamma_b(t_2)\rangle}_0=0
\qquad\forall\ a,\,b,\,t_1,\,t_2\ .
\end{equation}
We represent the propagator (\ref{Prgtg}) by a dashed oriented line,  see figure~\ref{AuxilProp}.
The (closed or open) partition function for the free dressed RW is now defined as
\begin{equation}
\label{Z0bdr}
\widetilde  Z_0(L)=\int \mathcal{D}[\rvec]\,\mathcal{D}[\gamma,\tilde\gamma]\,\emath^{-\cS_0[\rvec]-\cS_0[\gamma,\tilde\gamma]}\,
\mathbf{\Gamma}_{\mathrm{b}}
\end{equation}
where we insert the bilocal boundary operator
\begin{equation}
\label{GamBd}
\mathbf{\Gamma}_{\mathrm{b}}=\left({1\over N}\sum_a \tilde\gamma_a(0)\gamma_a(L)\right)
\end{equation}
with the proper boundary conditions for the path integral over $\rvec(t)$. 
 $\mathbf{\Gamma}_{\mathrm{b}}$  creates an auxiliary field at the initial point and annihilates it at the end point, thus still giving a contribution $1$ for the free RW.  
At that stage nothing changes for the expression of the closed and open RW partition functions
$Z_0^{\mathrm{closed}}$ and $Z_0^{\mathrm{open}}$, which are still given by (\ref{Z0c}) to (\ref{Zb0c2}) and by (\ref{Z0orre}) to (\ref{Z0obq}) respectively. However, the boundary operator is crucial for the correlation functions and the interacting theory.

The dressed RW partition functions are now graphically represented as a  ribbon (or fat line) with a full line for the RW and a dashed oriented line for the auxiliary field, see figure~\ref{AuxilProp}.

\subsubsection{Dressed contact operator}
\begin{figure}[h]
\begin{center}
\includegraphics[width=3.5in]{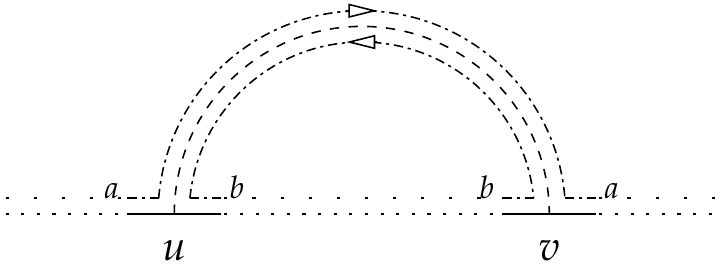}
\caption{Diagrammatic representation of the dressed contact operator $\Phi(u,v)$}
\label{PhiDressed}
\end{center}
\end{figure}

Now we can dress the contact operator $\Xi(u,v)$ with the auxiliary fields so that it becomes a ribbon arch with its topological features encoded by the auxiliary field color indices $a\in\{1,\,\cdots ,N\}$.
Let us define the dressed contact operator $\Phi$ as
\begin{equation}
\label{s3-6}
\Phi(u,v)={1\over N}\sum_{a,b}\gamma_a(u)\,\tilde\gamma_b(u)\,
{\tilde\delta}^d(\rvec(u)-\rvec(v))
\,\gamma_b(v)\,\tilde\gamma_a(v)
\quad.
\end{equation}
With the diagrammatic rules given above it is depicted as a ribbon arch (see figure~\ref{PhiDressed}).

\begin{figure}[t]
\begin{center}
\includegraphics[width=2.5in]{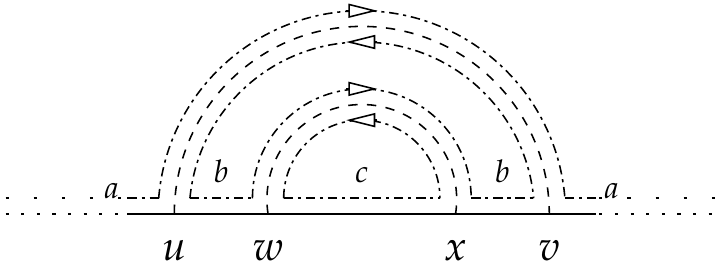}
\qquad
\includegraphics[width=2.5in]{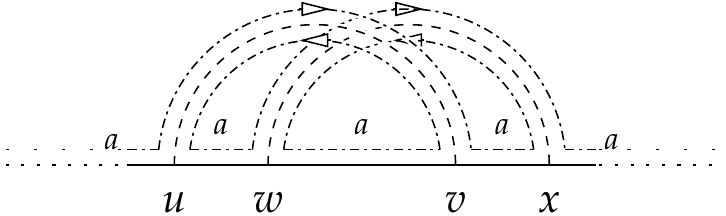}
\caption{Example of planar (left, $h=0$) and non planar (right, $h=1$) products of $\Phi$'s. The left one carries an additional factor of $N^2$ from the sums over $b$ and $c$.}
\label{PNPDressed}
\end{center}
\end{figure}

The partition function with the insertion of  operators $\Phi$, defined as in (\ref{KDelPF}) by
\begin{align}
\label{KPhiPF}
{\langle\Phi(u_1,v_1) \cdots  \Phi(u_K,v_K)  \rangle}_0^{\mathrm{closed}}&=
\int_{\rvec(0)=\rvec(L)=\mathbf{0}} \hskip -3.5em\mathcal{D}[\rvec]\,\mathcal{D}[\gamma,\tilde\gamma]\,
\emath^{-\cS_0[\rvec]-\cS_0[\gamma,\tilde\gamma]}
\,\mathbf{\Gamma}_{\mathrm{b}}
\,\Phi(u_1,v_1)\cdots\Phi(u_K,v_K)
\nonumber\\
&= N^{-2h}\,\, P_L[u_i,v_i]^{-d/2}
\end{align}
now contains a multiplicative color-counting factor $N^{-2h}$ where $h$ is the number of handles of the surface on which the planar arch system for the product of the $\Phi$'s can be drawn without crossings.  The argument is standard and requires counting the factors of $N$ which arise when summing over the internal  color indices $a$ carried by the auxiliary-field line, and by using the Euler relation for the system  of arch diagrams. This counting is illustrated on figure~\ref{PNPDressed}.

\subsubsection{The planar $N\to\infty$ limit}
If we take the planar $N\to\infty$ limit, only the contributions of the planar arch configurations (with  $h=0$ handles) survive. This shows that we can build the RNA perturbation theory in terms of a self-avoiding polymer model embedded in $d=3$. In the following, we discuss how this can be put to work for the random RNA model defined in Eq.~(\ref{Heff1}). The advantage of this formulation, and the only reason we have gone through this formal exercise, is that we can write perturbation theory with a polymer-like action ({\em microscopic free energy}), which allows us to apply the tools of non-local field theory, and especially the multilocal operator product expansion MOPE (see \cite{WieseHabil} for a review).

\subsection{Replicas, the interacting RW model and its diagrammatics}
\label{ss:RepRWdiag}
\subsubsection{The action}
\begin{figure}[t]
\begin{center}
\includegraphics[width=2.5in]{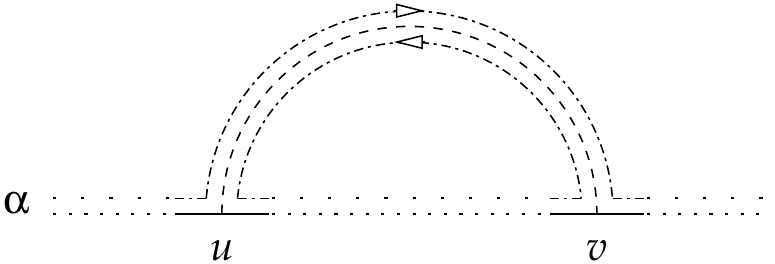}
\quad\qquad
\includegraphics[width=2.5in]{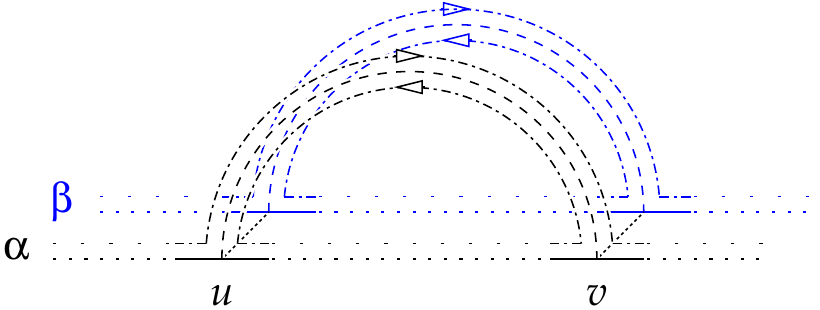}
\caption{Diagrammatic representation of the contact operator $\Phi_\alpha$ (left), and of the overlap operator $\Psi_{\alpha\beta}$ (right)}
\label{s3-8}
\end{center}
\end{figure}

To construct a random walk representation of the LW field theory, we introduce replicas and construct a RW representation for the effective interaction term (\ref{HintRep}) between replicas induced by the quenched disorder.
Consider $n$ replicas of the RW field,  $\rvec_\alpha(t)$, labeled by $\alpha\in\{1,\,\cdots, n\}$, and $n$ replicas for the auxiliary fields $\gamma_a^\alpha(t)$ and $\tilde\gamma_a^\alpha(t)$.
The action for the free replica system is
\begin{equation}
\label{S0DrRe}
\cS_0[\rvec,\gamma,\tilde\gamma]=\int_{0<t<L}{1\over 4}\sum_\alpha \dot\rvec_\alpha(t)^2+\sum_\alpha\sum_a \tilde\gamma_a^\alpha(t)\dot\gamma_a^\alpha(t)\ .
\end{equation}
The dressed contact operator $\Phi_\alpha$ for the replica $\alpha$ is
\begin{equation}
\label{PhiRep}
\Phi_\alpha(u,v)={1\over N}\sum_{a,b}\gamma_a^\alpha(u)\tilde\gamma_b^\alpha(u)\,\Xi_\alpha(u,v)\,\gamma_b^\alpha(v)\tilde\gamma_a^\alpha(v)
\end{equation}
with $\Xi_\alpha(u,v)$ the contact operator for the replica $\alpha$
\begin{equation}
\label{DelRep}
\Xi_\alpha(u,v)=(4\pi)^{d/2}\,\delta^d(\rvec_\alpha(u)-\rvec_\alpha(v))
\end{equation}
The dressed overlap operator between distinct replicas $\alpha\neq\beta$ is
\begin{equation}
\label{PsiDre}
\Psi_{\alpha\beta}(u,v)=\Phi_\alpha(u,v)\,\Phi_\beta(u,v)\ .
\end{equation}
The attractive replica interaction is
\begin{equation}
\label{SDreInt}
\cS_{\mathrm{{int}}}[\rvec,\gamma,\tilde\gamma]=-g\sum_{\alpha<\beta}\iint_{0\le u<v\le L} \Psi_{\alpha\beta}(u,v)\ ,
\end{equation}
so that the full action is
\begin{equation}
\label{SDreRep}
\cS=\cS_{{0}}+\cS_{\mathrm{{int}}}\ .
\end{equation}
We shall construct the perturbative expansion and its diagrammatics for this theory, starting from the diagrammatic representation of the interaction operators represented in figure~\ref{s3-8}.
We are interested in the double limit $N\to\infty$ (planar diagrams) and $n\to 0$ (quenched disorder). 
In the remaining discussion and in the calculations we shall take first the planar limit (the number of diagrams is thus greatly reduced), but keep the number  $n$ of replicas non-zero.  We take the $n\to 0$ limit at the end of the calculation.
Since $N=\infty$, we do not represent the auxiliary field propagators as dashed lines any more, but simply keep  planar diagrams.

\subsubsection{Closed-strand partition function}
The $n$-replica closed-strand partition function for the free model ($g=0$) is 
\begin{equation}
\label{Z0closed}
{\widetilde {\mathcal{Z}}}_0^{\mathrm{(closed)}}(L)
=\langle\mathbf{1}\rangle_0^{\mathrm{closed}}
=\int_{\rvec_\alpha(0)=\rvec_\alpha(L)=\mathbf{0}}\hskip -4.em
\mathcal{D}[\rvec,\gamma,\tilde\gamma]\,\emath^{-\cS_0}\,
\mathbf{\Gamma}_{\mathrm{b}}
=\left[\widetilde  Z_0(L)\right]^n
= L ^{-nd/2}\ ,
\end{equation}
where the measure and the boundary operators are
\begin{equation}
\label{MeaBouR}
\mathcal{D}[\rvec,\gamma,\tilde\gamma]=\prod_\alpha \mathcal{D}[\rvec_\alpha,\gamma_\alpha,\tilde\gamma_\alpha]
\ ,\quad 
\mathbf{\Gamma}_{\mathrm{b}}=\prod_\alpha \mathbf{\Gamma}_{\mathrm{b}}[\gamma_\alpha,\tilde\gamma_\alpha]
=\prod_\alpha\left( {1\over N} \sum_a\tilde\gamma_\alpha(0)\gamma_\alpha(L)\right)
\quad.
\end{equation} 
We represent it diagrammatically as a single (fat) line, but it is understood that this represents a bundle of $n$ lines.
\begin{figure}[t]
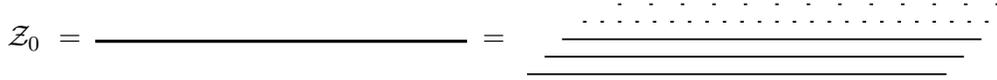

\begin{center}
$\mathcal{Z}_0\ =\ $\includegraphics[width=2in]{FatLine}$\ =\ $ 
\raisebox{-2.5 ex}{\includegraphics[width=2.5in]{BundleLine}}
\caption{The replica bundle.}
\label{s3-9}
\end{center}
\end{figure}

The closed-strand partition function for the interacting model is
\begin{equation}
\label{Zgclosed}
\widetilde {\mathcal{Z}}^{\mathrm{(closed)}}(L)=\langle\exp\left(-\cS_{\mathrm{int}}\right)\rangle_0^{\mathrm{closed}}
=\int_{\rvec_\alpha(0)=\rvec_\alpha(L)=\mathbf{0}}\hskip -4.em
\mathcal{D}[\rvec,\gamma,\tilde\gamma]\,\emath^{-\cS}\,
\mathbf{\Gamma}_{\mathrm{b}}
\end{equation}
It can be expanded in a perturbation  series in powers of $g$. The term of order $g^K$ is   of the form
\begin{equation}
\label{s3-10}
{1\over K!} \sum_{\cdots\alpha\cdots}\iiint {\langle\Psi\cdots\Psi\rangle}_0\ ,
\end{equation}
and can be represented diagrammatically in terms of double arch systems involving $P$ replicas with $2\le P\le 2K$, exactly as for the LW model.
In the following we represent as a line only the replicas coupled via $\Psi_{\alpha\beta}$ overlap operators. It is understood that the $n-P$ other replica are there and give a factor of $\big(\widetilde {Z}_0\big)^{n-P}$.

\begin{figure}[h]
\begin{align}
&\includegraphics[width=3.5in]{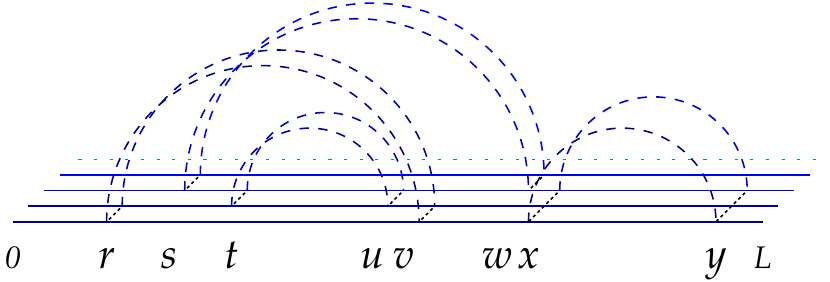}\nn\\
&=\left[(v-r)(y-x)(L-v+r-y+x) \right]^{-d/2}\nn\\
&\quad\  \times\left[ (u-t)(v-u+t-r)(L-v+r)\right]^{-d/2}\nn\\
&\quad\  \times\left[ (u-t)(w-u+t-s)(y-x)(L-y+x-w+s)\right]^{-d/2}\nn\\
&\quad\  \times\left[ (w-s) (L-w+s)\right]^{-d/2}\nn\\
&\quad\  \times L^{(4-n)d/2 }
\end{align}
\caption{Example of e.v.\ of $k=4$ $\Psi$'s as given in (\ref{PPsiL2}). The factors are grouped by replica lines, starting at the front.}
\label{FigPab}
\end{figure}

It is clear that for closed RWs the diagrammatics and the resulting integrals are equivalent to those of the LW theory. 
The amplitude ${\langle\Psi\cdots\Psi\rangle}_0$ is given as a product over each of the $P$ replicas $\alpha$ of products of the internal loop lengths to the power $-d/2$, {\em equivalent} to formula (\ref{PPsiL}):
\begin{equation}
\label{PPsiL2}
\langle\!\!\langle\underbrace{\Psi\cdots\Psi}_{k}  {\rangle\!\!\rangle}_0 =
\prod_{\alpha=1}^{P}\left(\prod_{j=0}^{p_\alpha}\ell_{j,\alpha}\right)^{-d/2}\ \left(L^{-d/2}\right)^{n-P}
\end{equation}
The combinatoric factor for each amplitude, which is a polynomial of degree $P$ in $n$, is also the same for the LW and the RW models.
An example of such an amplitude is given in figure~\ref{FigPab}.
The RW representation in the planar limit provides a (somewhat formal but systematic) functional integral representation of the LW model and a way to study its short distance structure (see next section). 

\subsection{Single open-strand partition function}
\label{ss:soRW}
As already explained above, it will be useful to consider other sectors and other observables in the RW model. These observables are associated to open random walks and have no interpretation in terms of the random pairing RNA model.

\begin{figure}[t]
\begin{center}
${\parbox{2.5in}{\includegraphics[width=2.5in]{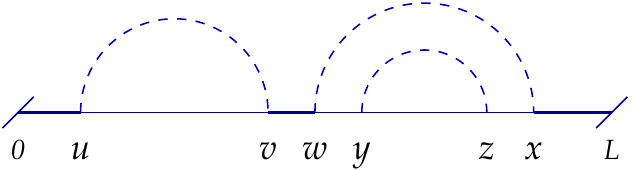}}}={\langle  \Delta(u_1,v_1)\cdots\Delta(u_p,v_p)\, \emath^{-\imath\qvec_1\rvec(L)} \rangle}_0^{{\mathrm{open}}}$
\caption{Diagrammatic representation for the open strand planar arch sub-system (\ref{OpDDE0}) for a single replica, and the corresponding amplitude. The open strand with length $\ell_0$ is depicted by a bold line. 
The cut at the ends of the strand indicate that we deal with a open strand with incoming momenta $\qvec_1$.}
\label{s3-14}
\end{center}
\end{figure}

By similarity with the partition function for an open RW (\ref{Z0oqq},\ref{Z0orr}), we consider now $n$ replicas of a dressed open RW 
$\rvec_\alpha(t)$ with $t\in[0,L]$, and with free boundaries $\rvec_\alpha(0)$ and $\rvec_\alpha(t)$. We attach to each endpoint momenta $\qvec_1$ and $\qvec_2$ which (for simplicity) are taken to be the same for the $n$ different replica. 
The open-strand partition function is defined as
\begin{equation}
\label{Z1oq1q2}
\cZo(\qvec_1,\qvec_2;L)=\iint \mathcal{D}[\rvec,\gamma,\tilde\gamma]\ \emath^{-\cS}\ \mathbf{\Gamma}_\mathrm{b}\ \emath^{\imath [\qvec_1\sum_\alpha \rvec_\alpha(0)+\qvec_2\sum_\alpha \rvec_\alpha(L)]}
\end{equation}
where $\cS$ is the action for the interacting dressed open RW model (\ref{SDreRep}) and $\mathbf{\Gamma}_\mathrm{b}$ the boundary operator, as defined by (\ref{MeaBouR}).
The index $^{(1)}$ added to the partition function
(instead of the index $^{\mathrm{(open)}}$ used in Sect.~\ref{s:3B}-\ref{s:3C})
 indicates: (\textit{i}) that we deal with open strands, (\textit{ii}) that we deal with one (1) single bundle of $n$ replica of the same open RW.
Later on we shall consider the partition functions $\cZp$ for $p>1$ (bundles of $n$ replica of) open RWs  interacting via the disorder-induced 2-replica contact-interaction term $\Psi_{\alpha\beta}$.
Using translational invariance in $\mathbb{R}^d$ for each replica, we can factor out a momentum conservation term for each replica, thus defining, by analogy with the free case (\ref{Z0oqqe})
\begin{equation}
\label{Z1oq}
\cZo(\qvec_1,\qvec_2;L)=\left[(2\pi)^{d}\delta^d(\qvec_1+\qvec_2)\right]^n\ \tcZo(\qvec_1;L)\ .
\end{equation}
The IR finite function $\tcZo(\qvec_1;L)$ can be expanded in perturbation theory in a power series in $g$,  as 
\begin{align}
\label{OpPPE0}
\tcZo(\qvec_1;L)&={\langle \emath^{-\imath \qvec_1 \sum_\alpha[\rvec_\alpha(0)-\rvec_\alpha(L)]}\rangle}^{{\mathrm{open}}}
={\langle \emath^{-\cS_{\mathrm{int}}}\,\emath^{\imath \qvec_1\sum_\alpha[\rvec_\alpha(0)-\rvec_\alpha(L)]}\rangle}_0
^{{\mathrm{open}}}
\nonumber\\
&=\iint_{\rvec_\alpha(0)=0} \mathcal{D}[\rvec,\gamma,\tilde\gamma]\ \emath^{-\cS}\ \mathbf{\Gamma}_\mathrm{b}\ \emath^{-\imath\qvec_1\sum\limits_\alpha \rvec_\alpha(L)}
\end{align}
with $\cS_{\mathrm{int}}$ given by (\ref{SDreInt}).
The rules to compute the perturbative expansion are a simple generalization of those for the closed RW model.
The term of order $g^K$ can be expanded into a sum (over the various distributions of $(\alpha,\beta)$) of integrals (over the $u$ and $v$) of expectation values of the form
\begin{equation}
\label{s3-12}
\langle \Psi_{\alpha_1\beta_1}(u_1,v_1)\cdots\Psi_{\alpha_k\beta_k}(u_k,v_k)\  \emath^{-\imath\qvec_1\sum\limits_\alpha \rvec_\alpha(L)}{\rangle}_0^{{\mathrm{open}}}\ .
\end{equation}
The integral over the auxiliary fields selects the arch configurations which are planar for each replica, and give zero for the others. 
We end up with a product for each replica of an e.v.\ for the open RW model of the form
\begin{equation}
\label{OpDDE0}
{\langle  \Delta(u_1,v_1)\cdots\Delta(u_p,v_p)\, \emath^{-\imath\qvec_1\rvec(L)} \rangle}_0^{{\mathrm{open}}}
\end{equation}
with the planar arch sub-system $\Omega=\Delta(u_1,v_1)\cdots\Delta(u_p,v_p)$ extracted from the planar double arch system $ \Psi_{\alpha_1\beta_1}(u_1,v_1)\cdots\Psi_{\alpha_k\beta_k}(u_k,v_k)$ for each replica $\alpha$.
The e.v.~(\ref{OpDDE0}) is easily calculated. 
The $p$ arches form $p$ internal loops with backbone lengths $\ell_1,\cdots,\ell_p$. At variance with the closed RW model, the remaining  segments of the strand that are not under an arch form an open sub-strand with total length $\ell_0=\ell_{\mathrm{free}}$ (with $\ell_0+\ell_1+\cdots+\ell_p=L$).
The e.v.\ (\ref{OpDDE0}) is 
\begin{equation}
\label{s3-13}
{\langle  \Delta(u_1,v_1)\cdots\Delta(u_p,v_p)\, \emath^{-\imath\qvec_1\rvec(L)} \rangle}_0^{{\mathrm{open}}}= \left(\prod_{\text{closed loops}\atop j=1,...,p}\hskip-.5em\ell_j\right)^{-d/2}\,\emath^{-\qvec_1^2 \ell_0}
\end{equation}
and is represented in figure~\ref{s3-14}.

\begin{figure}[h]

$\raisebox{-3.0 ex}{\includegraphics[width=3.in]{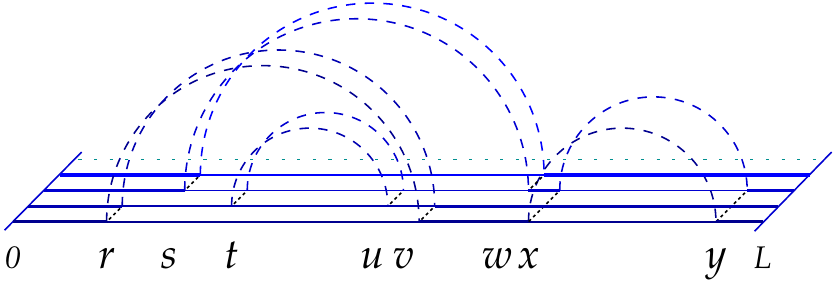}}
=\begin{array}{c}
 \left[(v-r)(y-x) \right]^{-\frac{d}{2} } \times\left[ (u-t)(v-u+t-r)\right]^{-\frac{d}{2} }\\
\times\left[ (u-t)(w-u+t-s)(y-x)\right]^{-\frac{d}{2} }\left[ (w-s)\right]^{-\frac{d}{2}}
\end{array}
$
\caption{Diagrammatic representation for a single planar open-strand $\Psi$ system, and the corresponding amplitude.
The bars at the endpoints of the bundle indicate that we deal with a open strand with incoming momenta $\qvec_1$ and $-\qvec_1$.}
\label{s3-16}
\end{figure}

Each e.v.\ (\ref{OpPPE0}) is the product over $P$ amplitudes of the form (\ref{OpDDE0}) (for the $P$ replicas coupled by the $\Psi_{\alpha\beta}$), times the $n-P$ remaining free-strand amplitudes. Hence it is given by an amplitude of the form
\begin{equation}
\label{s3-15}
\prod_{\alpha=1}^P\left(\prod_{ j=1}^{p_\alpha}\ell_{j,\alpha}\right)^{-d/2}\,\emath^{-\qvec_1^2 \left(\sum\limits_{\alpha=1}^P\ell_{0,\alpha}+(n-P)L\right)}
\end{equation}
and is represented as in figure~\ref{s3-16}.

\subsection{Multiple open-strand partition function}
\label{s:3F}
\begin{figure}[t]
\begin{center}
\includegraphics[width=2.5in]{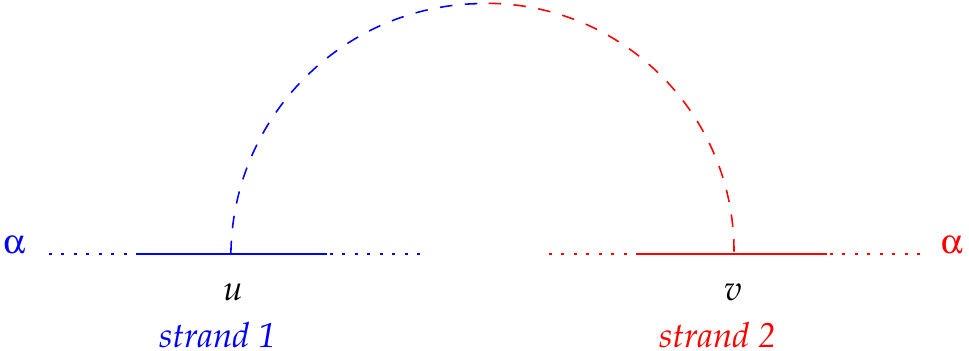}
\quad
\includegraphics[width=2.5in]{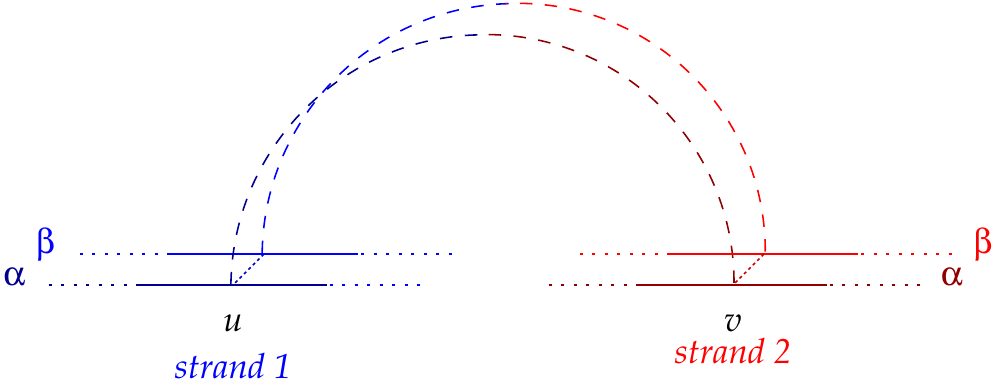}
\caption{The new contact vertex $\Phi^\ssst{1,2}_{\alpha\beta}(u,v)$  and the interaction vertex $\Psi^\ssst{1,2}_{\alpha\beta}(u,v)$ between two different strands (labeled  $1$ and $2$).}
\label{Fpsi12}
\end{center}
\end{figure}

Finally we  consider partition functions for $M>1$ strands interacting via the disorder-induced contact operator $\Psi$. Let us restrict ourselves to the two-strand case $M=2$.
These two open strands are described by the RW's $\rvec_{1}$ and $\rvec_{2}$, and more precisely by
\begin{align}
\label{s3-17}
&{\rvec_\ssst{1}}_\alpha(t_1),\ 
{\gamma_\ssst{1}^a}_\alpha(t_1) ,\ 
{\tilde{\gamma}_\ssst{1}^a{}}_\alpha(t_1)
\qquad \text{(strand 1)}\\
&{\rvec_\ssst{2}}_\alpha(t_2),\ 
{\gamma_\ssst{2}^a}_\alpha(t_2) ,\ 
{\tilde{\gamma}_\ssst{2}^a{}}_\alpha(t_2)
\qquad \text{(strand 2)}
\end{align}
For simplicity the two strands have the same length $L_1=L_2=L$, so that $t_I\in[0,L]$, $t_2\in[0,L]$.
The action is taken to be the sum of the action for strands one and two, plus an interaction term between the two strands
\begin{align}
\label{2StAct}
\cS=&
\sum_\alpha\left(
\int_{t_1}{1\over 4}({\dot\rvec_\ssst{1}{}}_\alpha)^2+\sum_a\int_{t_1}{\tilde\gamma_\ssst{1}^a{}}_\alpha{\dot\gamma_\ssst{1}^a{}}_\alpha
+
\int_{t_2}{1\over 4}({\dot\rvec_\ssst{2}{}}_\alpha)^2+
\sum_a\int_{t_2}{\tilde\gamma_\ssst{2}^a{}}_\alpha{\dot\gamma_\ssst{2}^a{}}_\alpha
\right)
\nonumber\\
-&g\sum_{\alpha<\beta} \left(\iint_{u_1<v_1}\Psi^\ssst{1}_{\alpha\beta}(u_1,v_1)+
\iint_{u_2<v_2}\Psi^\ssst{2}_{\alpha\beta}(u_2,v_2)
+\iint_{u_1,v_2}\Psi^\ssst{1,2}_{\alpha\beta}(u_1,v_2)\right)
\quad.
\end{align}
The first line in (\ref{2StAct}) is the free action for the two strands.
In the second line 
$\Psi^\ssst{1}_{\alpha\beta}(u_1,v_1)$ and $\Psi^\ssst{2}_{\alpha\beta}(u_2,v_2)$ are the overlap operators (\ref{PsiDre}) for the two strands 1 and 2.
The new operator $\Psi^\ssst{1,2}_{\alpha\beta}(u_1,v_2)$ is the overlap for the contact between strands 1 and 2, defined as
\begin{equation}
\label{s3-18}
\Psi^\ssst{1,2}_{\alpha\beta}(u_1,v_1)=\Phi^\ssst{1,2}_\alpha(u_1,v_2)\,\Phi^\ssst{1,2}_\beta(u_1,v_2)
\end{equation}
\begin{align}
\label{s3-19}
\Phi^\ssst{1,2}_\alpha(u_1,v_2)&={1\over N}\sum_{a,b}
{\gamma_\ssst{1}^\alpha{}}_{\! a}(u_1){\tilde\gamma_\ssst{1}^\alpha{}}_{\! b}(u_1)
\tilde \delta^d({\rvec_\ssst{1}}_\alpha(u_1)-{\rvec_\ssst{2}}_\alpha(v_2))
{\gamma_\ssst{2}^\alpha{}}_{\! b}(v_2){\tilde\gamma_\ssst{2}^\alpha{}}_{\! a}(v_2)
\end{align}
and depicted in figure~\ref{Fpsi12}. The model can be generalized by assigning different coupling constants $g_1$, $g_2$ and $g_{1,2}$ to the operators $\Psi^\ssst{1}_{\alpha\beta}$, $\Psi^\ssst{2}_{\alpha\beta}$ and $\Psi^\ssst{1,2}_{\alpha\beta}$. For simplicity we keep $g_1=g_2=g_{1,2}=g$.

\begin{figure}[h]
\begin{center}
\raisebox{3.ex}{$\qvec_1\ \to$}
\raisebox{20.ex}{\hskip -3em  $\qvec_4\ \to$}
\includegraphics[width=4in]{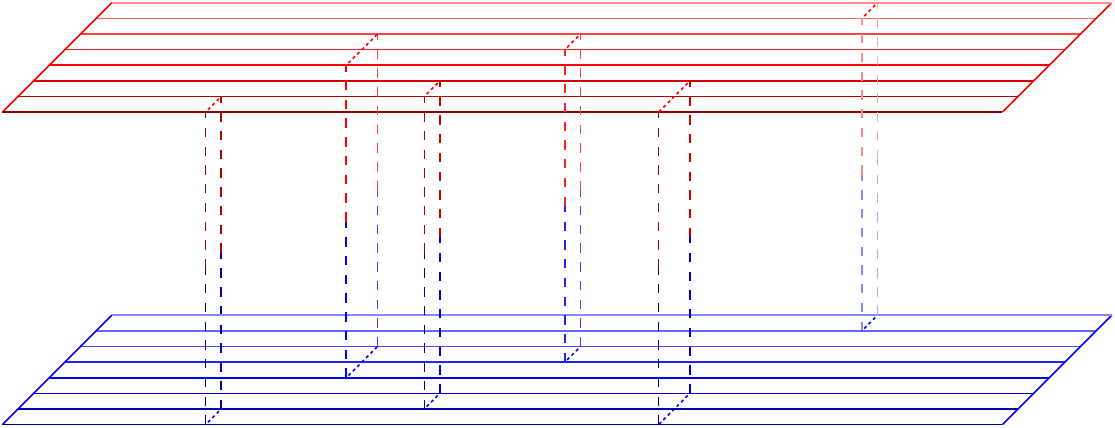}
\raisebox{3.ex}{$\leftarrow\ \qvec_2$}
\raisebox{20.ex}{\hskip -3em  $\leftarrow\ \qvec_3$}
\caption{An example of a two-strand diagram, representing a $\Psi^\ssst{1,2}_\mathnormal{1,1} \Psi^\ssst{1,2}_\mathnormal{4,6} \Psi^\ssst{1,2}_\mathnormal{2,3} \Psi^\ssst{1,2}_\mathnormal{5,6} \Psi^\ssst{1,2}_\mathnormal{1,3} \Psi^\ssst{1,2}_\mathnormal{7,8}$ term. Note that on the lower strand $0$ is left and $L$ right, whereas on the upper strand $L$ is left and $0$ right.}
\label{TwoStrandPartition}
\end{center}
\end{figure}

The two open-strand partition function is defined as
\begin{align}
\label{Z2odef}
&\cZt (\qvec_1,\qvec_2,\qvec_3,\qvec_4;L)=
\nonumber\\
&\iint \mathcal{D}[\rvec_\ssst{1},\gamma_\ssst{1},\tilde\gamma_\ssst{1}]\ 
\mathcal{D}[\rvec_\ssst{2},\gamma_\ssst{2},\tilde\gamma_\ssst{2}]
\ \emath^{-\cS}\ {\mathbf{\Gamma}_{\!\ssst{1}}}_\mathrm{b}\,  {\mathbf{\Gamma}_{\!\ssst{2}}}_\mathrm{b}\,
\emath^{\imath\left(\qvec_1\sum\limits_\alpha {\rvec_\ssst{1}}_\alpha(0)
+
\qvec_2\sum\limits_\alpha {\rvec_\ssst{1}}_\alpha(L)
+
\qvec_3\sum\limits_\alpha {\rvec_\ssst{2}}_\alpha(0)
+
\qvec_4\sum\limits_\alpha {\rvec_\ssst{2}}_\alpha(L)
\right)
}
\end{align}
It is calculated in perturbation theory as a power series in $g$. The term of order $g^K$  is a sum of expectation values of products of $\Psi^\ssst{1}_{\alpha\beta}$, $\Psi^\ssst{2}_{\alpha\beta}$ and $\Psi^\ssst{1,2}_{\alpha\beta}$ operators. 
Each term is represented diagrammatically as a system of two bundles (one for each strand), with a planar system of double arches on each strand (for the $\Psi^\ssst{1}$ and $\Psi^\ssst{2}$) and of double ribbons between the two strands (for the $\Psi^\ssst{1,2}$). An example of such a diagram (with only $\Psi^\ssst{1,2}_{\alpha\beta}$ operators)  is depicted on figure~\ref{TwoStrandPartition}.

When computing $\cZt$ there is a subtle technical point when dealing with translational invariance to factor out the $(2\pi)^d\delta^d(\Sigma\qvec)$ terms for each replica.
For each replica $\alpha$ one has to compute a term of the form
$$
\langle\, {\underbrace{\Phi^\ssst{1}_\alpha\cdots \Phi^\ssst{1}_\alpha}_{K_1}\,
 \underbrace{\Phi^\ssst{2}_\alpha\cdots \Phi^\ssst{2}_\alpha}_{K_2}\,
\underbrace{\Phi^\ssst{1,2}_\alpha\cdots \Phi^\ssst{1,2}_\alpha}_{K_{1,2}} 
\,\emath^{\imath(\qvec_1{\rvec_\ssst{1}}_\alpha(0)+\qvec_2{\rvec_\ssst{1}}_\alpha(L) +\qvec_3{\rvec_\ssst{2}}_\alpha(0)+\qvec_4{\rvec_\ssst{2}}_\alpha(L))}
}{\rangle}_0
 $$
Either there is no $\Psi_\ssst{1,2}$ operator ($K_{1,2}=0$) and the two strands are decoupled, so that from translational invariance we factor out a term
\begin{equation}\label{99}
(2\pi)^d\delta^d(\qvec_1+\qvec_2)\ (2\pi)^d\delta^d(\qvec_3+\qvec_4)\ ,
\end{equation}
or there is at least one $\Psi_\ssst{1,2}$ operator ($K_{1,2}>0$) and the two strands are coupled, so that from translational invariance we factor out a term
\begin{equation}\label{100}
(2\pi)^d\delta^d(\qvec_1+\qvec_2+\qvec_3+\qvec_4)\ .
\end{equation}
We already note that this subtlety will become fully manifest  when under renormalization the field $\bf r$ goes to ${\bf r} \sqrt{\mathbb Z}$, and correspondingly the momenta $\bf q$ go to ${\bf q}/ \sqrt{\mathbb Z}$, thus inducing different additional powers of $\mathbb Z$ in (\ref{99}) and (\ref{100}).

We must treat separately the contributions to $\cZt$ according to the number $Q$ of replica such that the strands $1$ and $2$ are coupled. More precisely, a term of order $g^K$ involves $K_1$ $\Psi^\ssst{1}$, $K_2$ $\Psi^\ssst{2}$ and $K_{1,2}$ $\Psi^\ssst{1,2}$,  with $K_1+K_2+K_{1,2}=K$, and may have $2\le Q\le 2 K_{1,2}$ replicas with strands $1$ and  $2$ coupled.
Thus we can decompose uniquely $\cZt$ into the contribution of these ``$Q$-sectors",  ${\tilde{\mathcal{Z}}}  ^{\ssst{(2,Q)}}$, as
\begin{align}
\label{Z2Qodef}
\cZt(\qvec_1,\qvec_2,\qvec_3,\qvec_4;L)=\sum_{Q}&\left[(2\pi)^{2d}\,\delta^d(\qvec_1+\qvec_2)\,\delta^d(\qvec_3+\qvec_4)\right]^{n-Q}
\nonumber\\
&\times \left[ (2\pi)^d\,\delta^d(\qvec_1+\qvec_2+\qvec_3+\qvec_4)\right]^Q \,
{\widetilde{\mathcal{Z}}}  ^{\ssst{(2,Q)}}(\qvec_1,\qvec_2,\qvec_3;L)\ .
\end{align}
The $Q=0$ term is the disconnected contribution
\begin{equation}
\label{s3-20}
{\widetilde{\mathcal{Z}}}  ^{\ssst{(2,0)}}(\qvec_1,\qvec_2,\qvec_3;L)={\widetilde{\mathcal{Z}}}  ^{\ssst{(1)}}(\qvec_1;L) \,{\widetilde{\mathcal{Z}}}  ^{\ssst{(1)}}(\qvec_3;L) \ .
\end{equation}
Since the interaction is a 2-replica interaction, the $Q=1$ term is  zero
\begin{equation}
\label{s3-21}
{\widetilde{\mathcal{Z}}}  ^{\ssst{(2,1)}}(\qvec_1,\qvec_2,\qvec_3;L)=0\ .
\end{equation}
The details of the calculations of these two-strand partition functions will be given in  section \ref{s:2loops}. 
To identify all necessary renormalisations,  we only need to compute the $Q=2$ sector contribution ${\widetilde{\mathcal{Z}}}  ^{\ssst{(2,2)}}$ at zero external momenta $\qvec=0$.

 \clean

\section{UV divergences and Renormalisation}
\label{s:renormalization}
\subsection{Introduction: dimensional analysis}
\label{ss:UVdimanalysis}

The LW field theory suffers from short-distance (UV) divergences when $\rho_0$, taken as an analytic regularisation parameter, is greater than or equal to 1. Using
\begin{equation}
\label{epsilonrho0}
\varepsilon=2\rho_0-2
\end{equation}
as a Wilson-Fisher expansion parameter, it was shown in \cite{LaessigWiese2005} that at one loop (first non-trivial order in $g$), these UV divergences appear as poles in $1/\varepsilon$, and can be absorbed into a renormalization  of the coupling constant $g$ (strength of the disorder) and of the strand length $L$.
In this renormalization  framework, for $\epsilon>0$, the one-loop calculation shows that there is a physically relevant non-trivial  UV fixed point $g^*=\mathcal{O}(\varepsilon)$ (with $g^*>0$), which controls the continuum limit of the LW theory. This UV fixed point separates a weak-disorder phase ($g<g^*$), where disorder is irrelevant at large scales, from a strong disorder phase ($g>g^*$), where disorder is strongly relevant at large distances.

In this section, we consider the UV divergences in the RW representation of the model. 
We show that the divergences can be analyzed via the short-distance behavior of the RW model, through a multilocal operator product expansion (MOPE). This MOPE is similar to the MOPE for polymers (the Edwards model) and for self-avoiding polymerized membranes (SAM), and is a generalization of the well-known operator product expansion (OPE) for local quantum field theories.

The RW model in $d$ dimensions is equivalent (for closed RW) to the LW model with analytic regularisation where the contact exponent is $\rho_0=d/2$. 
Hence we denote
\begin{equation}
\label{epsilond}
\varepsilon=d-2
\ .
\end{equation}
Using units of the strand length $L$, in the action $\mathcal{S}$ given by   (\ref{S0DrRe}--\ref{SDreRep})  the bare scaling dimensions of the base position $t$, of the position vector $\rvec$ and of the auxiliary fields $\gamma$ and $\tilde\gamma$ are
\begin{equation}
\label{dimtrgamma}
[t]=1\quad,\qquad [\rvec]=1/2\quad,\qquad [\gamma]=[\tilde\gamma]=0\ .
\end{equation}
The dimensions of the  elasticity $\dot\rvec^2$, of the contact operator $\Phi$, of the overlap operator $\Psi$, and of the coupling constant $g$ are
\begin{equation}
\label{dimr2phipsig}
[\dot\rvec^2]=-1 \quad,\qquad [\Phi]=-d/2 \quad,\qquad [\Psi]=-d \quad,\qquad [g]=d-2=\varepsilon\ .
\end{equation}
If the theory behaves as an ordinary quantum field theory, it is natural to expect that it has UV divergences for $d=2$, is perturbatively renormalizable for $d=2$ and non-renormalizable for $d>2$.
This will be true if the short-distance singularities are proportional to the operators already present in the theory, and if no new terms are generated under renormalization.

\subsection{The MOPE}
\label{ss.mope}
The short-distance singularities can be analyzed via a multi-local operator-product expansion (MOPE). The importance of this MOPE was already recognized by L\"assig-Wiese, \cite{LaessigWiese2005,LaessigWieseToBePublished} (see especially \cite{LaessigWieseToBePublished}). 

The fact that the short-distance singularities for products of $\Psi$ operators in our RW model are described by a MOPE is easy to understand, without much explicit calculations. Indeed, the operator $\Psi_{\alpha\beta}(u,v)$  is a product of two bilocal contact operators $\Phi_\alpha(u,v)$ for two independent replicas $\alpha$ and $\beta$
\begin{equation}
\label{Psiagain}
\Psi_{\alpha\beta}(u,v)=\Phi_\alpha(u,v)\Phi_\beta(u,v)\ .
\end{equation}
Each contact operator is the product of the standard bilocal contact operator $\Xi(u,v)$ for the plain RW, times two local operators $\Upsilon(u)$ and $\Upsilon(v)$ for the auxiliary fields $\gamma$ and $\tilde\gamma$
\begin{equation}
\label{Phiagain}
\Phi(u,v)={1\over N}\sum_{a,b}\Upsilon_{ab}(u)\,\Xi(u,v)\,\Upsilon_{ba}(v)
\end{equation}
with
\begin{equation}
\label{Xiagain}
\Xi(u,v)=\tilde \delta^d(\rvec(u)-\rvec(v)) = (4\pi)^{d/2}\,\delta^d(\rvec(u)-\rvec(v))
\ ,
\qquad \Upsilon_{ab}(u)=\gamma_a(u)\tilde\gamma_b(u)
\ .
\end{equation}
It is thus sufficient to analyze the short-distance behavior separately for each replica $\alpha$; and further separately for the $\Phi_\alpha$, functions of  $\rvec_\alpha$, and for the local operators $\Upsilon_\alpha$  involving only the auxiliary fields $\gamma_\alpha$ and $\tilde\gamma_\alpha$.

When dealing with open strands, one must be careful  to take into account the boundary operators at $u=0$ and $u=L$.  One has to write the MOPE at the boundaries for products involving a boundary operator and bulk operators, when some of the points go to the boundary.
As we shall see, boundary operators are important for the renormalization  of the model.

\subsubsection{MOPE for the $\rvec$ operators}
\label{sss:mopeggb}
\label{sss:moper}
The MOPE for the $\Xi$ operators is nothing but the standard MOPE for the operators of the Edwards model for a SAW,  and of the general $D>1$ polymerized (or tethered) membrane (SAM).
This MOPE was studied  extensively in
\cite{DDG1,DDG2,DDG3,DDG4,WieseDavid1995,DavidWiese1996,WieseDavid1997,WieseHabil}.
It is obtained by expressing the contact operator as
\begin{equation}
\label{XiFour}
\Xi(u,v)=\int {d^d\qvec \over \pi^{d/2}}\ \emath^{\imath \qvec [\rvec(u)-\rvec(v)]}\ .
\end{equation}
One then writes the short-distance expansion of products of vertex operators $\exp(\imath\qvec\rvec(u))$  in terms of normal products, and the short-distance OPE for the massless free field in 1D (the quantum free particle), using the explicit form of the 1D propagator.

The 1D massless propagator $G_0(u,v)\simeq\langle r(u)r(v)\rangle_0$ for the scalar field is the solution of
\begin{equation}
\label{LaplEqu}
-{1\over 2}{\partial^2\over \partial u^2}G_0(u,v)=\delta(u-v)-{1\over L}
\end{equation} 
with periodic b.c.\ for closed strands, and Neumann b.c. for open strands.
The $1/L$ term takes care of the zero mode.
The propagator is explicitly, for the closed strand
\begin{equation}
\label{ClosedProp}
G_0^{{\mathrm{closed}}}(u,v)=
(|u-v|-L/2)^2/L
\end{equation}
and for the open strand
\begin{equation}
\label{OpenProp}
G_0^{{\mathrm{open}}}(u,v)=
\left( (u-L/2)^2+(v-L/2)^2-L|u-v|\right)/L\ .
\end{equation}
We only need the difference propagators, see equation (\ref{rprop}), 
\begin{equation}
\frac 1{2d} \left< [\rvec(u)-\rvec(v)]^2\right>_0 = \left\{ \begin{array}{ccc}
 |u-v| &   \qquad \mbox{for open b.c.} &    \\
  &   &   \\
  \displaystyle \frac{|u-v|(L-|u-v|)}L &   \qquad \mbox{for closed b.c.} & 
\end{array}\right.
\end{equation}
The product of multilocal operators has a short-distance expansion in terms of other multilocal operators, of the general form
\begin{equation}
\label{MOPEGen}
\mathbf{O}_1(u_1^1,u_1^2,...)\,\mathbf{O}_2(u_2^1,u_2^2,...)\cdots\mathbf{O}_p(u_p^1,u_p^2,...)\ \mathop{=}_{\{u_j^i\}\to \{s^k_l\}}\ \sum_{\mathbf{O}}\ C^{\mathbf{O}_1\mathbf{O}_2\cdots\mathbf{O}_p}_{\mathbf{O}}(\{ u_j^i{-}s^k_l\}) \,\mathbf{O}(\{s^k_l\})
\end{equation}
This expansion generates an algebra containing all multilocal operators of the form
\begin{equation}
\label{GenMOp}
\mathbf{O}(s_1,\cdots,s_k)=\int \mathrm{d}^d{\rvec_0}\prod_{i=1}^{k} \left(\prod_{j=1}^{m_i}\left({\partial^{n_{ij}}\rvec(s_i)}\right)\nabla^{k_i}\delta^d(\rvec(s_i)-\rvec_0)\right)\ ,
\end{equation}
where $\partial$ and $\nabla$ are the derivative in internal and external space.
\begin{equation}
\label{PartNabl}
\partial={\partial\over\partial s}
\quad,\qquad
\nabla={\partial\over\partial\rvec}
\end{equation}
The coefficients $C^{\mathbf{O}_1\mathbf{O}_2\cdots\mathbf{O}_p}_{\mathbf{O}}(\{ u_j^i{-}s^k_l\})$ of the MOPE are homogeneous functions (or rather distributions) of the relative distances $u_j^i{-}s^k_l$. Except when $s^k_l$ is a boundary point, they do not depend on $s^k_l$, but on the differences of the $u_j^i$.

For $k=1$ (local operators) the most relevant operators (with the highest canonical dimension) are
\begin{equation}
\label{1&dr2}
\mathbf{1}\qquad\text{and}
\qquad
\partial\rvec\partial\rvec=(\dot\rvec)^2
\end{equation}
with canonical dimension $0$ and $-1$ respectively (remember that the dimensions of $s$ and $\rvec$ are $1$ and $1/2$).
For $k=2$ (bilocal operator) the most relevant operator is the contact operator
\begin{equation}
\label{Xi2def}
\Xi(s_1,s_2)= \tilde \delta^d(\rvec(s_1)-\rvec(s_2))=\int d^d\rvec_0\,\tilde\delta^d(\rvec(s_1)-\rvec_0)\,\tilde\delta^d(\rvec(s_2)-\rvec_0)
\end{equation}
with canonical dimension $-d/2$.
For $k=3$ (tri-local operator) the leading operator is
\begin{eqnarray}
\label{Xi3def}
\Xi^{(3)}(s_1,s_2,s_3)&=& \tilde\delta^d(\rvec(s_1)-\rvec(s_2))\,\tilde \delta^d(\rvec(s_1)-\rvec(s_3))\\
&=&\int d^d\rvec_0\, \tilde\delta^d(\rvec(s_1)-\rvec_0)\,\tilde \delta^d(\rvec(s_2)-\rvec_0)\,\tilde \delta^d(\rvec(s_3)-\rvec_0)
\end{eqnarray}
with canonical dimension $-d$,  etc.

\begin{figure}[h]
\begin{center}
\begin{align*}
\label{ak2}
\raisebox{-1ex}{\includegraphics[width=1.7in]{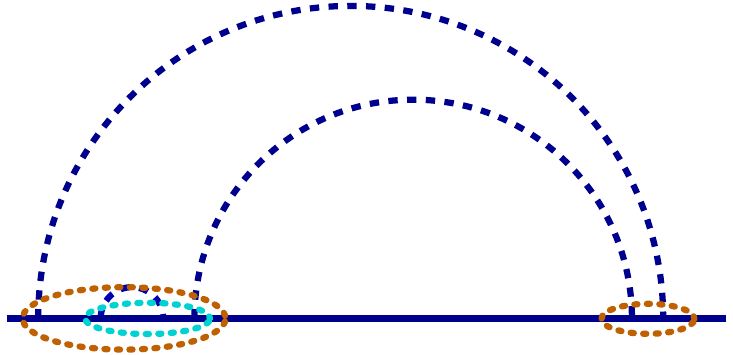}}
\ \to\ 
\raisebox{-1ex}{\includegraphics[width=1.7in]{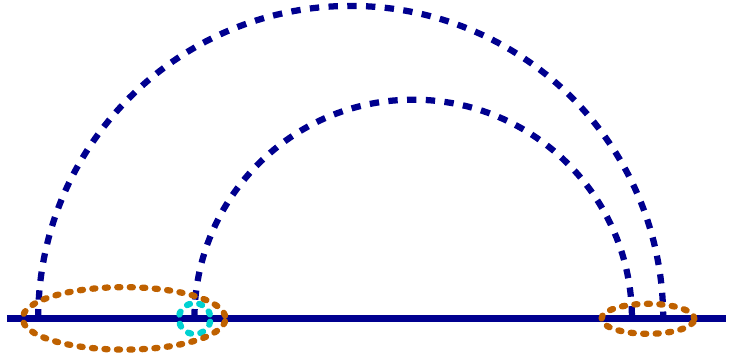}}
\ \to\ 
\raisebox{-1ex}{\includegraphics[width=1.7in]{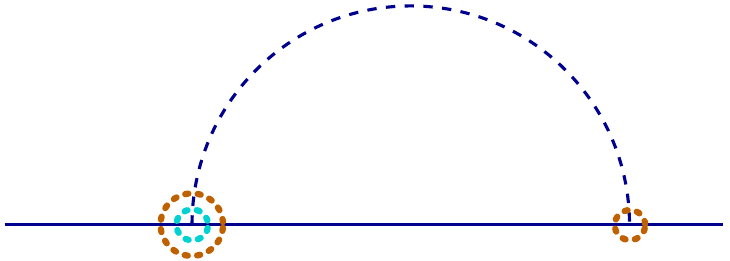}}
\end{align*}
\end{center}
\caption{Example of nested MOPE}
\label{NestMOPE}
\end{figure}

We give here the explicit form for the MOPE of the relevant operators at leading order, as calculated for instance in \cite{DDG4,WieseDavid1997,WieseHabil}
\begin{equation}
\label{Xi2I}
\Xi(u_1,u_2)\mathop{=}_{u_1\to u_2}|u_1-u_2|^{-d/2}\,\mathbf{1}(u)\,-\,\frac14\, |u_1-u_2|^{1-d/2}\,\dot\rvec(u)^2\,+\,\cdots
\end{equation}
with $u=(u_1+u_2)/2$.
\begin{equation}
\label{XiXi2Xi}
\Xi(u_1,v_1)\Xi(u_2,v_2)\ \mathop{=}_{{u_1\to u_2}\atop{v_1\to v_2}}\ \left(|u_1-u_2|+|v_1-v_2|\right)^{-d/2}\,\Xi(u,v)\ +\ \cdots
\end{equation}
with $u=(u_1+u_2)/2$, $v=(v_1+v_2)/2$.
\begin{equation}
\label{XiXi2Xi2}
\Xi(u,v)\,\Xi(u_1,u_2)\ \mathop{=}_{{u_1\to u}\atop{u_2\to u}}  \left| u_1-u_2 \right|^{-d/2}\,\Xi(u,v)
\ +\ \cdots
\end{equation}
\begin{equation}
\label{dr222dr2}
\dot\rvec^2(u_1)\,\dot\rvec^2(u_2)\ \mathop{=}_{u_1\to u_2} \ 8\ \delta(u_1-u_2)\ \dot\rvec^2(u)
\ +\ \cdots
\end{equation}
with $u=(u_1+u_2)/2$.
\begin{equation}
\label{Xidr22dr2}
\Xi(u_1,u_2)\ \dot\rvec^2(u)\ \mathop{=}_{{u_1\to u}\atop{u_2\to u}}  \  \left| u_1-u_2 \right|^{-d/2}\,\dot\rvec^2(u)
\end{equation}
And for the boundary operator
\begin{equation}
\label{XiB2B}
\Xi(u_1,u_2)\ \mathbf{1}_\ssst{\mathrm{b}}(0)\ \mathop{=}_{{u_1\to 0}\atop{u_2\to 0}}\ \left|u_1-u_2 \right|^{-d/2}\ \mathbf{1}_\ssst{\mathrm{b}}(0)\ +\ \cdots
\end{equation}
There is another potential term, but it vanishes because of the Neumann boundary condition for the open RW
\begin{equation}
\label{dr2B2B}
\dot\rvec^2(u)\ \mathbf{1}_\ssst{\mathrm{b}}(0)\ \mathop{=}_{{u\to 0}} \ 0
\ +\ \cdots
\end{equation}
Similar MOPEs can be written for the product of three or more operators. The coefficients $C$ of the MOPE  are homogeneous functions of the relative distance between the points involved. They have themselves a singular behavior when some of the points coalesce. These nested singularities are also given by a MOPE, with coefficients which have themselves nested sub-singularities, etc.  
This corresponds to the standard concept of nested sub-divergences (associated to Zimmerman forests) in the field theory. 
It is this nested MOPE structure which ensures the renormalisability of the self-avoiding polymer and membrane models studied in \cite{DDG1,DDG2,DDG3,DDG4,WieseDavid1995,DavidWiese1996,WieseDavid1997,WieseHabil,Wiese1997a,Wiese1997b,LeDoussalWiese1997,WieseLedoussal1998,Wiese1999,DavidWiese1998,Wiese1996a}.
An example is given on figure \ref{NestMOPE}.

\subsubsection{MOPE for $\gamma$ and $\tilde\gamma$}
The propagator for the auxiliary fields is a step function.
The OPE for the $\Upsilon$ operators is very simple and is exactly given by
\begin{equation}
\label{Uu2U}
\Upsilon_{ ab}(u)\Upsilon_{cd}(v)\   \mathop{=}_{u\to v} \ \theta(v-u)\,\delta_{bc}\ \Upsilon_{ad}(v)\ +\ \theta(u-v)\,\delta_{ad}\ \Upsilon_{cb}(v)
\end{equation}
Similarly, for the boundary operators of the open strand one has
\begin{equation}
\label{UB2B}
\tilde\gamma_a(0)\,\Upsilon_{bc}(v)\ \mathop{=}_{u\to 0_+}\ \delta_{ab}\,\tilde\gamma_c(0)
\qquad
\gamma_a(L)\,\Upsilon_{bc}(v)\ \mathop{=}_{u\to L_-}\ \delta_{ac}\,\gamma_b(L)
\end{equation}
This ensures that we keep track of the topology (planar structure) of the diagrams at short distance. It is clear that while (\ref{Uu2U}) and (\ref{UB2B}) eliminate some of the UV-divergences completely, they ``go along'' for the remaining ones, thus do not complicate the analysis. 

\subsubsection{MOPE for $\Phi$ in the planar limit}
\label{sss:mopephi}
Using (\ref{Phiagain})  it is easy to obtain the MOPE for the dressed contact operators $\Phi$ in each replica sector (we omit here the replica index $\alpha$ for simplicity of notation), and to take the planar large-$N$ limit ($N$ being the number of ``color" indices $a$ for the auxiliary fields $\gamma_a$ and $\tilde\gamma_a$) to obtain the MOPE for the planar pairing operators.
At leading order this MOPE   involves only $\Phi$,   and the local operators $\mathbf{1}$ and $\dot\rvec^2$, or rather their ``dressed versions"
\begin{equation}
\label{1dress}
\textrm{``}{\,\mathbf{1}}\textrm{''}(u)\ =\ {1\over N}\sum_a\tilde\gamma_a(u)\gamma_a(u)\ \mathbf{1}(u)
\ =\ {1\over N}\sum_a \Upsilon_{aa}(u) \ \mathbf{1}(u)
\end{equation}
\begin{equation}
\label{dr2dress}
\textrm{``}\,{\dot\rvec^2}\,\textrm{''}(u)\ =\ {1\over N}\sum_a\tilde\gamma_a(u)\gamma_a(u)\ \dot\rvec^2(u)
\ =\ {1\over N}\sum_a \Upsilon_{aa}(u) \ \dot\rvec^2(u)
\end{equation}
(we shall omit the dressing ``\ ''  for local operators in the rest of this section).

The first MOPE (\ref{Xi2I}) is unchanged 
\begin{equation}
\label{Phi21}
\Phi(u_1,u_2)\mathop{=}_{u_1\to u_2}|u_1-u_2|^{-d/2}\,{\mathbf{1}}(u)\,-\, \frac14 |u_1-u_2|^{1-d/2}\,{\dot\rvec^2}(u)\,+\,\cdots
\end{equation}
The two MOPE's involving two $\Phi$'s become
\begin{align}
\label{2Phi2Phi}
\Phi(u_1,v_1)\Phi(u_2,v_2)\ \mathop{=}_{{u_1,u_2\to u}\atop{v_1,v_2\to v}}\ &
\left[\theta(u_2-u_1)\theta(v_1-v_2)+\theta(u_1-u_2)\theta(v_2-v_1)\right]
\nonumber\\
&
\times \left(|u_1-u_2|+|v_1-v_2|\right)^{-d/2}\,\Phi(u,v)\ +\ \cdots
\end{align}
and
\begin{equation}
\label{2Phi2Phi2}
\Phi(u,v)\,\Phi(u_1,u_2)\ \mathop{=}_{{u_1\to u}\atop{u_2\to u}}  
\left[{\theta(u-u_2)+\theta(u_1-u)}\right]\ 
\left| u_1-u_2 \right|^{-d/2}\,\Phi(u,v)
\ +\ \cdots
\end{equation}
The remaining MOPE's (\ref{dr222dr2}-\ref{dr2B2B}) for the bulk and boundary operators are unchanged at this order.

One important remark is in order, about the MOPE's (\ref{XiXi2Xi2}) and (\ref{2Phi2Phi2}).
For the simple RW contact operator $\Xi$,  (\ref{XiXi2Xi2}) is just the product of the MOPE for $\Xi\to\mathbf{1}$, times an independent $\Xi$, which can be viewed as a ``spectator''.
As a consequence, the sum of the  diagrams 
\begin{equation}
\label{3diagrams}
\raisebox{-2ex}{\includegraphics[width=2in]{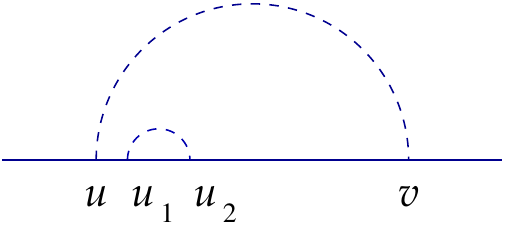}}
\ +\ 
\raisebox{-2ex}{\includegraphics[width=2in]{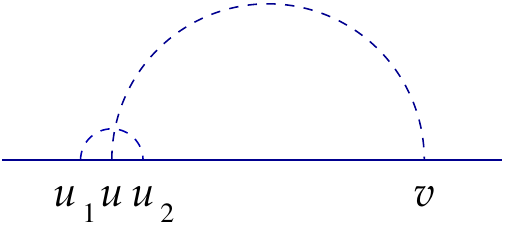}}
\ +\ 
\raisebox{-2ex}{\includegraphics[width=2in]{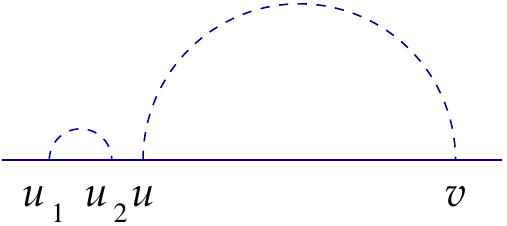}}\ ,
\end{equation}
which carry a potential UV divergence in the SAW model, are canceled by the counter-term for the leading divergence in (\ref{Phi21}), and no counterterm is associated to these diagrams  in the Edwards Model.
However the corresponding MOPE for the planar pairing model is different, since the term associated to a non planar configuration (the second diagram in (\ref{3diagrams}), is absent. 
As a consequence, and as we will see later, the MOPE (\ref{2Phi2Phi2}) gives a non-trivial UV singularity in the replica-interacting model, and requires an additional renormalization , which is absent in the SAW model.

\subsubsection{MOPE for $\Psi$}
It is clear that there is an analogous MOPE for the $\Psi_{\alpha\beta}$, since these operators are products of $\Phi$ operators on two independent replicas $\alpha$ and $\beta$. More generally, one can write a MOPE for any product of multilocal-multireplica operators of the form\ ,
\begin{equation}
\label{MOPErepl}
\mathbf{O}_{\alpha_1\alpha_2\cdots\alpha_p}(\{s^i_j\})=\mathbf{O}^1_{\alpha_1}(\{s^1_j\}) \, \mathbf{O}^2_{\alpha_2} (\{s^2_j\})\cdots \mathbf{O}^p_{\alpha_p} (\{s^p_j\})\
\end{equation}
where each $\mathbf{O}^i_{\alpha_i}$ is a multilocal operator $\mathbf{O}^i$ of the form (\ref{GenMOp}) for a single replica $\alpha_i$.
Of course, the MOPE is non-trivial only if some of the replica indices $\alpha$ are common to the different multilocal-multireplica operators.
We give here the MOPE for the $\Psi$ operators which are of interest for the renormalization  of the model at one loop.
For a single $\Psi$ we have
\begin{equation}
\label{Psi21}
\Psi_{\alpha\beta}(u_1,u_2)\mathop{=}_{u_1\to u_2}|u_1-u_2|^{-d}\,{\mathbf{1}}(u)\,-\,\frac 14\, |u_1-u_2|^{1-d}\,
\left[\dot\rvec^2_\alpha(u) + \dot\rvec^2_\beta(u)\right]\,+\,\cdots
\end{equation}
\begin{align*}
\raisebox{-3ex}{\includegraphics[width=2in]{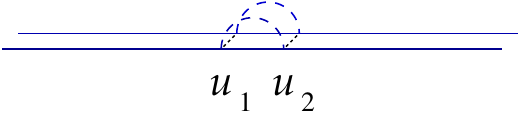} }\   \longrightarrow&\ \  
\raisebox{-2ex}{\includegraphics[width=2in]{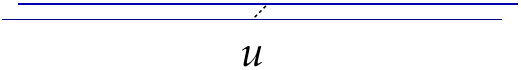}}\ +\   \\ \  &\ \\
    &  \hskip -6em +\ \raisebox{-2ex}{\includegraphics[width=2in]{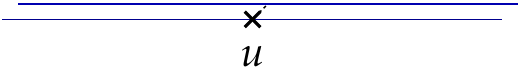}}\ +\  \raisebox{-2ex}{\includegraphics[width=2in]{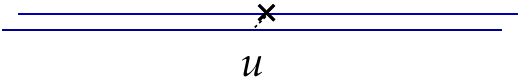}} \ +\ \cdots
\end{align*}
For two $\Psi$ we have
\begin{align}
\label{2Psi2Psi}
\Psi_{\alpha\beta}(u_1,v_1)\Psi_{\alpha\beta}(u_2,v_2)\ \mathop{=}_{{u_1,u_2\to u}\atop{v_1,v_2\to v}}\ &
\left[\theta(u_2-u_1)\theta(v_1-v_2)+\theta(u_1-u_2)\theta(v_2-v_1)\right]
\nonumber\\
&
\times \left(|u_1-u_2|+|v_1-v_2|\right)^{-d}\,\Psi_{\alpha\beta}(u,v)\ +\ \cdots
\end{align}
\begin{align*}
\raisebox{-3ex}{\includegraphics[width=2in]{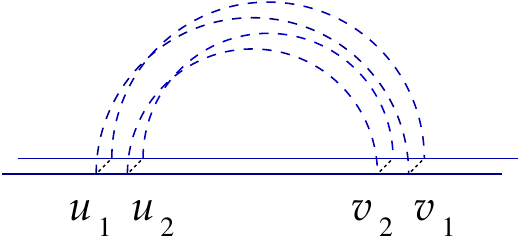} }\   \longrightarrow&\ \  
\raisebox{-2ex}{\includegraphics[width=2in]{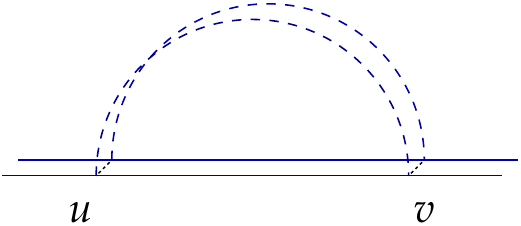}}\  +\ \cdots
\end{align*}
and
\begin{equation}
\label{2Psi2Psi2}
\Psi_{\alpha\beta}(u,v)\,\Psi_{\alpha\beta}(u_1,u_2)\ \mathop{=}_{{u_1\to u}\atop{u_2\to u}}  
\left[{\theta(u-u_2)+\theta(u_1-u)}\right]\ 
\left| u_1-u_2 \right|^{-d}\,\Psi_{\alpha\beta}(u,v)
\ +\ \cdots
\end{equation}
\begin{align*}
\label{ak6}
\raisebox{-3ex}{\includegraphics[width=2in]{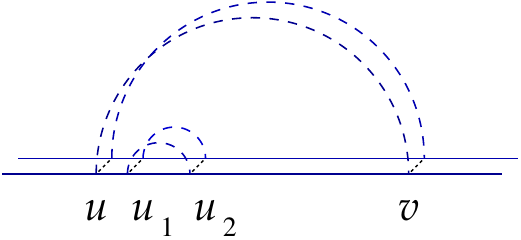} }\   \longrightarrow&\ \  
\raisebox{-2ex}{\includegraphics[width=2in]{MOPE-Psi-2}}\  +\ \cdots
\end{align*}
Let us note that the product (\ref{2Psi2Psi}) of two $\Psi$ which share only a single replica $\alpha$  generates a 3-replica operator of the form
\begin{equation}
\label{ak7}
\Psi_{\alpha\beta}\,\Psi_{\alpha\gamma}\to\Psi^{(3)}_{\alpha\beta\gamma}
\ ,\quad
\Psi^{(3)}_{\alpha\beta\gamma}(u,v)=\Phi_\alpha(u,v) \Phi_\beta(u,v) \Phi_\gamma(u,v) 
\end{equation}
which turns out to be irrelevant. But the product (\ref{2Psi2Psi2}) involving 3 replicas is relevant
\begin{equation}
\label{2Psi2Psi23}
\Psi_{\alpha\beta}(u,v)\,\Psi_{\alpha\gamma}(u_1,u_2)\ \mathop{=}_{{u_1\to u}\atop{u_2\to u}}  
\left[{\theta(u-u_2)+\theta(u_1-u)}\right]\ 
\left| u_1-u_2 \right|^{-d}\,\Psi_{\alpha\beta}(u,v)
\ +\ \cdots
\end{equation}
\begin{align*}
\label{ak8}
\raisebox{-3ex}{\includegraphics[width=2in]{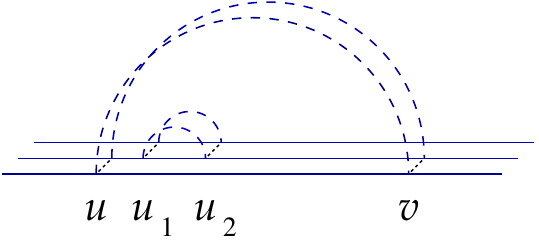} }\   \longrightarrow&\ \  
\raisebox{-2ex}{\includegraphics[width=2in]{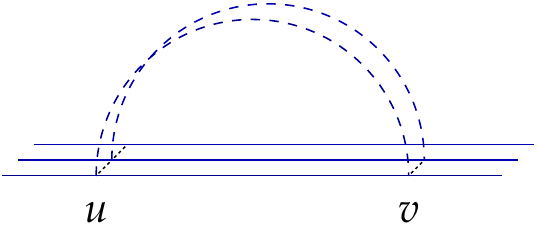}}\  +\ \cdots
\end{align*}
For the bulk operators and local boundary operators the MOPE is similar to (\ref{dr222dr2})-(\ref{dr2B2B}).
For the bulk we have
\begin{equation}
\label{dr222dr2a}
\dot\rvec_\alpha^2(u_1)\,\dot\rvec_\alpha^2(u_2)\ \mathop{=}_{u_1\to u_2} \ 8\,\delta(u_1-u_2)\ \dot\rvec_\alpha^2(u)
\ +\ \cdots
\end{equation}
\begin{align*}
\raisebox{-3ex}{\includegraphics[width=2in]{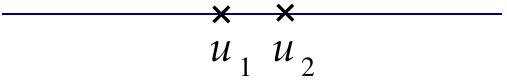} }\   \longrightarrow&\ \  
\raisebox{-2ex}{\includegraphics[width=2in]{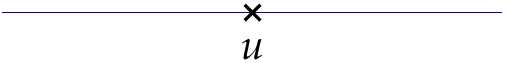}}\  +\ \cdots
\end{align*}
and \cite{WieseDavid1997}
\begin{eqnarray}
\label{Xidr22dr2}
\Psi_{\alpha\beta}(u_1,u_2)\left[ \dot\rvec^2_\alpha(u)+\dot\rvec^2_\beta(u)\right] &\displaystyle\mathop{=}_{{u_1\to u}\atop{u_2\to u}}& -4d \,\Theta(u_1<u<u_2)\,|u_1-u_2|^{-d-1} \,{\mathbf{1}}(u)\\
&& +\left[1+(d-1) \Theta(u_1<u<u_2)\right]  \left| u_1-u_2 \right|^{-d}\,\left[ \dot\rvec^2_\alpha(u)+\dot\rvec^2_\beta(u)\right] \nn
\end{eqnarray}
\begin{align*}
\label{ak10}
\raisebox{-3ex}{\includegraphics[width=2in]{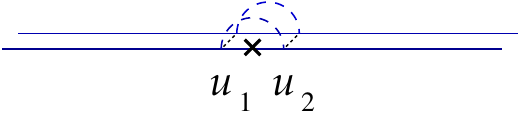} }\   \longrightarrow&\ \  
\raisebox{-2ex}{\includegraphics[width=2in]{MOPE-Psi-6}}\  +\ \cdots
\end{align*}
The four terms in the square bracket in (\ref{Xidr22dr2}) are obtained by not contracting  $ \dot\rvec^2_\alpha(u)$ (first term), contracting twice (second term), or once (third and fourth term). 

For the boundary operator we have
\begin{equation}
\label{XiB2B}
\Psi_{\alpha\beta}(u_1,u_2)\ \mathbf{1}_\ssst{\mathrm{b}}(0)\ \mathop{=}_{{u_1\to 0}\atop{u_2\to 0}}\ \left|u_1-u_2 \right|^{-d}\ \mathbf{1}_\ssst{\mathrm{b}}(0)\ +\ \cdots
\end{equation}
\begin{align*}
\label{ak11}
\raisebox{-3ex}{\includegraphics[width=2in]{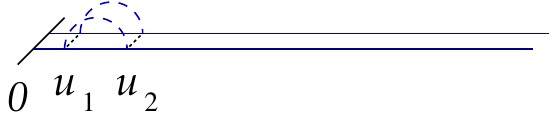} }\   \longrightarrow&\ \  
\raisebox{-3ex}{\includegraphics[width=2in]{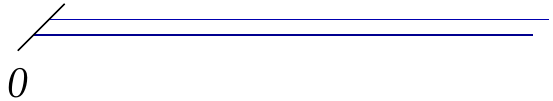}}\  +\ \cdots
\end{align*}
The trivial one is
\begin{equation}
\label{dr2B2B}
\dot\rvec^2_\alpha(u)\ \mathbf{1}_\ssst{\mathrm{b}}(0)\ \mathop{=}_{{u\to 0}} \ 0
\ +\ \cdots
\end{equation}
\begin{align*}
\label{ak12}
\raisebox{-2ex}{\includegraphics[width=2in]{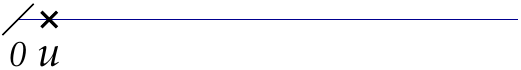} }\   \longrightarrow&\ \  
\raisebox{-2ex}{\includegraphics[width=2in]{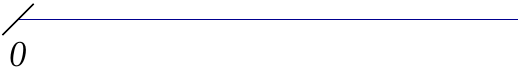}}\  +\ \cdots
\end{align*}
This MOPE structure for the multilocal-multireplica operators can be shown to hold at higher order, and to have a nested structure for its sub-singularities, as for the single replica case. It is this nested structure of the MOPE which ensures that the model is renormalizable.

\subsection{Renormalizability}
\label{ss:Renblty}
We can now analyze the short-distance singularities of the full interacting theory with $n$ replicas for $\varepsilon$ close to $0$. The analysis is very similar to the general proof of the renormalisability of the SAW and SAM models \cite{DDG1,DDG2,DDG3,DDG4,WieseHabil}, thus we can constrain ourselves to an outline.

We compute the partition function and the correlation functions of the model for open strands and generic $n$, as defined in section \ref{s:RW}, with $d$ or  $\varepsilon$ as an analytic regularization parameter. 
Using the standard rules of analytic (or dimensional)  regularisation, the integrals are calculated for $d\neq 2$, treating short-distance divergences via a finite-part prescription. Within this framework, UV divergences appear as poles in the complex $d$ plane. There are no long-distance IR divergences since the strand length $L$ is kept finite and acts as an IR regulator.

The UV poles at $\varepsilon=0$ are associated to the terms of the MOPE with the correct power counting. They give the superficial UV divergences of the theory. These divergences can be subtracted by adding counterterms to the action if they are proportional to the original operators in the action and if no new terms are generated under renormalization .

According to the MOPE  derived above, the dangerous UV terms come from the MOPE for products of operators of the following form:

Firstly
\begin{equation}
\label{mopepsi2one}
\prod_i\Psi_{\alpha\beta}(u_i,u'_i)
\prod_{j} \dot\rvec^2_{\alpha}(u''_j)
\prod_{k} \dot\rvec^2_{\beta}(u'''_k)
\mathop{\longrightarrow}_{
 \begin{matrix}
 \scriptscriptstyle{
 u_i,u'_i,u''_j,u'''_k\to u}
 \end{matrix}}
\mathbf{1}(u)+\ \left(\dot\rvec^2_{\alpha}(u)+\dot\rvec^2_{\beta}(u)\right)\ ,
\end{equation}
which describes the MOPE for a product of local and bilocal operators into a local operator.
The coefficient for the relevant identity operator $\mathbf{1}(u)$ gives dimension-full UV divergences, i.e.\ poles in the complex $\varepsilon$ plane for negative values of $\varepsilon=-2/p$, $p$ integer.
These do not give a pole at $\varepsilon=0$ and thus do not require a renormalization  in the minimal subtraction scheme.
Physically, this term gives a dimension-full UV divergence of the form $\Lambda^{-1} L$, where $\Lambda$ is a physical UV regulator (with dimension of a mass) and $L$ is the strand length. Since the strand length $L$ is kept fixed in our scheme, this  divergence is the same for all partition functions and does therefore cancel in correlation functions. 
The coefficient for the marginal operators $\dot\rvec^2_{\alpha}(u)$ and $\dot\rvec^2_{\beta}(u)$ gives a pole at $\varepsilon=0$ and will be subtracted by a wave-function counterterm proportional to the free action $\mathcal{S}_0$.

Secondly
\begin{equation}
\label{mopepsi2psi}
\prod_i\Psi_{\alpha\beta}(u_i,v_i)
\left[\prod_{j}\Psi_{..}(u'_j,u''_j)\prod_{k} \dot\rvec^2_.(u'''_k)\right]
\left[\prod_{l}\Psi_{..}(v'_l,v''_l)\prod_{m} \dot\rvec^2_.(v'''_m)\right]
\mathop{\longrightarrow}_{
 \begin{matrix}
 \scriptscriptstyle{
 u_i,u'_i,u''_j,u'''_j\to u}\\
\scriptscriptstyle{v_i,v'_k,v''_l,v'''_m\to v}
\end{matrix}}
\Psi_{\alpha\beta}(u,v)
\end{equation}
which describes the MOPE for a product of local and bilocal operators into a single bilocal operator. This gives also poles at $\varepsilon=0$ which are subtracted by a coupling-constant counterterm proportional to the interaction term $\mathcal{S}_{\mathrm{int}}$ in the action.

Thirdly, for the open strand, there is a divergence coming from the MOPE for the boundary operator $\mathbf{1}_b$, here located at $u=0$
\begin{equation}
\label{mopepsiboun}
\prod_i\Psi_{\alpha\beta}(u_i,u'_i)
\prod_{j} \dot\rvec^2_{\alpha}(u''_j)
\prod_{k} \dot\rvec^2_{\beta}(u'''_k)
\mathop{\longrightarrow}_{
 \begin{matrix}
 \scriptscriptstyle{
 u_i,u'_i,u''_j,u'''_k\to 0}
 \end{matrix}}
\mathbf{1}_b(0)
\end{equation}
This divergence is subtracted by a new counterterm proportional to the boundary operator $\mathbf{1}_b$. It is of course not present for closed strands.

There are no UV divergences associated to the auxiliary fields $\tilde\gamma^a_{\ \alpha}\dot\gamma^a_{\ \alpha}$.
This is not surprising since these auxiliary fields are introduced to organize the topological expansion of the perturbative expansion and to construct the planar limit for $N\to\infty$.

The analysis of the subdivergences and the fact that the UV poles can be recursively subtracted in perturbation theory is similar to the one of \cite{DDG1,DDG2,DDG3,DDG4,DavidWiese1996,WieseDavid1997,WieseHabil} and shall not be repeated here.

Finally let us stress that the arguments for perturbative renormalizability are valid for any value of $n$ (the number of replicas) and of $1/N$ (the parameter for the topological expansion of perturbation theory). 

We can now study the renormalized theory, and write the corresponding renormalization group equations.

\subsection{Bare and renormalized action and observables}
\label{ss:rsBsRO}
Renormalizability of the theory means that, when expressed in terms of the renormalized field $\rR$ and of the dimensionless renormalized coupling constant $\gR$, the theory defined with the renormalized action $\SR[\rR;\gR,\mu]$ for the open strand
 with
\begin{equation}
\label{Sren}
\SR[\rR]=\sum_{\alpha}\int_{0<s<L}\,
\left({\mathbb{Z}\over 4} \dot \rR^2 \ +\ \tilde\gamma\dot\gamma\right)
-\gR \mathbb{Z}_g\mu^{-\varepsilon}\sum_{\alpha<\beta}\iint_{0<u<v<L}\Psi^R_{\alpha\beta}(u,v)
+\sum_\alpha 2\mathbb{Z}_1
\end{equation}
with
\begin{equation}
\label{PsiR}
\Psi^R_{\alpha\beta}(u,v)=(4\pi)^{d/2}\delta^d(\rR_\alpha(u)-\rR_\alpha(u))\delta^d(\rR_\beta(u)-\rR_\beta(u))
\end{equation}
is UV finite when $\varepsilon=0$.
$\mathbb{Z}$ and $\mathbb{Z}_g$ are respectively the wave function counterterm and the coupling constant counterterms. They will correspond to a renormalization  of the operators $\dot\rvec^2$ and $\Psi$.
$\mu$ is the renormalization  mass scale, it has dimension $L^{-1}$.
$\mathbb{Z}_1$ is a new boundary counterterm, it is associated to a renormalization  of the boundary identity operator $\mathbf{1}_b$.
The counterterms are defined as a perturbative series in $\gR$, of the form 
\begin{align}
\label{ZCT}
    \mathbb{Z}& = 1+\sum_{k>0} \gR^k\,A_k(\varepsilon)  \\
    \mathbb{Z}_g& = 1+\sum_{k>0} \gR^k\,B_k(\varepsilon)  \label{ZgCT}\\  
    \mathbb{Z}_1& = \sum_{k>0} \gR^k\,C_k(\varepsilon) \ .\label{Z1CT}
\end{align}
The counterterm coefficients $A_k(\varepsilon)$, $B_k(\varepsilon)$ and  $C_k(\varepsilon)$ contain poles at $\varepsilon=0$ which cancel the UV poles of the bare theory. For the general case, these coefficients depend on $n$ and $N$. No renormalization is required for the auxiliary fields.

As usual, defining the bare field $\rB$ and the (dimensionfull) bare coupling constant $\gB$ as
\begin{equation}
\label{rBgBdef}
\rB=\rR\mathbb{Z}^{1/2}\quad,
\qquad
\gB=\gR \mathbb{Z}_g\mathbb{Z}^{d}\mu^{-\varepsilon}\quad,
\qquad\varepsilon=d-2\ ,
\end{equation}
we can rewrite the renormalized theory as a bare theory, defined by the bare action $\mathcal{S}_{\mathrm B}[\rB;\gB]$ given by
\begin{equation}
\label{Sbare}
S_{\mathrm B}[\rB;\gB]=\sum_{\alpha}\int_{0<s<L}\,
\left({1\over 4} \dot \rB^2\ +\ \tilde\gamma\dot\gamma\right)
-\gB\sum_{\alpha<\beta}\iint_{0<u<v<L}\Psi^{\mathrm B}_{\alpha\beta}(u,v) \ .
\end{equation}
Similarly, for an open strand, this amounts to renormalizing the boundary operator $\mathbf{1}_{\sss{b}}$ through
\begin{equation}
\label{ }
\mathbf{1}_{\sss{b}}^{\mathrm{B}}=\emath^{-\mathbb{Z}_1}\mathbf{1}_{\sss{b}}^{\mathrm{R}}
\end{equation}
When computing the counterterms and UV-finite observables for the renormalized theory from the UV divergent bare theory, one must be very careful with the field renormalization and the treatment of the translational zero modes.
To see this, let us first consider the partition function for the closed strand, $\widetilde{\mathcal{Z}}^{\mathrm{closed}}(g,L)$, as defined by (\ref{Zgclosed}).
Since the (infinite) contribution of the translational zero modes 
\begin{equation}
\label{intrdelq}
\int \prod_\alpha d^d\rvec_\alpha=\left[(2\pi)^d\delta^d(\qvec_\alpha=0)\right]^n 
\end{equation}
has already been factored out, this partition function has dimension
\begin{equation}
\label{dimZclosed}
\widetilde{\mathcal{Z}}^{\mathrm{closed}}(g;L)\simeq \left[\rvec\right]^{-d n}\ .
\end{equation}
Taking into account the renormalization of the field $\rvec$ given by  (\ref{rBgBdef}), the relation between the bare partition function, computed with the bare action $\mathcal{S}_{\mathrm B}[\rB;\gB]$, and the renormalized partition function is
\begin{equation}
\label{ZBcZRc}
\widetilde{\mathcal{Z}}_{\mathrm B}^{\mathrm{closed}}(\gB,L)\, \mathbb{Z}^{{nd\over 2}}=\widetilde{\mathcal{Z}}_{\mathrm R}^{\mathrm{closed}}(\gR,L)
\end{equation}
with $\mathbb{Z}$ the wave-function renormalization  factor.

The partition function for one open strand $\tcZo(\qvec;g,L)$ is defined by (\ref{Z1oq}), with $\qvec$ 
 the external momentum.
It has scaling dimension zero
\begin{equation}
\label{dimZ1o}
\tcZo(\qvec;g,L)\simeq  1\ .\end{equation}
Since the external momentum $\qvec$ is conjugate to the field $\rvec$, it is renormalized as
\begin{equation}
\label{qBqR}
\qB=\qR\,\mathbb{Z}^{-1/2}\ .
\end{equation}
The relation between the bare partition function for a single open strand (computed with the bare action $\SB[\rB;\gB]$, and expressed as a function of the bare coupling $\gB$ and the bare momentum $\qB$) and the UV finite renormalized partition function (computed with the renormalized action $\SR[\rR;\gR]$ and expressed as a function of the renormalized coupling $\gR$ and the renormalized momentum $\qR$) is\begin{equation}
\label{ZB1oZR1o}
\widetilde{\mathcal{Z}}^{(1)}_{\mathrm B}(\qB;\gB,L)\, \emath^{-2 n \mathbb{Z}_1}=\widetilde{\mathcal{Z}}^{(1)}_{\mathrm R}(\qR;\gR,L)\ .
\end{equation}
Note the additional multiplicative boundary renormalization  factor $\exp(-2 n \mathbb{Z}_1)$ which is of course not present for closed strands, and which is important for the calculations.

Finally we consider the partition function for two open strands in interaction $\mathcal{Z}^{(2)}(\qvec_1\cdots\qvec_4;g,L)$, defined by (\ref{Z2odef}).
We have seen that it must be decomposed into terms associated to the contribution of the so-called ``$Q$-sectors''  where $Q$ replicas interact amongst the $n$ available replicas, $\widetilde{\mathcal{Z}}^{(2,Q)}(\qvec_1\cdots\qvec_3;g,L)$, defined by
(\ref{Z2Qodef}).
Each term $\widetilde{\mathcal{Z}}^{(2,Q)}(\qvec_1\cdots\qvec_3;g,L)$ has a different scaling dimension
\begin{equation}
\label{Zt2Qscal}
\widetilde{\mathcal{Z}}^{(2,Q)}(\qvec_1\cdots\qvec_3;g,L)\simeq \left[\rvec\right]^{Qd}\ .
\end{equation}
Therefore the relation between the bare 2-strand partition function and the renormalized one is
\begin{equation}
\label{Z2Qscale}
\widetilde{\mathcal{Z}}_{\mathrm B}^{(2,Q)}(\qB\cdots;\gB,L)
\,\mathbb{Z}^{-{Q d\over 2}}\,\emath^{-4n\mathbb{Z}_1}
=
\widetilde{\mathcal{Z}}_{\mathrm R}^{(2,Q)}(\qR\cdots;\gR,L)\ .
\end{equation}
Note that the boundary renormalization factor is now $\exp(-4 n \mathbb{Z}_1)$ since we have two strands, hence four end points.

The argument is a bit sketchy, but can be made more rigorous by taking into account the functional measure $\mathcal{D}[\rvec]$ in the definition of the RW model, and using the fact that this functional measure is dimensionless for closed strands, but dimensionfull for open strands, with dimension
\begin{equation}
\label{Z2QRZ2QB}
\mathcal{D}[\rvec]^{\mathrm{closed}}\simeq 1
\quad,\qquad
\mathcal{D}[\rvec]^{\mathrm{(open)}}\simeq \left[\rvec\right]^{nd}\ .
\end{equation}
The difference can be viewed as the insertion of a contact operator $\tilde \delta^d(\rvec(0)-\rvec(L))$ for each replica in the open ensemble, which closes the strand and thus leads to the closed ensemble.

\subsection{Beta function and anomalous dimensions}
\label{MSanDim}
We use the minimal subtraction scheme (MS scheme), i.e.\ define the counterterms $\mathbb{Z}$, $\mathbb{Z}_g$ and $\mathbb{Z}_1$ in such a way that they contain only   poles at $\varepsilon=0$, but no finite part with analytic terms in $\varepsilon$.\begin{align}
\label{ZMS}
 \mathbb{Z}   =&  1+\gR  {a_1\over\varepsilon}+\gR ^2\left({a_2\over\varepsilon^2}+{a'_2\over\varepsilon}\right)+\mathcal{O}(\gR ^3)\\
 \label{ZgMS}
\mathbb{Z}_g   =&  1+\gR  {b_1\over\varepsilon}+\gR ^2\left({b_2\over\varepsilon^2}+{b'_2\over\varepsilon}\right)+\mathcal{O}(\gR ^3)\\
\label{Z1MS}
\mathbb{Z}_1   =&  \gR  {c_1\over\varepsilon}+\gR ^2\left({c_2\over\varepsilon^2}+{c'_2\over\varepsilon}\right)+\mathcal{O}(\gR ^3)\ .
\end{align}
The  $\beta_g$ function for the coupling is defined  as the variation of the renormalized coupling with the renormalization  scale $\mu$. Using (\ref{rBgBdef}) we get 
\begin{equation}
\label{betag}
\beta_g(\gR ):=\left.-\mu{d \gR \over d\mu}\right|_{\gB}=-\varepsilon\left[{1\over \gR }+{d\log\mathbb{Z}_g\over d\gR }+(2+\varepsilon){d\log\mathbb{Z}\over d\gR }\right]^{-1}
\quad.
\end{equation}
The function $\beta_g$ is  the Wilson flow function considered in L\"assig-Wiese \cite{LaessigWiese2005}, and our short paper \cite{DavidWiese2006}. 
Its derivative $\beta'_g$ gives the scaling dimension of $\gR$ (in the sense of Wilson, hence in units of mass $\mu$).
\begin{equation}
\label{ }
\Delta_g=\frac{\rmd }{\rmd \gR}\beta_g(\gR)
\quad.
\end{equation}
Since the renormalized coupling $\gR$ is the scaling field associated to the bilocal overlap operator $\PsiR(u,v)$, the scaling dimension of the operator $\Psi$  (in units of $\mu$) is 
\begin{equation}
\label{DeltaPsi}
\Delta_{\Psi}=2 - \frac{\rmd }{\rmd \gR}\beta_g(\gR)
\quad.
\end{equation}
In \cite{DavidWiese2006} we considered the dimension of  $\rvec$ (in units of length $L\sim 1/\mu$),  $\chi_\rvec(\gR)$, defined as 
\begin{equation}
\label{Deltar}
\chi_\rvec(\gR)= {1\over 2}\left[ 1+\beta_g(\gR ){d\log\mathbb{Z}(\gR)\over d \gR } \right]
\equiv {1\over 2}\left[1+\gamma(g_R)\right]
\end{equation}
with
\begin{equation}
\label{gammadef1}
\gamma(g_R):=\beta_g(\gR ){d\log\mathbb{Z}(\gR)\over d \gR } 
\quad.
\end{equation}
$\chi_\rvec(\gR)$ equals minus the scaling dimension of $\rvec$ considered as a local operator (hence in units of mass $\mu$)
\begin{equation}
\label{ }
\Delta_{\rvec}=-\chi_\rvec(\gR)
\quad.
\end{equation}
Finally  the scaling dimension for the boundary operator $\mathbf{1}_b$ is (still in units of mass $\mu$)
\begin{equation}
\label{m19}
\Delta_{\mathbf{1}_b}=\gamma_1(\gR)=\beta_g(\gR ){d\mathbb{Z}_1\over d \gR }
\quad.
\end{equation}
These formulas will be used to derive the anomalous dimensions and the renormalization -group flow of the RNA model from the two-loop calculation of the counterterms presented in the next section.

It is also possible to study the renormalization  of the contact operator $\Phi_\alpha(u,v)=\tilde\delta^d(\rvec_\alpha(u)-\rvec_\alpha(v))$ and to compute its scaling dimension $\Delta_\Phi$. This is done in section \ref{ss:RenPhi}.

\subsection{MS and $\protect\overline{\protect\mbox{MS}}$ subtraction schemes}
\label{ss:MSandMSbar}
We use two slightly different subtraction schemes. 
The first one stated in (\ref{ZMS})-(\ref{Z1MS}) is the standard minimal subtraction scheme (MS scheme), where the counterterms $\mathbb{Z}$, $\mathbb{Z}_g$ and $\mathbb{Z}_1$ are chosen such that they contain only   poles in $\varepsilon$, but no term analytic in $\varepsilon$.
It is useful to make the connection with the renormalization  in the L\"assig-Wiese formulation \cite{LaessigWiese2005}, see subsection \ref {ss:LSscheme}.

The second one is denoted $\overline{\mbox{MS}}$ and is defined as follows: Remarking that the relation between the bare  and the renormalized coupling constant is $\gB=\gR\mathbb{Z}_g\mathbb{Z}^{2+\varepsilon}\mu^{-\varepsilon}$, its total renormalization  factor is
\begin{equation}
\label{barZg}
\overline{\mathbb{Z}}_g=\mathbb{Z}_g\mathbb{Z}^{2+\varepsilon}\ .
\end{equation}
It contains also analytic terms of order $\varepsilon^n$, $n\ge 0$. If we choose  $\overline{\mathbb{Z}}_g$ to  have pure poles in $\varepsilon$, together with $\mathbb{Z}$ and $\mathbb{Z}_1$, we obtain the $\overline{\mbox{MS}}$ renormalization  scheme where
\begin{equation}
\label{gBgRMSbar}
\rB=\rR\mathbb{Z}^{1/2}\quad,\qquad \gB=\gR\overline{\mathbb{Z}}_g\mu^{-\varepsilon}\ .
\end{equation}
$\mathbb{Z}$ and $\mathbb{Z}_1$ are still of the form (\ref{ZMS}) and (\ref{Z1MS}) (but with a priori different coefficients $a'_2$ and $c'_2$), and 
\begin{equation}
\label{barZgMSbar}
\overline{\mathbb{Z}}_g   = 1+\gR  {\overline{b}_1\over\varepsilon}+\gR ^2\left({\overline{b}_2\over\varepsilon^2}+{\overline{b}'_2\over\varepsilon}\right)+\mathcal{O}(\gR ^3) \ .
\end{equation}
The two schemes MS and $\overline{\mbox{MS}}$ differ  by a finite redefinition of the renormalized coupling constant $\gR$.
In the $\overline{\mbox{MS}}$ scheme the definition (\ref{betag}) for the new beta function $\overline{\beta}_g$ reads
\begin{equation}
\label{betag'}
\overline{\beta}_g(\gR )=-\left.\mu{d \gR \over d\mu}\right|_{\gB}=-\varepsilon\left[{1\over \gR }+{d\log\overline{\mathbb{Z}}_g\over d\gR }\right]^{-1}\ .
\end{equation}

\subsection{Renormalisation and anomalous dimension of the operator $\Phi$}
\label{ss:RenPhi}
We have seen that the local field $\rvec$ and the overlap operators $\Psi_{\alpha\beta}$ are renormalized and get dimensions $\chi_\rvec$ and $\Delta_\Psi$.
Similarly the bilocal contact operator $\Phi_\alpha$, involving a single replica $\alpha$, must be renormalized and gets an anomalous dimension $\Delta_\Phi$. This dimension plays an important role in the analysis of the model, in particular in the ``locking'' mechanism presented in \cite{LaessigWiese2005} and discussed below.

\subsubsection{Contact operator $\Phi$ for a single strand}
\label{sss:Phi1}
To compute the anomalous dimension of $\Phi$, we must compute correlation functions involving this operator. We  first consider the contact operator between two points $u$ and $v$ for a single strand with replica index $\alpha$. It reads, from  (\ref{s3-6}),
\begin{equation}
\label{defPhi1}
\Phi_{\alpha}(u,v)=\tilde\delta^d(\rvec_\alpha(u)-\rvec_\alpha(v))\ .
\end{equation}
(For simplicity of notation, we omit the dressing by the auxiliary fields which is necessary to make it a planar operator).
It is depicted by a single arch as in Fig.~\ref{FigPhi}.
Its engineering dimension in units of $\mu$ is 
\begin{equation}
\label{DimPhi0}
\Delta_\Phi^{{0}}={d\over 2}=1+{\varepsilon\over 2}\ .
\end{equation}
This operator allows to define the ``height operator'' $h(s)$ for a single closed strand
\begin{equation}
\label{hsPhi}
h(s):=\iint_{u<s<v}\Phi(u,v)\ .
\end{equation}
Since the theory is renormalizable the operator  $\Phi$ can be renormalized as 
 \begin{equation}
\label{PhiRen}
\PhiR=\mathbb{Z}_\Phi\PhiB\ ,
\end{equation}
where $\PhiB$ is the bare operator (\ref{defPhi1})   involving the bare field $\rB$, and $\mathbb{Z}_\Phi$ is the renormalization  factor, computed in either the MS or $\overline{\mbox{MS}}$ scheme.  Its coefficients depend on $n$ (the replica number) and contains ultraviolet poles in $\varepsilon$.
The standard RG analysis shows that the scaling dimension of $\Phi$ in the renormalized theory is
\begin{equation}
\label{DimPhi1}
\rho(\gR)\equiv\Delta_\Phi(\gR)=\Delta^{ 0}_\Phi+\gamma_\Phi(\gR)\quad,\qquad
\gamma_\Phi(\gR)=\beta_g(\gR ){d\log\mathbb{Z}_\Phi\over d \gR }\ ,
\end{equation}
where $\beta_g(\gR)$ is the coupling-constant $\beta$-function given by (\ref{betag}).
Its value at the UV fixed point $\gR^*$, i.e.\ at the freezing transition, gives the contact exponent $\rho$ at the transition
\begin{equation}
\label{rhostarDPhi}
\rho^*=\Delta_\Phi(\gR^*)\ .
\end{equation}

\subsubsection{Two-strand contact operator $\Phi^{\scriptscriptstyle{1,2}}$}
\label{sss:Phi2}
It is more convenient to consider the contact operator between two different interacting (open) strands, in the same spirit as in 
subsection~\ref{s:3F}. 
The two strands are labeled $1$ and $2$ and described by the RW ${\rvec_1}_\alpha$ and ${\rvec_2}_\alpha$ respectively (and the associated auxiliary fields).
The inter-strand contact operator $\Phi_{\alpha}^{1,2}(u,v)$ is 
\begin{equation}
\label{DefPhi12}
\Phi_{\alpha}^{\scriptscriptstyle{1,2}}(u,v)=\tilde \delta^d\big({\rvec_1}_\alpha(u)-{\rvec_2}_\alpha(v)\big)\end{equation}
times the appropriate dressing to ensure planarity.
This operator is represented diagrammatically by a single line between the two strands.
\begin{figure}[t]
\begin{center}
\includegraphics[width=2in]{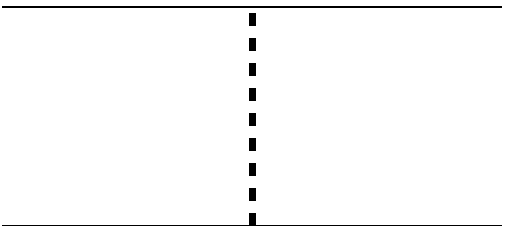}
\caption{Contact operator $\Phi_{\alpha}^{\scriptscriptstyle{1,2}}$ between two strands}
\label{f:Phi12}
\end{center}
\end{figure}
Repeating the analysis of the UV divergences and the renormalization  for this operator, it is easy to see that the singularities come from the very same terms in the MOPE as for $\Phi$, that is from the singularities when a bunch of  $\Psi$ coalesces on the endpoints of $\Phi$.
This implies that $\Phi_{\alpha}^{\scriptscriptstyle{1,2}}$ is  renormalized exactly in the same way as $\Phi$, namely that the renormalized operator is
\begin{equation}
\label{ Phi12Ren}
\Phi_\mathrm{R}^{\scriptscriptstyle{1,2}}=\mathbb{Z}_\Phi\Phi_{\mathrm{B}}^{\scriptscriptstyle{1,2}}\ ,
\end{equation}
with $\mathbb{Z}_\Phi$ containing the same counterterms as in (\ref{PhiRen}).

We consider in perturbation theory the following partition function
\begin{eqnarray}
\label{ZPhi2}
\widetilde{\mathcal{Z}}_\Phi^{(2)}&=&\largediagram{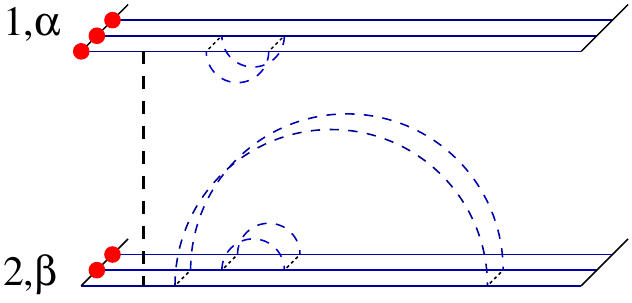}\nn\\
&=& \int\limits_{{\rvec_1}_\alpha(0)=\mathbf{0}}\mathcal{D}[{\rvec_1}_\alpha]
\int\limits_{{\rvec_2}_\beta(0)=\mathbf{0};\,\beta\neq\alpha_0}\mathcal{D}[{\rvec_2}_\beta]\ \mathrm{e}^{-(\mathcal{S}[{\mathbf{r}_1}_\alpha]+\mathcal{S}[{\mathbf{r}_2}_\beta])}\int_u\int_v\Phi^{\scriptscriptstyle{1,2}}_{\alpha_{\scriptscriptstyle{0}}}(u,v)
\end{eqnarray}
which contains all the diagrams which contribute to the renormalization  of $\Phi$.
This partition function is the partition function of 2 strands (replica bundles), with interactions between replicas inside each strand $1$ and $2$, no interactions between these two strands (this would result in higher replica operators, not $\Phi$), and constrained to be in contact via the replica $\alpha_0$.
We have omitted the auxiliary fields and the boundary operators $\mathbf{\Gamma}_1$ and $\mathbf{\Gamma}_2$ which are present in the definition of the partition functions of open strands (\ref{Z1oq1q2}).
The constraints on the path-integrals $\rvec_{1\alpha}(0)=0$ and $\rvec_{2\beta}(0)=0$ take care of the zero modes,  except for $\rvec_{2\alpha_0}$, whose zero-mode is fixed through $\rvec_{1\alpha_0}(0)=0$, and the operator $\Phi^{\scriptscriptstyle{1,2}}_{\alpha_{\scriptscriptstyle{0}}}(u,v)$. 

This partition function is renormalized as 
\begin{equation}
\label{m32}
\left.{  {   \widetilde{\mathcal{Z}}_\Phi  }^{(2)}  }\right._{\!\!\mathrm{R}}(\gR )
=\mathbb{Z}^{-d/2}\,\mathbb{Z}_\Phi\,\mathrm{e}^{-4n\mathbb{Z}_1}\,
\left.{  {   \widetilde{\mathcal{Z}}_\Phi  }^{(2)}  }\right._{\!\!\mathrm{B}}(\gB )
\end{equation}
(compare with the renormalization  of the ${\widetilde{\mathcal{Z}}}^{(2,Q)}$ partition functions in (\ref{Z2QRZ2QB}), where there are $Q$ replica interactions between the two strands).
The diagrams and the corresponding amplitudes are discussed in section~\ref{s:Diagr4Phi}, and the corresponding counterterm $\mathbb{Z}_\Phi$ and scaling dimension $\Delta_\Phi$ are calculated in section \ref{s:Diagr4Phi}.
However, we shall now derive directly a scaling relation for $\Delta_\Phi$ at the fixed point $g^*$, which  implies a scaling relation between the exponents $\rho^*$ (the contact exponent) and $\zeta^*=\zeta(g^*)$ (the full dimension of the $\rvec$ field).

\subsubsection{A scaling relation between $\Delta_\Phi$ and $\Delta_\rvec$}\label{sss:scalPhi}
We remark that $\widetilde{\mathcal{Z}}_\Phi^{(2)}$ can be written as
\begin{equation}
\label{m36}
\widetilde{\mathcal{Z}}_\Phi^{(2)}={1\over n^2}\left[-\left.{\partial\over\partial \qvec^2}\widetilde{\mathcal{Z}}^{(1)}(\qvec)\right\vert_{\qvec=0}\right]^2\ .
\end{equation}
Indeed each $\left.-{\partial\over\partial \qvec^2}\widetilde{\mathcal{Z}}^{(1)}(\qvec)\right|_{\qvec=0}$ gives the partition function for a single open-strand bundle with a point marked on one of the $n$ replicas. Taking the square gives the partition function for two replicas.  The factor of $1/n^2$ takes care of the fact that the replica $\alpha_0$ is not summed over in the definition of $\widetilde{\mathcal{Z}}_\Phi^{(2)}$.
Using the renormalization of the  partition function for one open strand given in Eqs.\ (\ref{qBqR}) and (\ref{ZB1oZR1o}),
 $\widetilde{\mathcal{Z}}_\Phi^{(2)}$ is UV finite for
\begin{equation}
\label{PhiCT'}
\overline{\mathbb{Z}}_\Phi=\mathbb{Z}^{-2+d/2}\ .
\end{equation}
$\overline{\mathbb{Z}}_\Phi$ is not subtracted minimally, but this is enough to compute the full dimension of $\Phi$ at the fixed point, using (\ref{DimPhi1})
\begin{eqnarray}
\label{m38a}
\bar\rho(\gR)=\overline{ \Delta}_\Phi(\gR )&=&1+{\varepsilon\over 2}-\left(1-{\varepsilon\over 2}\right)\beta_g(\gR ){\partial\log(\mathbb{Z})\over\partial \gR }\nn\\
&=&1+{\varepsilon\over 2}-\left(1-{\varepsilon\over 2}\right)\left(2\chi_\rvec(\gR)-1 \right)\nn\\
&=& 2 -\left(2- \varepsilon \right) \chi_\rvec(\gR)
\end{eqnarray}
with $\chi_\rvec(\gR)$ defined in (\ref{Deltar}).
At a  fixed point $g^*$ we get the scaling relation
\begin{equation}
\label{m39a}
\rho^* + \left(2 -{\varepsilon}\right)\chi^* = 2\ .
\end{equation}
Using the scaling relation $\rho^* + \zeta^* =2 $ from Eq.~(\ref{rho+zeta=2}), 
this implies the relation between the roughness exponent  $\zeta^*$ of the height $h$ and the dimension of $\rvec$
\begin{equation}
\label{m40b}
\zeta^*=
(2-\epsilon)\chi^* \equiv (4-d)\chi^*
\end{equation}
For $d=3$ ($\varepsilon=1$)
\begin{equation}
\label{m41a}
\zeta^*=\chi^*\ ,
\end{equation}
as expected,
since $h$ is the height variable, equivalent to the radial coordinate $h=|\rvec|$ for the random walk.

\subsection{Relation with the length-renormalization framework of L\"assig-Wiese}
\label{ss:LSscheme}
In the original one-loop renormalization-group calculation of \cite{LaessigWiese2005} by L\"assig and Wiese the random RNA folding model is not formulated in terms of a dressed random walk model with auxiliary fields $\rvec$. The model is defined as a perturbative expansion in the disorder strength $g$ in terms of the overlap operator $\Psi$. The renormalization s at 1-loop order are extracted from the  behavior of the free energy at 2-loop order.
Since there is no $\rvec$ field in this formulation, there is no renormalization  factor $\mathbb{Z}$ for  this field. 
Also in their scheme two renormalization s are  required: the first one
is a renormalization  of the coupling constant $g$ as here, but the second one is a renormalization  of the length $L$ of the  RNA strand.

In order to compare our results with those of \cite{LaessigWiese2005}, and to check that our general argument for the renormalizability of the model implies the consistency of the renormalization scheme of \cite{LaessigWiese2005}, we must understand the relation between these two renormalization  schemes.
This relation is in fact easy to establish. 
\medskip

Let us start from our renormalization  scheme. The renormalized action is of the form (we omit the auxiliary fields, the associated indices $a,b,\ldots$, the replica indices $\alpha,\beta,\ldots$, the boundary counterterms, and consider only closed strands)
\begin{equation}
\label{ak13}
\SR[\rR]= {\mathbb{Z}(\gR)\over 4}\int_{0<u<L} \drR(u)^2 - \gR\mathbb{Z}_g(\gR)\mu^{-\varepsilon}\iint_{0<u<v<L}\Psi_{\mathrm{R}}(u,v)
\end{equation}
This renormalized action can be rewritten as a bare action $\SB$ in terms of the bare field $\rB$ and the bare coupling constant $\gB$
\begin{equation}
\label{ak14}
\SR[\rR]=\SB[\rB]={1\over 4} \int_{0<u<L}\drB(u)^2-\gB \iint_{0<u<v<L}\Psi_{\mathrm{B}}(u,v)
\end{equation}
provided that we renormalize the field $\rvec$ and the coupling constant $g$ as
\begin{equation}
\label{ak15}
\rB(u)=\mathbb{Z}(\gR)^{1/2}\rR(u)
\quad,\qquad
\gB=\gR\mathbb{Z}_g(\gR) \mathbb{Z}(\gR)^d \mu^{-\varepsilon}\ .
\end{equation}
The overlap operator $\Psi$ is defined as usual as
\begin{equation}
\label{ak16}
\Psi_{\alpha\beta}(u,v)=\delta^d(\rvec_\alpha(u)-\rvec_\alpha(v))\delta^d(\rvec_\beta(u)-\rvec_\beta(v))\ ,
\end{equation}
so that we have
\begin{equation}
\label{ak17}
\Psi_{\mathrm{B}}(u,v)=\mathbb{Z}(\gR)^{-d}\,\Psi_{\mathrm{R}}(u,v)\ .
\end{equation}
We now reconstruct the L\"assig-Wiese renormalization  scheme from our scheme. 
We can consider the strand length $L$ and the base coordinates  $u\in[0,L]$ as renormalized length parameters
\begin{equation}
\label{ak18}
L=\LR\quad,\qquad u=\uR\in[0,\LR]
\end{equation}
and in our scheme the length $L$ is \emph{not} renormalized, i.e.\ the length in the bare action $\SB$ is the same as the length in the renormalized action $\SR$.

Instead let us construct another renormalization  scheme (hereafter indicated by ``tilde" superscripts)  where the coupling constant $g$ and the length $L$ are renormalized, but where the field $\rvec$ is \emph{not} renormalized.
We define the base length parameter $\tuB$ and the renormalized length $\tLB$ as
\begin{equation}
\label{ak19}
\tLB=\tZ\LR\quad,\qquad\tuB=\tZ\,\uR\ \in\ [0,\tLB]
\end{equation}
with $\tZ$ a renormalization  length factor (to be determined).
Similarly the bare coupling constant $\tgB$ is defined as
\begin{equation}
\label{ak20}
\tgB=\gR\,\tZg\,\mu^{-\varepsilon}
\end{equation}
with $\tZg$ the coupling constant renormalization  factor. The $\rvec$ field is not renormalized, the bare field is simply
\begin{equation}
\label{ak21}
\trB(\tuB)=\rR(\uR)
\end{equation}
so that
\begin{equation}
\label{ak22}
\tilde\Psi_\mathrm{B}(\tuB,\tvB)=\Psi_\mathrm{R}(\uR,\vR)\ .
\end{equation}
It is a simple exercise to check that if we choose as counterterms
\begin{equation}
\label{tildeCT}
\tZ=\mathbb{Z}(\gR)^{-1}\quad,\qquad\tZg=\mathbb{Z}_g(\gR)\,\mathbb{Z}^{2}(\gR)
\end{equation}
we have renormalized the theory in the L\"assig-Wiese scheme, since we
can rewrite our renormalized theory, given by our renormalized action
$\SR$ in eq.\ (\ref{ak13}), which gives an UV finite theory, as a bare LW theory, with the bare action $\tSB$ of the form
\begin{equation}
\label{ak23}
\SR[\rR]=\tSB[\trB]={1\over 4}\int_{0<\tuB<\tLB}{\dot{\tilde\rvec}}_{\scriptscriptstyle{\mathrm{B}}}^2-\tgB\iint_{0<\tuB<\tvB<\tLB}\tilde\Psi_\mathrm{B}(\tuB,\tvB)
\end{equation}
We have thus proven that our renormalization  scheme implies that the L\"assig-Wiese renormalization  scheme exists and is consistent at all orders.
The reader interested in the precise correspondence between the one-loop calculations of 
 \cite{LaessigWiese2005} and our formalism may check that in \cite{LaessigWiese2005} the calculations are done in a minimal subtraction scheme where the renormalization  scale is set by the renormalized strand length. This corresponds in our formalism to set $\mu=\LR^{-1}$ and to chose a minimal subtraction scheme where the counterterms for the coupling constant $\tilde g$, that is $\tilde{\mathbb{Z}}_g$, contain pure poles in $\epsilon$. 
The relation between the $\beta$-functions is obtained as follows. Insert $\mu^{-1}=\LR =\tLB/\tZ= \tLB \mathbb{Z}(\gR)$ in (\ref{ak20}), and use (\ref{tildeCT}) to obtain
\begin{equation}\label{216}
\tgB = \gR \mathbb{Z}_g \mathbb{Z}^{2+\epsilon} (\gR) \tLB^\epsilon \ .
\end{equation}
Comparing  (\ref{ak15}) and (\ref{216}), we conclude that 
\begin{equation}
 \tLB \frac{d}{d \tLB} \gR \Big|_{\tgB} \equiv -\mu \frac{d }{d \mu} \gR \Big|_{\gB}
\end{equation}
The l.h.s.\ is the $\beta$-function denoted $\beta(u)$ (with $u=\gR$) in LW \cite{LaessigWiese2005}, whereas the r.h.s.\ is our $\beta$-function $\beta_g(\gR)$ in eq.\ (\ref{betag}).
 According to (\ref{tildeCT}) this corresponds to our minimal subtraction scheme MS. 
The precise dictionary between our notations and results and those of
\cite{LaessigWiese2005} are given in table \ref{ak24}.
\begin{table}[h]
\begin{tabular}{|c|c|c|}
\hline
   this work &  \cite{LaessigWiese2005} by LW &  denomination\\
   \hline 
   $\mu$ & $\mu=L_R^{-1}$ &renormalization  scale\\
   \hline
  $\gR$ &  $u$ & dimensionless renormalized coupling \\
  $\tgB$ & $g_0$ &   bare coupling \\
  $\tLB$ & $L_0$ & bare length \\
  $L=\LR$ & $L$ & renormalized length \\
  $\tilde {\mathbb{Z}}^{-1}=\mathbb{Z}$ & $Z_L$ & length renormalization  factor \\
  $\tilde {\mathbb{Z}}_g^{-1}=\mathbb{Z}_g^{-1}\mathbb{Z}^{-2}$ & $Z_g$ & coupling renormalization  factor \\
{ $\beta_g(\gR)$} &{$\beta(u)$} &  beta function \\
\hline
\end{tabular}
\caption{Correspondence between our renormalization scheme and the LW
scheme \cite{LaessigWiese2005}.}\label{ak24}
\end{table}

 \clean
\newcommand{\LL}{L}
\newcommand{\TT}{\tau}

\section{2-loop diagrams}
\label{s:2loops}
\nopagebreak[4]
\subsection{Presentation of the calculation, fixed-length scheme}
\label{ss:2Lpresent}
This section is devoted to the explicit calculation of the 2-loop diagrams which are required to evaluate the renormalization  factors and the renormalization-group functions at second order.
The main calculation will be performed for open strands in the fixed-length framework which is presented and discussed in the previous sections.
For completeness and comparison with the calculations of
\cite{LaessigWiese2005} we shall also discuss the so-called
grand-canonical scheme where an integration over the length $L$ of the
strands is performed. This scheme has the advantages that
it allows for an independent check, and that some integrals are much
simpler. 

We first recall the definition of the amplitudes in the fixed-length framework.
The partition function for $p$ strands,
${\mathcal{Z}}^{(p)}(\qvec\cdots; \gB,L)$ is decomposed into a sum of
the contribution of the $Q$ sectors where $Q$ is the number of
interacting replicas between two strands, according to
(\ref{Z2odef}). Each sector gives  the partition function
$\widetilde{\mathcal{Z}}^{(p,Q)}(\qvec\cdots; \gB,L)$ which is a
function of the $2p-1$ independent external momenta $\qvec$. The
$2n-Q$ delta functions for the conservation of external momenta are
already extracted and taken into account in (\ref{Z2odef}).
To extract the necessary renormalization  factors, in fact we only need
two cases: First, 
$p=1$, and  $Q=0$, with the partition function  denoted
$\widetilde{\mathcal{Z}}^{(1)}(\qvec; \gB,L)$. And second 
$p=2,Q=2$.  Note that $p=2,Q=0$ reduces to the previous one since
$\widetilde{\mathcal{Z}}^{(2,0)}=\widetilde{\mathcal{Z}}^{(1)}\widetilde{\mathcal{Z}}^{(1)}$
and the case $p=2,Q=1$ is trivial since
$\widetilde{\mathcal{Z}}^{(2,1)}=0$. 

\subsubsection{One-strand diagrams}
\label{sss:1stdiag}
The perturbative expansion of the one-strand partition function $\widetilde{\mathcal{Z}}^{(1)}(\qvec; g,L)$ is written as a sum 
over one-strand planar diagrams $\mathcal{E}_i$
\begin{equation}
\label{Z1exp}
\widetilde{\mathcal{Z}}^{(1)}(\qvec; \gB,L)=\sum_{k=0}^\infty \gB^k\sum_{\begin{matrix}
     \scriptstyle{ \mathrm{diagrams}}    \\       \scriptstyle{\mathcal{E}_i}
\end{matrix}} c(\mathcal{E}_i)\, \tilde E_i(\qvec;L)\ .
\end{equation}
$c(\mathcal{E}_i)$ is the combinatorial  factor of the planar diagram $\mathcal{E}_i$. This combinatorial (or symmetry) factor depends on $n$ (the number of replicas).
$\tilde E_i(\qvec;L)$ is the amplitude of the one-strand diagram $\mathcal{E}_i$. It is given by an integral over the position of the internal arch points of amplitudes of the general form given by (\ref{OpDDE0}).
For a diagram of order $k$ (with $k$ double arches), by homogeneity this amplitude is of the form
\begin{equation}
\label{E1bar}
\tilde E_i(\qvec;L)=L^{k(2-2\theta)} E_i(\qvec L^{1/2})\,\emath^{-n \qvec^2 L}\ ,
\end{equation}
where for historical reasons and to make an easy comparison with \cite{LaessigWiese2005} we keep the notation
\begin{equation}
\label{thetadef}
\theta={d\over 2}=1+{\varepsilon\over 2}\ .
\end{equation}
Each dimensionless amplitude  $E_i(\qvec')$ depends only on the dimensionless momentum $\qvec'=\qvec L^{1/2}$ and of course on $\theta$. 
The UV divergences appear in our dimensional-regularisation scheme as poles at $\theta=1$ (that is $\varepsilon=0$). Power counting shows that the primary divergences (needed to compute the RG functions) are contained in the first two coefficients of the amplitudes $E_i(\qvec)$ in a small-momentum expansion in $\qvec$ around $\qvec=0$. 
We  denote these two coefficients $A_i$ and $B_i$
\begin{equation}
\label{AbarAB}
E_i(\qvec)=A_i+\qvec^2B_i+\mathcal{O}(\qvec^4)\ ,
\end{equation}
(they depend on $\theta$).
Finally, since we can express everything in terms of dimensionless quantities, we set 
\begin{equation}
\label{Lto1}
L=1
\end{equation}
in the rest of the calculations in the fixed-length scheme. In this case, the calculation of the amplitudes $E_i(\qvec)$ and of the coefficients $A_i$ and $B_i$ for each diagram $\mathcal{E}_i$ involves integrals over the relative successive distances between arch end points on the strand of the form
\begin{equation}
\label{muint1}
\int_0^1 \rmd u_1\int_0^{u_1}\rmd u_2\cdots\int_0^{u_{2k-1}}\rmd u_{2k} \cdots\end{equation}
Setting $u_0=0$, and  $u_{2k+1}=1$ and denoting $v_j=u_{j+1}-u_{j}$ the relative distance between ordered end points  we denote such integrals by
\begin{equation}
\label{muint2}
\int_{v_1,v_1,\cdots v_{2k}}\ =\ \int_0^1\cdots\int_0^1 \rmd v_0\cdots
\end{equation}

\subsubsection{Two-strand diagrams}
\label{sss:2stdiag}
The perturbative expansion for the two-strand partition function $\widetilde{\mathcal{Z}}^{( 2,2)}(\qvec\cdots,g,L)$ is written as a sum over the connected planar two-strand diagrams $\mathcal{C}_i$ with  $Q=2$ replica interactions between the two strands,
\begin{equation}
\label{Z22exp}
\widetilde{\mathcal{Z}}^{(2,2)}(\qvec\cdots; \gB,L)=\sum_{k=0}^\infty \gB^k\sum_{\begin{matrix}
     \scriptstyle{ \mathrm{diagrams}}    \\\hskip -1ex
      \scriptstyle{\mathcal{C}_i}
\end{matrix}} c(\mathcal{C}_i)\, \tilde C_i(\qvec\cdots;L)\ .
\end{equation}
The two-strand diagrams are now denoted by $\mathcal{C}_i$, and $c(\mathcal{C}_i)$ is the combinatorial factor for the diagram 
$\mathcal{C}_i$. It depends on the number of replicas $n$.
$\tilde C_i(\qvec\cdots;L)$ denotes the amplitude of the diagram $\mathcal{C}_i$. It depends on the three independent external momenta 
$\qvec=\{\qvec_1,\qvec_2,\qvec_3\}$, on the length of the two strands (which are taken to be equal for simplicity $L_1=L_2=L$), and on $\theta$ (or $d$ or $\varepsilon$).
Following the discussion of the previous sections, the amplitudes $\tilde C_i(\{ \qvec\};L)$ have a scaling form similar to (\ref{E1bar}), 
\begin{equation}
\label{C1bar}
\tilde C_i(\{\qvec\};L)=L^{2\theta+2k(1-\theta)}\  C_i(\{\qvec L^{1/2}\})\ .
\end{equation}
The UV divergences which are needed to compute the RG functions are
contained in the first coefficient of the 2-strand amplitude at small
momenta  $\qvec$. Similarly we can deal with dimensionless amplitudes
by setting as in (\ref{Lto1})  $L=1$. 
Thus we denote the dimensionless zero-momenta 2-strand amplitude coefficient for the diagram $\mathcal{C}_i$ as
\begin{equation}
\label{Ciq0def}
C_i=\left.C_i(\{\qvec\};L)\right|_{\qvec=0,L=1}\ .
\end{equation}
These coefficients still depend on $\theta$ (or $d$ or $\varepsilon$) and have UV poles at $\theta=1$.

\subsection{Grand-canonical scheme}
\label{sss:GCanS}
We had defined above the canonical, i.e.\ fixed-length,    partition
function for open strands, 
$\widetilde{\mathcal{Z}}^{(p,Q)}(\qvec\cdots; \gB,L)$. 
We now define the partition function for open strands in the
grand-canonical scheme. This scheme leads to several important simplifications
in the calculations, and therefore offers a valuable check.

\subsubsection{1-strand diagrams}\label{nk1}
We start with the 1-strand diagrams. The partition function is  defined as
\begin{equation}\label{nk2}
\widetilde{\mathcal{Z}}^{(1)}_{\tau }(\qvec\cdots; \gB,\tau ) :=
\int_{0}^{\infty}\rmd L\, \rme^{-L \tau} \widetilde{\mathcal{Z}}^{(1)}(\qvec\cdots; \gB,L)\ .
\end{equation}
As an example, the free (non-interacting) single-strand partition
function for the $n$-times replicated open RNA is  
\begin{equation}\label{nk3}
\widetilde {\mathcal{Z}}^{(1)} (\qvec,L) = \rme^{-n \qvec^2 L}\ .
\end{equation}
Thus, its Laplace transform reads 
\begin{equation}\label{nk4}
\widetilde {\mathcal{Z}}^{(1)}_{\tau } (\qvec,\tau) =
\int_{0}^{\infty}\rmd L\, \rme^{-L \tau}  {\mathcal{Z}}^{(1)}
(\qvec,L)  = \frac{1}{\tau +n \qvec^2}\ .
\end{equation}
According to (\ref{E1bar}), the contribution of diagram $i$ to the 1-strand partition function at order
$g^k$ is of the form 
\begin{equation}\label{nk5}
\tilde E_i(\qvec;L)=L^{-k\epsilon } E_i(\qvec L^{1/2})\,\emath^{-n
\qvec^2 L} = L^{-k\epsilon }\,\emath^{-n
\qvec^2 L}  \left[ A_i +B_{i} \qvec^2 L + \dots \right]\ .
\end{equation} 
Its Laplace transform reads 
\begin{eqnarray}\label{nk6}
\tilde E_{i}^{\tau} (q,\tau) &=& \int_{0}^{\infty}\rmd L\, \rme^{-L
\tau} \tilde E_i(\qvec;L) \nonumber \\
&=& \frac{1}{\tau +n \qvec^2} \left[ \Gamma
(1-k \epsilon) (\tau +n \qvec^2)^{k \epsilon} A_{i} +\qvec^2 \Gamma
(2-k \epsilon) (\tau +n \qvec^2)^{k \epsilon-1} B_{i}+\dots \right]\ .
\end{eqnarray}
Thus we can write for diagram $i$ at order $\gB^k$
\begin{equation}\label{nk7}
\tilde E_{i}^{\tau} (q,\tau) \gB^k = \frac{\left[\gB (\tau +n
\qvec^2) \Gamma (1-\epsilon)\right]^{k} }{\tau +n \qvec^2}
\left[\frac{\Gamma (1-k\epsilon)}{\Gamma (1-\epsilon)^{k}} 
A_{i}+ \frac{\Gamma (2-k\epsilon)}{\Gamma (1-\epsilon)^{k}}
\frac{B_{i} \qvec^2}{\tau +n\qvec^{2}}  \right]\ .
\end{equation}
To simplify calculations, in the next subsections we will evaluate all
amplitudes\emph{ with the combination $\tau
+n\qvec^{2}=1$ kept fixed}. We define $A_{i}^{\tau}$ and $B_{i}^{\tau}$ to be the coefficients in the
square brackets of the above expansion, i.e.
\begin{equation}\label{nk8}
A_{i}^{\tau} : = \frac{\Gamma (1-k\epsilon)}{\Gamma (1-\epsilon)^{k}}
A_{i}\ , \qquad B_{i}^{\tau} : = \frac{\Gamma (2-k\epsilon)}{\Gamma (1-\epsilon)^{k}}
B_{i} \ .
\end{equation}  
These are the coefficients which we calculate in the next
subsection. 
Note that the division by $1/\Gamma (1-\epsilon)^{k}$ was introduced to
have the simplest possible coefficients, especially no derivatives of
the $\Gamma$-function. This is achieved by replacing $\gB$ by
$\gB /\Gamma(1-\epsilon)$ in the bare action (\ref{Sbare}). As we will see from the explicit calculation,
this replacement leads to 
{\em identical} RG-functions up to 2-loop order in the both schemes.  

\subsubsection{2-strand diagrams}\label{nk9}
We now discuss 2-strand diagrams, for which the differences are more
important. In principle, there are two possible definitions: One could
either take 2 strands of the same length $L$, and then perform the
Laplace transform, or one can take a length $L_{1}$ for the first and
a length $L_{2}$  for the second strand, and then perform a
Laplace-transform for each of the two strands separately, using the
same chemical potential. The first method will give essentially the
same results as in the fixed-length framework, and will not allow for
any simplifications. On the other hand, the latter method, which we
shall adopt in the following, leads to important simplifications, see
diagram $C_{17}$, eq.\ (\ref{a37}) to (\ref{i26}), and eq.\
(\ref{i27}). 
Formally, if the contribution of diagram $i$ to $\widetilde Z^{(2,2)}
(\qvec\dots ;\gB;L_{1}, L_{2})$ (with  length $L_{1}$ for strand
1 and  length $L_{2}$ for strand 2) is for the only needed
contribution at $\qvec=0$ given by $C_{i} (L_{1},L_{2}) g_{0}^{k}$, then
we define 
\begin{equation}\label{nk10}
C_{i}^{\tau} := \frac{1}{\Gamma (1-\epsilon)^{k}}\int_{0}^{\infty}\rmd L_{1} \int_{0}^{\infty}\rmd
L_{2}\, \rme^{-\tau (L_{1}+L_{2})} C_{i} (L_{1},L_{2}) \ .
\end{equation}
These are the coefficients which we will calculate in the next
subsection. Of course, we will not calculate them first for fixed
lengths, and then do the Laplace transform, but take advantage of the
fact that after Laplace-transform many integrals factorize quite
nicely. 
Note that we have again consistently divided each diagram by $\Gamma
(1-\epsilon)^{k}$, as was done for the single-strand diagrams, due to
the replacement of $\gB$ by $\gB/\Gamma(1-\epsilon)$ in the bare
action (\ref{Sbare}). The coefficients $C_{i}^{\tau}$ will be
calculated in the next subsection.

To keep notations compact, we will indicate the corresponding  integrals in the fixed length scheme and in the grand canonical scheme respectively with superscript $L$ (for the length-cutoff) and with superscript $\tau$ (for the chemical potential). Equivalent integrals are denoted by ``$\equiv$'' to distinguish them from the same diagram, so e.g.\ $C_{1}\equiv C_{2}$ means that diagrams $C_{1}$ and $C_{2}$ have the same integrals in {\em both schemes}, although their combinatorial  factors, denoted  $c(C_{1})$ and $c(C_{2})$ may be different. Writing $C_{1}^{L }\equiv C_{2}^{L}$ denotes that this is true only for the integral in the fixed length scheme.

\subsection{Explicit diagrams: single-strand}
\label{ss:1stdiag}

\subsubsection{Order $k=1$:}
\label{sss:1sk1}
\noindent There is only one diagram.
\begin{align}
\mathcal{E}_2&=\largediagram{R1-2} \label{ee2a}\\
& \hskip 3em r \hskip 2.5em s \hskip 2.5em t \nonumber\\
  c({\mathcal{E}_2})&=    1 \times    n(n-1)/2 
  \qquad,\qquad{E}_2(\qvec)=\int_{r,s,t}  \rme^{2s\,\qvec^2} \,s^{-2\theta} \label{ee2b}\\
  A_2^{\LL}&=\int_{r+s+t=1}\,s^{-2\theta}=\int_0^1 \rmd s\,s^{-2\theta}(1-s)=
{\Gamma(1-2\theta)\Gamma(2)\over\Gamma(3-2\theta)}={1\over (2-2\theta)(1-2\theta)} \nn\\
&=\frac1\varepsilon-1+\mathcal{O}(\varepsilon)\label{ee2c}\\
B_2^{\LL}&= \int_{r+s+t} \,2\,s^{1-2\theta}=2{\Gamma(2-2\theta)\Gamma(2)\over\Gamma(4-2\theta)}={2\over (3-2\theta)(2-2\theta)}=-\frac2{\varepsilon}-2+\mathcal{O}(\varepsilon) \label{ee2d}\\
A_{2}^{\TT}&=\frac{\Gamma (1-2\theta)}{\Gamma (1-\varepsilon)}= 
\frac{1}{\varepsilon } -1 +{\cal O}(\epsilon) \label{ee2e}\\
B_{2}^{\TT}&= -\frac{2}{\varepsilon }\label{ee2f}
\end{align}

\subsubsection{Order $k=2$}
\label{sss:1sk2}
\noindent There are 5 inequivalent diagrams.
\begin{align}
\mathcal{E}_3&=\largediagram{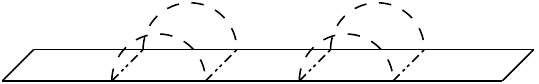}\label{ee3a}\\
& \hskip 2.5em r \hskip 1.2em s \hskip 1.2 em t \hskip 1.2 em u \hskip 1.2em v \nonumber\\
c(\mathcal{E}_{3})&=   1   \times  n(n-1)/2    
\qquad,\qquad {E}_2(\qvec)=  \int_{r,s,t,u,v} \rme^{2(s+u)Zq^2} \,s^{-2\theta}\,u^{-2\theta}\label{ee3b}\\ 
A_3^{\LL}&=  \int_{r,s,t,u,v} s^{-2\theta}u^{-2\theta}=\frac12 \int_{s+u<1}s^{-2\theta}u^{-2\theta}(1-s-u)^2\nn\\
&\nn =\frac12\int_0^1 dx\, x^{1-4\theta}(1-x)^2 \int_{s+u=1}s^{-2\theta}\,u^{-2\theta}\\
&={1\over 2}{\Gamma(2-4\theta)\Gamma(3)\over\Gamma(5-4\theta)}{\Gamma(1-2\theta)\Gamma(1-2\theta)\over\Gamma(2-4\theta)}
={\Gamma(1-2\theta)^2\over\Gamma(5-4\theta)}= \frac{1}{\varepsilon^2}-\frac{2}{\varepsilon}+\left(3-\frac{\pi^2}6\right)+\mathcal{O}(\varepsilon)\label{ee3e}\\
\nn
B_3^{\LL}&=  \int_{r+s+t+u+v=1}2(s+u) s^{-2\theta}u^{-2\theta}=\int_0^1 dx\, x^{2-4\theta}(1-x)^2\int_{s+u=1}s^{-2\theta}\,u^{-2\theta}\\
&= {\Gamma(3-4\theta)\Gamma(3)\over\Gamma(6-4\theta)}{\Gamma(1-2\theta)\Gamma(1-2\theta)\over\Gamma(2-4\theta)}\equiv {2(2-4\theta)\over(5-4\theta)}A_3\nn\\
&=-\frac{4}{\varepsilon^{2}}-\frac{4}{\varepsilon}+\left(-12+\frac{2\pi^2}3\right)+\mathcal{O}(\varepsilon)\label{ee3f}\\
A_{3}^{\TT} &\equiv ( A_{2}^{\tau})^{2}\label{ee3g}\\
B_{3}^{\TT} &\equiv B_{2}^{\tau}A_{2}^{\tau}\label{ee3h}\\
\ \nn\\
\mathcal{E}_{4}&=\largediagram{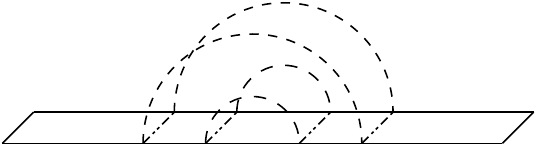} 
\quad,\qquad E_4(q)=\int_{r,s,t,u,v}  \rme^{2(s+t+u)Zq^2} \,(s+u)^{-2\theta}\,t^{-2\theta} \   \label{ee4a} \\  
 c(\mathcal{E}_{4})&= 1\times      n(n-1)/2 
\label{ee4a1}\\\nn
A_4^{L}&=\int_{r+s+t+u+v=1} (s+u)^{-2\theta}t^{-2\theta}=\int_{s+u+t<1}(1-(s+u+t))(s+u)^{-2\theta}t^{-2\theta}\\
&=\int_{x+t<1}x(1-(x+t))x^{-2\theta}t^{-2\theta}=\int_0^1dy\,y^{2-4\theta}(1-y) \int_{x+t=1} x^{1-2\theta}t^{-2\theta}\nn \\
&={\Gamma(3-4\theta)\Gamma(2)\over\Gamma(5-4\theta)}{\Gamma(2-2\theta)\Gamma(1-2\theta)\over\Gamma(3-4\theta)}={\Gamma(2-2\theta)\Gamma(1-2\theta)\over\Gamma(5-4\theta)} \nn \\
&=-\frac1{\varepsilon^{2}}+\frac1{\varepsilon}+\left(\frac{\pi^2}6-1\right)+\mathcal{O}(\varepsilon)\label{ee4b}\\ 
A_{4}^{\tau}&= \frac{\Gamma (-\varepsilon)\Gamma (-1-\varepsilon)}{\Gamma
(1-\varepsilon)^{2}} =\frac{-1}{\varepsilon^{2}+\varepsilon^{3}}=
-\frac{1}{\varepsilon ^2}+\frac{1 }{\varepsilon } -1
+{\cal O}(\epsilon) \label{ee4c}\\
B_4^{L}&= \int_{r+s+t+u+v=1} 2(s+t+u)(s+u)^{-2\theta}t^{-2\theta} =2\int_0^1dy\,y^{3-4\theta}(1-y) \int_{x+t=1} x^{1-2\theta}t^{-2\theta}\label{ee4d}\\
&=2{\Gamma(4-4\theta)\Gamma(2)\over\Gamma(6-4\theta)}{\Gamma(2-2\theta)\Gamma(1-2\theta)\over\Gamma(3-4\theta)}\equiv{2(3-4\theta)\over(5-4\theta)}A_4^{L}\nn\\
&=\frac2{\varepsilon^2}+\frac6{\varepsilon}+\left(10-\frac{\pi^2}3\right)+\mathcal{O}(\varepsilon)\label{ee4e}\\ 
B_{4}^{\tau} &=  \frac{2+4\varepsilon}{\varepsilon^{2}+\varepsilon^{3}} = 
\frac{2}{\varepsilon ^2}+\frac{2}{\varepsilon }-2+{\cal O}(\epsilon)\label{ee4f}\\
\ \nn\\
\label{a5}
\mathcal{E}_{5}&=\largediagram{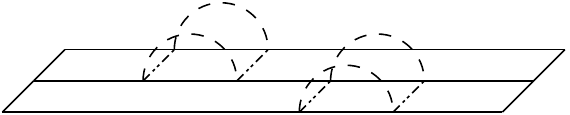} \ ,\quad E_5(q)=\int_{r,s,t,u,v}  \rme^{2(s+u)Zq^2} \,s^{-2\theta}\,u^{-2\theta}\equiv E_3(q)\\
  c(\mathcal{E}_{5})&=    1 \times      n(n-1)(n-2)  \ , \qquad 
A_5\equiv A_3\ ,
\qquad
B_5\equiv B_3  \label{ee5b}\\
\ \nn\\
\label{a6}
\mathcal{E}_{6}&=\largediagram{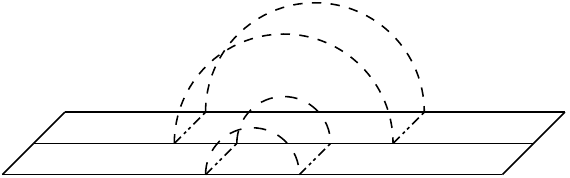} 
\\
\label{a7}
c(\mathcal{E}_6)&=  1 \times     n(n-1)(n-2)
\ ,\ E_6(q)=\int_{r,s,t,u,v} \!\!\!\!\!\!\!\!\!\rme^{(2s+3t+2u)Zq^2} (s+t+u)^{-\theta}\,(s+u)^{-\theta}\,t^{-2\theta}\\
A_6^{L}&=\int_{r+s+t+u+v=1}(s+t+u)^{-\theta}\,(s+u)^{-\theta}\,t^{-2\theta}\nn\\
&=\int_{s+t+u<1}(1-(s+t+u))(s+t+u)^{-\theta}(s+u)^{-\theta}t^{-2\theta} \nn\\
&=\int_0^1dy\,y^{2-4\theta}(1-y) \int_{x+t=1} x^{1-\theta}t^{-2\theta}\nn\\
&={\Gamma(3-4\theta)\Gamma(2)\over\Gamma(5-4\theta)}{\Gamma(2-\theta)\Gamma(1-2\theta)\over\Gamma(3-3\theta)}= -\frac{3}{4\varepsilon}+\frac{9}{4}+\mathcal{O}(\varepsilon)\label{a7a}\\
\label{a8}
A_{6}^{\tau}&= \frac{\Gamma (-2 \varepsilon -1) \Gamma (-\varepsilon -1)
\Gamma \left(1-\frac{\varepsilon }{2}\right)}{\Gamma (1-\varepsilon )^2
\Gamma \left(-\frac{3 \varepsilon }{2}\right)} = -\frac{3}{4 \varepsilon
}+\frac{9}{4}+{\mathcal{O}}\left(\varepsilon ^1\right) 
\\
\label{a9}
B_6^{L}&=\int_{r+s+t+u+v=1}(2s+3t+2u)(s+t+u)^{-\theta}\,(s+u)^{-\theta}\,t^{-2\theta}\nn \\
&=\int_0^1dy\,y^{3-4\theta}(1-y) \int_{x+t=1} (2x+3t)x^{1-\theta}t^{-2\theta}\nn\\
&={\Gamma(4-4\theta)\Gamma(2)\over\Gamma(6-4\theta)}{\Gamma(2-\theta)\Gamma(1-2\theta)\over\Gamma(3-3\theta)}{(2(2-\theta)+3(1-2\theta))\over (3-3\theta)}\nn\\
&={(3-4\theta)(7-8\theta)\over(5-4\theta)(3-3\theta)}A_6\nn\\
&=\frac{1}{2 \varepsilon^{2}}+\frac{5}{2\varepsilon}+\left(\frac72-\frac{\pi^2}{24}\right)+\mathcal{O}(\varepsilon)\\ 
\label{a10}
B_{6}^{\tau}&= -\frac{2^{\varepsilon } \sqrt{\pi } (4 \varepsilon +1) \Gamma
(-2 \varepsilon )}{\varepsilon (\varepsilon +1) \Gamma \left(1-\frac{3 \varepsilon
}{2}\right) \Gamma \left(\frac{1}{2}-\frac{\varepsilon }{2}\right)}
=\frac{1}{2 \varepsilon ^2}+\frac{3}{2 \varepsilon }+\frac{1}{24}
\left(-36+\pi ^2\right)+{\cal O}(\epsilon)\\
\ \nn\\
\label{a11}
\mathcal{E}_{7}&= \largediagram{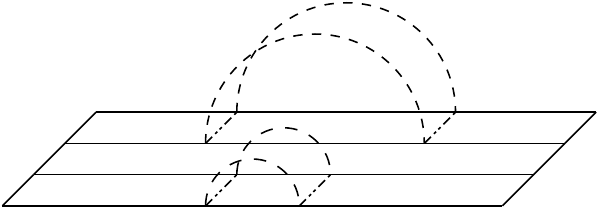}\ ,\ \ E_7(q)= \left[\int_{r+s+t=L}\hskip-2em \rme^{2sZq^2} \,s^{-2\theta}\right]^2 =A_2(q)^2\\
  c( \mathcal{E}_{7}) & = 1 \times    n(n-1)(n-2)(n-3)/8   \ , \qquad
   A_7\equiv A_2^2\ ,\qquad B_7\equiv 2A_2B_2\qquad \qquad\label{a11a}
\end{align}

\subsection{Explict diagrams: 2 strands}
\label{ss:2stdiag}
We now consider 2-strand partition functions. For simplicity both strands have the same length $L$, or the same chemical potential $\tau$, respectively.
Note that in order to transform from one ensemble to the other, one would need to know quantities for different lengths, or different chemical potentials, which are more difficult to calculate. This can  be seen as follows: If one inverse-Laplace transforms results for fixed chemical potential, one obtains results where the sum of the lengths on both RNA-strands equals $L$, and not their individual lengths. 

We recall that there are subtleties for defining and computing multi-strand correlation functions, which are discussed above.
We only need the $Q=2$ diagrams, i.e. the connected 2-strand diagrams with inter-strand interactions between $Q=2$ replicas. It is also sufficient to evaluate the amplitudes at zero external momenta $\qvec=0$.
At two loops we need the diagrams with $k\le 3$ interactions.

\subsubsection{Order $k=1$}
\label{sss:2sk1}
\begin{align}\label{a13}
\mathcal{C}_{1}&=\largediagram{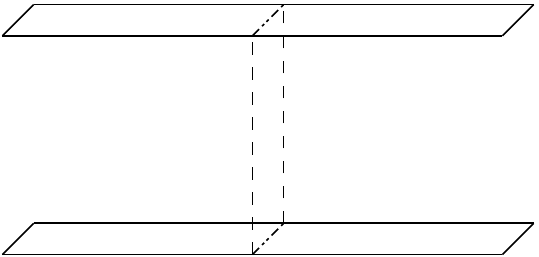} \ ,\quad C_1=1\ ,\qquad c(\mathcal{C}_1)=    1 \times {n(n-1)\over 2}
\end{align}

\subsubsection{Order $k=2$}
\label{sss:2sk2}
\begin{align}\label{a14}
\mathcal{ C}_{2}&=\largediagram{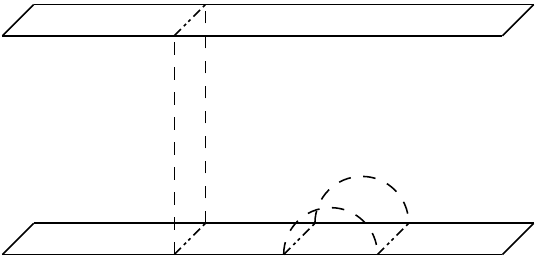}
\ ,\ \ C_2=\int_{r,s,t,u}t^{-2\theta}
 \ ,\qquad c( \mathcal{C}_2) =     4 \times {n(n-1)\over 2}\\
C_2^{L}&=\int_{r+s+t+u=1}t^{-2\theta}\equiv\int_{r+v=1}v^{2-2\theta}A_2^{L}={\Gamma(3-2\theta)\Gamma(1)\over\Gamma(4-2\theta)}A_2^{L}\nn\\
&={1\over(3-2\theta)(2-2\theta)(1-2\theta)}=\varepsilon^{-1}+\mathcal{O}(\varepsilon)\label{a14a}\\
C_{2}^{\tau} &\equiv A_{2}^{\tau}\label{a14b}
\\ \ \nn\\
\label{a15}
\mathcal{C}_{3}&=\largediagram{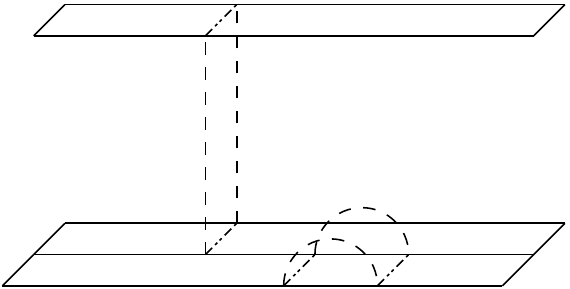}
\ ,\ \ C_3= \int_{r,s,t,u}t^{-2\theta}\equiv C_{2} \ ,\qquad c(\mathcal{C}_{3})= 4 \times     n(n-1)(n-2)  
\\ \ \nn\\
\label{a16}
\mathcal{C}_{4}&=\largediagram{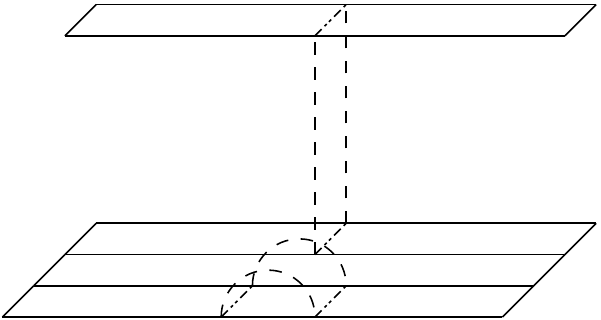}
\ ,\qquad c(\mathcal{C}_{4})=  2 \times    {n(n-1)(n-2)(n-3)\over 4} 
\\
C_4^{L}&\equiv C_1^{L}A_2^{L}  \label{a16a}\\ 
\label{a17}
C_{4}^{\tau} & \equiv \frac{\Gamma (4-2\theta )}{\Gamma
(1-\varepsilon)} A_{2}^{L} = \frac{1-\varepsilon}{\varepsilon (1+\varepsilon)}
= \frac{1}{\varepsilon} -2 + 2 \varepsilon +{\cal O}(\varepsilon^{2})
\\ \ \nn\\
\mathcal{C}_{5}&=\largediagram{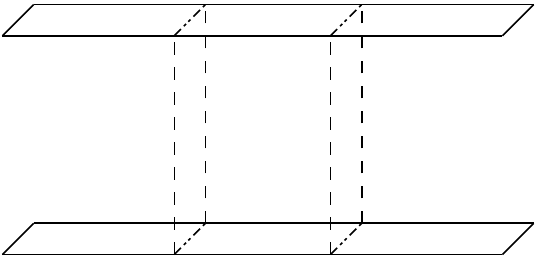}
\ ,\ \ C_5=\int_{r,s,t}\int_{u,v,w}(s+v)^{-2\theta}\nn
\label{a17a}
\\
c(\mathcal{C}_{5})&=   1 \times     {n(n-1)\over 2}      \\ \label{a18}
\nn
C_5^{L}&=\int_{r+s+t=1}\int_{u+v+w=1}(s+v)^{-2\theta}=\int_0^1\rmd s\int_0^1dv\,(1-s)(1-v)(s+v)^{-2\theta}\\
&={8\cdot
4^{-\theta}+2\theta-5\over(2-\theta)(1-\theta)(3-2\theta)(1-2\theta)}{1\over
2}=-\frac{1}\varepsilon+\left[\frac 1 2-2\ln(2)\right]+\mathcal{O}(\varepsilon)\\ \label{a19}
C_{5}^{\tau} &= -\frac{1}{\varepsilon} 
\end{align}

\subsubsection{Order $k=3$}
\label{sss:2sk3}
\begin{align}
\mathcal{C}_{8}&=\largediagram{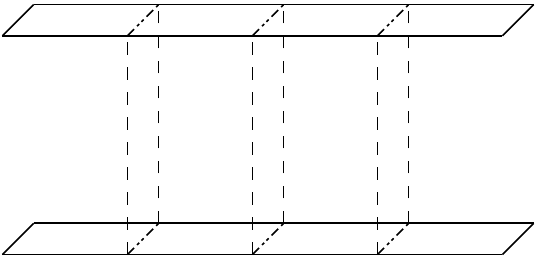} 
\ ,\ \ C_8=\int_{r,s,t,u}\int_{v,w,x,y}(s+w)^{-2\theta}(t+x)^{-2\theta} 
\label{a20a}\\
c(C_{8}) &=    1 \times    {n(n-1)\over 2}  
\label{a20b} \\ 
\nn
C_8^{L}&=\int_{r+s+t+u=1}\int_{v+w+x+y=1}(s+w)^{-2\theta}(t+x)^{-2\theta}\\ 
\nn
&=\int_{s+t<1}\int_{w+x<1}(1-(s+t))(1-(w+x))(s+w)^{-2\theta}(t+x)^{-2\theta}\\ 
&={1\over 4 {(1-\theta)}^2{(1-2\theta)}^2}\int_0^1ds\int_0^1dx\,K(s,x)K(x,s)\\ 
K(s,x)&=s^{2-2\theta}-{(1+s-x)}^{2-2\theta}+2(1-x)(1-\theta)s^{1-2\theta} \\ \label{a22}
C_8^{L} &=\frac{1}{\varepsilon^2}+\frac{-1+\ln(16)}\varepsilon+\left[2-\frac{\pi^2}3-4\ln(2)^2+8\ln(2)\right]+\mathcal{O}(\varepsilon) \\
C_{8}^{\tau}& \equiv ( C_{5}^{\tau})^{2}
\label{a20f}\\
\ \nn\\
\mathcal{C}_{9}&=\largediagram{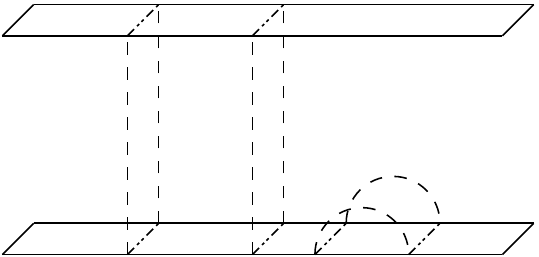} 
\ ,\ \ C_{9}=\int_{r,s,t}\int_{u,v,w,x,y} (s+v)^{-2\theta}x^{-2\theta}
\label{a22a}\\ 
\label{a23}
c(\mathcal{C}_{9})&=  4 \times    {n(n-1)\over 2}  \\ \label{a24}
C_9^{L}&\equiv D_9^{L} A_2^{L} \\ \nn
D_9^{L}&=\int_{r+s+t}\int_{u+v+w}(s+v)^{-2\theta}w^{2-2\theta}=\int_0^1ds \,(1-s)\int_{v+w<1}(s+v)^{-2\theta}w^{2-2\theta}\\
&={1\over (1-2\theta)(2-2\theta)(3-2\theta)}\int_0^1dv\, \left[(1+v)^{2-2\theta}-v^{1-2\theta}(2-2\theta+v)\right](1-v)^{3-2\theta}
\end{align}
{The last integral is  a hypergeometric function, finite when $\theta\to 1$. Therefore}
\begin{align}
D_{9}^{L}&={1\over (1-2\theta)(2-2\theta)(3-2\theta)}\left[{-(7-4\theta)\Gamma(3-2\theta)\Gamma(4-2\theta)\over\Gamma(7-4\theta)}+{_2F_1(1,2\theta-2,5-2\theta,-1)\over(4-2\theta)}\right]\nn \\ \label{a26}
 &=-\frac{1}\varepsilon-\frac12+\ln(4)+\mathcal{O}(\varepsilon)\\ \label{a27}
C_9^{L}&=-\frac{1}{\varepsilon^2}+\left(\frac{1}{2}-2\ln(2)\right)\frac{1}{\varepsilon}+\left(-2+\frac{\pi^2}{12}+2\ln(2)^2-2\ln(2)\right)+\mathcal{O}(\varepsilon)\\ \label{a28}
C_{9}^{\tau} &\equiv C_{5}^{\tau} A_{2}^{\tau }
\end{align}

\begin{align}
\mathcal{C}_{10}&=\largediagram{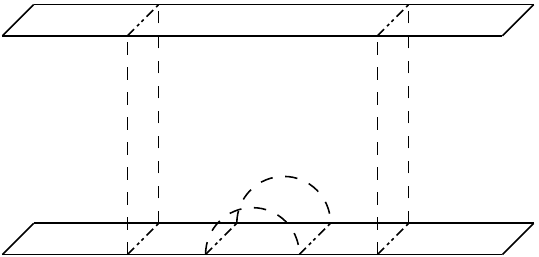}
\ ,\ \ C_{10}=\int_{r,s,t}\int_{u,v,w,x,y} (s+v+x)^{-2\theta}w^{-2\theta}
\label{a28a}\\ 
\label{a29}
c(\mathcal{C}_{10}) &=  2 \times    {n(n-1)\over 2}    \\ 
\nn
C_{10}^{L}&= \int_{r+s+t=1}\int_{u+v+w+x+y=1} (s+v+x)^{-2\theta}w^{-2\theta}\\ \nn
&=\int_0^1ds\,(1-s)\int_{z+w<1}z(1-z-w)(s+z)^{-2\theta}w^{-2\theta}\nn \\ 
&=\int_{0}^{1} \frac{(1-z )^{-\varepsilon } \left((\varepsilon
-z)z^{-\varepsilon} + (z+1)^{-\varepsilon } z \right)}{\varepsilon ^2 (\varepsilon
+1)^2} \nn \\ \label{a32}
 &= \frac{1}{2 \varepsilon }+\left(-\frac{1}{2}+\frac{\pi ^2}{12}\right)+\left(1-\frac{\pi ^2}{6}+\zeta (3)\right)
   \varepsilon +O\left(\varepsilon ^2\right)\\ 
C_{10}^{\tau}&\equiv \frac{A_{2}^{\tau}}{\Gamma (1-\epsilon)} \int_{s,v,x} \left(s+v+x
\right)^{-2\theta} \rme^{-s-v-x} = \frac{A_{2}^{\tau}}{\Gamma
(1-\epsilon)} \int_{s,v} v \left(s+v
\right)^{-2\theta} \rme^{-s-v}\nonumber \\ \label{a34}
&= \frac{\Gamma (3-2 \theta )}{2 \Gamma (1-\epsilon )} = \frac{1}{2 \epsilon }-\frac{1}{2}+{\cal O} (\epsilon )
\\
\ \nn\\
\mathcal{C}_{11}&=\largediagram{{R2-11}} \ ,\ \ C_{11}= \int_{r,s,t,u,v,w}t^{-2\theta}v^{-2\theta}\nn \\ \label{a57}
c(\mathcal{C}_{11})&= 4 \times    {n(n-1)\over 2}\\ \label{a58}
C_{11}^{L}&\equiv {1\over
5-4\theta}A_3=\frac 1{\varepsilon^{2}}+\left(3-\frac{\pi^2} 6\right)+\mathcal{O}(\varepsilon)\\ \label{a59}
C_{11}^{\tau} &\equiv  ( A_{2}^{\tau})^{2}\\ 
\ \nn\\
\mathcal{C}_{12}&=\largediagram{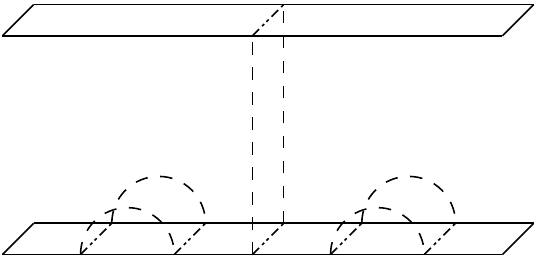}\ , \quad C_{12}=\int_{r,s,t,u,v,w}s^{-2\theta}w^{-2\theta}\label{a60a}\\ 
\label{a61}
c(\mathcal{C}_{12})&= 2\times  {n(n-1)\over 2}
\\ 
\label{a62}
C_{12}^{L}&=\int_{r+s=1}r^{2-2\theta}s^{2-2\theta}A_2^2={\Gamma(3-2\theta)^2\over\Gamma(6-4\theta)}A_2^2=\frac{1}{\varepsilon^2}+\left(3-\frac{\pi^2}6\right)+\mathcal{O}(\varepsilon)\\ \label{a63}
C_{12}^{\tau} &= \frac{1}{\left(\varepsilon ^2+\varepsilon \right)^2} =
\frac{1}{\varepsilon ^2}-\frac{2}{\varepsilon }+3+{\cal O}(\varepsilon)
\\
\ \nn\\
\mathcal{C}_{13}&= \largediagram{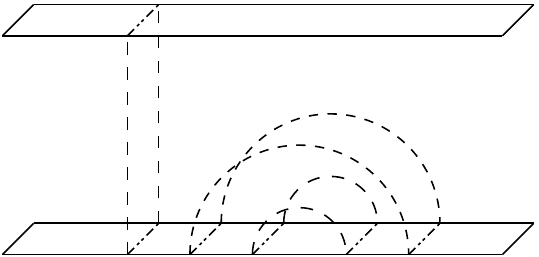}
\label{a63a}\\ 
\label{a64}
c(\mathcal{C}_{13}) &= 4 \times  {n(n-1)\over 2}
\\ 
\label{a65}
C_{13}^{L}&\equiv {1\over
5-4\theta}A_4^{L}=-\frac{1}{\varepsilon^2}-\frac1{\varepsilon^1}+\left(-3+\frac{\pi^2}6\right)+\mathcal{O}(\varepsilon)\\ \label{a66}
C_{13}^{\tau}&\equiv A_{4}^{\tau}
\\
\ \nn\\
\mathcal{C}_{14}&= \largediagram{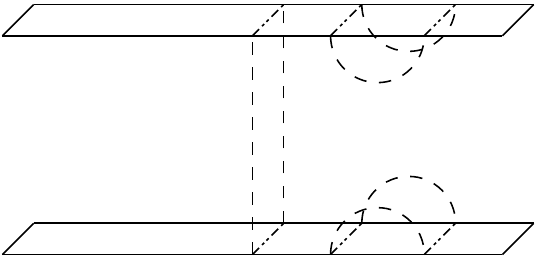}\ , \qquad 
c(\mathcal{C}_{14}) = 2  \times {n(n-1)\over 2} \ , \qquad 
 C_{14}\equiv C_{2}^2 
 \label{a67}\\
\ \nn\\
\mathcal{C}_{15}&= \largediagram{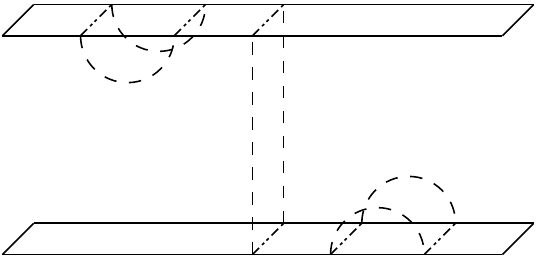}\ , \qquad
c(\mathcal{C}_{15})= 2 \times  {n(n-1)\over 2}\ , \qquad   C_{15}\equiv C_{14}\equiv C_2^2
\label{a67a}\\
\ \nn\\
\mathcal{C}_{16}&=\largediagram{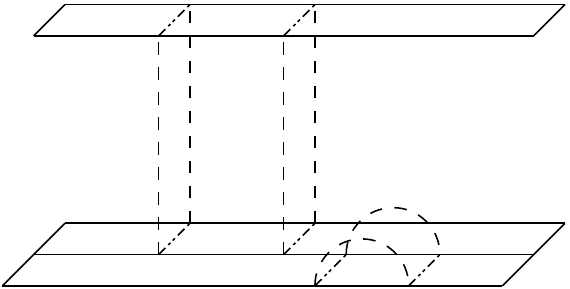} \ , \qquad C_{16} = 
\int_{r,s,t}\int_{u,v,w,x,y} (s+v)^{-2\theta}x^{-2\theta} 
\label{a67b}\\ 
\label{a35}
c(\mathcal{C}_{16})&=    4 \times n(n-1)(n-2)    \ , \qquad C_{16}\equiv C_{9}
\\
\ \nn\\
\mathcal{C}_{17}&=\largediagram{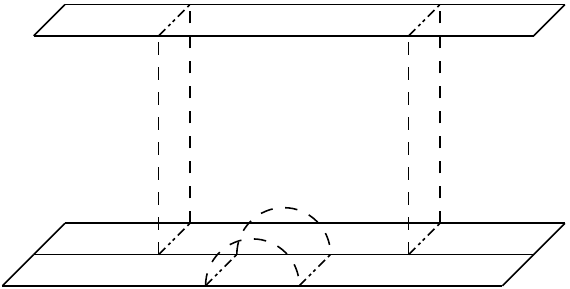} C_{17}=\int_{r,s,t}\int_{u,v,w,x,y} (s+v+w+x)^{-\theta}(s+v+x)^{-\theta}w^{-2\theta} 
\label{a35a} 
\\ \label{a36}
c(\mathcal{C}_{17})&= 2 \times     n(n-1)(n-2)    
\end{align}
The integral $C_{17}^{L}$ is hard to calculate. In order to simplify matters, we calculate instead its derivative $C_{17}'$ w.r.t.\ $L$. Using that $C_{17}^{L}\sim L^{6-4\theta}$, this gives  
\begin{align}
C_{17}^{L}&= \left.\int_0^L ds\,(L-s)\int_{z+w<L} z\,(L-z-w)(s+z+w)^{-\theta}(s+z)^{-\theta}w^{-2\theta}\right|_{L=1}\nn \\ \label{a37}
&= {1\over 6-4\theta} C_{17}'  \\ 
C_{17}'&= \left.\int_0^L ds\,\int_{z+w<L}z(2L-s-z-w)(s+z+w)^{-\theta}(s+z)^{-\theta}w^{-2\theta}\right|_{L=1} \nn \\ 
&= {1\over 5-4\theta}\left(E+F+G\right) \ ,\label{a40}
\intertext{where we have again taken the derivative w.r.t.\ $L$. The different contributions, named $E$, $F$, and $G$ are}
E&=\left.\int_{0<z+w<L}z(L-z-w)(L+z+w)^{-\theta}(L+z)^{-\theta}w^{-2\theta}\right|_{L=1}
=E_1+E_2 \label{a41}
\intertext{which we integrate by part w.r.t.\ $w$. The boundary term is $E_{1}$, the remaining one $E_{2}$:}
E_1&={1\over 1-2\theta}\int_0^1dz\ \left[{z(1-z-w)(1+z+w)^{-\theta}(1+z)^{-\theta}w^{1-2\theta}}\right]_{w=0}^{w=1-z}=0 \\
E_2&={1\over 1-2\theta}\int_0^1dz\int_0^{1-z}dw\,z\left[(1+z+w)+\theta(1-z-w)\right](1+z)^{-\theta}(1+z+w)^{-1-\theta}w^{1-2\theta}\nn \\ \label{a44}
&=E'_2+E''_2\\ \label{a45}
E_2'&={1\over 1-2\theta}\int_0^1dz\int_0^{1-z}dw\,z((1+z)+\theta(1-z))(1+z)^{-1-2\theta}w^{1-2\theta}
\intertext{and $E_{2}''=E_{2}-E_{2}'$. The decomposition is such that $E_{2}'$ can be integrated analytically over $w$, whereas $E_{2}''$ is finite, s.t.\ it can be expanded in $\epsilon$.}
E'_2&={1\over 1-2\theta}{1\over 2-2\theta}\int_0^1dz\,z((1+z)+\theta(1-z))(1+z)^{-1-2\theta}(1-z)^{2-2\theta}\nn\\
&=\frac{1}{4 \epsilon }+\left[-\frac{3}{4}+\ln (2)\right]+{\cal O}(\epsilon ) \label{a46}\\
E_{2}''&=\frac{1}{8}+{\cal O}(\epsilon ) \label{a46b}
\end{align}
The next contribution is 
\begin{align}
F&=\left.\int_{0<s<L}\int_{z+w=L}z(L-s)(L+s)^{-\theta}(s+z)^{-\theta}w^{-2\theta}\right|_{L=1} = F_{1}+F_{2}\ .\label{a47}
\end{align}
Again we split into a contribution $F_{1}$, containing the divergent contribution, but integrable analytically, and $F_{2}$, the finite rest to be evaluated at $\epsilon=0$:
\begin{align}
F_{1}&=\int_{0<s<1}\int_{0<w<1} w^{-2 \theta } \left[(1-s) w (-s+\theta -1) (s+1)^{-2 \theta -1}+(1-s) (s+1)^{-2
   \theta }\right] \nn\\
&=\frac{4^{-\theta -1} \left[-2 \left(4^{\theta } (\theta -1)+1\right) \theta  (2
   \theta -3)-4^{\theta }+2\right]}{(\theta  (2 \theta -3)+1)^2} \nn\\
   &=\frac{\frac{3}{4}-\ln (2)}{\epsilon }+\frac{1}{4} \left[-9+2 \ln ^2(2)+\ln
   (512)\right] +{\cal O}(\epsilon )  \label{a47}\\
F_{2}&= \int_{0<s<1}\int_{0<w<1} \frac{(s-1) s}{(s+1)^3 (s-w+1)}+{\cal O} (\epsilon )\nn\\
&= \frac{1}{24} \left[2 \pi ^2-3 \left(5-4 \ln ^2(2)+\ln (64)\right)\right]+{\cal O} (\epsilon )
\end{align} 
The last contribution is 
\begin{align}
G&=\left.\int_{0<s<L}\int_{0<z+w<L}2z(s+z+w)^{-\theta}(s+z)^{-\theta}w^{-2\theta}\right|_{L=1}\label{a48}
\end{align}
which is the most divergent one, since both the integrals over $w$ as over the global scale will give a pole in $1/\epsilon$. To isolate the global pole, we again derive w.r.t.\ $L$, reducing this time the number of integrations by 1. Since $G\sim L^{4-4\theta}$, we obtain 
\begin{align}
G&={1\over 4-4\theta}(H+K) \label{90}\\
H&=\left.\int_{0<z+w<L}2z(L+z+w)^{-\theta}(L+z)^{-\theta}w^{-2\theta}\right|_{L=1}=H_{1}+H_{2}+H_{3} \label{a49}\\
K&=\left.\int_{0<s<L}\int_{z+w=L}2z(s+L)^{-\theta}(s+z)^{-\theta}w^{-2\theta}\right|_{L=1}=K_{1}+K_{2}\ .
 \label{a50}
\end{align}
From both $H$ and $K$, we need also the finite term, since it is multiplied by $1/\epsilon$ from (\ref{90}), but we will even calculate the next sub-leading term. We use again the technique to split $H$ into  parts which contain the divergent contributions $H_{1}$ and $H_{2}$ and which are doable analytically, and a part  for the rest, which is finite and can be expanded in $\epsilon$.
\begin{align}
H_{1} &=  \int_{0}^{1}\rmd z \int_{0}^{1-z}\rmd w\, 2z (L+z)^{-2\theta}w^{-2\theta} = \int_{0}^{1}\rmd z\, \frac{2 (z-1) z \left(1-z^2\right)^{-2
   \theta }}{2 \theta -1} =\frac{2-\frac{\sqrt{\pi } \Gamma (-\epsilon
   )}{\Gamma \left(\frac{1}{2}-\epsilon
   \right)}}{2 (\epsilon +1)^2}  \label{eH1a}\\
H_{2}&= -2 \theta \int_{0}^{1}\rmd z \int_{0}^{1-z}\rmd w\,  w^{1-2 \theta } z (z+1)^{-2 \theta -1}
      = \frac{\epsilon +2} {\epsilon } \int_{0}^{1}\rmd z\, {(1-z)^{-\epsilon } z (z+1)^{-\epsilon
   -3} }\label{eH2a}\\
\intertext{Since the last integral is convergent, we can expand it in $\epsilon$}   
H_{2}&= \frac{\epsilon +2} {\epsilon } \int_{0}^{1}\rmd z\, \left[ \frac{z}{(z+1)^3}-\frac{z \ln \left(\frac{1-z}{z+1}\right) \epsilon
   }{(z+1)^3}+\frac{z \ln^2\left(\frac{1-z}{z+1}\right) \epsilon^2}{2 (z+1)^3}+{\cal O}(\epsilon ^3) \right]\label{eH2b}\nn\\
   & =  \frac{\epsilon +2} {\epsilon } \left[\frac{1}{8}+\frac{3 \epsilon }{16}+\frac{7
   \epsilon ^2}{32}+{\cal O}(\epsilon ^3)  \right] \\
H_{3}&=  2 \int_{0}^{1}\rmd z \int_{0}^{1-z}\rmd w\,  w^{-2 \theta } z (z+1)^{-2 \theta } \left[(z+1)^{\theta } (w+z+1)^{-\theta }+\frac{w
   \theta }{z+1}-1\right]\ .\label{eH3a}\\
\intertext{Having calculated separately the most diverging terms in $w$, $H_{1}$ and $H_{2}$, the remainder $H_{3}$ is finite, since the  square bracket is of order $w^{2}$. Thus the whole expression can be expanded in $\epsilon$, and then integrated analytically. This is straightforward, but since the intermediate expressions are rather cumbersome, we only give the final result:}
H_{3}&= \left[-\frac{5}{8}+\ln (2)\right]+\left[\frac{11 \pi ^2}{96}+\frac{1}{4} (9-\ln (2)
   (18+\ln (2)))\right] \epsilon +{\cal O}(\epsilon ^2 )
   \label{eH3b}
\end{align}
We now treat $K$, which we split into a diverging integral $K_{1}$, doable analytically, and a convergent rest $K_{2}$, which we expand in $\epsilon$:
\begin{align}
K_{1}&= \int_{0<s<1}\int_{0<w<1} w^{-2 \theta } \left(2 (s+1)^{-2 \theta }-2 (s+1)^{-2 \theta -1} w (s-\theta +1)\right) \nn\\
&=\frac{1 -\left(1-2^{-\epsilon -2}\right) \epsilon  (\epsilon +2)-3\cdot 2^{-\epsilon
   -2}}{\epsilon  (\epsilon +1)^2} \nn\\
   &= \frac{1}{4 \epsilon }+\left[\frac{3 \ln (2) }{4}-2\right]+\left[3-2 \ln (2)-\frac{3 \ln
   ^2(2)}{8}\right] \epsilon +{\cal O}(\epsilon ^2)\label{eK1a}\\
K_{2}&= 2 \int_{0<s<1}\int_{0<w<1} (s+1)^{-2 \theta } w^{-2 \theta } \left[-(s+1)^{\theta } (w-1) (s-w+1)^{-\theta
   }+w-\frac{w \theta }{s+1}-1\right] \nn\\
   &= \left[-\frac{1}{8}-\frac{\ln (2)}{4}\right]+\left[2-\frac{23 \pi ^2}{96}+\ln (2)-\frac{7 \ln
   ^2(2)}{8}\right] \epsilon +{\cal O}(\epsilon ^2)\ ,\label{eK2a}
\end{align}
where in order to integrate $K_{2}$, we have again expanded in $\epsilon$, which is justified since the square bracket is of order $w^{2}$. 

Taking all terms together, the final result for $C_{17}^L$ is 
\begin{align}
\label{i26}
 C_{17}^{L}   &= -{1\over 4 \varepsilon^{2}}+\left[{5\over 8}-\ln(2)\right]\frac1\varepsilon+\left[-{5\over 4}+{\pi^2\over 16}-{1\over 2}\ln(2)+\ln(2)^2\right]+\mathcal{O}(\varepsilon) \intertext{The diagram in the grand-canonical scheme is much simpler, and all integrals can be done analytically, using the standard Schwinger representation. The result is:}
\label{i27}
C_{17}^{\tau}& = \frac{2^{-2 (\varepsilon +1)} \sqrt{\pi } \csc \left(\frac{\pi  \varepsilon }{2}\right) \Gamma \left(\frac{1}{2}-\varepsilon
   \right)}{\varepsilon ^2 (\varepsilon +1) \Gamma \left(1-\frac{3 \varepsilon
}{2}\right) \Gamma \left(\frac{\varepsilon }{2}-1\right)}=
 -\frac{1}{4 \varepsilon ^2}+\frac{3}{8 \varepsilon }+\left(-\frac{3}{8}-\frac{\pi ^2}{48}\right)+{\cal O}(\epsilon)
\intertext{The remaining diagrams are:}
\mathcal{C}_{18}&= \largediagram{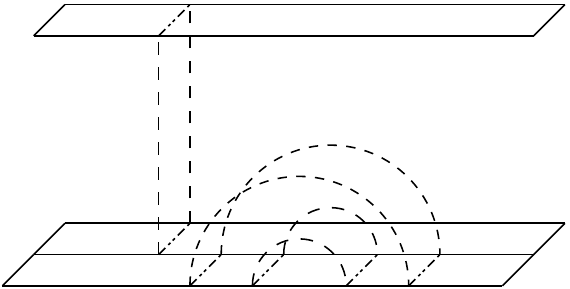} \ ,  \qquad c(\mathcal{C}_{18})=      4   \times   n(n-1)(n-2) \ ,  \qquad C_{18}\equiv C_{13}
\label{eC18}\\ \label{a70}
\mathcal{C}_{19}&= \largediagram{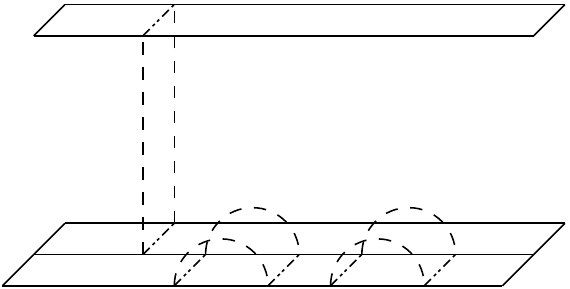} \ , \qquad c(\mathcal{C}_{19}) = 4\times  n(n-1)(n-2)  \ ,  \qquad  C_{19}\equiv C_{11}\\ \label{a71}
\mathcal{C}_{20}&= \largediagram{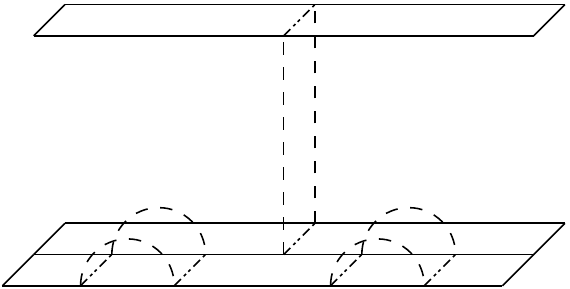} \ , \qquad c(\mathcal{C}_{20}) = 2 \times  n(n-1)(n-2) \ , \qquad C_{20}\equiv C_{12}\\ \label{a72}
\mathcal{C}_{21}&= \largediagram{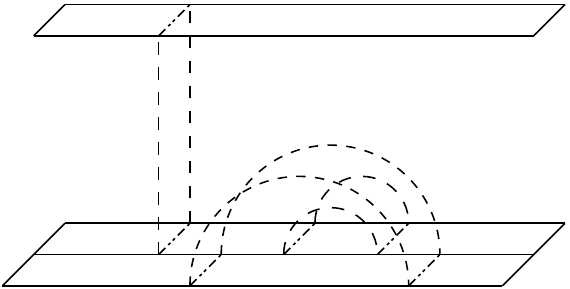} \ , \qquad c(\mathcal{C}_{21}) = 4 \times  n(n-1)(n-2) \\ \label{a73}
C_{21}^{L}&\equiv   \int_{r,s}s^{4-4\theta}A_{6}^{L} = {1\over 5-4\theta}A_6^{L}=-{3\over 4 \varepsilon}+{3\over 4}+\mathcal{O}(\varepsilon)
\ , \qquad C_{21}^{\tau}\equiv A_{6}^{\tau}\\ \label{a74}
\mathcal{C}_{22}&= \largediagram{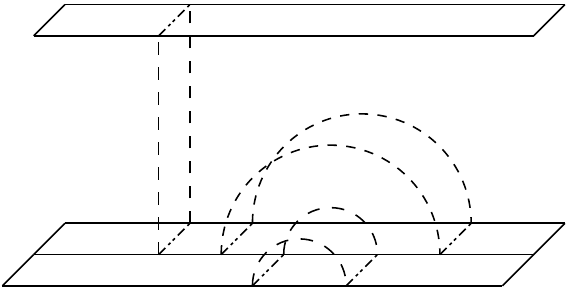}  \ ,  \qquad c(\mathcal{C}_{22})=       4   \times   n(n-1)(n-2) \ ,  \qquad  C_{22}\equiv C_{21}\\ \label{a75}
\mathcal{C}_{23}&=\largediagram{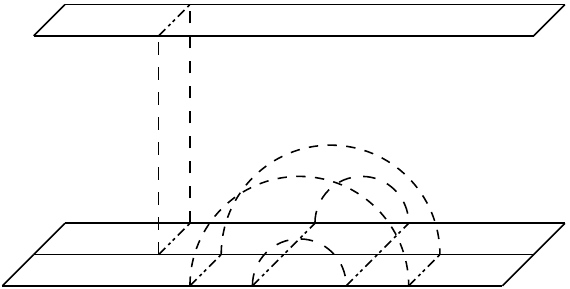} \ ,  \qquad c(\mathcal{C}_{23}) =    4 \times   n(n-1)(n-2)\ ,   \qquad    C_{23}\equiv C_{21}  \\ \label{a76}
\mathcal{C}_{24}&=\largediagram{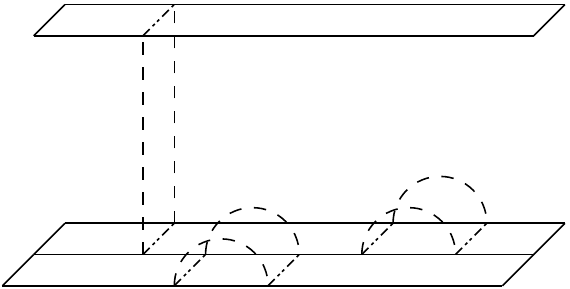}  \ ,   \qquad c(\mathcal{C}_{24})=   4   \times   n(n-1)(n-2)\\ \label{a77}
C_{24}^{L}&\equiv \int_{r,s}s^{4-4\theta}A_{5}^{L}\equiv C_{11}^{L}  \ , \qquad C_{24}^{\tau}\equiv   (A_{2}^{\tau})^{2}\\ \label{a78}
\mathcal{C}_{25}&=\largediagram{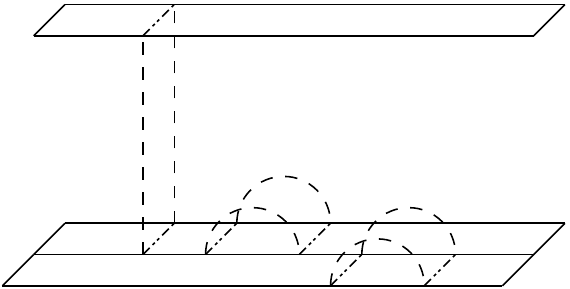}  \ , \qquad c(\mathcal{C}_{25})    =   4   \times   n(n-1)(n-2) \ ,  \qquad 
C_{25}\equiv C_{24}  \\ \label{a79}
\mathcal{C}_{26}&=\largediagram{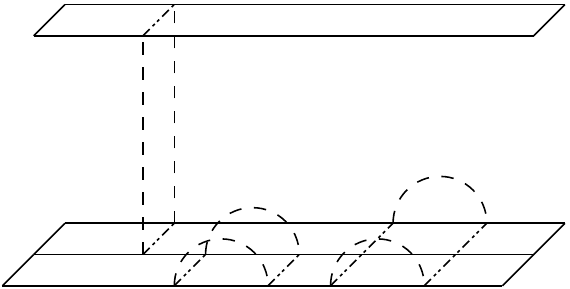} \ ,  \qquad c(\mathcal{C}_{26})    =   4   \times   n(n-1)(n-2)  \ , \qquad   C_{26}\equiv C_{24}   \\ \label{a80}
\mathcal{C}_{27}&=\largediagram{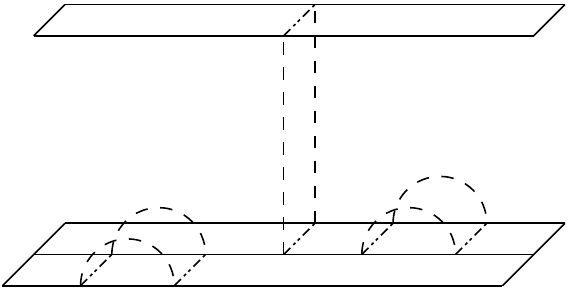}  \ ,  \qquad c(\mathcal{C}_{27})    =   4   \times  n(n-1)(n-2)  \ , \qquad  C_{27}\equiv C_{12}    \\ \label{a81}
 \mathcal{C}_{28}&=\largediagram{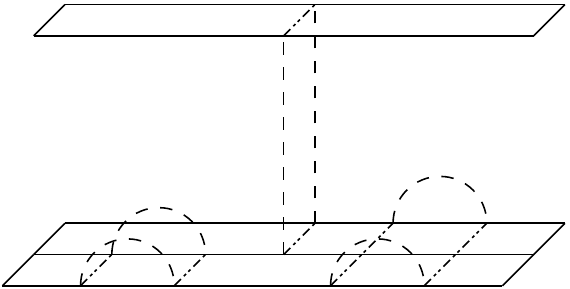}   \ ,  \qquad c(\mathcal{C}_{28})    =   2   \times   n(n-1)(n-2) \ ,  \qquad 
C_{28}\equiv C_{12}  \\ \label{a82}
 \mathcal{C}_{29}&=\largediagram{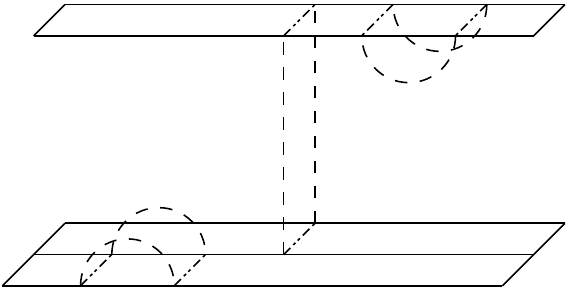}   \ ,  \qquad c(\mathcal{C}_{29})    =   4   \times   n(n-1)(n-2) \ ,  \qquad  
C_{29}\equiv C_{14} \\ \label{a83}
 \mathcal{C}_{30}&=\largediagram{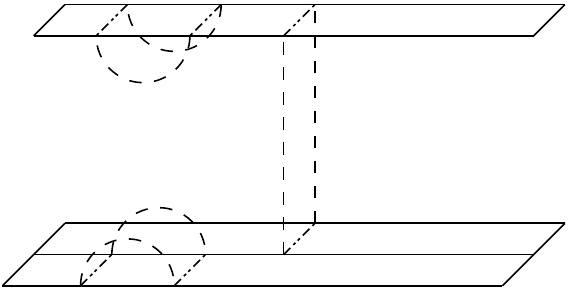} \ ,  \qquad c(\mathcal{C}_{30})    =   4   \times     n(n-1)(n-2)  \ , \qquad   
C_{30}\equiv C_{14} \\ \label{a84}
 \mathcal{C}_{31}&=\largediagram{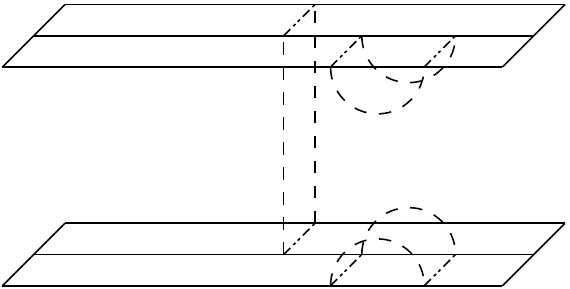} \ ,  \qquad c(\mathcal{C}_{31})    =    2   \times   n(n-1)(n-2)^2  \ , \qquad
  C_{31}\equiv C_{14}  \\ \label{a85}
 \mathcal{C}_{32}&=\largediagram{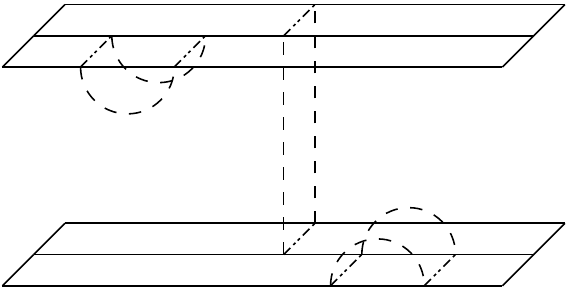}  \ ,  \qquad c(\mathcal{C}_{32})    =   2   \times   n(n-1)(n-2)^2  \ , \qquad  
C_{32}\equiv C_{14}\\ \label{a86}
 \mathcal{C}_{33}&=\largediagram{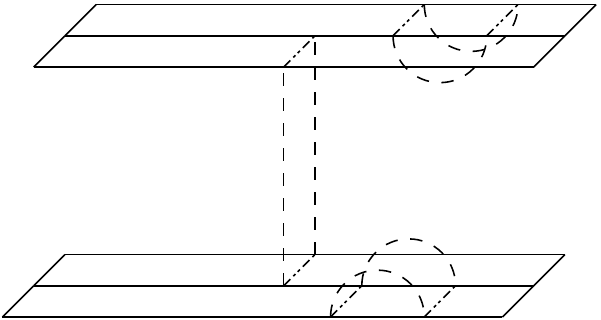}  \ ,  \qquad c(\mathcal{C}_{33})    =  2   \times    n(n-1)(n-2)^2 \ ,  \qquad  
C_{33}\equiv C_{14} \\ \label{a87}
 \mathcal{C}_{34}&=\largediagram{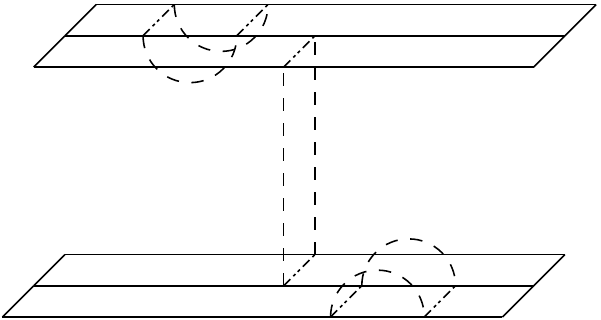}   \ ,  \qquad c(\mathcal{C}_{34})    =   2   \times    n(n-1)(n-2)^2 \ ,  \qquad 
C_{34}\equiv C_{14}\\ \label{a88}
\mathcal{C}_{35}&=\largediagram{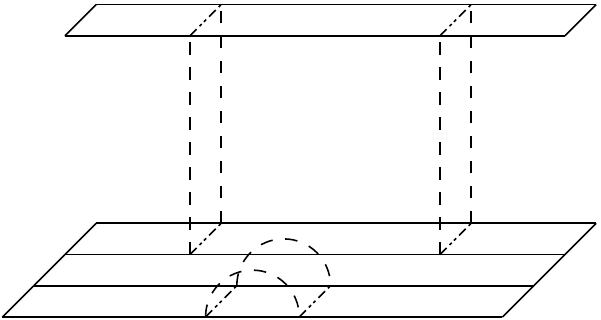}\\ \label{a52}
c(\mathcal{C}_{35}) &=   2 \times    n(n-1)(n-2)(n-3)/4 \ , \qquad
C^{L}_{35}\equiv C^{L}_{5}A^{L}_2  \\ 
C_{35}^{\tau} &\equiv  \frac{A_{2}}{\Gamma (1-\varepsilon)^{2}} \int_{L_{1}}
\int_{L_{2}} \rme^{-L_{1}-L_{2}} C_{5} (L_{1},L_{2}) L_{2}^{2-2\theta}
\nn \\ 
&= 
\frac{(\varepsilon -2) \left(2 \Gamma (3-\varepsilon )^2+(\varepsilon -2)
\Gamma (3-2 \varepsilon )\right)}{2 \left(\varepsilon ^2+\varepsilon \right)^2
\Gamma (3-\varepsilon )^2}\nn\\ \label{a55}
 &= -\frac{1}{\varepsilon ^2}+\frac{2}{\varepsilon }+\left(-4+\frac{\pi
^2}{6}\right)+{\cal O}(\epsilon)
\\ 
\ \nn\\
\mathcal{C}_{36}&=\largediagram{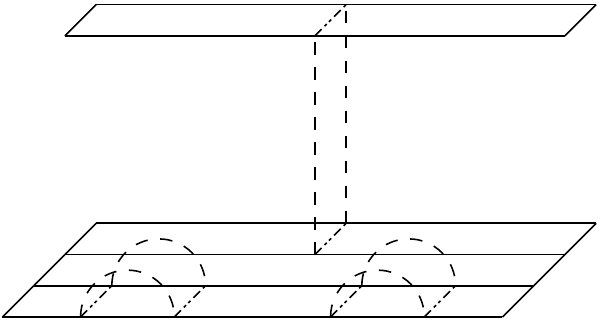}  \ , \qquad c(\mathcal{C}_{36})=   2  \times  {n(n-1)(n-2)(n-3)\over 4}  
\label{a55a}\\ 
\label{a89}
C_{36}^{L}&\equiv C_{1}^{L} A_3^{L} \ , \qquad 
C_{36}^{\tau}\equiv \frac{C_{36}^{L} \Gamma (2-2\varepsilon)}{\Gamma
(1-\varepsilon)^{2}}=\frac{1- 2 \varepsilon }{\varepsilon ^2 (\varepsilon +1)^{2}}
= \frac{1}{\varepsilon ^2}-\frac{4}{\varepsilon }+7+{\cal O}(\epsilon)\\ 
\ \nn\\
\label{a90}
\mathcal{C}_{37}&=\largediagram{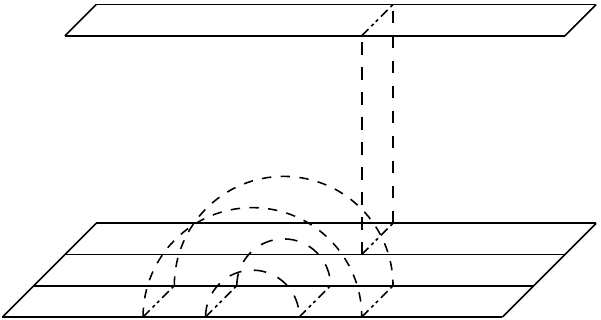} \ ,  \qquad c(\mathcal{C}_{37})=  2  \times   {n(n-1)(n-2)(n-3)\over 4}  \\ \label{a91}
C_{37}^{L}&\equiv C_{1}^{L} A_4^{L} \ ,\qquad 
C_{37}^{\tau}\equiv \frac{C_{37}^{L} \Gamma (2-2\varepsilon)}{\Gamma
(1-\varepsilon)^{2}}=\frac{-1+2 \varepsilon }{\varepsilon ^2 (\varepsilon +1)} =
- \frac{1}{\varepsilon ^2}+\frac{3}{\varepsilon }-3+{\cal O}(\epsilon)
\end{align}
\begin{align}
\mathcal{C}_{38}&=\largediagram{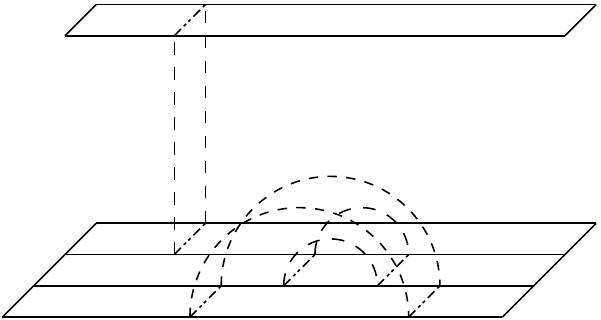} C_{38}= \int_{r,s,t,u,v}(s+t+u)^{-\theta}(s+u)^{-\theta} t^{-2\theta} (r+s)
\label{a91a}\\ 
\label{a92}
c(\mathcal{C}_{38})&=   4   \times    n(n-1)(n-2)(n-3)    \\ 
C_{38}^{L}&= \int_{r+s+t+u+v=1}
(s+t+u)^{-\theta}(s+u)^{-\theta} t^{-2\theta} (r+s) \label{a94}
\intertext{Symmetrizing $\{r,s\}\leftrightarrow \{v,u\}$ gives:}
&= \int_{r+s+t+u+v=1}(s+t+u)^{-\theta}(s+u)^{-\theta} t^{-2\theta} (1-t)/2\nn \\ 
&=\int_{x+t<1} {1\over 2}x (1-t-x)(1-t)(x+t)^{-\theta}x^{-\theta}t^{-2\theta}\nn \\ 
&=\int_0^1 dy\,y^{2-4\theta}(1-y)\int_{t+x=1}{1\over 2}(1-yt)x^{1-\theta}t^{-2\theta}\nn\\
&={1\over 2}{\Gamma(3-4\theta)\Gamma(2)\over\Gamma(5-4\theta)}\left({\Gamma(2-\theta)\Gamma(1-2\theta)\over\Gamma(3-3\theta)}-{(3-4\theta)\over (5-4\theta)}{\Gamma(2-\theta)\Gamma(2-2\theta)\over\Gamma(4-3\theta)}\right)\nn \\ \label{a98}
C_{38}^{L}&=-{1\over 4 \varepsilon^2}-{7\over 8\varepsilon}+{1\over 48}(6+\pi^2)+\mathcal{O}(\varepsilon) \\ \label{a99}
C_{38}^{\tau} &\equiv \frac{\Gamma (2-2\epsilon)
C_{38}^{L}}{\Gamma(1-\epsilon)^{2}} = -\frac{1}{4 \epsilon ^2}-\frac{3}{8 \epsilon }+\left(\frac{15}{8}-\frac{\pi ^2}{48}\right)+{\cal O}\left(\varepsilon\right)\\
\ \nn\\
\mathcal{C}_{39}&= \largediagram{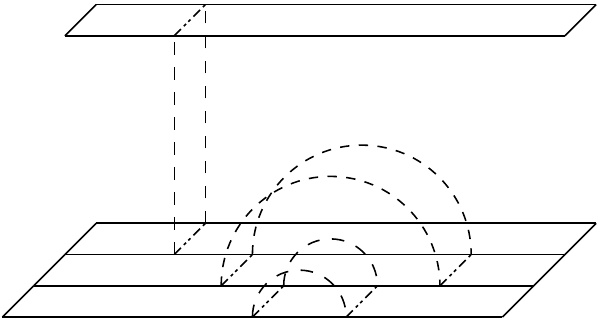}  \ ,\qquad c(\mathcal{C}_{39})= 4 \times  n(n-1)(n-2)(n-3)\nn \\ \label{a101}
C_{39}&\equiv C_{21}  \\ 
\ \nn\\
\mathcal{C}_{40}&=\largediagram{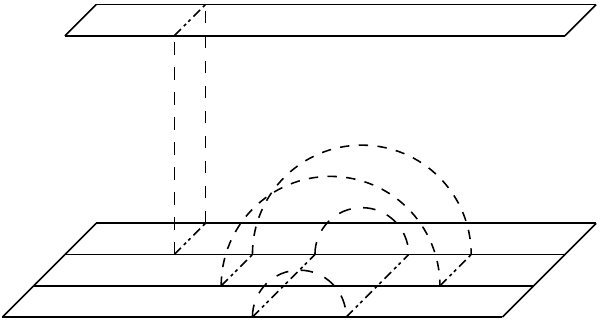} \ ,  \qquad c(\mathcal{C}_{40})=      4   \times     {n(n-1)(n-2)(n-3)}\nn \\ \label{a103}
C_{40}&\equiv C_{21}\\ 
\ \nn\\
\mathcal{C}_{41}&=\largediagram{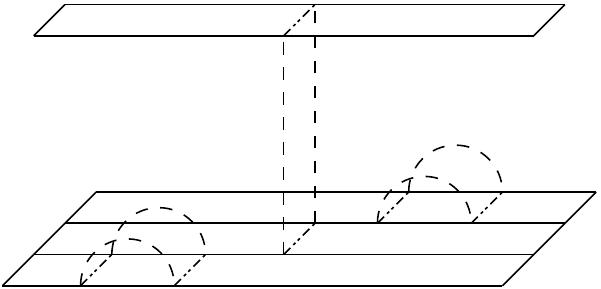}  \ ,  \qquad c(\mathcal{C}_{41})=   2   \times   {n(n-1)(n-2)(n-3)} \nn \\ \label{a105}
C_{41}&\equiv C_{12}  \\ 
\ \nn\\
\mathcal{C}_{42}&=\largediagram{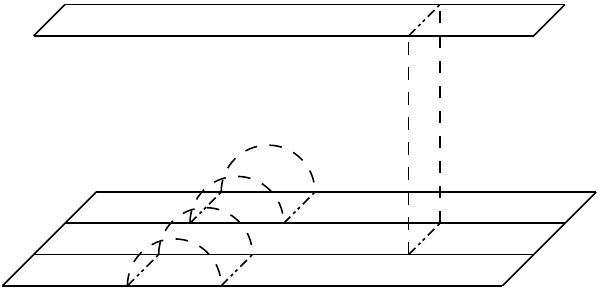}   \ ,  \qquad c(\mathcal{C}_{42})=   4   \times   {n(n-1)(n-2)(n-3)\over 2}  \nn   \\ \label{a107} 
C_{42}^{L}&\equiv  \int_{r+s=1}r^{4-4\theta}(A_2)^2={1\over 5-4\theta}(A_2)^2={1\over (5-4\theta)(2-2\theta)^2(1-2\theta)^2}=\frac{1}{\varepsilon^2}+3+\mathcal{O}(\varepsilon)\nonumber \\
\\ \label{a108}
C_{42}^{\tau} &\equiv  \frac{\Gamma (2-2\epsilon)}{\Gamma (1-\epsilon)^{2}}
C_{42}^{L} = \frac{\Gamma (2-2 \epsilon )}{(2 \epsilon -1) \left(\epsilon
^2+\epsilon \right)^2 \Gamma (1-\epsilon )^2} = \frac{1}{\epsilon
^2}-\frac{2}{\epsilon }+\left(3+\frac{\pi
^2}{6}\right)+{\cal O}(\varepsilon)\\ 
\ \nn\\
\mathcal{C}_{43}&=\largediagram{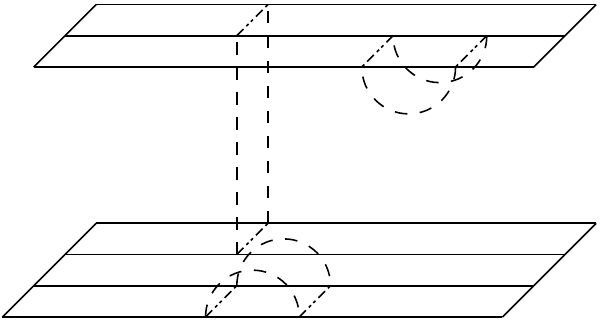}   \ ,  \qquad c(\mathcal{C}_{43})=    4   \times    {n(n-1)(n-2)^2(n-3)\over 2} \nn\\ \label{a110}
C_{43}^{L}&\equiv C_3^{L }A_2^{L} \ , \qquad C_{43}^{\tau}\equiv C_4^{\tau  }A_2^{\tau }\qquad \\ 
\ \nn\\
\mathcal{C}_{44}&=\largediagram{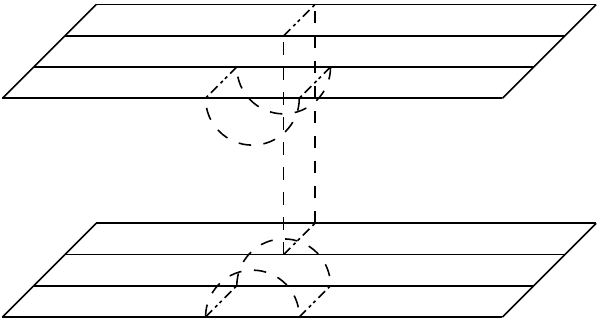}  \ ,  \qquad c\mathcal{(C}_{44})=    1   \times    {n(n-1)(n-2)^2(n-3)^2\over 8} \nn \\ \label{a112}
C_{44}^{L}&\equiv  ( A_2^{L})^2\ , \qquad C_{44}^{\tau} \equiv  (
A_{2}^{L})^{2}\frac{\Gamma (4-2\theta)^{2}}{\Gamma (1-\epsilon)^{2}} =
\frac{1}{\epsilon^{2}}-\frac{4}{\epsilon}+8+{\cal O}(\epsilon)\\ 
\ \nn\\
\mathcal{C}_{45}&=\largediagram{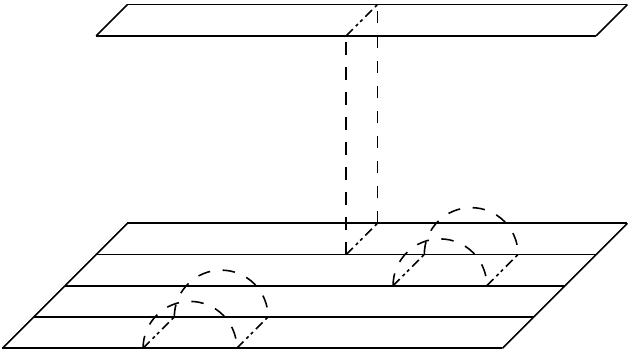}  \ ,  \qquad c(\mathcal{C}_{45})=   4   \times     {n(n-1)(n-2)(n-3)(n-4)\over 2} \nn \\ \label{a114}
C_{45}&\equiv C_3^{L}A_2^{L} \ ,\qquad C_{45}^{\tau}\equiv  C_{2}^{L}A_{2}^{L}\frac{\Gamma (6-4\theta)}{\Gamma (1-\epsilon)^{2}} = \frac{1}{\epsilon^{2}}-\frac{3}{\epsilon}+\left(\frac{\pi^{2}}{6}+4 \right)+{\cal O}(\epsilon) \\ 
\ \nn\\
\mathcal{C}_{46}&=\largediagram{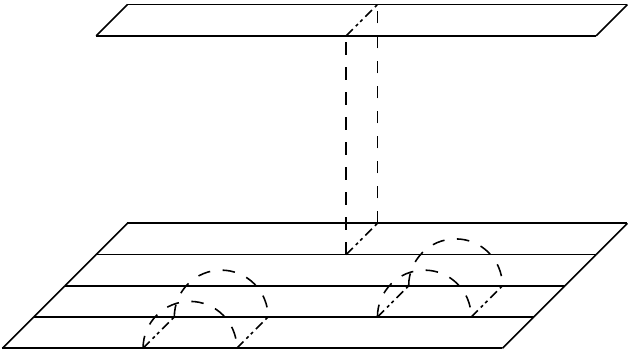}  \ ,  \qquad c(\mathcal{C}_{46})=    2   \times    {n(n-1)(n-2)(n-3)(n-4)\over 2} \nn\\ \label{a116}
C_{46}&\equiv  C_1^{L}A_5^{L}\ ,\qquad C_{46}^{\tau} \equiv  A_{3}^{L} \frac{\Gamma (6-4\theta)}{\Gamma (1-\epsilon)^{2}} =  \frac{1}{\epsilon^{2}}-\frac{4}{\epsilon}+7+{\cal O} (\epsilon)    \\ 
\ \nn\\
\mathcal{C}_{47}&=\largediagram{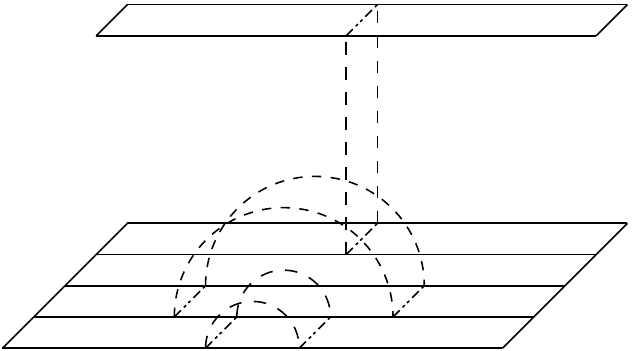}  \ ,  \qquad c(\mathcal{C}_{47})=  2   \times   {n(n-1)(n-2)(n-3)(n-4)\over 2} \nn\\ \label{a118} 
C_{47}^{L}&\equiv C_1^{L}A_6^{L} \ ,\qquad C_{47}^{\tau}\equiv  A_{6}^{L} \frac{\Gamma (6-4\theta)}{\Gamma (1-\epsilon)^{2}} = -\frac{3}{4\epsilon}+\frac{15}{4}+{\cal O} (\epsilon)  \\ 
\mathcal{C}_{48}&=\largediagram{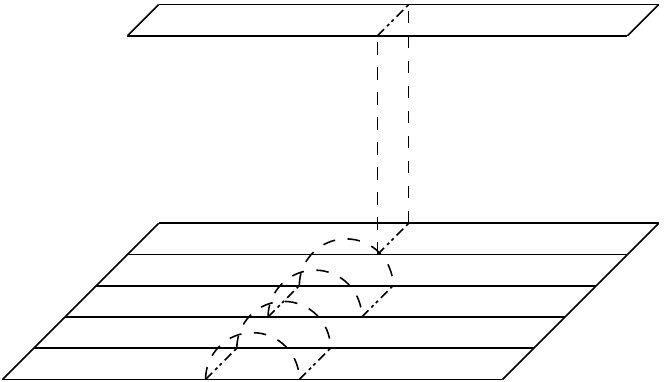}  \nn\\  
c(\mathcal{C}_{48})&=   2   \times    {n(n-1)(n-2)(n-3)(n-4)(n-5)\over 16} \nn \\ \label{a121}
C_{48}^{L}&\equiv  C_{1}^{L}( A_2^{L})^2 \ , \qquad C_{48}^{\tau} =  \frac{\Gamma (6-4\theta)}{\Gamma (1-\epsilon)^{2}} = \frac{1}{\epsilon^2}-\frac{4}{\epsilon}+\left( \frac{\pi^{2}}{6}+7\right)+{\cal O} (\epsilon)\\ 
\ \nn\\
\mathcal{C}_{49}&=\largediagram{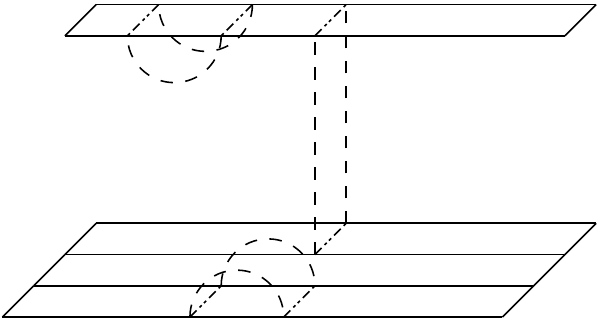}   \ ,  \qquad c(\mathcal{C}_{49})=   4   \times    {n(n-1)(n-2)(n-3)\over 4} \nn\\ \label{a123}
C_{49}^{L}&\equiv  C_2^{L}A_2^{L}\ , \qquad C_{49}^{\tau} \equiv A_{2}^{L}A_{2}^{\tau} \frac{\Gamma (4-2\theta)}{\Gamma (1-\epsilon)} \equiv \frac{\Gamma (2-2\epsilon)}{\Gamma (1-\epsilon)^{2}} C_{49}^{L}= \frac{1}{\epsilon^{2}} -\frac{3}{\epsilon} +{\cal O}(\epsilon )   \\ 
\ \nn\\
\mathcal{C}_{50}&=\largediagram{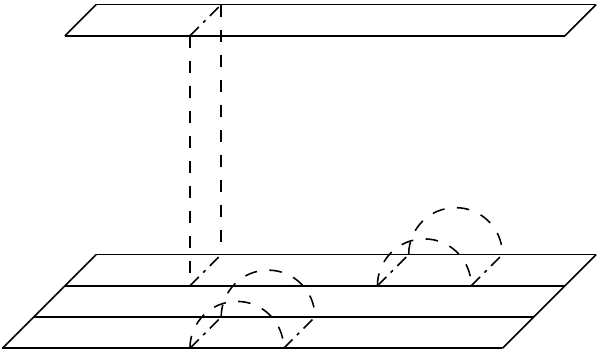}    \ ,  \qquad c(\mathcal{C}_{50})=   4   \times    {n(n-1)(n-2)(n-3)\over 4} \nn\\ \label{a125}
C_{50}&\equiv C_{49}   \\ 
\ \nn\\
\mathcal{C}_{51}&=\largediagram{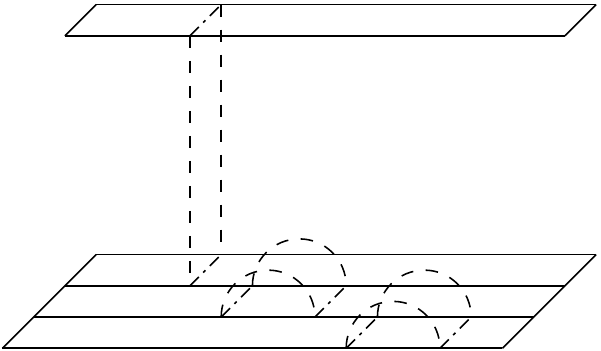}    \ ,  \qquad c(\mathcal{C}_{51})=    4   \times    {n(n-1)(n-2)(n-3)}  \nonumber \\ \label{a127}
  C_{51}&\equiv C_{11}   \\ 
  \ \nn\\
\mathcal{C}_{52}&=\largediagram{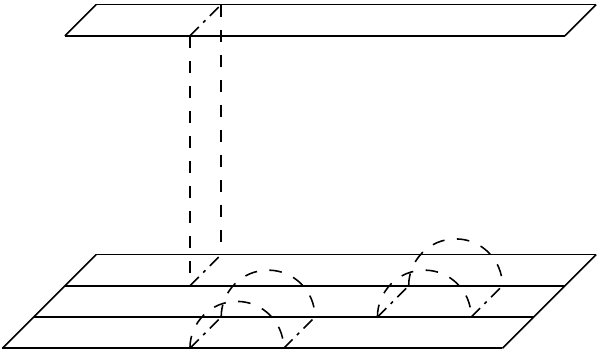}   \ ,  \qquad c(\mathcal{C}_{52})=   4   \times    {n(n-1)(n-2)(n-3)} \nn\\ 
C_{52}^{L}&\equiv  A_2^{L}\int_{r+s+t+u=1}t^{-2\theta}(r+s)^{2-2\theta}=A_2^{L}\int_{x+t+u=1}t^{-2\theta}x^{3-2\theta} \nn\\ \nn &=A_2^{L}\int_{u+y=1}y^{4-4\theta}\int_{x+t=1}t^{-2\theta}x^{3-2\theta}=A_2^{L}{1\over 5-4\theta}{\Gamma(1-2\theta)\Gamma(4-2\theta)\over\Gamma(5-4\theta)}\\ \label{a131}
&=\frac{1}{\varepsilon^2}-\frac1{\varepsilon}+3-{\pi^2\over
6}+\mathcal{O}(\varepsilon)\\ 
\label{a132}
C_{52}^{\tau} &\equiv  \frac{C_{52}^{L}\Gamma (2-2\epsilon)}{\Gamma
(1-\epsilon)^{2} } = \frac{1}{\epsilon ^2}-\frac{3}{\epsilon
}+5+{\cal O}(\epsilon ) \\ 
\ \nn\\
\mathcal{C}_{53}&=\largediagram{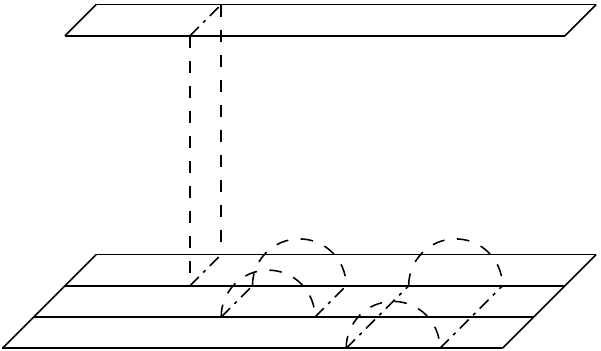}  \ ,  \qquad c(\mathcal{C}_{53})=    4   \times    {n(n-1)(n-2)(n-3)}  \nn\\ \label{a134}
C_{53}&\equiv C_{11}   \\ 
\ \nn\\
\mathcal{C}_{54}&=\largediagram{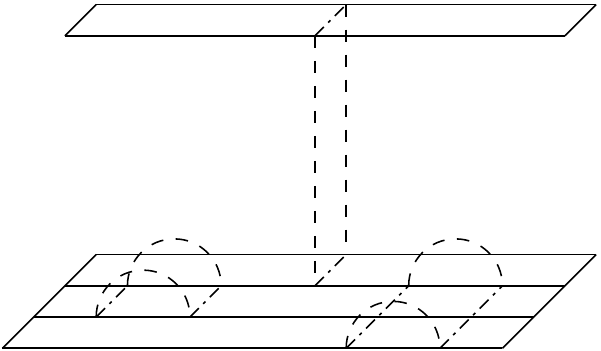}   \ ,  \qquad c(\mathcal{C}_{54})=  2   \times    {n(n-1)(n-2)(n-3)}  \nn\\ \label{a136}  
C_{54}&\equiv C_{12} 
\end{align}

\subsection{Diagrams and amplitudes for $\Phi$}\label{s:Diagr4Phi} 
\label{ss:PhiDiag}

We compute in perturbation theory up to two loops (order $g^2$) the partition function
$\widetilde{\mathcal{Z}}_\Phi^{(2)}(g,L)$ defined in (\ref{ZPhi2}).
$g$ is the (bare) coupling constant and $L$ the length of the open strands. 
It is written as a sum over the diagrams (labeled by $\mathcal{D}_i$) involving two-strand bundles with a single contact line (the $\Phi$ operator) between the two strands and arbitrary (planar system of) double arches within each strand bundle (the $\Psi$ interaction operators).
The perturbative expansion thus reads
\begin{equation}
\label{ZPhi2exp}
\widetilde{\mathcal{Z}}_\Phi^{(2)}(g,L)=\sum_{k=0}^\infty g^k\, L^{2+2k(1-\theta)} \sum_{\mathcal{D}_i} c(\mathcal{D}_i)\, D_i
\end{equation}
$\mathcal{D}_i$ are the diagrams, $c(\mathcal{D}_i)$ their symmetry factor and $D_i$ the amplitude.
The amplitudes $D_i$ can be deduced from the amplitudes $C_j$ already calculated for the 2-strands diagrams $\mathcal{C}_j$ considered for the renormalization of $g$.

At order $k=0$ there is only the trivial diagram

\begin{align}
\label{m31a}
\mathcal{D}_{1} &= \largediagram{P2-1}   \ ,  &   c(\mathcal{D}_{1})&=  1 \ , &   D_1&\equiv 1 \end{align}
\subsubsection{order $k=1$}
At order 1 there are 2 diagrams
\begin{align}
\mathcal{D}_{2} &= \largediagram{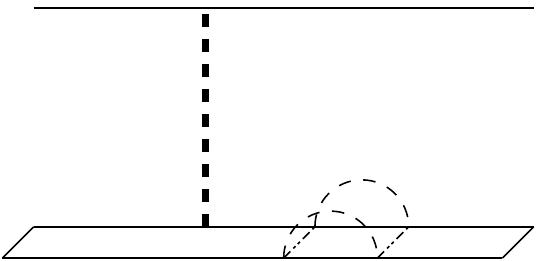}   \ ,  &   c(\mathcal{D}_{2})&=  4(n-1)\ , &   D_2&\equiv C_3 \label{eD2}\\
&&&\medskip\nonumber\\
\mathcal{D}_{3} &= \largediagram{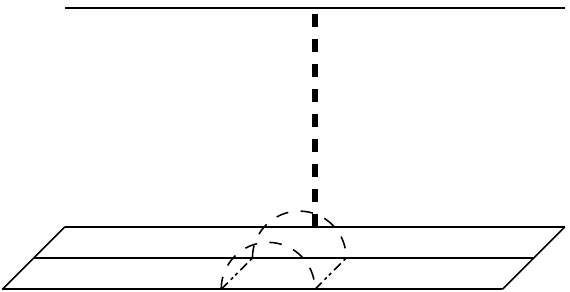}  \ ,  &   c(\mathcal{D}_{3})&=  (n-1)(n-2)\ , &     D_3&\equiv C_4 \label{eD3}\end{align}
\subsubsection{order $k=2$}
At order 2 there are 19 diagrams
\begin{align}
\mathcal{D}_{4} &= \largediagram{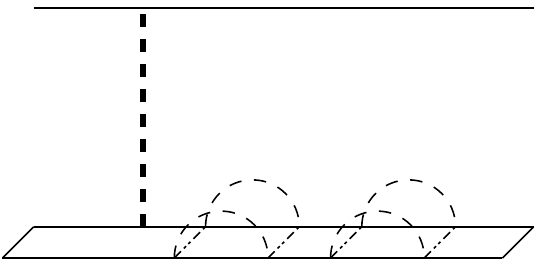}   \ ,  &   c(\mathcal{D}_{4})&=  4(n-1)\ ,    &  D_4&\equiv C_{19} \label{eD4}\\
&&&\medskip\nonumber\\
\mathcal{D}_{5} &= \largediagram{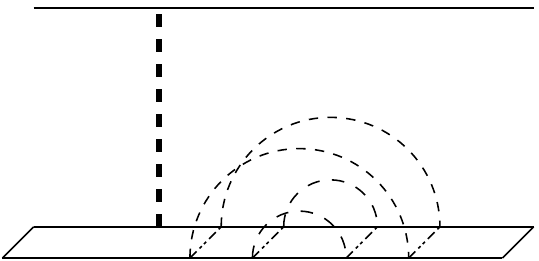}  \ ,  &   c(\mathcal{D}_{5})&=   4(n-1)\ ,    &  D_5&\equiv C_{18} \label{eD5} \\
&&&\medskip\nonumber\\
\mathcal{D}_{6} &= \largediagram{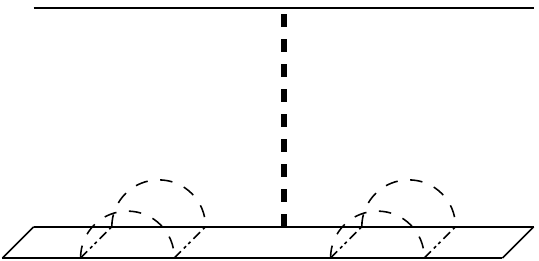}   \ ,  &   c(\mathcal{D}_{6})&=   2(n-1)\ ,     & D_6&\equiv C_{20} \label{eD6}\\
&&&\medskip\nonumber\\
\mathcal{D}_{7} &= \largediagram{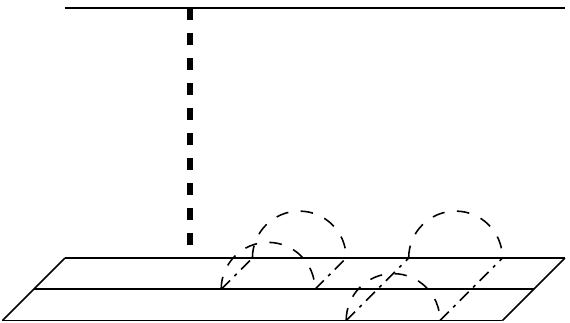}   \ ,  &  c(\mathcal{D}_{7})&=  4(n-1)(n-2) \ , &    D_7&\equiv C_{53} \label{eD7}\\
&&&\medskip\nonumber\\
\mathcal{D}_{8} &= \largediagram{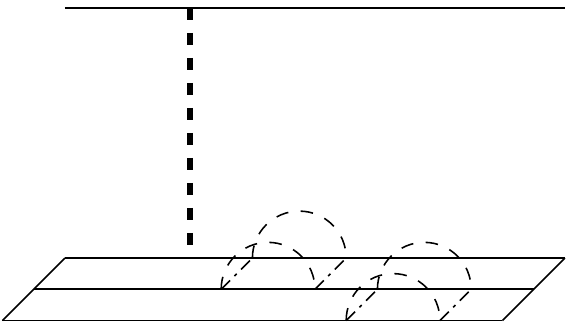}  \ ,  &    c(\mathcal{D}_{8})&=   4(n-1)(n-2) \ , &    D_8&\equiv C_{51} \label{eD8} \\
&&&\medskip\nonumber\\
\mathcal{D}_{9} &=\largediagram{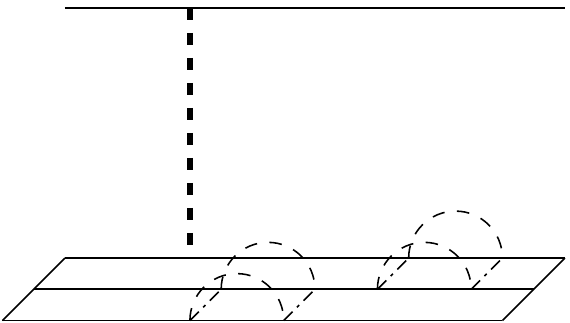}   \ ,  &   c(\mathcal{D}_{9})&=  4(n-1)(n-2) \ ,   &  D_9&\equiv C_{52} \label{eD9} \\
&&&\medskip\nonumber\\
\mathcal{D}_{10} &=\largediagram{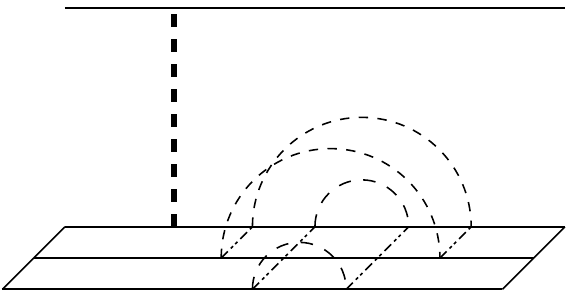}   \ ,  &  c(\mathcal{D}_{10})&=  4(n-1)(n-2)\ ,  &    D_{10}&\equiv C_{40} \label{eD10}\\
&&&\medskip\nonumber\\
\mathcal{D}_{11} &=\largediagram{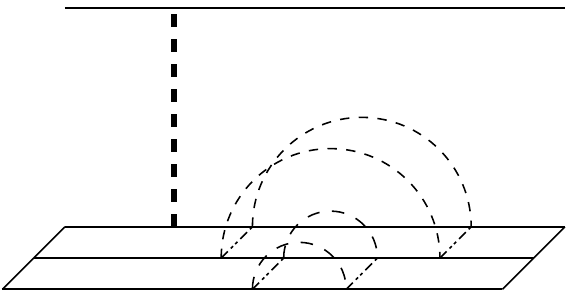}   \ ,  &  c(\mathcal{D}_{11})&=  4(n-1)(n-2)\ ,   &   D_{11}&\equiv C_{39} \label{eD11} \\
&&&\medskip\nonumber\\
\mathcal{D}_{12} &=\largediagram{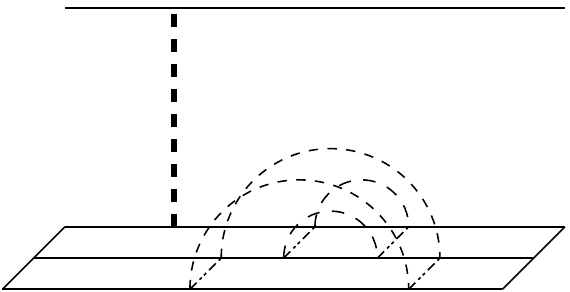}   \ ,  &  c(\mathcal{D}_{12})&= 4(n-1)(n-2)\ ,     & D_{12}&\equiv C_{38} \label{eD12}\\
&&&\medskip\nonumber\\
\mathcal{D}_{13} &=\largediagram{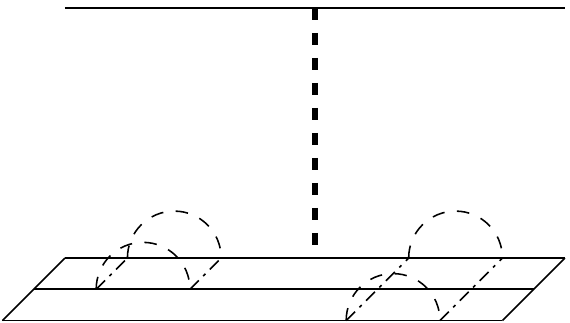}  \ ,  &   c(\mathcal{D}_{13})&= 2(n-1)(n-2)\ ,     & D_{13}&\equiv C_{54} \label{eD13}\\
&&&\medskip\nonumber\\
\mathcal{D}_{14} &=\largediagram{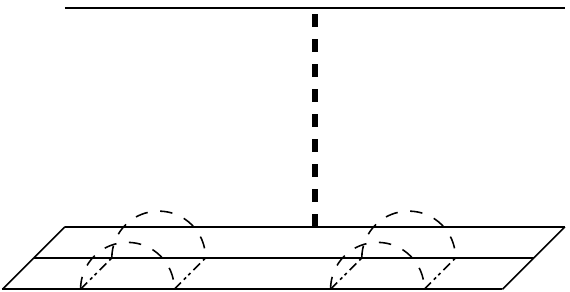}  \ ,  &   c(\mathcal{D}_{14})&=  (n-1)(n-2)\ ,  &    D_{14}&\equiv C_{36} \label{eD14}\\
&&&\medskip\nonumber\\
\mathcal{D}_{15} &=\largediagram{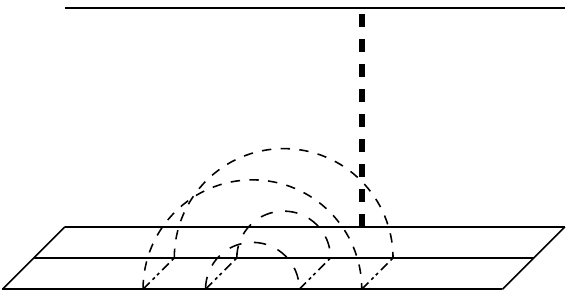}   \ ,  &  c(\mathcal{D}_{15})&= (n-1)(n-2) \ ,   &  D_{15}&\equiv C_{37} \label{eD15}\\
&&&\medskip\nonumber\\
\mathcal{D}_{16} &=\largediagram{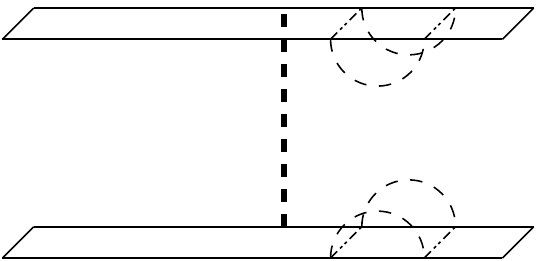}  \ ,  &   c(\mathcal{D}_{16})&= 4(n-1)^2 \ ,   &  D_{16}&\equiv C_{31} \label{eD16}
\end{align}
\begin{align}
\mathcal{D}_{17} &=\largediagram{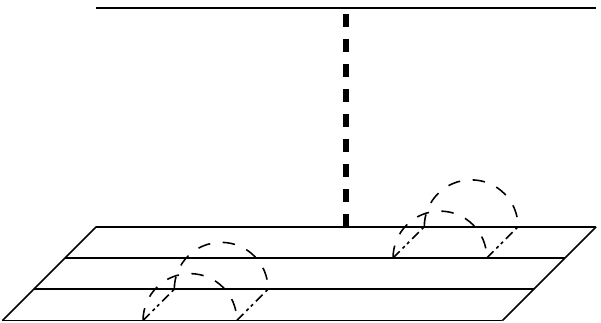}  \ ,  &   c(\mathcal{D}_{17})&= 2(n-1)(n-2)(n-3)\ , &     D_{17}&\equiv C_{45}\label{eD17} \\
\mathcal{D}_{18} &=\largediagram{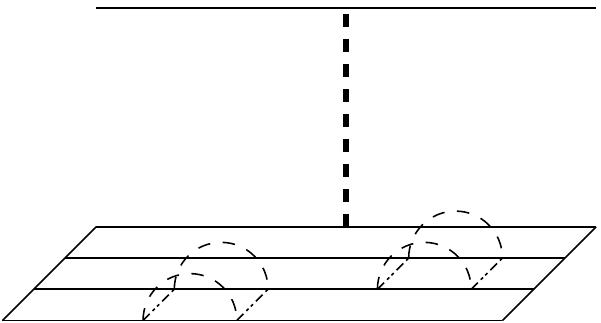}  \ ,  &   c(\mathcal{D}_{18})&=   2(n-1)(n-2)(n-3)\ , &     D_{18}&\equiv C_{46} \label{eD18}\\
\mathcal{D}_{19} &=\largediagram{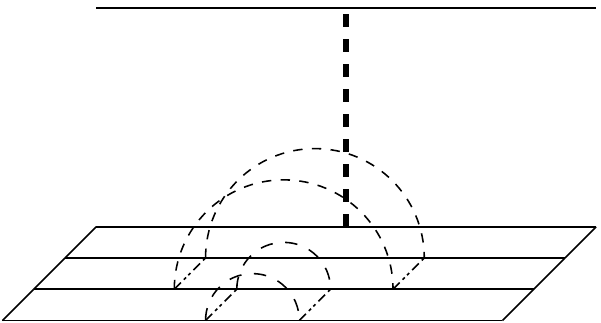}  \ ,  &   c(\mathcal{D}_{19})&= 2(n-1)(n-2)(n-3) \ ,  &   D_{19}&\equiv C_{47} \label{eD19}\\
\mathcal{D}_{20} &=\largediagram{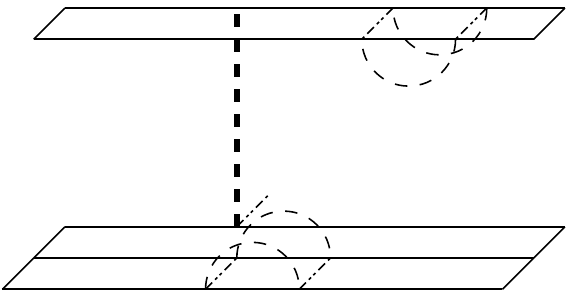}  \ ,  &   c(\mathcal{D}_{20})&= 2(n-1)^2(n-2)  \ ,  &  D_{20}&\equiv C_{43} \label{eD20} \\
\mathcal{D}_{21} &=\!\!\!\!\!\!\largediagram{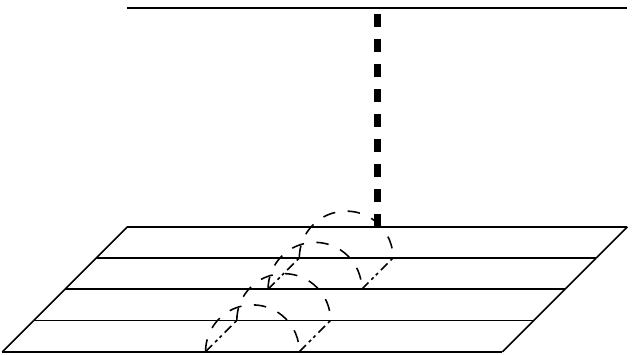}   \ ,  &  c(\mathcal{D}_{21})&= \frac{(n-1)(n-2)(n-3)(n-4)}4\ , &   D_{21}&\equiv C_{48} \label{eD21}\\
\mathcal{D}_{22} &=\largediagram{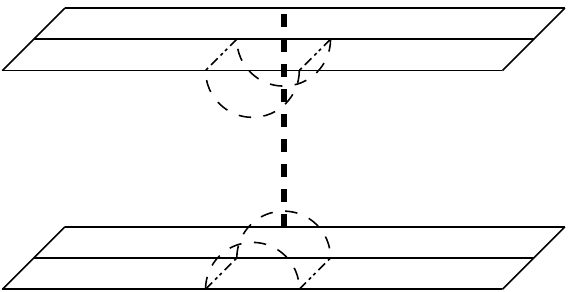}  \ ,  &   c(\mathcal{D}_{22})&= \frac{(n-1)^2(n-2)^2}4 \ ,  &   D_{22}&\equiv C_{44}  \label{eD22}
\end{align}

 \clean
\newcommand{\qvecB}{{\qvec_{\sss{B}}}}

\section{RG-functions and results}
\label{s:RG2L}
\subsection{Calculation of the UV poles and finite parts at 2-loop order}
\label{ss:Poles2L}

In the last section, we have computed the first two terms in the small-momentum expansion (in powers of the external momenta $\qvec_{\sss{B}}$) of the bare (unrenormalized) single-open-strand partition function $\widetilde{\mathcal{Z}}^{(1)}_{B}(\qvecB;\gB,L)$, as well as the  two-open-strand  partition function in the $Q=2$ replica sector $\widetilde{\mathcal{Z}}^{(2,2)}_{B}(\qvecB_i;\gB,L)$
at zero momenta ($\qvecB_i=0$). We consider here the fixed length $L$ ensemble and normalize the strand length $L$  to
\begin{equation}
\label{L=1sec6 }
L=1
\end{equation}
to simplify the calculations.
This is equivalent to deal with the dimensionless momenta $\qvec '=\qvec\, L^{1/2}$ (see section \ref{ss:2Lpresent}).
To order $\gB^2$ and order $\qvecB^{\!\!\!\!2}$ this single-strand partition function reads
\begin{equation}
\label{Z1Bexpq}
\widetilde{\mathcal{Z}}^{(1)}_{B}(\qvecB;\gB)=\mathbb{A}(\gB)+\qvecB^{\!\!\!\!2}\ \mathbb{B}(\gB)+\mathcal{O}(\qvecB^4)
\end{equation}
with
\begin{equation}
\label{AexpBexp}
\mathbb{A}(\gB)=\mathbb{A}_0+\gB\,\mathbb{A}_1+\gB^2\,\mathbb{A}_2+\mathcal{O}(\gB^3)
\ ,\quad
\mathbb{B}(\gB)=\mathbb{B}_0+\gB\,\mathbb{B}_1+\gB^2\,\mathbb{B}_2+\mathcal{O}(\gB^3)
\end{equation}
The two-strand partition function is
\begin{equation}
\label{Z2Bexpq}
\widetilde{\mathcal{Z}}^{(2,2)}_{B}(\qvecB_i;\gB)=\mathbb{C}(\gB)+\mathcal{O}(\qvecB^2)
\end{equation}
with
\begin{equation}
\label{Cexpbb}
\mathbb{C}(\gB)=\gB \mathbb{C}_1+\gB^2\,\mathbb{C}_2+\gB^3\,\mathbb{C}_3+\mathcal{O}(\gB^4)\ .
\end{equation}
The amplitudes $\mathbb{A}_{k}$, $\mathbb{B}_{k}$ and $\mathbb{C}_{k}$ depend on the dimension $d$, and have UV poles at $\varepsilon=0$.
They also depend on the number of replica $n$.
They are given by the sum of the diagrams calculated in section \ref{s:2loops}, more precisely they are given by
\begin{equation}
\label{ ABCsumDiag}
\mathbb{A}_{k}=\hskip-1.em\sum_{\text{order $k$ diagrams}} \hskip-1.emc(\mathcal{E}_j) A^L_j
\quad,\qquad
\mathbb{B}_{k}=\hskip-1.em\sum_{\text{order $k$ diagrams}}\hskip-1.em c(\mathcal{E}_j) B^L_j
\quad,\qquad
\mathbb{C}_{k}=\hskip-1.em\sum_{\text{order $k$ diagrams}}\hskip-1.em c(\mathcal{C}_j) C^L_j
\end{equation}
We only need the poles and the finite part of these amplitudes. We find
\begin{align}\label{m3}
\mathbb A_0=&1\ ,\qquad \mathbb B_0=-n\\
\label{m4}
\mathbb A_1=&{n(n-1)\over 2}{1\over \varepsilon}-{n(n-1)\over 2}+\mathcal{O}(\varepsilon)\\
\mathbb B_1=&-{n(n-1)(n+2)\over 2}{1\over \varepsilon}+{n(n-1)(n-2)\over 2}+\mathcal{O}(\varepsilon)
\label{m4b1}\\
\label{m5}
\mathbb    A_2=&   {n(n-1)(n-2)(n+5)\over 8}{1\over\varepsilon^2}+{n(n-1)(-n^2-6n+14)\over 4}{1\over\varepsilon} \nonumber \\
& +{n(n-1)(3n^2+(111-4\pi^2)n-210+8\pi^2)\over 24}+\mathcal{O}(\varepsilon)
\\
\label{m6}
\mathbb    B_2=&   {n(n-1)(n+2)(-n^2-5n+12)\over 8}{1\over\varepsilon^2}+{n(n-1)(n^3+6n^2-20n+16)\over 4}{1\over\varepsilon}\nonumber\\
&+{n(n-1)[-3n^3+(-99+4\pi^2)n^2-54n+(456-26\pi^2)]\over 24}+\mathcal{O}(\varepsilon)\\
 \mathbb C_1=&{n(n-1)\over 2}
\label{m4c1}\\
\label{m7}\mathbb C_2=& \frac{\left( -1 + n \right) \,n\,\left( -7 + 3\,n +
n^2 \right) }{2}{1\over\varepsilon} -\frac{ \left( -1 + n \right)
\,n\,\left[ 11 - 10\,n + 2\,n^2 + \ln (16) \right]
}{4}+\mathcal{O}(\varepsilon) \\
\label{m8} 
\mathbb C_3=&\frac{n (n-1)}{4}\times \bigg\{ ( -7 + 3 n + n^2 ) (
-12 + 5 n + n^2)
{1\over\varepsilon^2} \nonumber  \\
      &- \left[  -104 + 142 n - 58 n^2 + 3 n^3 + 2 n^4 + n^2 \ln (16)
- 8 \ln (64) + 2 n\, \ln (1024) \right] {1\over\varepsilon} \nonumber
\\
&+ \frac{1}{6} \Bigl[ 2238 - 2091 n + 231 n^2 + 102 n^3 + 18 n^4 - 140
{\pi }^2 +
      137 n \pi ^2 - 26 n^2 \pi ^2  - 4 n^3 {\pi }^2 \nonumber\\
      & 
      + 504 \ln (2) - 276 n\, \ln (2) - 
      360 \ln (2)^2 + 180 n \ln (2)^2 + 12 n^2 \ln (2)^2 + n^2 \ln (4096) \Bigr]
      \nonumber \\
      &+\mathcal{O}(\varepsilon)   \bigg\}   
\end{align}
Finally, the partition function $\widetilde{\mathcal{Z}}_\Phi^{(2)}$ with one $\Phi$ operator defined by (\ref{ZPhi2}) is given in the bare theory by
\begin{equation}
\label{6s1}
\left.{\widetilde{\mathcal{Z}}_\Phi^{(2)}}\right._{\!\!\mathrm{B}}=\mathbb{D}(\gB)=\mathbb{D}_0+\mathbb{D}_1\,\gB+\mathbb{D}_2\,\gB^2+\mathcal{O}(\gB^3)\ ,
\end{equation}
with
\begin{align}
    \mathbb{D}_0=&  1 \label{equD0}\\
    \mathbb{D}_1=& {(n-1)(n+2) } {1 \over \,\varepsilon}+{(n-1)(2-n)} + \mathcal{O}(\varepsilon)  \label{equD1}\\
    \mathbb{D}_2=&  \frac{1}{2} (n-1) (n+2) \left(n^2+3 n-7\right) {1\over \varepsilon^2}
    -\frac{1}{2} (n-1) \left(2 n^3+5 n^2-24 n+20\right){1\over \varepsilon}\nn\\
    &+\frac{1}{12} (n-1) \left(18 n^3-4 \pi ^2 n^2+96 n^2-7 \pi ^2 n-6 n+26 \pi^2-348\right)
    +\mathcal{O}(\varepsilon)
    \label{equD2}
\end{align}

\subsection{Counterterms and RG functions in the MS scheme}
\label{ss:CT2L}
\subsubsection{Counterterms}
We use the minimal subtraction (MS) scheme described in sections \ref{ss:rsBsRO} and \ref{MSanDim}.
The one- and two-strand partition functions are renormalized according to (\ref{ZB1oZR1o}) and (\ref{Z2Qscale}) into $\widetilde{\mathcal{Z}}^{(1)}_{\mathrm R}(\qR;\gR)$ and $\widetilde{\mathcal{Z}}_{\mathrm R}^{(2,Q)}(\qR\cdots;\gR)$, with the coupling constant $g$ and the momenta $\qvec$ renormalized by (\ref{rBgBdef}) and (\ref{qBqR}) and the renormalisation factors for the field $\mathbb{Z}$, the coupling constant $\mathbb{Z}_g$ and the boundary $\mathbb{Z}_1$ of the form (\ref{ZCT}), (\ref{ZgCT}) and (\ref{Z1CT}).
This means that the renormalized coupling constant is related to the bare coupling constant by
\begin{equation}
\label{gBgRagain}
\gB=\gR\, \mathbb{Z}_g(\gR)\,\mathbb{Z}(\gR)^{2+\varepsilon}\ \mu^{-\varepsilon}\ ,
\end{equation}
$\mu$ being the renormalization mass scale, and that the three amplitudes
\begin{equation}
\label{ABCren}
\mathbb{A}(\gB)\,\emath^{-2n\mathbb{Z}_1(\gR)}
\ ,\quad
\mathbb{B}(\gB)\,\mathbb{Z}(\gR)\,\emath^{-2n\mathbb{Z}_1(\gR)}
\ ,\quad
\mathbb{C}(\gB)\,{\mathbb{Z}(\gR)}^{-(2+\varepsilon)}\,\emath^{-4n\mathbb{Z}_1(\gR)}
\end{equation}
considered as series in the renormalized coupling constant $\gR$, must be UV finite at $\varepsilon=0$.
In the minimal subtraction scheme (MS) the counterterms $\mathbb{Z}$, $\mathbb{Z}_g$ and $\mathbb{Z}_1$ are chosen to have only poles in $1/\epsilon$.
Putting together (\ref{AexpBexp}), (\ref{Cexpbb}), (\ref{m3})-(\ref{m8}), (\ref{gBgRagain}) and (\ref{ABCren}), we obtain for the counterterms
\begin{align}\label{m9}
{\mathbb{Z}}(\gR)&=1+\gR  \frac{(n-1)}{\epsilon} +
\gR ^{2}\frac{(n-1)}{2}\left(\frac{4-n}{\epsilon^{2}}+\frac{2-n}{\epsilon}
\right) + {\cal O} (\gR ^{3}) \\
{\mathbb{Z}}_{g}(\gR)&= 1+ \gR \frac{(7-4n)}{\epsilon}+ \gR ^{2}(n-2)
(4n-7)\left(\frac{3}{\epsilon^{2}}+\frac{1}{2\epsilon } \right)+ {\cal
O} (\gR ^{3})
\label{m10}\\
\label{m11} {\mathbb Z}_{1}(\gR)&= \gR  \frac{(n-1)}{4}- \gR ^{2}
\frac{(n-1)}{8}\left(\frac{2n-5}{\epsilon^{2}}+ \frac{n-2}{\epsilon }
\right)+{\cal O} (\gR ^{3})
\end{align}
Similarly, the amplitude 
\begin{equation}
\label{DampRen}
\mathbb{D}(\gB)\,\mathbb{Z}(\gR)^{-(1+\varepsilon/2}\,\mathbb{Z}_\Phi(\gR)\,\emath^{-4n\mathbb{Z}_1(\gR)}
\end{equation}
must be UV finite.
This fixes the $\mathbb{Z}_\Phi$ counterterms to be
\begin{equation}
\label{ZPhiCT2L}
\mathbb{Z}_\Phi(\gR) = 1+\frac{(1-n)}{\epsilon }\gR+\left(\frac{3 (n-2)
   (n-1)}{2 \epsilon ^2}+\frac{(n-1) }{4\epsilon }\right)
   \gR^2+\mathcal{O}\left(\gR^4\right)
\end{equation}

\subsubsection{RG functions in the MS scheme}
\label{sss:RGfunMS}
We now use the definition (\ref{betag}) of the $\beta_g$ function to obtain
\begin{equation}
\label{m13}
\beta_g(\gR)=-\varepsilon \gR+\left[(5-2n)+\varepsilon(n-1)\right]\gR^2+
(3-2n)\bigl[(5-2n)+\varepsilon(n-1)\bigr] \gR^3
+\mathcal{O}(\gR^4)
\end{equation}
Using (\ref{Deltar}) and (\ref{m19}) we obtain  the anomalous dimension for the $\rvec^2$ field
\begin{equation}
\label{gamma2L}
\gamma(\gR)=(1-n)\gR+(1-n)(3-2n)\gR^2+\mathcal{O}(\gR^3)\ ,
\end{equation}
and the anomalous dimension for the boundary operator $\mathbf{1}_b$
\begin{equation}
\label{gamma12L}
\gamma_1(\gR)={(1-n)\over 4}\gR+{(1-n)(3-2n)\over 4}\gR^2+\mathcal{O}(\gR^3)\ .
\end{equation}
Finally, using (\ref{DimPhi1}) we obtain the anomalous dimension for the $\Phi$ operator
\begin{equation}
\label{gammaPhi2L}
\gamma_\Phi(\gR)=\beta_g(\gR)\,{d\ln\mathbb{Z}_\Phi\over d \gR}=\gR (n-1)-\gR ^2{(n-1)(2n-1)\over 2}+\mathcal{O}(\gR ^3)\ .
\end{equation}
We remark that we still have at two loop
\begin{equation}
\label{gamgam1rel}
\gamma(\gR)=4\,\gamma_1(\gR)+\mathcal{O}(\gR^3)\ ,
\end{equation}
but we do not know if this is a general relation valid at all orders.

\subsection{Counterterms and RG functions in the $\overline{\mathrm{MS}}$ scheme}
\label{ss:CTMS'}
We now compute the RG functions in the $\overline{\mathrm{MS}}$ scheme.
The counterterms are $\mathbb{Z}$, $\overline{\mathbb{Z}}_g$ and $\mathbb{Z}_1$ with $\gB=\gR\overline{\mathbb{Z}}_g\mu^{-\varepsilon}$ and they are chosen to make the three amplitudes in (\ref{ABCren}) finite at $\varepsilon=0$.
We obtain
\begin{eqnarray}
\label{m79}
\mathbb{Z}&=&1+\frac{(n-1)}{\epsilon }\gR
-\left(
\frac{(n-4) (n-1)}{2\, \epsilon ^2}
+
\frac{(3 n-4)(n-1)}{2\, \epsilon }
\right)
\gR^2+\mathcal O(\gR^3)\\
\label{m80}
\overline{\mathbb{Z}}_{g}&=&1+\frac{(5-2 n) }{\epsilon }\gR
+\left(
\frac{(2 n-5)^2}{\epsilon^2}
+\frac{(2n-5)(2n-3)}{2\,\epsilon }
\right)
\gR^2+\mathcal O(\gR^3)\\
\label{m82}
\mathbb{Z}_{1}&=&\frac{(n-1)}{4 \epsilon }\gR
-\left(
\frac{(n-1)(2n-5)}{8\, \epsilon ^2}
+\frac{(n-1)(3n-4)}{8\, \epsilon }
\right)
   \gR^2+\mathcal O (\gR^3)\ .
\end{eqnarray}
We find for the beta function and the anomalous dimension in the $\overline{\mathrm{MS}}$ scheme
\begin{align}
\label{6s4}
\overline{\beta}_g(\gR)&=-\varepsilon\gR +(5-2n)\gR^2+(5-2n)(3-2n)\gR^3+\mathcal{O}(\gR^4) \\
\label{6s5}
\overline{\gamma}(\gR)&=(1-n)\gR+(1-n)(4-3n)\gR^2+\mathcal{O}(\gR^3) \\
\label{6saz}
\overline{\gamma}_{1} (\gR)&= \frac{1}{4} \gR (1-n)+\frac{1}{4} \gR^2 (1-n) (4-3n)+\mathcal O(\gR^3)\\
\label{6ser}
\overline{\gamma }_{\Phi} (\gR) &= \gR (n-1)-\frac{1}{2} \gR^2 (n-1)
(4 n-3)+\mathcal O(\gR^3)\ .
\end{align}

\subsection{Counter-terms and RG functions in the grand-canonical scheme }\label{ss:CTGrCa}
In section \ref{sss:GCanS}, we have seen that the perturbation
expansion in the grand-canonical scheme is in terms of the
dimensionless combination 
\begin{equation}\label{}
\tgB := \gB (\tau +n \qvec_{\mathrm{\tiny R}}^{2} /\mathbb{Z})^{\epsilon }
\end{equation}
Using this, we find in the grand-canonical scheme
\begin{eqnarray}\label{}
{\mathbb{A}}^{\tau } (\tgB) = 1&+&\tgB \left[\frac{(n-1) n}{2 \epsilon
   }-\frac{1}{2} (n-1) n+O(\epsilon
   )\right]\nn\\ &+& \tgB^2 \bigg[\frac{(n-2)
   (n-1) n (n+5)}{8 \epsilon ^2}-\frac{(n-1)
   n \left(n^2+6 n-14\right)}{4 \epsilon
   }\nn\\
&&\qquad +\frac{1}{48} (n-2) (n-1) n \left(\pi ^2
   n+18 n-3 \pi ^2+102\right)+\mathcal O(\epsilon
   )\bigg]+\mathcal O (\tgB^{3})\qquad \end{eqnarray}
\begin{eqnarray}
{\mathbb{B}}^{\tau } (\tgB) = -n &+&  \tgB \left[-\frac{(n-1) n}{\epsilon
   }+\mathcal O(\epsilon )\right]\nn\\&+&\tgB^2
   \bigg[-\frac{(n-1) n \left(n^2+2
   n-6\right)}{2 \epsilon ^2}+\frac{(n-1) n
   \left(2 n^2+n-4\right)}{2 \epsilon
   }\nn\\&&\qquad-\frac{1}{24} (n-1) n \left(2 \pi ^2
   n^2+12 n^2-11 \pi ^2 n-24 n+14 \pi
   ^2+24\right)+\mathcal O(\epsilon
   )\bigg]\nn\\&+& \mathcal O(\tgB^{3})
\end{eqnarray}
\begin{eqnarray}\label{}
{\mathbb{C}}^{\tau } (\tgB) &=& \frac{1}{2} \tgB (n-1) n+\tgB^2 \left[\frac{(n-1) n \left(n^2+3 n-7\right)}{2 \epsilon }-(n-1)^2
   n^2+\mathcal O(\epsilon )\right]\nn\\&&
+\tgB^3 \bigg[\frac{(n-1) n \left(n^2+3 n-7\right)
   \left(n^2+5 n-12\right)}{4 \epsilon ^2}\nonumber \\
&&\qquad -\frac{(n-1) n \left(4 n^4+19 n^3-63 n^2+15
   n+28\right)}{4 \epsilon }\nonumber \\
&&\qquad+\frac{1}{48} (n-1) n \Big(\pi ^2 n^4+90 n^4+2 \pi ^2 n^3+24
   n^3-49 \pi ^2 n^2-102 n^2+140 \pi ^2 n \nonumber \\
&&\qquad \qquad-504 n-116 \pi ^2+192\Big)+ \mathcal O(\epsilon
   )\bigg]+\mathcal O(\tgB^{4})\ .
\end{eqnarray}
Since this scheme will mostly serve as a check for the integrals in
the fixed-length scheme, we restrict ourselves to one possible
subtraction scheme, namely the  grand-canonical $\overline{\mbox{MS}}$ scheme. 
This gives the RG factors, the index $\tau$ indicating the
grand-canonical scheme, 
\begin{eqnarray}\label{}
\mathbb{Z}^{\tau }&=& 1 + \gR \frac{n-1}{\epsilon
   }+\gR^2
   \left[-\frac{(n-4) (n-1)}{2 \epsilon
   ^2}-\frac{(n-1) (3 n-4)}{2 \epsilon
   }\right] +\mathcal O(\gR^{3}) \\
\overline{\mathbb{Z}}_{g}^{\tau } &=& 1 + \gR \frac{5-2 n}{\epsilon
   }+\gR^2
   \left[\frac{(2 n-5)^2}{\epsilon
   ^2}+\frac{(2 n-5) (2 n-3)}{2 \epsilon
   }\right] +\mathcal O(\gR^{3}) \\
\mathbb{Z}_{1}^{\tau } &=& \gR\frac{n-1}{4 \epsilon
   }+\gR^2
   \left[-\frac{(n-1) (2 n-5)}{8 \epsilon
   ^2}-\frac{(n-1) (3 n-4)}{8 \epsilon
   }\right] + \mathcal O(\gR^{3})\ .
\end{eqnarray}
The RG functions read
\begin{eqnarray}\label{}
{{\bar\beta}}^{\tau} (\gR)&=&- \gR \epsilon-\gR^2 (2 n-5)+\gR^3 (2 n-5) (2
   n-3)+\mathcal O (\gR^4)\\
\bar\chi^{\tau } (\gR) &=& \frac{1}{2}+\frac{1}{2} \gR (1-n)+\frac{1}{2} \gR^2 (n-1)
(3 n-4)+\mathcal O(\gR^3) \\
\bar\gamma_{1}^{\tau } (\gR) &=& \frac{1}{4} \gR (1-n)+\frac{1}{4} \gR^2 (1-n) (4-3n)+\mathcal O(\gR^3) \ .
\end{eqnarray}
They are identical to (\ref{6s4})-(\ref{6ser}),  confirming our calculations in
the fixed-length scheme.

\subsection{Fixed point and critical exponents}
\label{FPexp2L}
We now analyze the RG flow within an $\varepsilon$ expansion as in LW \cite{LaessigWiese2005} and in our letter \cite{DavidWiese2006}. First, we give all results at arbitrary $n$, before specifying to $n=0$ in the next subsection. 
For small $\varepsilon>0$ and $n>-5/2$, thus especially for the limit of interest $n\to 0$, there is an IR unstable fixed point $g^*$.
In the $\overline{\mathrm{MS}}$ scheme this fixed point is at
\begin{equation}
\label{m15}
g^*={1\over 5-2n}\,\varepsilon
+{2n -3 \over (5-2n)^2}\,\varepsilon^2
+\mathcal{O}(\varepsilon^3)\ .
\end{equation}
This  {\em freezing transition} separates the molten (weak coupling) phase with $0<g<g^*$, where disorder is irrelevant, from the (strong coupling) glass phase with $g>g^*$, where disorder is strongly relevant.  
The anomalous dimensions at the fixed point give the critical exponents at the glass (freezing) transition.

The correlation-length exponent $\nu$ defined in eq.\ (\ref{a35b}) is 
\begin{equation}
\label{nu2Loop}
\nu^*=\frac{1}{\bar \beta_g'(g^*)}={1\over\varepsilon}+{2n-3\over 5-2n}+\mathcal{O}(\varepsilon)\ .
\end{equation}
The scaling dimension of the overlap operator, which gives the decay of the pair correlator, is
\begin{equation}
\label{m16}
\theta^*=2-\bar \beta_g'(g^*)=2-{1\over\nu^*}=2-\varepsilon
+{2n-3\over 5-2n}\,\varepsilon^2
+\mathcal{O}(\varepsilon^3)\ .
\end{equation}
It is in agreement with our result announced in \cite{DavidWiese2006} (note that \cite{DavidWiese2006} contains a sign typo in the intermediate identity of Eq.~(20), but the final result at order $\epsilon^2$ for $\theta^*$ is correct).
The anomalous dimensions of the $\rvec^2$ and contact operator $\Phi$ are respectively
\begin{equation}
\label{m17}
\gamma^*=\bar \gamma(g^*)={n-1\over 5-2n}\varepsilon
-\left({n-1\over 5-2n}\right)^2\,\varepsilon^2
+\mathcal{O}(\varepsilon^3)        
\end{equation}
\begin{equation}
\label{6s6}
\gamma_{\Phi}^*=\bar \gamma_\Phi(g^*)={n-1\over 5-2n}\varepsilon+{3 \over 2}{1-n\over (5-2 n)^2}\varepsilon^2+\mathcal{O}(\varepsilon^3)
\end{equation}
This gives the scaling dimension for $\rvec$,
\begin{equation}
\label{6s7}
\chi^*={1\over 2}-{\gamma^*\over 2}=\frac{1}{2}+\frac{n-1  }{4 n-10} \varepsilon+\frac{(n-1)^2
 }{2 (5-2 n)^2}\varepsilon ^2+\mathcal O(\varepsilon ^3)
\end{equation}
and the dimension of the contact operator $\Phi$,
\begin{equation}
\label{6s8}
\rho^*=1+\frac{\varepsilon}2+\gamma^*_\Phi=1+\frac{3 }{10-4 n}\varepsilon-\frac{3 (n-1) }{2 (5-2 n)^2} \varepsilon
   ^2+\mathcal O(\epsilon ^3)\ .
\end{equation}
Finally, our explicit calculation shows that the scaling relation  (\ref{m39a}), $\zeta^*=(2-\varepsilon)\chi^*$, holds at two loops. Indeed,
\begin{equation}
\label{6s9}
\zeta^* + \rho^* \to (2-\varepsilon)\chi^* + \rho^* =2 +{\cal O}(\epsilon^3)\ ,
\end{equation}
in agreement with the exact scaling relation  (\ref{rho+zeta=2}).

\subsection{Results for the disordered system ($n=0$)}
The physically relevant case is the zero-replica limit $n=0$. There we find
\begin{align}
\label{6s10}
    \nu^*& = {1\over\varepsilon}-{3 \over 5}+\mathcal{O}(\varepsilon)  \\
   \theta^* & = 2 -\varepsilon-{3\over 5}\,\varepsilon^2+\mathcal{O}(\varepsilon^3)\\
   \rho^* & =  1+{3\over 10}\varepsilon+{3\over 50}\varepsilon^2+\mathcal O(\varepsilon ^3 )\\
   \chi^* & =  \frac{1}{2}+\frac{1
   }{10}\varepsilon+\frac{1}{50}\varepsilon^2+\mathcal O(\epsilon ^3)
   \\
   \zeta^* &= 1-\frac{3 }{10} \epsilon -\frac{3 }{50} \epsilon ^2+{\cal O}(\epsilon
   ^3)\ .
\end{align}
The physical case for the RNA folding model corresponds to $\varepsilon=1$.

\subsection{Physical interpretation}\label{s:phys-interprete}

\begin{figure}[t]
\begin{center}\includegraphics[width=0.4\textwidth]{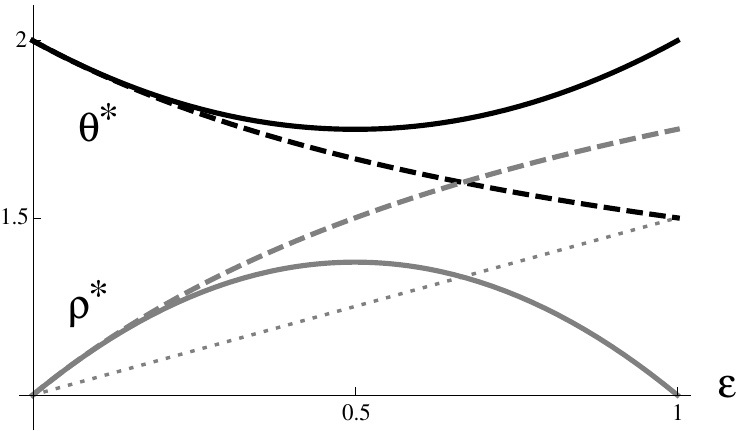} 
\includegraphics[width=0.4\textwidth]{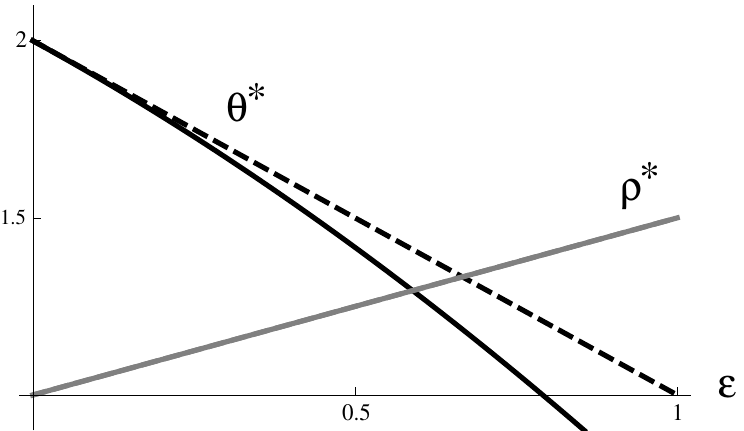} \\
{$n=2$\hspace*{0.38\textwidth}$n=1$}\medskip\\
\includegraphics[width=0.4\textwidth]{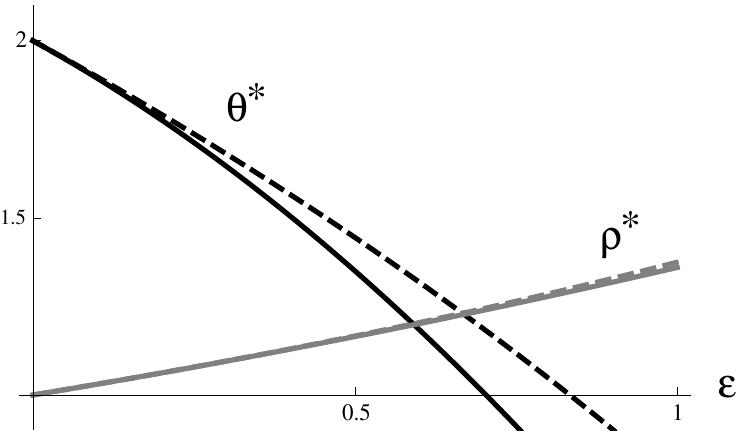} \\
$n=0$
\caption{Results for
$\theta^{*}$ (black) and $\rho^{*}$ (grey) at 1- (dashed) and 2-loop
(solid) order, for $n=2$, $n=1$, and $n=0$. The dotted line for $n=2$ is the value of $\rho_0$, which is an upper bound for $\rho^*$. For interpretation see main text.} 
\label{fig23}
\end{center}
\end{figure}
In fig.\ \ref{fig23}, we have plotted, for $n=2$, $n=1$ and $n=0$,  $\theta^*$ and $\rho^*$, as obtained through our 2-loop calculation (solid lines), as well as the resummed 1-loop results of \cite{LaessigWiese2005} (dashed lines)
\begin{equation}
\theta^*_{\mbox{\scriptsize LW}} = 2-\frac{\epsilon }{1+\frac{(n-1) \epsilon }{5-2 n}}\ , \qquad \rho^*_ {\mbox{\scriptsize LW}} =\frac{1+\frac{\epsilon }{2}+\frac{2 (n-1) \epsilon }{5-2 n}}{1+\frac{(n-1) \epsilon }{5-2 n}}\ .
\end{equation}
The latter are interesting, since they are exact for $n=2$, where they can be obtained by an exact solution of the corresponding self-consistency equations \cite{LaessigWiese2005, LaessigWieseToBePublished}. For $n\neq 2$, they are expected to be better than the pure 1-loop results, which are obtained by extrapolating $\theta^*$ and $\rho^*$ linearly around $\epsilon=0$. 

Consider now fig.\ \ref{fig23} for $n=2$. The dashed lines (resummed 1-loop results) are exact. However since $\rho^*>\rho_0 = 1+\frac{\epsilon}2$ (dotted line), the exponent $\rho^*$ gives only the subleading corrections to the contact-probability, and the dominant contribution comes from  the mean field exponent $\rho_0=1+{\epsilon/2}$  {(represented by the dotted line on fig.~\ref{fig23}, first graph)}, instead of $\rho^*$. 
Since base-pairings common to two replicas are a subset of base-pairings in one replica, the pairing probabilities satisfy
\begin{equation}
\left<\psi_{\alpha \beta}(s,t)\right> \le \left<\phi_\alpha(s,t)\right> 
\end{equation}
and as a consequence 
\begin{equation}\label{ineq}
\theta^*\ge \rho^*\ .
\end{equation}
Note that $\theta^* \ge \rho_0$, consistent with (\ref{ineq}), and the inequality is attained for $\epsilon=1$. One can show \cite{LaessigWieseToBePublished} that for $\epsilon>1$ both exponents take the value $\frac32$, and that there are logarithmic corrections at $\epsilon=1$. We also see that the 2-loop corrections are rather large, and a resummation appropriate. However, as we will see below, for the smaller $n$ we are interested in, this will not be necessary.

Consider now $n=1$. This case is peculiar, since the disorder  lives on pairs of distinct replicas, thus ``does not exist'' for $n=1$, as can be seen e.g.\ from the factor of $(n-1)$ in the contribution to the anomalous dimension $\gamma_\Phi$ in eq.\ (\ref{6s6}). As a consequence, $\rho^*=\rho_0$ exactly, which is respected by our perturbation theory. The $\beta$ function can nevertheless be defined, as the limit of $n\to 1$, resulting in  the exponent $\theta^*$ plotted in fig.~\ref{fig23}. In order to respect (\ref{ineq}), one has to set $\theta^*$ to $\rho^*=\rho_0$ as soon as the two curves cross, i.e.\ for $\epsilon>\epsilon_c=0.589$. We call $\epsilon_c$ the upper critical dimension, i.e.\ the dimension at which at the freezing-transition the two replicas get locked \cite{LaessigWiese2005}. 

We now arrive at the physically relevant case $n=0$. The first observation is that $\rho^*<\rho_0$, thus disorder makes the pairing probability more long-range correlated, and $\rho^*$ is the leading exponent for the latter  {(this is in fact true as soon as $n>1$)}.
As in the case $n=1$,  {beyond the  upper critical dimension}, i.e.\ $\epsilon>\epsilon_c = 0.592$, the inequality (\ref{ineq}) is violated. Following \cite{LaessigWiese2005}, we assume that the value of $\rho^*$  can again be extrapolated analytically beyond $\epsilon_c$, and that $\theta^*$ is given by $\rho^*$. We have no proof that this assumption is correct, but expect that due to the very character of the $\epsilon$-expansion, which states that the topology of fixed points and their corresponding exponents evolves smoothly upon a change in parameters (here $n$ from $n=1$ to $n=0$), the possible error is at least small.  
Last not least, we remark that 2-loop corrections (especially for $\rho^*$) are small, thus no resummation is necessary. These arguments lead to the following value for $\theta^*$ and $\rho^*$ at the transition
\begin{equation}
\theta^*=\rho^*=1.36\ .
\end{equation}
Our discussion above implies that at the transition the two replicas are already locked together, i.e.\ for a given disorder one possible fold dominates the partition function. This is not what we would expect. Indeed, at high temperatures, in the so-called molten phase, all possible folds are equally probable; and one expects that lowering the temperature, at the transition only a subset remains relevant, whereas at zero temperature a single configuration dominates. This naturally leads to three different sets of critical exponents: above, at, and below the transition. 
However, since our theory suggests that at the transition as in the low-temperature (glass) phase the same single configuration dominates the partition function, we are led to conjecture that in the glass phase
\begin{equation}\label{lock}
\theta^{\mathrm{glass}}=\rho^{\mathrm{glass}}=\theta^*=\rho^*\ .
\end{equation}
This ``locking scenario'' was first proposed by L\"assig and Wiese \cite{LaessigWiese2005}.
Our final result for   $\rho$ and $\zeta$ at the transition and in the glass phase is
\begin{equation}
\label{o1}
\rho_{\mathrm{glass}}=\rho^*= 1.37\pm 0.01, \quad \zeta_{\mathrm{glass}}=2{-}\rho^*= 0.63\mp 0.01\ ,
\end{equation}
where the central value and the reported error is an {\em estimate based on the three possible Pad\'es} for $\zeta^*(\epsilon)$,  the Pad\'es (2,0),  (1,1), and (0,2). The error due to the neglect of higher-order corrections is difficult to estimate, and not given here. 
Numerical results obtained by Krzakala et
al.~\cite{KrzakalaMezardMueller2002} in agreement with Bundschuh et
al.~\cite{BundschuhHwa2002a,BundschuPrivateCommunication} give
\begin{equation}\label{o2}
\rho_{\mathrm{glass}}\simeq 1.34\pm 0.003,\quad \zeta_{\mathrm{
glass}}\simeq0.67 \pm 0.02\ .
\end{equation}
These numerical results compare favorably well with those from the renormalization group, using the locking argument.

 \clean
\newcommand{\kvec}{\mathbf{k}}

\section{Random RNA under tension}
\label{s:tension}
An interesting question is what happens when a RNA molecule is pulled at its both ends. This problem has been studied in \cite{GerlandBundschuhHwa2001,KrzakalaMezardMueller2002,GerlandBundschuhHwa2003} both  numerically and analytically in the molten phase, and experimentally in \cite{LiphardtOnoaSmithTinocoBustamante2001}. 
In \cite{DavidHagendorfWiese2007a} a generalization of the field
theory was proposed for {\em random RNA} under tension, and the force-induced denaturation transition. 
In this model the pulling force $f$ is treated as a small perturbation. A RG calculation to first order in the force was performed to study the denaturation transition and its interplay with the disorder-induced freezing transition. 

In this section we recall the definition of the model,
show that it is renormalizable for $\varepsilon=0$ as for the tension-less
random RNA model, and give the details of the calculation of the RG
flow and of the critical exponents at two loops. For a
more detailed justification we refer to \cite{DavidHagendorfWiese2007a}, where  some of the results
were already announced.

\subsection{The model}
\label{ss:TenModel}

\begin{figure}[t]
\medskip
  \begin{center}
        \begin{tabular}{ccc}
      \includegraphics[scale=.75]{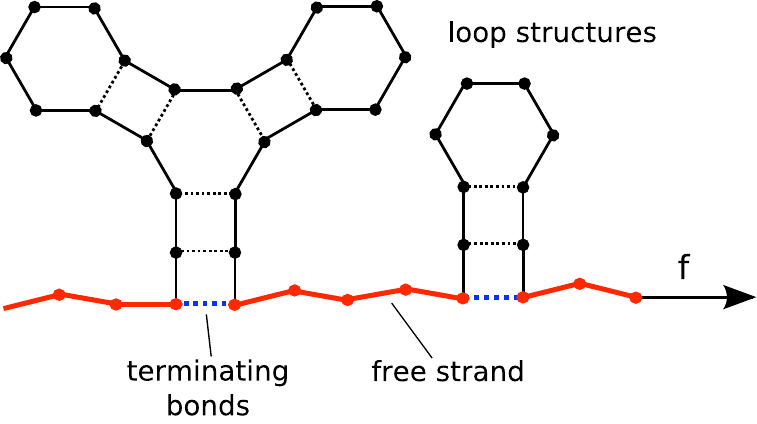}&~
      \raisebox{8.8mm}{\includegraphics[scale=.75]{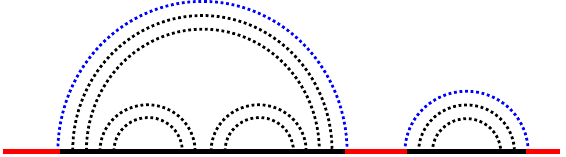}} &~
            \raisebox{8mm}{\includegraphics[scale=.75]{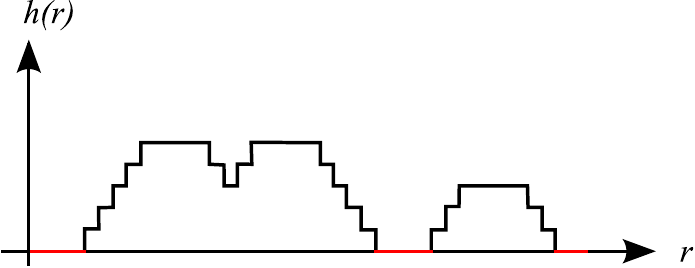}}\\
      (a) &~(b)&~ (c)
    \end{tabular}
  \end{center}
  \caption{(a) An open  planar RNA structure, (b) arch
  diagram and (c) corresponding height relief. The free part of the
  strand is in red, corresponding to height $h=0$.}
  \label{fig:structures}
\end{figure}

The discrete model for random RNA under tension is obtained by adding
to the disorder Hamiltonian $\mathcal{H}[{\Phi}]$ given by (\ref{Heffr2}) a new term proportional to the external force $f$ of the form
\begin{equation}
\label{Hforcei}
\mathcal{H}_{\sss{\mathrm{force}}}[\Phi] =-f\sum_\alpha\sum_i {\Delta}_\alpha(i)
\ .
\end{equation}
Here, the ``free-strand operator'' $\Delta_\alpha(i)$ is defined for each replica $\alpha$ and for each site $i$ as
\begin{equation}
\label{contacti}
{\Delta}_\alpha(i)=\begin{cases}
      1& 
      \text{if}\  h(i)=0
             \\
      0 & \text{otherwise}\ ,
\end{cases}
\end{equation}
where the height $h(i)$ of the site $i$ is defined via the
height representation (\ref{ph2}) of the planar pairing system,
\begin{equation}
\label{heighti}
h(i)=\sum_{j\le i}\sum_{k>i} \Phi(j,k)\ .
\end{equation}
The graphical interpretation of this operator is depicted on fig.~\ref{fig:structures}.

\begin{figure}[h]
\begin{center}
\includegraphics[width=1.25in]{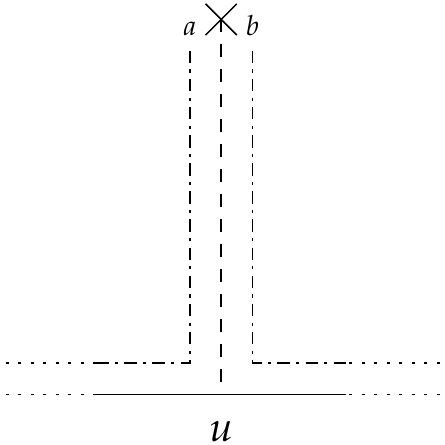}
\caption{Diagrammatic representation of the dressed force insertion operator $\boldsymbol{\Delta}_\alpha(u)$}
\label{f:forceInsV}
\end{center}
\end{figure}

In the continuum limit, this operator corresponds in the planar random-walk representation to the 1-point contact operator 
\begin{equation}
\label{conticontu}
\Delta_\alpha(i)\quad\to\quad
\boldsymbol{\Delta}_\alpha(u)={1\over N}\sum_{a,b} \gamma^\alpha_a(u)\,\tilde\delta^d\!\left(\rvec_\alpha(u)\right)\,\tilde\gamma^\alpha_b(u)
= {1\over N}\sum_{a,b}\Upsilon_{ab}(u)\,\tilde\delta^d\!\left(\rvec_\alpha(u)\right)\,
\end{equation}
which represents the interaction of the random walk $\rvec(s)$ with an impurity fixed at the origin $\rvec=\mathbf{0}$.
The dressing by the auxiliary fields $\gamma$, $\tilde\gamma$ (which are the same fields as in the previous sections) is introduced in order to eliminate  non-planar diagrams.
The force-insertion operator $\boldsymbol{\Delta}_\alpha(u)$ is represented diagrammatically on fig.\ \ref{f:forceInsV} by a single dashed line connecting the random walk (represented by the full line) to the impurity (represented by a cross).
The planar dressing by the auxiliary fields is represented as in the
force-free case by additional lines carrying  color indices $a$ ($a=1,\dots ,N$). Their role is to replace the lines by fat lines so that the planarity constraint is implemented in the limit of $N\to\infty$.

\begin{figure}[h]
\begin{center}
\begin{tabular}{ccc}
  \includegraphics[width=2in]{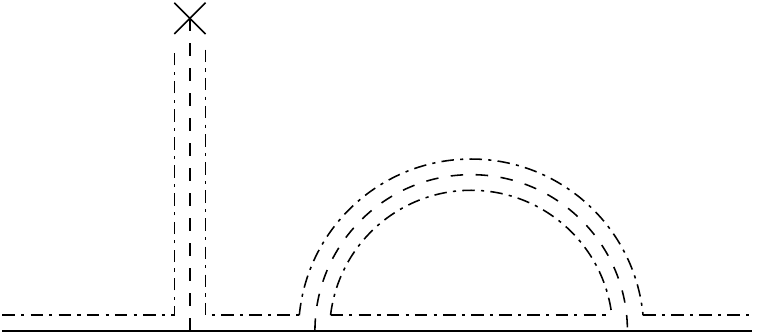} & & \includegraphics[width=2in]{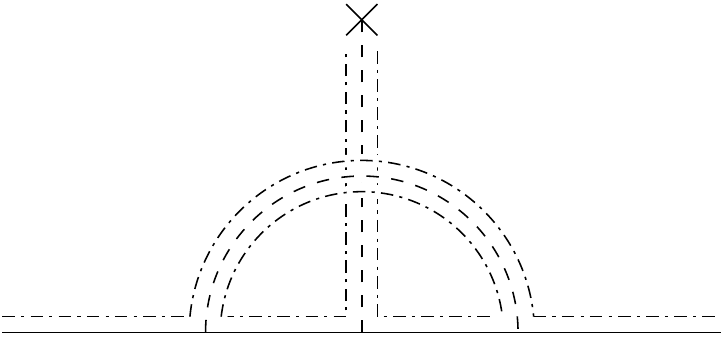}  \\
 (a)  & \qquad &  (b)
\end{tabular}
\caption{ (a) A force insertion on the free strand is represented by a planar diagram, of order $\mathcal{O}(1)$. (b) A non-planar diagram represents a force insertion on a non-free part of the strand and is of order $\mathcal{O}(N^{-1})$.}
\label{ForceDiagPlanar}
\end{center}
\end{figure}

This planarity constraint is required since the force term only acts on the free part of the RNA strand. 
It is easy to see that once we include  the disorder induced
interaction between replica, non-planar diagrams such as those of type
(b) in fig.~\ref{ForceDiagPlanar} represent force terms acting on the
non-free part of the RNA strand, since they are inside an interaction
arch. Only planar diagrams such as those of type (a) in
fig.~\ref{ForceDiagPlanar} are to be taken into account.

The force is thus taken into account in the action by the additional interaction term
\begin{equation}
\label{Sforce}
\mathcal{S}_{\sss{\mathrm{force}}}[\rvec,\gamma,\tilde\gamma]=-f_0\,\sum_\alpha\int _{0\le u\le L} \boldsymbol{\Delta}_\alpha(u)
\end{equation}
In the RW picture  it corresponds to  the addition of  an \emph{attractive short-ranged pinning potential} at the origin $\rvec=0$ for each polymer.
The potential is attractive because of the minus sign in the action (the pulling force is $f_0>0$).

The full action of the model with \emph{disorder and force} is 
\begin{equation}
\label{SS0SiSf}
\mathcal S=\mathcal{S}_{\sss{0}}+\mathcal{S}_{\sss{\mathrm{int}}}+\mathcal{S}_{\sss{\mathrm{force}}}
\end{equation}
and depends now on two coupling constants, the bare disorder-induced
coupling constant  $g_0>0$, and  the bare pulling force $f_0>0$.
The naive (engineering) dimension (in units of $L$) of the force-insertion operator $\boldsymbol{\Delta}_\alpha(u)$ and of the bare force $f_{\sss{0}}$ are respectively
\begin{equation}
\label{EngDimDelF}
\Delta^{\!\sss{0}}_{\boldsymbol{\Delta}}={d\over 2}={1+{\varepsilon\over 2}}
\quad,\qquad
\Delta^{\!\sss{0}}_{f}={-{\varepsilon\over 2}}
\end{equation}
The force $f$ is thus relevant   if $\varepsilon>0$, exactly as  the disorder coupling $g_0$.

Finally we consider  observables in the force model.
The length of the free open  strand in the discrete model (for a
single replica) is
\begin{equation}
\label{Lfree}
L_{\sss{\mathrm{free}}}=\sum_{i}\delta_{h(i),0}=\sum_i\Delta(i)\ .
\end{equation}
Therefore in the continuum model  it is
\begin{equation}
\label{ellfree}
\ell_{\sss{\mathrm{free}}}=\int_{0\le u\le L}\boldsymbol{\Delta}(u)\ ,
\end{equation}
and it has engineering dimension ${\varepsilon/2}$.
Secondly we consider partition functions for open strands, that we describe in the next section.

\subsection{Perturbation theory}
\label{ss:PertTen}
We can now build a systematic perturbation theory by expanding observables as a power series in both $g_0$ (the disorder strength) and $f_0$ (the pulling force). The diagramatics involves  both the two-replica bi-local vertex of fig.~\ref{s3-8}, representing the overlap operator $\Psi_{\alpha\beta}$ in 
$\mathcal{S}_{\sss{\mathrm{int}}}$, and the single-replica force-insertion vertex of fig.~\ref{f:forceInsV}, representing the operator $\boldsymbol{\Delta}_\alpha$ in  $\mathcal{S}_{\sss{\mathrm{force}}}$.
This last term breaks the translational symmetry of the $f=0$ model, since  it corresponds to the addition of an attractive short-ranged pinning potential at the origin $\rvec=0$ for each polymer.
Therefore one must be careful when dealing with the zero modes in perturbation theory.

\subsubsection{One strand partition function: Zero-mode decomposition}
We  consider the partition function for a single (bundle of $n$ replicas of an) open strand with length $L$ and external momenta $\qvec_1^\alpha$, $\qvec_2^\alpha$ for each replica $\alpha$. It is defined as
\begin{equation}
\label{m49}
\Xi_{(1)}(\qvec_1,\qvec_2;\qvec_0,f_0)=\int \mathcal{D}[\rvec_\alpha]\
\emath^{-\mathcal{S}[\rvec_\alpha]}\ \emath^{\imath\sum\limits_\alpha
\qvec_1^\alpha \rvec_\alpha(0)}\ \emath^{\imath\sum\limits_\alpha
\qvec_2^\alpha \rvec_\alpha(L)} \ .
\end{equation}
In the perturbative expansion of (\ref{m49}) in powers of $g_0$ and $f_0$, 
we must be careful how many different replicas get at least one force insertion, since as soon as a replica gets a force insertion vertex, no $\delta^d(\qvec_1+\qvec_2)$ occurs (since there are no translational zero modes any more). 
Therefore we decompose the partition function $\Xi_{(1)}(\qvec_1,\qvec_2;\qvec_0,f_0)$ in a sum over sectors labeled by the number $k$ of different replicas with at least one force insertion, and factor out for the $n-k$ remaining replicas the $\tilde\delta(\qvec_1+\qvec_2)$ factor coming out of the translational zero mode, so that
$0\le k\le n$:
\begin{equation}
\label{m50}
\Xi_{(1)}(\qvec_1,\qvec_2;g_0,f_0)=\sum_{k=0}^n\,\tilde\delta^d(\qvec_1+\qvec_2)^{(n-k)}\ \tilde\Xi_{(1,k)}(\qvec_1,\qvec_2;g_0,f_0)\ .
\end{equation}
Each sector $(k)$ gives a finite partition function 
$\tilde\Xi_{(1,k)}(\qvec_1,\qvec_2;g_0,f_0)$
and the translational zero modes are taken into account in the factors of $\tilde\delta^d(\qvec_1+\qvec_2)$.

One should keep in mind that, as in the force-free case discussed in the previous sections, each replica may carry different incoming external momenta $(\qvec_1^\alpha,\qvec_2^\alpha)$; for simplicity we take these momenta independent of the replica index $\alpha$.
This is why we must keep the dependence on the two momenta $\qvec_1,\qvec_2$ for  $\tilde\Xi_{(1,k)}$.
Note also that for the sector $k=0$ (no force insertions) we recover the partition function for the single free strand (no force)
\begin{equation}
\label{m51}
\tilde\Xi_{(1,k=0)}(\qvec_1,-\qvec_1;g_0,f_0)=\overline Z_{(1)}((\qvec_1;g_0)\ .
\end{equation}
As we shall see, only the sector $(k=1$) is in fact needed to study
the renormalisation of the model. It is of course possible to define
and study multi-strand partition functions in the presence of a force, and to treat properly the zero modes. Fortunately this is not necessary either.

\subsubsection{Diagrammatics and UV divergences}
We now compute the partition function $\tilde\Xi_{(1,k)}$ up to second order in the couplings $g_0$ (disorder) and $f_0$ (force).
The corresponding diagrams are detailed below. Their amplitudes are calculated in dimensional regularization using the same methods as in the force-free case. 
As discussed above, the force-insertion vertex is associated with a delta function in position space, which is represented by an external momentum-vertex insertion
\begin{equation}
\label{m47}
\boldsymbol{\Delta}_\alpha(u)\ \to\  \tilde\delta^d(\rvec_\alpha(u))= \int \frac{\rmd^d \kvec_\alpha}{\pi^{d/2}}\,\mathrm{e}^{\imath \kvec_\alpha r_\alpha(u)} \ .
\end{equation} 
As explained in \cite{DavidHagendorfWiese2007a}, using this representation for the force insertion vertex, together with that for the bilocal contact operator
\begin{equation}
\label{fi1}
\boldsymbol{\Xi}_\alpha(u,v)\ \to\ \tilde\delta^d(\rvec_\alpha(u)-\rvec_\alpha(v))= \int \frac{d^d \kvec_\alpha}{\pi^{d/2}}\,\mathrm{e}^{\imath \kvec_\alpha [\rvec_\alpha(u)-\rvec_\alpha(v)]} 
\end{equation}
taking the e.v. w.r.t. the free field $\rvec$, and then integrating over the momenta $\kvec$, we obtain an integral representation for the amplitudes of the diagrams in terms of the relative positions of vertices (the $u$'s and $v$'s) on the line $[0,L]$.

Let us look at a few simple cases to see how this works and where the UV divergences occur. The complete list of diagrams is given in section \ref{alsdk}.

The diagrams without disorder ($g=0$) and ($k=1$) (force insertions on one replica) are the diagrams  for the theory of a free polymer pinned by a single impurity. They are represented with the corresponding amplitude. $q_1$ and $q_2$ are the momenta entering at the both ends, and we drop the combinatorial factors (given in section \ref{alsdk}). The points are labeled $x$, $y$ and $z$ (if appropriate) from left to right.
\begin{align}
\label{g0fdiagrams}
\largediagram{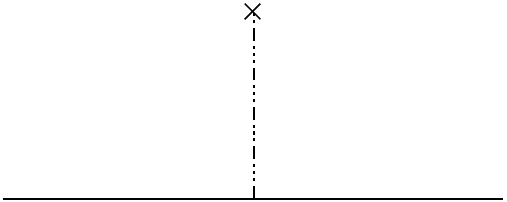}   &=  \int_{0<x<L}\emath^{-q_1^2x} \emath^{-q_2^2(L-x)}   \\
 \label{g0f2diag}\largediagram{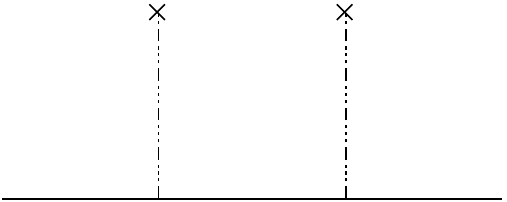}   &=  \int_{0<x<y<L}\emath^{-q_1^2x} \emath^{-q_2^2(L-y)}|y-x|^{-d/2} \\
\label{g0f3diag} \largediagram{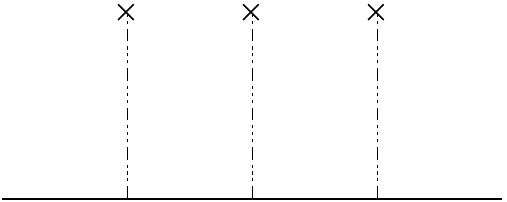} & = \int_{0<x<y<z<L}\emath^{-q_1^2x} \emath^{-q_2^2(L-z)}|y-x|^{-d/2}|z-y|^{-d/2}
\end{align}
The diagrams of order $f\times g$ are with the same convention for the points $x$, $y$ and $z$:
\begin{align}
\label{m53}
&\includegraphics[width=2in]{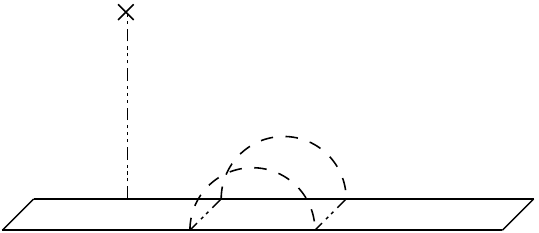}\nn\\
&\qquad= \int_{0<x<y<z<L} \emath^{-q_1^2x}\emath^{-q_2^2(L-z+y-x)}\emath^{-q_1^2y}\emath^{-q_2^2(L-z)}|z-y|^{-d}\delta^d(q_1+q_2)\\&\includegraphics[width=2in]{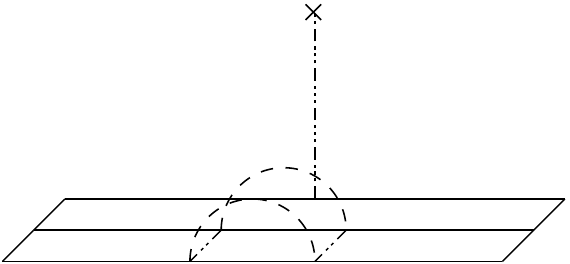}  \nn\\   & \qquad = \int_{0<x<L}\emath^{-q_1^2x} \emath^{-q_2^2(L-x)} \int_{0<y<z<L} \emath^{-2q_1^2y}\emath^{-2q_1^2(L-z)}|z-y|^{-d} \left[\delta^d(q_1+q_2)\right]^2\
\end{align}
(Here $y$ and $z$ are the points on the first two replicas, and $x$ is the point on the third replica.)

These amplitudes can be computed explicitly, and contain short-distance UV divergences (UV pole) at $\epsilon=0$.
For instance (\ref{g0f2diag}) contains a pole at $\epsilon=0$ coming from $y\to x$, which is proportional to the amplitude of (\ref{g0fdiagrams}). Therefore it is proportional to the insertion of a force operator $\boldsymbol{\Delta}(x)$, and can be absorbed into a renormalisation of the force $f$. 

Similarly, (\ref{m53}) contains (in addition to the poles coming from $z\to y$ and $\{y,\,z\}\to L$), a pole coming from $\{ y,\,z\}\to x$, which is proportional to (\ref{g0fdiagrams}), therefore also to a renormalisation of $f$.

These new UV divergences can be analyzed by the Multilocal Operator Product Expansion techniques already used for the tension-free model. One must consider the short-distance behavior of products involving both replica-interaction operators $\Psi_{\alpha\beta}(u,v)$ and force-insertion operators $\boldsymbol{\Delta}_\alpha(u)$.
We do not discuss this MOPE in more detail, but  mention that at first order, in addition to the standard MOPE,
\begin{equation}
\label{fi2}
\Psi\times\dot\rvec^2\to \mathbf{1} +\dot\rvec^2+\cdots
\ ,\quad
\Psi\times\Psi\to \mathbf{1} +\dot\rvec^2+\cdots
\ ,\quad
\Psi\times\Psi\to \Psi+\cdots
\ ,\quad
\Psi\times\mathbf{1}_b\to \mathbf{1}_b+\cdots
\end{equation}
the following additional terms appear:
\begin{equation}
\label{fi3}
\Psi\times\boldsymbol{\Delta}\to\boldsymbol{\Delta}+\cdots
\ ,\quad
\boldsymbol{\Delta}\times\boldsymbol{\Delta}\to\boldsymbol{\Delta}\ .
\end{equation}
This implies that there is a force renormalisation induced both by the force $f$ and by the disorder $g$, but that the force $f$ does not renormalise the disorder $g$.

\subsection{Renormalization}
\label{ss:RenTen}
\subsubsection{Renormalized action and beta functions}
The renormalized action for the model of open RNA strands under tension is thus (we omit the auxiliary fields $\gamma$ and $\bar\gamma$)
\begin{equation}
\label{m70}
\SR[\rR]=\sum_\alpha\int_L {\mathbb{Z}\over 4}\left.{\drR}\!\right._\alpha^2-\sum_{\alpha<\beta}\gR\,\mathbb{Z}_g\,\mu^{-\epsilon}\!\iint_L\Phi_{\alpha\beta}^\mathrm{R}+\sum_\alpha 2\ \mathbb{Z}_1-\sum_\alpha \fR\ \mathbb{Z}_f\,\mu^{-\epsilon/2}\int_L \boldsymbol{\Delta}_\alpha^{\mathrm{\!R}}\ .
\end{equation}
The counterterms $\mathbb{Z}=\mathbb{Z}(\gR)$, $\mathbb{Z}_g=\mathbb{Z}_g(\gR)$ and $\mathbb{Z}_1=\mathbb{Z}_1(\gR)$ are the same as those for the force-free model, since  they are not changed when $f>0$.
$\fR$ is the renormalized force and $\mathbb{Z}_f=\mathbb{Z}_f(\fR,\gR)$ is the new force-renormalisation factor. It depends both on $\fR$ and on $\gR$. 
For the disorder-free model $\gR=0$, one recovers the renormalized theory for the model of a random walk pinned by an impurity \cite{DDG1,DDG2,PinnowWiese2001,PinnowWiese2002a,PinnowWiese2004}.

The renormalized action can be written as a bare action $\SR[\rR]=\SB[\rB]$, with the same bare fields and bare couplings $\rB$ and $\gB$ as before, but with a bare force $\fB$ given by
\begin{equation}
\label{m72}
\gB=\gR\mathbb{Z}_g\mathbb{Z}^{d}\mu^{-\epsilon}\ ,
\qquad
\fB=\fR\mathbb{Z}_f\mathbb{Z}^{d/2}\mu^{-\epsilon/2}
\end{equation}
The renormalized  functions $\Xi^{\mathrm{R}}_{(1,k)}(\qvec_1,\qvec_2;\gR,\fR)$ calculated with the renormalized action are UV finite. They are given in terms of the bare functions $\Xi^{\mathrm{B}}_{(1,k)}(\qvec_1,\qvec_2;\gB,\fB)$ (calculated with the bare action) by 
\begin{equation}
\label{m71}
\Xi^R_{(1,k)}(\qvec_1,\qvec_2;\gR,\fR)=\mathbb{Z}^{-kd/2}\,\emath^{-2n\mathbb{Z}_1}\,\Xi^B_{(1,k)}(\mathbb{Z}^{-1/2}\qvec_1,\mathbb{Z}^{-1/2}\qvec_2;\gB,\fB)
\end{equation}
The factor $\mathbb{Z}^{-kd/2}$ comes from the zero modes. 

The RG beta function for the disorder coupling $\gR$, $\beta_g(\gR)$, is unchanged. In addition there is a RG beta function for the force, which is defined as
\begin{eqnarray}
\label{m73}
\beta_f(\fR,\gR):=-\mu\left.{\partial \fR\over\partial\mu}\right|_{\gB,\fB}&=& \fR
\left[-\frac\epsilon 2  -\frac d2 \beta_g(\gR) \frac{d \ln Z}{d \gR}-\beta_g(\gR)\frac{d \ln \mathbb Z_f}{d \gR}\right]\nn\\
&\equiv&
\fR\left[1-(2+\epsilon)\chi_\rvec(\gR)-\beta_g(\gR){d\ln \mathbb{Z}_f\over d\gR}\right]
\end{eqnarray}
\subsubsection{MS versus $\overline{\mathrm{MS}}$ schemes}
As for the force-free model, we consider the two different subtraction schemes MS and $\overline{\mathrm{MS}}$.
In the MS scheme, the counterterms $\mathbb{Z}$, $\mathbb{Z}_g$, $\mathbb{Z}_1$ and $\mathbb{Z}_f$ are chosen such that they contain only pure poles in $\epsilon$, and no finite or analytic term at $\epsilon=0$.
It is more convenient to consider the $\overline{\mathrm{MS}}$ scheme, where the coupling-constant renormalisation factor $\overline{\mathbb{Z}}_g=\mathbb{Z}_g\mathbb{Z}^{2+\epsilon}$ has only pure poles.
This scheme is easily extended to RNA under tension:
\begin{equation}
\label{fi4}
\overline{\mathbb{Z}}_f=\mathbb{Z}_f\mathbb{Z}^{1+\epsilon/2}
\qquad\text{such that}\quad\fB=\fR\,\overline{\mathbb{Z}}_f\,\mu^{-\epsilon/2}\ .
\end{equation}
In the  $\overline{\mathrm{MS}}$ scheme, it is $\overline{\mathbb{Z}}_f$, together with $\overline{\mathbb{Z}}_b$, which has only pure poles and no finite analytic part at $\epsilon=0$.
The $\overline{\beta}_f$ function becomes 
\begin{eqnarray}
\label{m73+}
\overline{\beta}_f(\fR,\gR):=-\mu\left.{\partial \fR\over\partial\mu}\right|_{\gB,\fB}&=& \fR
\left[-\frac\epsilon 2  -\beta_g(\gR)\frac{d \ln \overline{\mathbb Z}_f}{d \gR}\right]\ .
\end{eqnarray}

\subsection{Two-loop calculations}
\label{ss:2LTen}
\subsubsection{Principle}
Since we have already computed the boundary and the wavefunction counterterms (they are independent of $f_0$), it is enough to compute the bare $\Xi^B_{(1,k)}(\qvec_1,\qvec_2;\gB,\fB)$ function for $k=1$ (force insertions on a single replica) and at zero momenta $\qvec_1=\qvec_2=0$, and to isolate its UV poles at $\epsilon=0$, in order to compute the force counterterm $\mathbb{Z}_f$ and the RG functions for the force.  

Further simplifications  occur, since the contribution of the diagrams with a single force insertion  are related to the contributions of the corresponding diagrams with no force insertion at first orders in $\qvec$, which we have already calculated.
For instance
\begin{equation}
\label{m54}
\left.\includegraphics[width=1in]{G1-3a}+\includegraphics[width=1in]{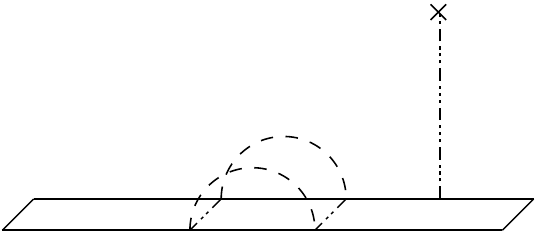}\right|_{q=0}=\left.-{1\over 2}{\partial\over\partial q^2}\includegraphics[width=1in]{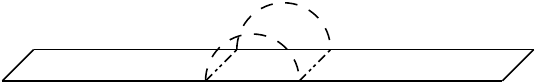}\right|_{q=0}\ .
\end{equation}
Similar identities hold for the two-loop diagrams.

\subsubsection{Diagrams}\label{alsdk}
With these remarks we now collate the diagrams $\mathcal{G}_i$, their symmetry factors $c(\mathcal{G}_i)$ and their amplitudes $G_i$ for the bare function $\Xi_{1,k=1}(\qvec_1=\qvec_2=0;\gB,\fB)$ (with the strand length set to $L=1$) up to two loops
\begin{equation}
\label{fi5}
\Xi_{(1,k=1)}(\qvec_1=\qvec_2=0;g,f)=\sum_{\mathrm{diagrams}\,\mathcal{G}_i}\gB^m\,\fB^n\,c(\mathcal{G}_i)\, G_i\ .
\end{equation}
When possible,
we express the new amplitudes $G_i$ in terms of the already calculated amplitudes $A_i$ and $B_i$ of the force-free model. We do not give results for the grand-canonical scheme.

\medskip

\leftline{\underline{Order $f_0$:}}
\begin{align}
\label{m55}
\mathcal{G}_{1}&=\largediagram{G1-1} \ ,\qquad c(\mathcal{G}_{1})=n\ , \qquad G_1= 1
\end{align}
\leftline{\underline{Order $f_0^{2}$:}}
\begin{align}
\label{m56}
\mathcal{G}_{2}=\largediagram{G1-2} \ ,\qquad c(\mathcal{G}_{2})=n\ , \qquad G_2&\equiv\int_u (1-u) u^{-d/2} \nonumber\\
&={1\over(2-d/2)(1-d/2)}
\end{align}
\leftline{\underline{Order $f_0 g_0$:}}
\begin{align}
\label{m57}
\mathcal{G}_{3}&=\largediagram{G1-3a}+\largediagram{G1-3b}\nn \\
c(\mathcal{G}_{3})&=n(n-1) \ ,\qquad G_3\equiv A_2-{1\over 2}B_2
\\
\mathcal{G}_{4}&=\largediagram{G1-4}\  ,\qquad c(\mathcal{G}_{4})=\frac{n(n-1)(n-2)}2\ ,\qquad G_{4}=G_1 A_2
\end{align}
\leftline{\underline{Order $f_0^3$:}}
\begin{align}
\label{m58}
\mathcal{G}_{5}&=\largediagram{G1-5}\ ,\qquad c(\mathcal{G}_{5})=n\ , \qquad G_5\equiv{\Gamma(1-d/2)^2\over\Gamma(4-d)}
\end{align}
\leftline{\underline{Order $f_0^2\times g_0$:}}
\begin{align}
\label{m59}
\mathcal{G}_{6a}&=\!\!\!\largediagram{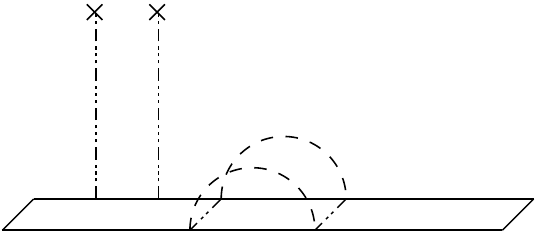}\!\!\ , \quad
c(\mathcal{G}_{6a})= n(n-1)\ , &G_{6a}\equiv& \frac12\int_{r+s+t=1}r^{-d}s^{-d/2}(1-r-s)^2\nonumber \\
& &  =& \frac{\Gamma (-1-\epsilon)\Gamma (-\epsilon /2)}{\Gamma (2-\frac{3\epsilon}{2})}\\
\mathcal{G}_{6b}&=\!\!\!\largediagram{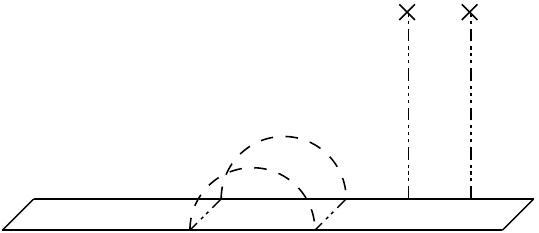}\!\!\ , \quad
c(\mathcal{G}_{6b}) = n(n-1)\ ,  &G_{6b}\equiv&G_{6a}\\
\mathcal{G}_{6c}&=\!\!\!\largediagram{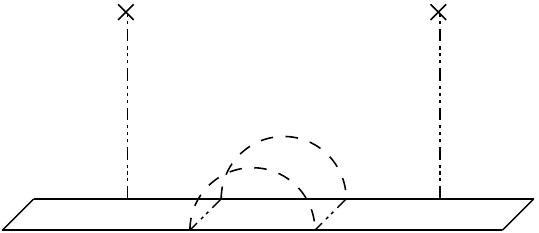} \!\!\ , \quad
c(\mathcal{G}_{6c})= n(n-1)\ , &G_{6c}\equiv& \int_{t+w+x=1}t^{-d}w^{1-d/2}x \nonumber\\
& &  =& {\Gamma(1-d)\Gamma(2-d/2)\over \Gamma(5-3d/2)}
\end{align}
\begin{align}
\mathcal{G}_{7}&=\!\!\! \largediagram{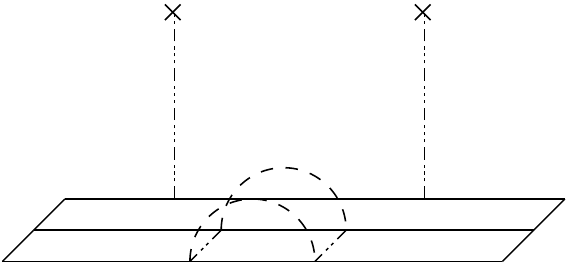}\ ,    & c(\mathcal{G}_{7})=\frac{n(n-1)(n-2)}2\ ,\qquad G_{7} \equiv G_2 A_2
\end{align}
\leftline{\underline{Order $f_0\times g_0^2$:}}
\begin{align}
\label{m60}
\mathcal{G}_{8}=& \largediagram{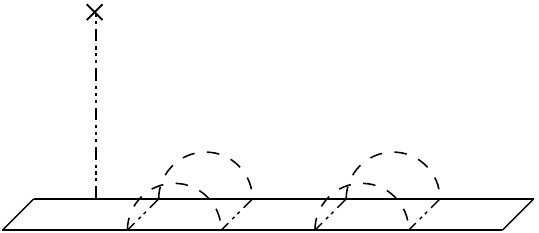}+
\largediagram{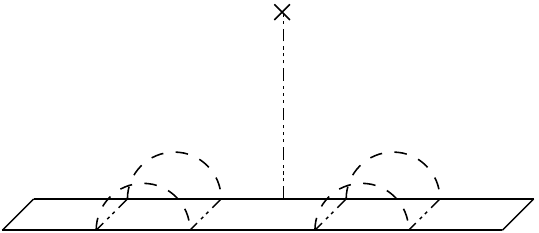}+
\largediagram{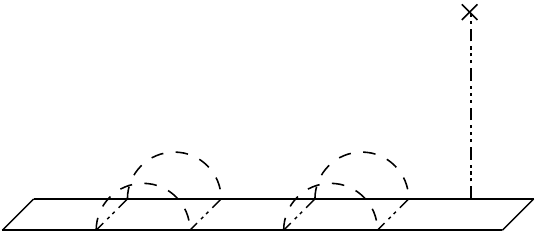} \nonumber\\
c(\mathcal{G}_{8})   &= n(n-1)\ ,\qquad G_8 \equiv A_3-B_3/2  \\
\label{m61}
\mathcal{G}_{9}=&\!\!\!\largediagram{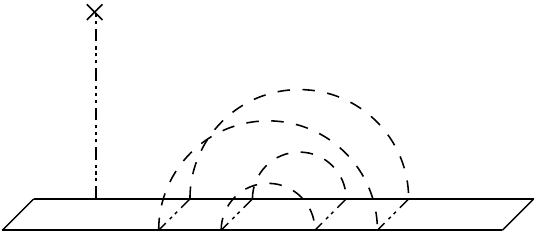}+
\largediagram{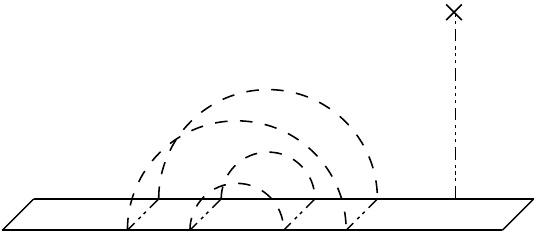} \nn \\
 c(\mathcal{G}_{9}) =&\, n(n-1)\ ,\qquad G_9\equiv A_4-B_4/2\\   
\label{m62}
\mathcal{G}_{10}=&~~\largediagram{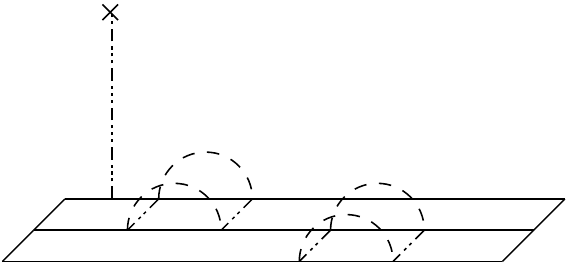}+
\largediagram{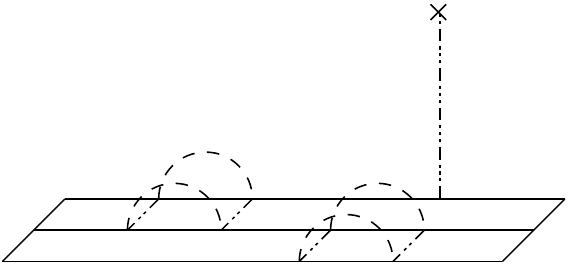}+
\largediagram{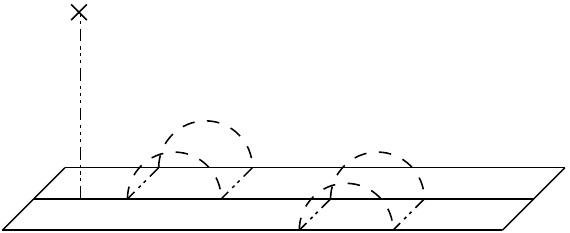}\nonumber\\ 
&+\!\!\!\largediagram{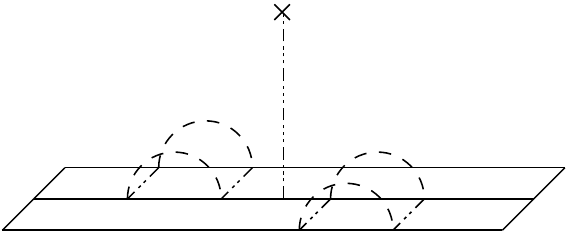}
+\largediagram{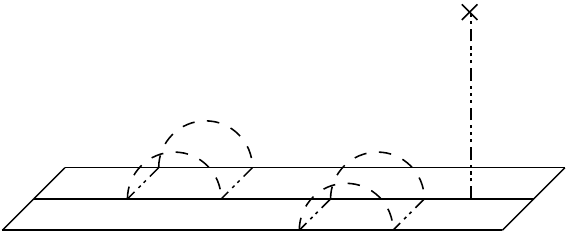}+
\largediagram{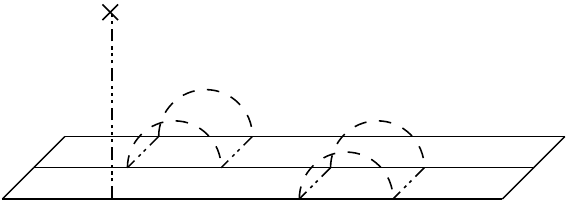}\nonumber \\
&+\!\!\!  
\largediagram{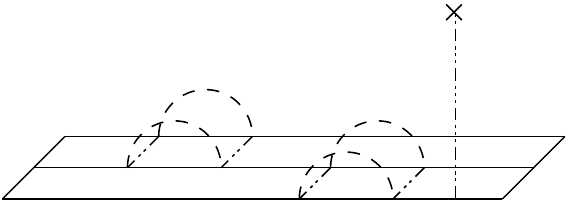} \ , 
\quad c(\mathcal{G}_{10})=n(n-1)(n-2)\ ,\quad  G_{10}\equiv 3A_5-B_5 
\\\label{m63}  
\mathcal{G}_{11}=&~~\largediagram{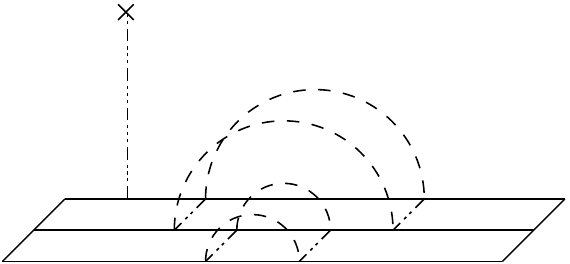}
+\largediagram{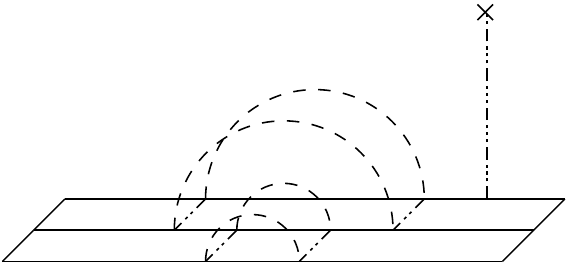}+
\largediagram{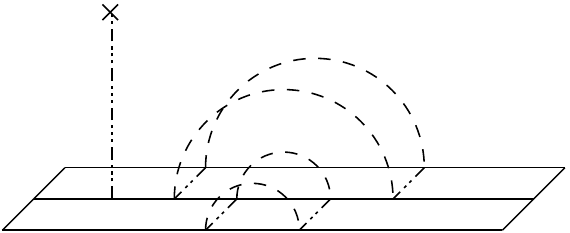}\\
&+\!\!\! \largediagram{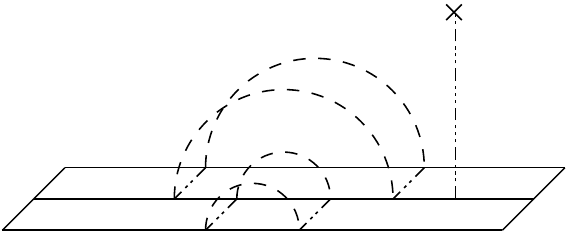}
+\largediagram{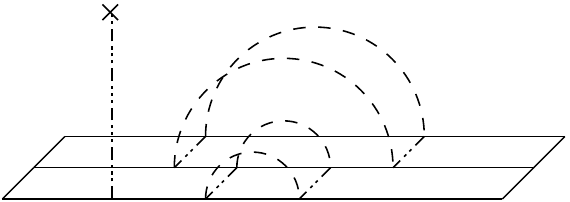} +
\largediagram{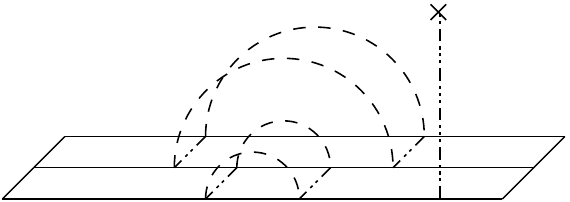} \nonumber  \\
c(\mathcal{G}_{11})=&\, n(n-1)(n-2)\ ,\qquad   G_{11}\equiv 3A_5-B_5 
\\\label{m64}
\mathcal{G}_{12}=&\largediagram{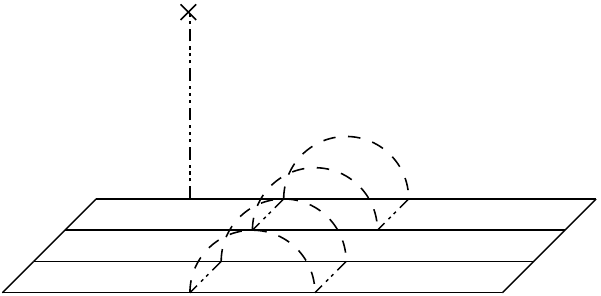}\ ,\quad  c(\mathcal{G}_{12})= \frac{n(n-1)(n-2)(n-3)}2\ ,  \quad
G_{12}\equiv A_2 G_3
\\\label{m65}
\mathcal{G}_{13}=&\largediagram{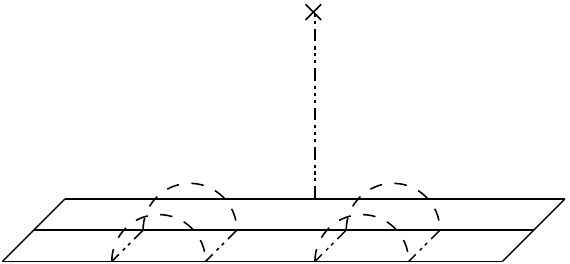}\ ,\quad  c(\mathcal{G}_{13})=\frac{n(n-1)(n-2)}2 \ ,\quad G_{13}\equiv G_1 A_3
\\
\label{m66}
\mathcal{G}_{14}=&\largediagram{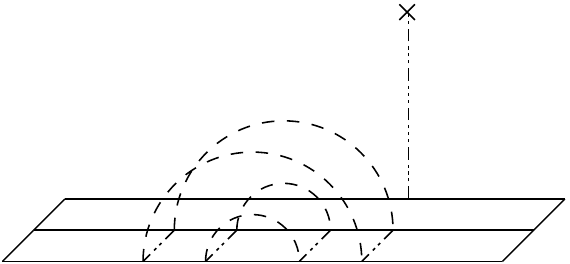}  \ ,\quad c(\mathcal{G}_{14})= \frac{n(n-1)(n-2)}2 \ , \quad G_{14}\equiv G_1 A_4
\\
\label{m67}
\mathcal{G}_{15}=&\largediagram{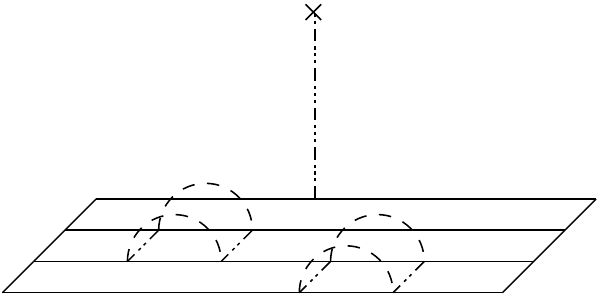}  \ ,\quad c(\mathcal{G}_{15})= n(n-1)(n-2)(n-3)  \ , \quad G_{15} \equiv G_1 A_5
\\
\label{m68}
\mathcal{G}_{16}=&\largediagram{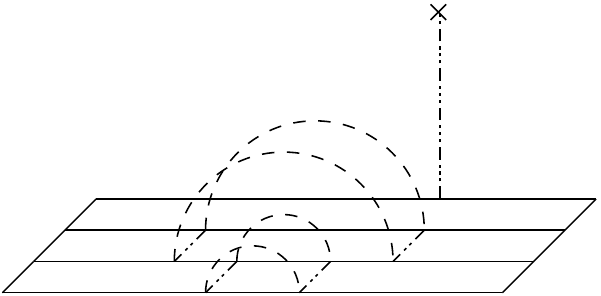}   \ ,\quad c(\mathcal{G}_{16})= n(n-1)(n-2)(n-3)  \ , \quad G_{16} \equiv G_1 A_6
\\
\label{m69}
\mathcal{G}_{17}=&\largediagram{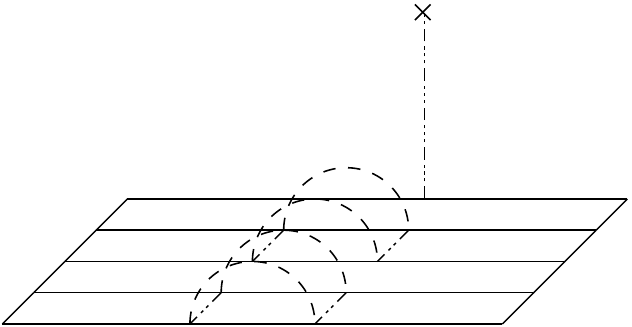} \ ,\quad c(\mathcal{G}_{17})=\frac{n(n-1)(n-2)(n-3)(n-4)}8  \nn \\
  G_{17}\equiv &G_1 A_7
\end{align}
Gathering the results we get
\begin{equation}
\label{fi6}
\Xi_{(1,k=1)}(\qvec_1{=}\qvec_2{=}0;\gB,\fB)=\mathbb{E}_{0,1}\fB+\mathbb{E}_{1,1}\gB \fB + \mathbb{E}_{0,2}\fB ^2
+\mathbb{E}_{1,2}\gB \fB^2 +\mathbb{E}_{2,1}\gB^1 \fB  + \mathbb{E}_{0,3}\fB ^3+\cdots
\end{equation}
with  (keeping the singular and finite part in $\epsilon$ for the amplitudes $\mathbb{E}_i$)
\begin{align}
\label{fi7}
  \mathbb{E}_{0,1}  &  =1\quad,\qquad \mathbb{E}_{0,2}=-{2\over\epsilon}-1+\cdots
 \quad,\qquad \mathbb{E}_{0,3}= \frac{4}{\epsilon^2}+\frac{4}{\epsilon}-\frac{\pi ^2}{6}+4+\cdots\\
 \mathbb{E}_{1,1}    & = \frac{n^2+n-2}{2 \epsilon}-\frac{1}{2} (n-2) (n-1)+\cdots\\\
 \mathbb{E}_{2,1}    &=\frac{(n-1) (n+2) (n (n+5)-12)}{8 \epsilon^2}-\frac{(n-1) (n (n (n+6)-20)+16)}{4
   \epsilon}\nonumber\\
   &+\frac{1}{24} (n-1) \left(3 n (n (3 n+31)-10)+\pi ^2 (26-n (4
   n+7))-312\right)+\cdots\\
  \mathbb{E}_{1,2}  &= -\frac{(n+2) (n-1)}{\epsilon^2}+\frac{(n-4) (n-1)}{2 \epsilon}-\frac{1}{6} \left(3 n-2
   \pi ^2+33\right) (n-1)+\cdots\
    \end{align}
    
\subsubsection{Beta functions}
Starting from the relation (\ref{m71}) between the bare and renormalized functions $\Xi_{(1,k=1)}$, using the already obtained two-loop expression for the counterterms $\mathbb{Z}$, $\mathbb{Z}_1$ and $\overline{\mathbb{Z}}_g$,  we are left with the determination of the force-renormalisation factor $\overline{\mathbb{Z}}_f$. This counterterm is determined by enforcing that the renormalized function $\Xi_{(1,k=1)}^{\mathrm{R}}$ has no poles in $\epsilon$.
In the $\overline{\mathbf{MS}}$ scheme this yields
\begin{equation}
\label{fi8}
\overline{\mathbb{Z}}_f(\gR,\fR)=1+\fR {2\over\epsilon} +\fR^2{4\over\epsilon^2} +\fR\gR {1-n\over \epsilon}+\gR^2 {(n-1)(2n-5)\over 4 \epsilon}+\cdots\ .
\end{equation}
We find for the 2-loop beta function for the force $\fR$ in the $\overline{\mathbf{MS}}$ subtraction scheme,
\begin{equation}
\label{m74}
\overline\beta_f(\gR,\fR)=-\frac{\epsilon}{2} \fR+ \fR^2 +\frac{3(1-n)}{2}\fR^2 \gR-\frac{(1-n)(5-2n)}{2}\fR \gR^2+\cdots\ .
\end{equation}
We recall that the beta function for the disorder coupling $\gR$ in this scheme is
\begin{equation}
\label{m75}
\overline\beta_g(\gR)=-\epsilon\,\gR +(5-2n)\gR^2+(5-2n)(3-2n)\gR^3+\cdots\ .
\end{equation}

\subsection{Freezing and denaturation transitions}
\label{ss:FrDenTen}
We now discuss the physical applications, following \cite{DavidHagendorfWiese2007a}.
To apply our calculation to the problem of random RNA under tension, we must take the $n=0$ limit, and set $\epsilon=1$. The RG functions become at $n=0$
\begin{eqnarray}
\label{m75n0}
\overline\beta_g(\gR)&=&-\epsilon\,\gR +5\gR^2+15\gR^3+\cdots
\\
\label{m74n0}
\overline\beta_f(\gR,\fR)&=&-\frac{\epsilon}{2} \fR+ \fR^2 +\frac{3}{2}\fR^2 \gR-\frac{5}{2}\fR \gR^2+\cdots
\end{eqnarray}
Note that there is no $\fR^3$ term in the $\overline{\mathbf{MS}}$ scheme.

\begin{figure}[t]
\begin{center}
\includegraphics[width=4in]{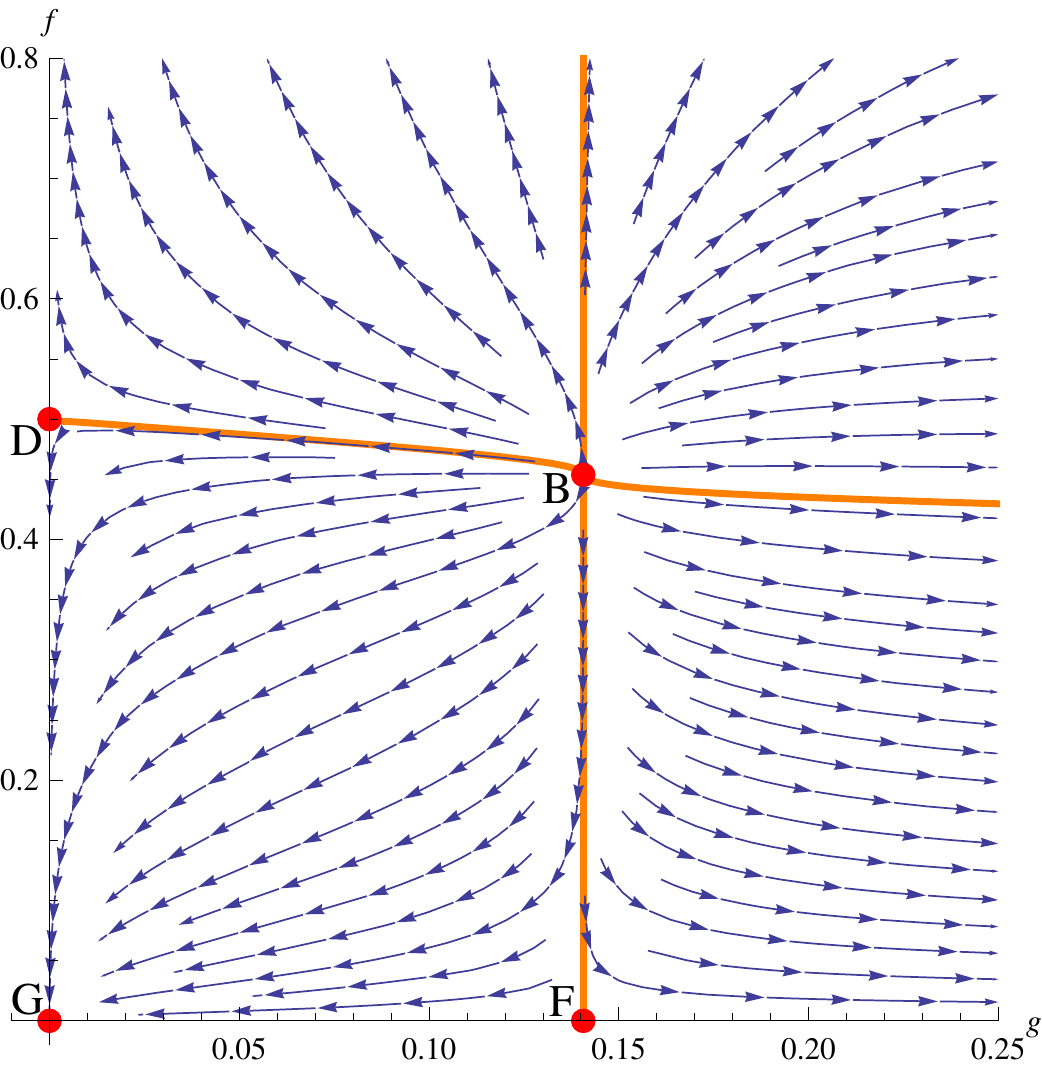}\caption{RG-flow as given by equations (\ref{m74n0}) and
(\ref{m75n0}), with fixed points (red) and separatrices (thick orange/grey lines).}
\label{flowdisfor}
\end{center}
\end{figure}

We recall, that equations (\ref{m75n0}) and (\ref{m74n0}) give the RG flow, when going to large scales (large RNA molecules).
For small $\gR$ and $\fR$, the flow is dominated by fixed points of order $\gR\sim \fR\sim \epsilon$, and can thus be studied in a Wilson-Fischer small $\epsilon$-expansion. This flow in the $(\gR,\fR)$ plane is depicted on figure~\ref{flowdisfor}.
Besides the unstable Gaussian fixed point $\mathbf{O}=(0,0)$, there are three non-trivial fixed points, which read at order $\epsilon^2$:
\begin{align}
\label{label1}
\text{freezing transition f.p.}\quad \mathbf{F}\qquad& g^*_{\mathrm{F}} ={\epsilon\over 5}-{3\,\epsilon^2\over 25}\ ,\quad f^*_{\mathrm{F}}=0  \\
\text{denaturation transition f.p.}\quad \mathbf{D}\qquad&  g^*_{\mathrm{D}}=0\ ,\quad f^*_{\mathrm{D}}= {\epsilon\over 2}  \\
\text{bicritical f.p.}\quad \mathbf{B}\qquad&  g^*_{\mathrm{B}}={\epsilon\over 5}-{3\,\epsilon^2\over 25}\ ,\quad f^*_{\mathrm{B}}=  {\epsilon\over 2}-{\epsilon^2\over 20}
\end{align}
$\mathbf{D}$ is  the fixed point for the tension-induced denaturation transition of homogeneous RNA (i.e.\ for homopolymers).
$\mathbf{F}$ is the fixed point for the force-free RNA freezing-transition.
$\mathbf{B}$  is the new unstable fixed point of \cite{DavidHagendorfWiese2007a}, corresponding to a bicritical freezing+denaturation point.

The denaturation exponent $\gamma$ at weak disorder is given by the derivative w.r.t.\ $\fR$ of $\beta_f(\gR,\fR)$ at the fixed point $\mathbf{D}$,  
\begin{equation}
\label{fi10}
\gamma=\left.{\partial\overline{\beta}_f(\gR,\fR) \over\partial \fR}\right|_{\mathbf{D}}={\epsilon\over 2}\ \mathop{=}_{\epsilon=1}\ 0 .5\ ,
\end{equation}
(this is in fact an exact result).
At the bicritical point $\mathbf{B}$ this denaturation exponent is modified to 
\begin{equation}
\label{m77}
\gamma'=\left.{\partial \overline\beta_f(\gR,\fR)\over\partial \fR}\right|_{\mathbf{B}}={\epsilon\over 2}+{1\over 10}\epsilon^2
\mathop{\simeq}_{\epsilon=1}0.6
\end{equation}
Assuming the locking scenario of L\"assig-Wiese, this exponent
should be equal to the critical exponent for the tension-induced
denaturation transition of RNA in the frozen phase ($g$ large).  Our
2-loop estimate $\gamma'_{\mathrm{2\mbox{\scriptsize -}loop}}=0.6$ is to be compared with the
result of numerical simulations of Krzakala et al.\ \cite{KrzakalaMezardMueller2002}, $\gamma_{\mathrm{sim}}\simeq
0.7$, and is definitely better than the 1-loop estimate $\gamma'_{\mathrm{1\mbox{\scriptsize -}loop}}=0.5$.

 \clean

\section{Conclusions and perspectives}
\label{s:conclusion}
In this article, we have considered the folding of RNA strands with random pairing energies, as a model for folding of random RNA sequences. We have established the existence of a phase transition from a high-temperature/low-disorder phase with a pair-contact exponent $\rho=\frac 32$ to an exponent of $\rho^*\approx 1.36$ at the transition. This was achieved via a designed field theory, modeled along the lines of self-avoiding polymers and membranes. It allowed us to use the tools of the multi-local operator product expansion, show renormalizability, and to obtain results  at 2-loop order.

Our considerations here are for RNA-molecules in thermal equilibrium. 
An important question is whether the dynamics of RNA folding \cite{FernandezAppignanesiMontani1997, XayaphoummineBucherIsambert2005}, especially after a quench can be modeled too,  or melting scenarios \`a la \cite{BundschuhBruinsma2008} explored. 

The order of the phase transition may depend delicately on the role of excluded-volume interactions. It is now possible to ask questions about the tertiary structure, using as a starting point our field theory and the ensuing statistics of branching, or equivalently the statistics of the RNA fold seen as a tree.  
%

{Our results suggests that steric interactions should be relevant. Indeed, the internal fractal dimension is $d_{\mathrm{f}}=1/\zeta\approx 1.56$. If the embedding of the tree in external space is dominated by entropic effects (mean field), so that the tree forms a ``blob'', the fractal dimension of the tree in bulk space should be $d_{\mathrm{blob}}=2 d_{\mathrm{f}}\approx 3.12$. Steric interactions are a priori relevant if the dimension of the embeding space is $d_{\mathrm{emb}}<2d_{\mathrm{blob}}$, thus are important for long strands, even though their effect may not be pronounced since $2d_{\mathrm{blob}}-d_{\mathrm{emb}}$ is small. The numerical results in eq.~(\ref{o2}) are indeed consistent with $2d_{\mathrm{blob}}-d_{\mathrm{emb}}=0$, which would lead to logarithmic corrections only.}

Finally, in view of the results of \cite{XayaphoummineBucherIsambert2005} it will be necessary to reanalyze the role of knots and pseudo-knots, even though their existence may not change the asymptotic scaling. Our field theory is capable of achieving this, by including corrections in $1/N$, i.e.\ corrections to the planar limit, systematically, in the spirit of what was done in \cite{OrlandZee2002} for homopolymers.

\acknowledgements
KW thanks M.\ L\"assig for the stimulating collaboration on RNA folding prior to this work, which raised many questions answered here. We acknowledge useful  discussions with C.~Hagendorf and H.\ Isambert. 
This research is supported by Agence Nationale de la Recherche
(France) under contracts  ANR-05-BLAN-0029-01 (F.D.), ANR-05-BLAN-0099-01 (K.W.), and the ENRAGE network,
MRTN-CT-2004-5616 (F.D.).

 \clean


\clean
\tableofcontents

 \end{document}